\documentclass[12pt, a4paper]{report}
\usepackage{graphicx}
\usepackage{latexsym}
\usepackage{amsmath}
\usepackage{amssymb}
\usepackage{epsfig}
\begin{document}
\begin{titlepage}

\title{From Primordial Quantum Fluctuations to the
\\Anisotropies of the Cosmic Microwave Background Radiation
\thanks{Based on lectures given at the \textit{Physik-Combo}, in
Halle, Leipzig and Jena, winter semester 2004/5. To appear in \emph{Ann. Phys.
(Leipzig)}.}}

\author{Norbert Straumann\\
Institute for Theoretical Physics University of Zurich,\\
CH--8057 Zurich, Switzerland}
\date{May, 2005}
\maketitle
\begin{abstract}
These lecture notes cover mainly three connected topics. In the first part we give a
detailed treatment of cosmological perturbation theory. The second part is devoted to
cosmological inflation and the generation of primordial fluctuations. In part three it will
be shown how these initial perturbation evolve and produce the temperature anisotropies of
the cosmic microwave background radiation. Comparing the theoretical prediction for the
angular power spectrum with the increasingly accurate observations provides important
cosmological information (cosmological parameters, initial conditions).
\end{abstract}
\end{titlepage}

\tableofcontents
\newpage
\section*{Introduction}

Cosmology is going through a fruitful and exciting period. Some of the developments are
definitely also of interest to physicists outside the fields of astrophysics and cosmology.

These lectures cover some particularly fascinating and topical subjects. A central theme
will be the current evidence that the recent ( $z<1$) Universe is dominated by an exotic
nearly homogeneous dark energy density with {\it negative} pressure. The simplest candidate
for this unknown so-called \textbf{Dark Energy} is a cosmological term in Einstein's field
equations, a possibility that has been considered during all the history of relativistic
cosmology. Independently of what this exotic energy density is, one thing is certain since
a long time: The energy density belonging to the cosmological constant is not larger than
the cosmological critical density, and thus \textbf{incredibly small by particle physics
standards}. This is a profound mystery, since we expect that all sorts of \textbf{vacuum
energies} contribute to the effective cosmological constant.

Since this is such an important issue it should be of interest to see how convincing the
evidence for this finding really is, or whether one should remain sceptical. Much of this
is based on the observed temperature fluctuations of the cosmic microwave background
radiation (CMB). A detailed analysis of the data requires a considerable amount of
theoretical machinery, the development of which fills most space of these notes.

Since this audience consists mostly of diploma and graduate students, whose main interests
are outside astrophysics and cosmology, I do not presuppose that you had already some
serious training in cosmology. However, I do assume that you have some working knowledge of
general relativity (GR). As a source, and for references, I usually quote my recent
textbook \cite{NS1}.

In an opening chapter those parts of the Standard Model of cosmology will be treated that
are needed for the main parts of the lectures. More on this can be found at many places,
for instance in the recent textbooks on cosmology  \cite{Cos1}, \cite{Cos2}, \cite{Cos3},
\cite{Cos4}, \cite{Cos5}.

In Part I we will develop the somewhat involved cosmological perturbation theory. The
formalism will later be applied to two main topics: (1) The generation of primordial
fluctuations during an inflationary era. (2) The evolution of these perturbations during
the linear regime. A main goal will be to determine the CMB power spectrum.

\chapter{Essentials of Friedmann-Lema\^{\i}tre models}

For reasons explained in the Introduction I treat in this opening chapter some standard
material that will be needed in the main parts of these notes.  In addition, an important
topical subject will be discussed in some detail, namely the Hubble diagram for Type Ia
supernovas that gave the first evidence for an accelerated expansion of the `recent' and
future universe. Most readers can directly go to Sect. 0.2, where this is treated.

\section{Friedmann-Lema\^{\i}tre spacetimes}

There is now good evidence that the (recent as well as the early) Universe\footnote{By
\textit{Universe} I always mean that part of the world around us which is in principle
accessible to observations. In my opinion the `Universe as a whole' is not a scientific
concept. When talking about \textit{model universes}, we develop on paper or with the help
of computers, I tend to use lower case letters. In this domain we are, of course, free to
make extrapolations and venture into speculations, but one should always be aware that
there is the danger to be drifted into a kind of `cosmo-mythology'.}is -- on large scales
-- surprisingly homogeneous and isotropic. The most impressive support for this comes from
extended redshift surveys of galaxies and from the truly remarkable isotropy of the cosmic
microwave background (CMB). In the Two Degree Field (2dF) Galaxy Redshift
Survey,\footnote{Consult the Home Page: http://www.mso.anu.edu.au/2dFGRS~.} completed in
2003, the redshifts of about 250'000 galaxies have been measured. The distribution of
galaxies out to 4 billion light years shows that there are huge clusters, long filaments,
and empty voids measuring over 100 million light years across. But the map also shows that
there are \textit{no larger structures}. The more extended Sloan Digital Sky Survey (SDSS)
has already produced very similar results, and will in the end have spectra of about a
million galaxies\footnote{For a description and pictures, see the  Home Page:
http://www.sdss.org/sdss.html~.}.

One arrives at the Friedmann (-Lema\^{\i}tre-Robertson-Walker) spacetimes by postulating
that for each observer, moving along an integral curve of a distinguished four-velocity
field $u$, the Universe looks spatially isotropic. Mathematically, this means the
following: Let $I\!so_x(M)$ be the group of local isometries of a Lorentz manifold $(M,g)$,
with fixed point $x\in M$, and let $SO_3(u_x)$ be the group of all linear transformations
of the tangent space $T_x(M)$ which leave the 4-velocity $u_x$ invariant and induce special
orthogonal transformations in the subspace orthogonal to $u_x$, then
\[ \{T_x\phi:~\phi\in I\!so_x(M),~\phi_\star u=u\}\supseteq SO_3(u_x) \]
($\phi_\star$ denotes the push-forward belonging to $\phi$; see \cite{NS1}, p. 550). In
\cite{NS2} it is shown that this requirement implies that $(M,g)$ is a Friedmann spacetime,
whose structure we now recall. Note that $(M,g)$ is then automatically homogeneous.

A \textit{Friedmann spacetime} $(M,g)$ is a warped product of the form $M= I\times \Sigma$,
where $I$ is an interval of $\mathbb{R}$, and the metric $g$ is of the form
\begin{equation}
g=-dt^2+a^2(t)\gamma,
\end{equation}
such that $(\Sigma,\gamma)$ is a Riemannian space of constant curvature $k=0,\pm 1$. The
distinguished time $t$ is the \textit{cosmic time}, and $a(t)$ is the \textit{scale factor}
(it plays the role of the warp factor (see Appendix B of \cite{NS1})). Instead of $t$ we
often use the \textit{conformal time} $\eta$, defined by $d\eta=dt/a(t)$. The velocity
field is perpendicular to the slices of constant cosmic time, $u=\partial/\partial t$.

\subsection{Spaces of constant curvature}

For the space $(\Sigma,\gamma)$ of constant curvature\footnote{For a detailed discussion of
these spaces I refer -- for readers knowing German -- to \cite{NS3} or \cite{NS4}.} the
curvature is given by
\begin{equation}
R^{(3)}(X,Y)Z=k\left[\gamma(Z,Y)X-\gamma(Z,X)Y\right];
\end{equation}
in components:
\begin{equation}
R^{(3)}_{ijkl}=k(\gamma_{ik}\gamma_{jl}-\gamma_{il}\gamma_{jk}).
\end{equation}
Hence, the Ricci tensor and the scalar curvature are
\begin{equation}
R^{(3)}_{jl}=2k\gamma_{jl}~~,~~ R^{(3)}=6k.
\end{equation}
For the curvature two-forms we obtain from (3) relative to an orthonormal triad
$\{\theta^i\}$
\begin{equation}
\Omega^{(3)}_{ij}=\frac{1}{2}R^{(3)}_{ijkl}~\theta^k\wedge\theta^l=k~\theta_i\wedge\theta_j
\end{equation}
($\theta_i=\gamma_{ik}\theta^k$). The simply connected constant curvature spaces are in $n$
dimensions the (n+1)-sphere $S^{n+1}$ ($k=1$), the Euclidean space ($k=0$), and the
pseudo-sphere ($k=-1$). Non-simply connected constant curvature spaces are obtained from
these by forming quotients with respect to discrete isometry groups. (For detailed
derivations, see \cite{NS3}.)

\subsection{Curvature of Friedmann spacetimes}

Let $\{\bar{\theta}^i\}$ be any orthonormal triad on $(\Sigma,\gamma)$. On this Riemannian
space the first structure equations read (we use the notation in \cite{NS1}; quantities
referring to this 3-dim. space are indicated by bars)
\begin{equation}
d\bar{\theta}^i+\bar{\omega}^i{}_j\wedge\bar{\theta}^j=0.
\end{equation}

On $(M,g)$ we introduce the following orthonormal tetrad:
\begin{equation}
\theta^0=dt,~~ \theta^i=a(t)\bar{\theta}^i.
\end{equation}
From this and (6) we get
\begin{equation}
d\theta^0=0,~~
d\theta^i=\frac{\dot{a}}{a}\theta^0\wedge\theta^i-a~\bar{\omega}^i{}_j\wedge\bar{\theta}^j.
\end{equation}
Comparing this with the first structure equation for the Friedmann manifold implies
\begin{equation}
\omega^0{}_i\wedge\theta^i=0,~~
\omega^i{}_0\wedge\theta^0+\omega^i{}_j\wedge\theta^j=\frac{\dot{a}}{a}
\theta^i\wedge\theta^0+a~\bar{\omega}^i{}_j\wedge\bar{\theta}^j,
\end{equation}
whence
\begin{equation}
\fbox{$\displaystyle\omega^0{}_i=\frac{\dot{a}}{a}~\theta^i,~~
\omega^i{}_j=\bar{\omega}^i{}_j.$}
\end{equation}

The worldlines of \textit{comoving observers} are integral curves of the four-velocity
field $u=\partial_t$. We claim that these are geodesics, i.e., that
\begin{equation}
\nabla_u u=0.
\end{equation}
To show this (and for other purposes) we introduce the basis $\{e_\mu\}$ of vector fields
dual to (7). Since $u=e_0$ we have, using the connection forms (10),
\[\nabla_u u=\nabla_{e_0}e_0=\omega^\lambda{}_0(e_0)e_\lambda=\omega^i{}_0(e_0)e_i=0.\]

\subsection{Einstein equations for Friedmann spacetimes}

Inserting the connection forms (10) into the second structure equations we readily find for
the curvature 2-forms $\Omega^\mu{}_\nu$:
\begin{equation}
\Omega^0{}_i=\frac{\ddot{a}}{a}\theta^0\wedge\theta^i,~~\Omega^i{}_j=
\frac{k+\dot{a}^2}{a^2}\theta^i\wedge\theta^j.
\end{equation}
A routine calculation leads to the following components of the Einstein tensor relative to
the basis (7)
\begin{eqnarray}
G_{00} &=& 3\left(\frac{\dot{a}^2}{a^2}+\frac{k}{a^2}\right),\\
G_{11} &=& G_{22}=G_{33}=-2\frac{\ddot{a}}{a}-\frac{\dot{a}^2}{a^2}-\frac{k}{a^2},\\
G_{\mu\nu} &=& 0~~ (\mu\neq\nu).
\end{eqnarray}

In order to satisfy the field equations, the symmetries of $G_{\mu\nu}$ imply that the
energy-momentum tensor \textit{must} have the perfect fluid form (see \cite{NS1}, Sect.
1.4.2):
\begin{equation}
T^{\mu\nu}=(\rho+p)u^\mu u^\nu+pg^{\mu\nu},
\end{equation}
where $u$ is the comoving velocity field introduced above.

Now, we can write down the field equations (including the cosmological term):
\begin{eqnarray}
\displaystyle 3\left(\frac{\dot{a}^2}{a^2}+\frac{k}{a^2}\right) &=& 8\pi G\rho+\Lambda,\\
-2\frac{\ddot{a}}{a}-\frac{\dot{a}^2}{a^2}-\frac{k}{a^2} &=& 8\pi Gp-\Lambda.
\end{eqnarray}

Although the `energy-momentum conservation' does not provide an independent equation, it is
useful to work this out. As expected, the momentum `conservation' is automatically
satisfied. For the `energy conservation' we use the general form (see (1.37) in \cite{NS1})
\begin{equation}
\nabla_u\rho=-(\rho+p)\nabla\cdot u.
\end{equation}
In our case we have for the \textit{expansion rate}
\[\nabla\cdot u=\omega^\lambda{}_0(e_\lambda)u^0=\omega^i{}_0(e_i),\]
thus with (10)
\begin{equation}
\nabla\cdot u=3\frac{\dot{a}}{a}.
\end{equation}
Therefore, eq. (19) becomes
\begin{equation}
\dot{\rho}+3\frac{\dot{a}}{a}(\rho+p)=0.
\end{equation}

For a given equation of state, $p=p(\rho)$, we can use (21) in the form
\begin{equation}
\frac{d}{da}(\rho a^3)=-3pa^2
\end{equation}
to determine $\rho$ as a function of the scale factor $a$. Examples: 1. For free massless
particles (radiation) we have $p=\rho/3$, thus $\rho\propto a^{-4}$. 2. For dust ($p=0$) we
get $\rho\propto a^{-3}$.

With this knowledge the \textit{Friedmann equation} (17) determines the time evolution of
$a(t)$.

------------

\textbf{Exercise}. Show that (18) follows from (17) and (21).

------------

As an important consequence of (17) and (18) we obtain for the acceleration of the
expansion
\begin{equation}
\ddot{a}=-\frac{4\pi G}{3}(\rho+3p)a+\frac{1}{3}\Lambda a.
\end{equation}
This shows that as long as $\rho+3p$ is positive, the first term in (23) is decelerating,
while a positive cosmological constant is repulsive. This becomes understandable if one
writes the field equation as
\begin{equation}
G_{\mu\nu} = \kappa (T_{\mu\nu} + T_{\mu\nu}^{\Lambda}) \quad\quad (\kappa =  8\pi G),
\end{equation}
with
\begin{equation}
T_{\mu\nu}^{\Lambda} = -\frac{\Lambda}{8\pi G} g_{\mu\nu}.
\end{equation}
This vacuum contribution has the form of the energy-momentum tensor of an ideal fluid, with
energy density $\rho_\Lambda = \Lambda/8\pi G$ and pressure $p_\Lambda = -\rho_\Lambda$.
Hence the combination $\rho_\Lambda+3p_\Lambda$ is equal to $-2\rho_\Lambda$, and is thus
negative. In what follows we shall often include in $\rho$ and $p$ the vacuum pieces.

\subsection{Redshift}

As a result of the expansion of the Universe the light of distant sources appears
redshifted. The amount of redshift can be simply expressed in terms of the scale factor
$a(t)$.

Consider two integral curves of the average velocity field $u$. We imagine that one
describes the worldline of a distant comoving source and the other that of an observer at a
telescope (see Fig. 1). Since light is propagating along null geodesics, we conclude from
(1) that along the worldline of a light ray $dt=a(t)d\sigma$, where $d\sigma$ is the line
element on the 3-dimensional space $(\Sigma,\gamma)$ of constant curvature $k=0,\pm 1$.
Hence the integral on the left of
\begin{equation}
\int_{t_e}^{t_o}\frac{dt}{a(t)}=\int_{source}^{obs.}d\sigma,
\end{equation}
between the time of emission ($t_e$) and the arrival time at the observer ($t_o$), is
independent of $t_e$ and $t_o$. Therefore, if we consider a second light ray that is
emitted at the time $t_e+\Delta t_e$ and is received at the time $t_o+\Delta t_o$, we
obtain from the last equation
\begin{equation}
\int_{t_e+\Delta t_e}^{t_o+\Delta t_o}\frac{dt}{a(t)}=\int_{t_e}^{t_o}\frac{dt}{a(t)}.
\end{equation}

For a small $\Delta t_e$ this gives
\[ \frac{\Delta t_o}{a(t_o)}=\frac{\Delta t_e}{a(t_e)}.\]
The observed and the emitted frequences $\nu_o$ and $\nu_e$, respectively, are thus related
according to
\begin{equation}
\frac{\nu_o}{\nu_e}=\frac{\Delta t_e}{\Delta t_o}=\frac{a(t_e)}{a(t_o)}.
\end{equation}
The redshift parameter $z$ is defined by
\begin{equation}
z:=\frac{\nu_e-\nu_o}{\nu_o},
\end{equation}
and is given by the key equation
\begin{equation}
\fbox{$\displaystyle 1+z=\frac{a(t_o)}{a(t_e)}.$}
\end{equation}
One can also express this by the equation $\nu\cdot a=const$ along a null geodesic.

\begin{figure}
\begin{center}
\includegraphics[height=0.3\textheight]{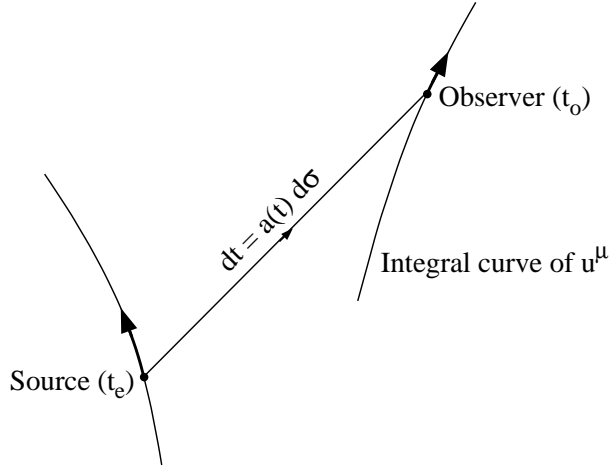}
\caption{Redshift for Friedmann models.} \label{Fig-1}
\end{center}
\end{figure}

\subsection{Cosmic distance measures}

We now introduce a further important tool, namely operational definitions of three
different distance measures, and show that they are related by simple redshift factors.

If $D$ is the physical (proper) extension of a distant object, and $\delta$ is its angle
subtended, then the \textit{angular diameter distance} $D_A$ is defined by
\begin{equation}
D_A:= D/\delta.
\end{equation}
If the object is moving with the proper transversal velocity $V_\bot$ and with an apparent
angular motion $d\delta/dt_0$, then the \textit{proper-motion distance} is by definition
\begin{equation}
D_M:=\frac{V_\bot}{d\delta/dt_0}.
\end{equation}
Finally, if the object has the intrinsic luminosity $\mathcal{L}$ and $\mathcal{F}$ is the
received energy flux then the  \textit{luminosity distance} is naturally defined as
\begin{equation}
D_L := (\mathcal{L}/4\pi\mathcal{F})^{1/2}.
\end{equation}
Below we show that these three distances are related as follows
\begin{equation}
\fbox{$D_L=(1+z)D_M=(1+z)^2D_A.$}
\end{equation}

It will be useful to introduce on $(\Sigma,\gamma)$ `polar' coordinates
$(r,\vartheta,\varphi)$, such that
\begin{equation}
\gamma=\frac{dr^2}{1-kr^2}+ r^2d\Omega^2,~~ d\Omega^2=d\vartheta^2+\sin^2\vartheta
d\varphi^2.
\end{equation}
One easily verifies that the curvature forms of this metric satisfy (5). (This follows
without doing any work by using in \cite{NS1} the curvature forms (3.9) in the ansatz (3.3)
for the Schwarzschild metric.)

To prove (34) we show that the three distances can be expressed as follows, if $r_e$
denotes the comoving radial coordinate (in (35)) of the distant object and the observer is
(without loss of generality) at $r=0$.
\begin{equation}
D_A=r_ea(t_e),~~ D_M=r_ea(t_0),~~ D_L=r_ea(t_0)\frac{a(t_0)}{a(t_e)}.
\end{equation}
Once this is established, (34) follows from (30).

From Fig. 2 and (35) we see that
\begin{equation}
D=a(t_e)r_e\delta,
\end{equation}
hence the first equation in (36) holds.

\begin{figure}
\begin{center}
\includegraphics[height=0.3\textheight]{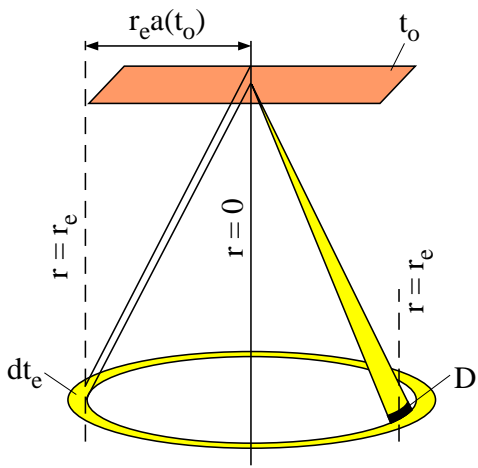}
\caption{Spacetime diagram for cosmic distance measures.} \label{Fig-2}
\end{center}
\end{figure}

To prove the second one we note that the source moves in a time $dt_0$ a proper transversal
distance
\[dD=V_\bot dt_e=V_\bot dt_0\frac{a(t_e)}{a(t_0)}.\]
Using again the metric (35) we see that the apparent angular motion is
\[d\delta=\frac{dD}{a(t_e)r_e}=\frac{V_\bot dt_0}{a(t_0)r_e}.\]
Inserting this into the definition (32) shows that the second equation in (36) holds. For
the third equation we have to consider the observed energy flux. In a time $dt_e$ the
source emits an energy $\mathcal{L}dt_e$. This energy is redshifted to the present by a
factor $a(t_e)/a(t_0)$, and is now distributed by (35) over a sphere with proper area
$4\pi(r_ea(t_0))^2$ (see Fig. 2). Hence the received flux (\textit{apparent luminosity}) is
\[\mathcal{F}=\mathcal{L}dt_e\frac{a(t_e)}{a(t_0)}\frac{1}{4\pi(r_ea(t_0))^2}\frac{1}{dt_0},\]
thus
\[\mathcal{F}=\frac{\mathcal{L}a^2(t_e)}{4\pi a^4(t_0)r_e^2}.\]
Inserting this into the definition (33) establishes the third equation in (36). For later
applications we write the last equation in the more transparent form
\begin{equation}
\fbox{$\displaystyle\mathcal{F}=\frac{\mathcal{L}}{4\pi(r_ea(t_0))^2}\frac{1}{(1+z)^2}.$}
\end{equation}
The last factor is due to redshift effects.

Two of the discussed distances as a function of $z$ are shown in Fig. 3 for two Friedmann
models with different cosmological parameters. The other two distance measures will be
introduced later (Sect. 3.2).

\begin{figure}
\begin{center}
\includegraphics[height=0.3\textheight]{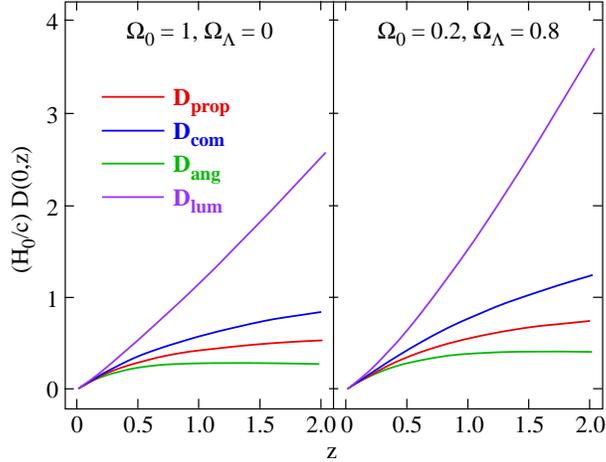}
\caption{Cosmological distance measures as a function of source
redshift for two cosmological models. The angular diameter distance
$D_{ang}\equiv D_A$ and the luminosity distance $D_{lum}\equiv D_L$
have been introduced in this section. The other two will be
introduced later.} \label{Fig-3}
\end{center}
\end{figure}

\section{Luminosity-redshift relation for Type Ia supernovas}

A few years ago the Hubble diagram for Type Ia supernovas gave, as a big surprise, the
first serious evidence for a currently accelerating Universe. Before presenting and
discussing critically these exciting results, we develop on the basis of the previous
section some theoretical background. (For the benefit of readers who start with this
section we repeat a few things.)

\subsection{Theoretical redshift-luminosity relation}

We have seen that in cosmology several different distance measures are in use, which are
all related by simple redshift factors. The one which is relevant in this section is the
{\it luminosity distance} $D_L$. We recall that this is defined by
\begin{equation}
D_L = (\mathcal{L}/4\pi\mathcal{F})^{1/2},
\end{equation}
where $\mathcal{L}$ is the intrinsic luminosity of the source and $\mathcal{F}$ the
observed energy flux.

We want to express this in terms of the redshift $z$ of the source and some of the
cosmological parameters. If the comoving radial coordinate $r$ is chosen such that the
Friedmann- Lema\^{\i}tre metric takes the form
\begin{equation}
g = -dt^2 + a^2(t) \left[ \frac{dr^2}{1-kr^2} + r^2d\Omega^2 \right ],\;\;\;  k = 0,\pm 1,
\end{equation}
then we have
\[\mathcal{F}dt_0 = \mathcal{L}dt_e \cdot\frac{1}{1+z}\cdot
\frac{1}{4\pi(r_e a(t_0))^2}. \]

The second factor on the right is due to the redshift of the photon energy; the indices $0,
e$ refer to the present and emission times, respectively. Using also $1+z=a(t_0)/ a(t_e)$,
we find in a first step:
\begin{equation}
D_L(z) = a_0 (1+z)r(z) \;\;\; (a_0\equiv a(t_0)).
\end{equation}

We need the function $r(z)$. From
\[ dz= -\frac{a_0}{a}\frac{\dot{a}}{a}dt,\;\;\;\; dt =
-a(t)\frac{dr}{\sqrt{1-kr^2}} \]
for light rays, we see that
\begin{equation}
\frac{dr}{\sqrt{1-kr^2}} = \frac{1}{a_0} \frac{dz}{H(z)}  \;\;\;\; (
H(z)=\frac{\dot{a}}{a}).
\end{equation}
Now, we make use of the Friedmann equation
\begin{equation}
H^2 + \frac{k}{a^2} = \frac{8\pi G}{3}\rho .
\end{equation}
Let us  decompose the total energy-mass density $\rho $ into nonrelativistic (NR),
relativistic (R), $\Lambda$, quintessence (Q), and possibly other contributions
\begin{equation}
\rho = \rho_{NR} + \rho_R + \rho_\Lambda + \rho_Q + \cdots .
\end{equation}
For the relevant cosmic period we can assume that the ``energy equation''
\begin{equation}
\frac{d}{da}(\rho a^3) = -3pa^2
\end{equation}
also holds for the individual components $X=NR,R,\Lambda,Q,\cdots$. If $w_X \equiv
p_X/\rho_X $ is constant, this implies that
\begin{equation}
\rho_Xa^{3(1+w_X)} = const.
\end{equation}
Therefore,
\begin{equation}
 \rho = \sum_X \left(\rho_X a^{3(1+w_X)}\right)_0 \frac{1}{a^{3(1+w_X)}} =
 \sum_X (\rho_X)_0 (1+z)^{3(1+w_X)}.
\end{equation}
Hence the Friedmann equation (43) can be written as
\begin{equation}
\frac{H^2(z)}{H_0^2} + \frac{k}{H_0^2 a_0^2 }(1+z)^2 = \sum_X \Omega_X (1+z)^{3(1+w_X)},
\end{equation}
where $\Omega_X$ is the dimensionless density parameter for the species $X$,
\begin{equation}
\Omega_X = \frac{(\rho_X)_0}{\rho_{crit}},
\end{equation}
where $\rho_{crit}$ is the critical density:
\begin{eqnarray}
\rho_{crit} &=& \frac{3H_0^2}{8\pi G} \nonumber \\
&=& 1.88\times 10^{-29}~ h_0^2 ~\textit {g}~\textit {cm}^{-3} \\
&=& 8\times10^{-47} h_0^2~ \textit{GeV}~^{4}. \nonumber
\end{eqnarray}
Here $h_0$ is the {\it reduced Hubble parameter}
\begin{equation}
h_0 = H_0/(100~\textit{km}~\textit{s}^{-1}~\textit{Mpc}^{-1})
\end{equation}
and is close to 0.7. Using also the curvature parameter $\Omega_K\equiv-k/H_0^2 a_0^2$, we
obtain the useful form
\begin{equation}
\fbox{$\displaystyle H^2(z) = H_0^2 E^2(z;\Omega_K,\Omega_X),$}
\end{equation}
with
\begin{equation}
E^2(z;\Omega_K,\Omega_X) = \Omega_K(1+z)^2 + \sum_X \Omega_X (1+z)^{3(1+w_X)}.
\end{equation}
Especially for $z=0$ this gives
\begin{equation}
\Omega_K + \Omega_0 = 1,\;\;\;\; \Omega_0 \equiv \sum_X \Omega_X.
\end{equation}
If we use (52) in (42), we get
\begin{equation}
\int_0^{r(z)} \frac{dr}{\sqrt{1-kr^2}} = \frac{1}{H_0 a_0}\int_0^z \frac{dz'}{E(z')}
\end{equation}
and thus
\begin{equation}
r(z) = \mathcal{S}(\chi(z)),
\end{equation}
where
\begin{equation}
\chi(z) = \frac{1}{H_0 a_0} \int_0^z \frac{dz'}{E(z')}
\end{equation}
and
\begin{equation}
\mathcal{S}(\chi) = \left\{\begin{array}{r@{\quad:\quad}l} \sin\chi & k=1\\ \chi & k=0\\
\sinh\chi & k=1 .\end{array} \right.
\end{equation}
Inserting this in (41) gives finally the relation we were looking for
\begin{equation}
D_L(z) = \frac{1}{H_0}\mathcal{D}_L(z;\Omega_K,\Omega_X) ,
\end{equation}
with
\begin{equation}
\mathcal{D}_L(z;\Omega_K,\Omega_X) = (1+z)\frac{1}{|\Omega_K|^{1/2}}\mathcal{S}
\left(|\Omega_K|^{1/2}\int_0^z \frac{dz'}{E(z')}\right)
\end{equation}
for $k=\pm 1$. For a flat universe, $\Omega_K =0$ or equivalently  $\Omega_0 =1 $, the
``Hubble-constant-free'' luminosity distance is
\begin{equation}
\mathcal{D}_L(z) = (1+z)\int_0^z\frac{dz'}{E(z')}.
\end{equation}

Astronomers use as logarithmic measures of $\mathcal{L}$ and $\mathcal{F}$ the {\it
absolute and apparent  magnitudes }\footnote{Beside the (bolometric) magnitudes $m,M$,
astronomers also use magnitudes $m_B,~m_V,~\ldots$ referring to certain wavelength bands
$B$ (blue), $V$ (visual), and so on.}, denoted by $M$ and $m$, respectively. The
conventions are chosen such that the {\it distance modulus} $m-M$ is related to $D_L$ as
follows
\begin{equation}
m-M = 5 \log \left( \frac{D_L}{1~Mpc}\right) + 25.
\end{equation}
Inserting the representation (59), we obtain the following relation between the apparent
magnitude $m$ and the redshift $z$:
\begin{equation}
m=\mathcal{M} + 5 \log \mathcal{D}_L(z;\Omega_K,\Omega_X),
\end{equation}
where, for our purpose, $\mathcal{M}=M-5\log H_0 + 25$ is an uninteresting fit parameter.
The comparison of this theoretical {\it magnitude redshift relation} with data will lead to
interesting restrictions for the cosmological $\Omega$-parameters. In practice often only
$\Omega_M$ and $\Omega_\Lambda$ are kept as independent parameters, where from now on the
subscript $M$ denotes (as in most papers) nonrelativistic matter.

The following remark about {\it degeneracy curves} in the $\Omega$-plane is important in
this context. For a fixed $z$ in the presently explored interval, the contours defined by
the equations $\mathcal{D}_L(z;\Omega_M,\Omega_\Lambda) = const $ have little curvature,
and thus we can associate an approximate slope to them. For $z=0.4$ the slope is about 1
and increases to 1.5-2 by $z=0.8$ over the interesting range of $\Omega_M $ and
$\Omega_\Lambda $. Hence even quite accurate data can at best select a strip in the
$\Omega$-plane, with a slope in the range just discussed. This is the reason behind the
shape of the likelihood regions shown later (Fig. 5).

In this context it is also interesting to determine the dependence of the {\it deceleration
parameter}
\begin{equation}
q_0 = - \Bigl( \frac{a\ddot{a}}{\dot{a}^2} \Bigr )_0
\end{equation}
on $\Omega_M$ and $\Omega_\Lambda$. At an any cosmic time we obtain from (23) and (47)
\begin{equation}
-\frac{\ddot{a}a}{\dot{a}^2} = \frac{1}{2}\frac{1}{E^2(z)}\sum_X \Omega_X
(1+z)^{3(1+w_X)}(1+3w_X).
\end{equation}
For $z=0$ this gives
\begin{equation}
q_0 = \frac{1}{2}\sum_X \Omega_X (1+3w_X) = \frac{1}{2} (\Omega_M - 2\Omega_\Lambda +
\cdots).
\end{equation}
The line $q_0=0 \; (\Omega_\Lambda = \Omega_M /2 )$ separates decelerating from
accelerating universes at the present time. For given values of $\Omega_M, \Omega_\Lambda$,
etc, (65) vanishes for $z$ determined by
\begin{equation}
\Omega_M(1+z)^3 - 2\Omega_\Lambda +\cdots = 0.
\end{equation}
This equation gives the redshift at which the deceleration period ends (coasting redshift).

\paragraph{Generalization for dynamical models of Dark Energy.}

If the vacuum energy constitutes the missing two thirds of the
average energy density of the \emph{present} Universe, we would be
confronted with the following \emph{cosmic coincidence} problem:
Since the vacuum energy density is constant in time -- at least
after the QCD phase transition --, while the matter energy density
decreases as the Universe expands, it would be more than surprising
if the two are comparable just at about the present time, while
their ratio was tiny in the early Universe and would become very
large in the distant future.  The goal of dynamical models of Dark
Energy is to avoid such an extreme fine-tuning. The ratio $p/\rho$
of this component then becomes a function of redshift, which we
denote by $w_Q(z)$ (because so-called quintessence models are
particular examples). Then the function $E(z)$ in (53) gets
modified.

To see how, we start from the energy equation (45) and write this as
\[\frac{d\ln(\rho_Q a^3)}{d\ln(1+z)}=3w_Q.\]
This gives
\[ \rho_Q(z)=\rho_{Q0}(1+z)^3\exp\left(\int_{0}^{\ln(1+z)} 3w_Q(z')d\ln(1+z')\right)\]
or
\begin{equation}
\rho_Q(z)=\rho_{Q0}\exp\left(3\int_{0}^{\ln(1+z)}
(1+w_Q(z'))d\ln(1+z')\right).
\end{equation}
Hence, we have to perform on the right of (53) the following substitution:
\begin{equation}
\Omega_Q(1+z)^{3(1+w_Q)}\rightarrow \Omega_Q
\exp\left(3\int_{0}^{\ln(1+z)} (1+w_Q(z'))d\ln(1+z')\right).
\end{equation}

\subsection{Type Ia supernovas as standard candles}

It has long been recognized that supernovas of type Ia are excellent
standard candles and are visible to cosmic distances \cite{Bad} (the
record is at present at a redshift of about 1.7). At relatively
closed distances they can be used to measure the Hubble constant, by
calibrating the absolute magnitude of nearby supernovas with various
distance determinations (e.g., Cepheids). There is still some
dispute over these calibration resulting in differences of about
10\% for $H_0$. (For recent papers and references, see \cite{STR}.)

In 1979 Tammann \cite{Tam} and Colgate \cite{Col} independently suggested that at higher
redshifts this subclass of supernovas can be used to determine also the deceleration
parameter. In recent years this program became feasible thanks to the development of new
technologies which made it possible to obtain digital images of faint objects over sizable
angular scales, and by making use of big telescopes such as Hubble and Keck.

There are two major teams investigating high-redshift SNe Ia, namely the `Supernova
Cosmology Project' (SCP) and the `High-Z Supernova search Team' (HZT). Each team has found
a large number of SNe, and both groups have published almost identical results. (For
up-to-date information, see the home pages \cite{SCP} and \cite{HZT}.)

Before discussing these, a few remarks about the nature and properties of type Ia SNe
should be made. Observationally, they are characterized by the absence of hydrogen in their
spectra, and the presence of some strong silicon lines near maximum. The immediate
progenitors are most probably carbon-oxygen white dwarfs in close binary systems, but it
must be said that these have not yet been clearly identified.\footnote{This is perhaps not
so astonishing, because the progenitors are presumably faint compact dwarf stars.}

In the standard scenario a white dwarf accretes matter from a nondegenerate companion until
it approaches the critical Chandrasekhar mass and ignites carbon burning deep in its
interior of highly degenerate matter. This is followed by  an outward-propagating nuclear
flame leading to a total disruption of the white dwarf. Within a few seconds the star is
converted largely into nickel and iron. The dispersed nickel radioactively decays to cobalt
and then to iron in a few hundred days. A lot of effort has been invested to simulate these
complicated processes. Clearly, the physics of thermonuclear runaway burning in degenerate
matter is complex. In particular, since the thermonuclear combustion is highly turbulent,
multidimensional simulations are required. This is an important subject of current
research. (One gets a good impression of the present status from several articles in
\cite{CEx}. See also the recent review \cite{HiN}.) The theoretical uncertainties are such
that, for instance, predictions for possible evolutionary changes are not reliable.

It is conceivable that in some cases a type Ia supernova is the result of a merging of two
carbon-oxygen-rich white dwarfs with a combined mass surpassing the Chandrasekhar limit.
Theoretical modelling indicates, however, that such a merging would lead to a collapse,
rather than a SN Ia explosion. But this issue is still debated.

In view of the complex physics involved, it is not astonishing that type Ia  supernovas are
not perfect standard candles. Their peak absolute magnitudes have a dispersion of 0.3-0.5
mag, depending on the sample. Astronomers have, however, learned in recent years to reduce
this dispersion by making use of empirical correlations between the absolute peak
luminosity and light curve shapes. Examination of nearby SNe showed that the peak
brightness is correlated with the time scale of their brightening and fading: slow
decliners tend to be brighter than rapid ones. There are also some correlations with
spectral properties. Using these correlations it became possible to reduce the remaining
intrinsic dispersion, at least in the average, to $\simeq 0.15 mag$. (For the various
methods in use, and how they compare, see \cite{Leib1}, \cite{Rie}, and references
therein.) Other corrections, such as Galactic extinction, have been applied, resulting for
each supernova in a corrected (rest-frame) magnitude. The redshift  dependence of this
quantity is compared with the theoretical expectation given by Eqs. (62) and (60).

\subsection{Results}

After the classic papers \cite{P99}, \cite{S98}, \cite{R98} on the Hubble diagram for
high-redshift type Ia supernovas, published by the SCP and HZT teams, significant progress
has been made (for reviews, see \cite{Leib2} and \cite{Fil04}). I discuss here the main
results presented in \cite{Rie}. These are based on additional new data for $z>1$, obtained
in conjunction with the GOODS (Great Observatories Origins Deep Survey) Treasury program,
conducted with the Advanced Camera for Surveys (ACS) aboard the Hubble Space Telescope
(HST).

The quality of the data and some of the main results of the analysis are shown in Fig. 4.
The data points in the top panel are the distance moduli relative to an empty uniformly
expanding universe, $\Delta(m-M)$, and the redshifts of a ``gold'' set of 157 SNe Ia. In
this `reduced' Hubble diagram the filled symbols are the HST-discovered SNe Ia. The bottom
panel shows weighted averages in fixed redshift bins.

\begin{figure}
\begin{center}
\includegraphics[height=0.35\textheight]{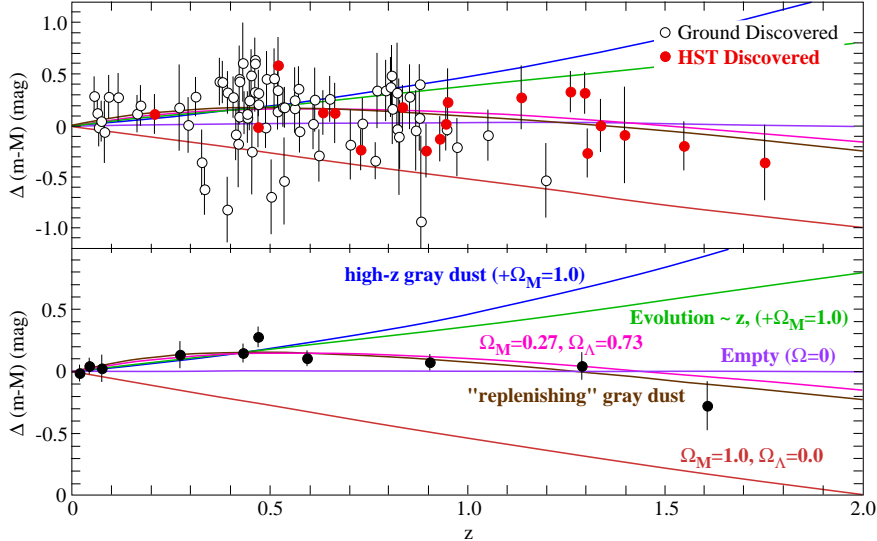}
\caption{Distance moduli relative to an empty uniformly expanding universe (residual Hubble
diagram)
 for SNe Ia; see text for further explanations. (Adapted from \cite{Rie}, Fig. 7.).} \label{Fig-4}
\end{center}
\end{figure}

These data are consistent with the ``cosmic concordance'' model
($\Omega_M=0.3,~\Omega_\Lambda=0.7$), with $\chi^2_{dof}=1.06)$. For a flat universe with a
cosmological constant, the fit gives $\Omega_M=0.29\pm_{0.19}^{0.13}$ (equivalently,
$\Omega_\Lambda=0.71$). The other model curves will be discussed below. Likelihood regions
in the ($\Omega_M,\Omega_\Lambda$)-plane, keeping only these parameters in (62) and
averaging $H_0$, are shown in Fig. 5. To demonstrate the progress, old results from 1998
are also included. It will turn out that this information is largely complementary to the
restrictions we shall obtain from the CMB anisotropies.

\begin{figure}
\begin{center}
\includegraphics[height=0.4\textheight]{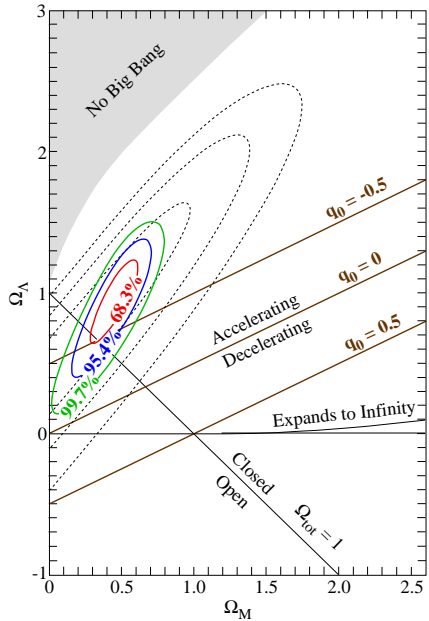}
\caption{Likelihood regions in the ($\Omega_M,\Omega_\Lambda$)-plane. The dotted contours
are old results from 1998. (Adapted from \cite{Rie}, Fig. 8.).} \label{Fig-5}
\end{center}
\end{figure}

In the meantime new results have been published. Perhaps the best
high-z SN Ia compilation to date are the results from the Supernova
Legacy Survey (SNLS) of the first year \cite{Leg}. The other main
research group has also published new data at about the same time
\cite{Clo}.

\subsection{Systematic uncertainties}

Possible systematic uncertainties due to astrophysical effects have been discussed
extensively in the literature. The most serious ones are (i) {\it dimming} by intergalactic
dust, and (ii) {\it evolution} of SNe Ia over cosmic time, due to changes in progenitor
mass, metallicity, and C/O ratio. I discuss these concerns only briefly (see also
\cite{Leib2}, \cite{Rie}).

Concerning extinction, detailed studies show that high-redshift SN Ia suffer little
reddening; their B-V colors at maximum brightness are normal. However, it can a priori not
be excluded that we see distant SNe through a grey dust with grain sizes large enough as to
not imprint the reddening signature of typical interstellar extinction. One argument
against this hypothesis is that this would also imply a larger dispersion than is observed.
In Fig. 4 the expectation of a simple grey dust model is also shown. The new  high redshift
data reject this monotonic model of astrophysical dimming. Eq. (67) shows that at redshifts
$z\geq (2\Omega_\Lambda /\Omega_M)^{1/3}-1 \simeq 1.2$ the Universe is {\it decelerating},
and this provides an almost unambiguous signature for $\Lambda$, or some effective
equivalent. There is now strong evidence for a transition from a deceleration to
acceleration at a redshift $z=0.46\pm 0.13$.

The same data provide also some evidence against a simple luminosity evolution that could
mimic an accelerating Universe. Other empirical constraints are obtained by comparing
subsamples of low-redshift SN Ia believed to arise from old and young progenitors. It turns
out that there is no difference within the measuring errors, {\it after} the correction
based on the light-curve shape has been applied. Moreover, spectra of high-redshift  SNe
appear remarkably similar to those at low redshift. This is very reassuring. On the other
hand, there seems to be a trend that more distant supernovas are bluer. It would, of
course, be helpful if evolution could be predicted theoretically, but in view of what has
been said earlier, this is not (yet) possible.

In conclusion, none of the investigated systematic errors appear to reconcile the data with
$\Omega_\Lambda = 0$ and $q_0\geq 0$. But further work is necessary before we can declare
this as a really established fact.

To improve the observational situation a satellite mission called SNAP (``Supernovas
Acceleration Probe'') has been proposed \cite{Snap}. According to the plans this satellite
would observe about 2000 SNe within a year and much more detailed studies could then be
performed. For the time being some scepticism with regard to the results that have been
obtained is still not out of place, but the situation is steadily improving.

Finally, I mention a more theoretical complication. In the analysis of the data the
luminosity distance for an ideal Friedmann universe was always used. But the data were
taken in the real inhomogeneous Universe. This may not be good enough, especially for
high-redshift standard candles. The simplest way to take this into account is to introduce
a filling parameter which, roughly speaking, represents matter that exists in galaxies but
not in the intergalactic medium. For a constant filling parameter one can determine the
luminosity distance by solving the Dyer-Roeder equation. But now one has an additional
parameter in fitting the data. For a flat universe this was recently investigated in
\cite{Kan}.

\section{Thermal history below 100 $MeV$}

\subsection*{A. Overview}

Below the transition at about 200 $MeV$ from a quark-gluon plasma to the confinement phase,
the Universe was initially dominated by a complicated dense hadron soup. The abundance of
pions, for example, was so high that they nearly overlapped. The pions, kaons and other
hadrons soon began to decay and most of the nucleons and antinucleons annihilated, leaving
only a tiny baryon asymmetry. The energy density is then almost completely dominated by
radiation and the stable leptons ($e^{\pm}$, the three neutrino flavors and their
antiparticles). For some time all these particles are in thermodynamic equilibrium. For
this reason, only a few initial conditions have to be imposed. The Universe was never as
simple as in this lepton era. (At this stage it is almost inconceivable that the complex
world around us would eventually emerge.)

The first particles which freeze out of this equilibrium are the weakly interacting
neutrinos. Let us estimate when this happened. The coupling of the neutrinos in the lepton
era is dominated by the reactions:
\[ e^{-}+e^{+}\leftrightarrow \nu+\bar{\nu},~~ e^{\pm}+\nu\rightarrow e^{\pm}+\nu,
~~e^{\pm}+\bar{\nu}\rightarrow e^{\pm}+\bar{\nu}. \] For dimensional reasons, the cross
sections are all of magnitude
\begin{equation}
\sigma\simeq G_F^2T^2,
\end{equation}
where $G_F$ is the Fermi coupling constant ($\hbar=c=k_B=1$). Numerically,
$G_Fm^2_p\simeq10^{-5}$. On the other hand, the electron and neutrino densities $n_e,n_\nu$
are about $T^3$. For this reason, the reaction rates $\Gamma$ for $\nu$-scattering and
$\nu$-production per electron are of magnitude $c\cdot v\cdot n_e\simeq G_F^2T^5$. This has
to be compared with the expansion rate of the Universe
\[H=\frac{\dot{a}}{a}\simeq (G\rho)^{1/2}.\] Since $\rho\simeq T^4$ we get
\begin{equation}
H\simeq G^{1/2}T^2
\end{equation}
and thus
\begin{equation}
\frac{\Gamma}{H}\simeq G^{-1/2}G_F^2T^3\simeq (T/10^{10}~K)^3.
\end{equation}
This ration is larger than 1 for $T>10^{10}~K \simeq 1~MeV$, and the neutrinos thus remain
in thermodynamic equilibrium until the temperature has decreased to about 1 $MeV$. But even
below this temperature the neutrinos remain Fermi distributed,
\begin{equation}
n_\nu(p)dp=\frac{1}{2\pi^2}\frac{1}{e^{p/T_\nu}+1}p^2dp~,
\end{equation}
as long as they can be treated as massless. The reason is that the number density decreases
as $a^{-3}$ and the momenta with $a^{-1}$. Because of this we also see that the neutrino
temperature $T_\nu$ decreases after decoupling as $a^{-1}$. The same is, of course true for
photons. The reader will easily find out how the distribution evolves when neutrino masses
are taken into account. (Since neutrino masses are so small this is only relevant at very
late times.)

\subsection*{B. Chemical potentials of the leptons}

The equilibrium reactions below 100 $MeV$, say, conserve several additive quantum
numbers\footnote{Even if $B,L_e,L_\mu,L_\tau$ should not be strictly conserved, this is not
relevant within a Hubble time $H_0^{-1}$.}, namely the electric charge $Q$, the baryon
number $B$, and the three lepton numbers $L_e,L_\mu,L_\tau$. Correspondingly, there are
five independent chemical potentials. Since particles and antiparticles can annihilate to
photons, their chemical potentials are oppositely equal: $\mu_{e^{-}}=-\mu_{e^{+}}$, etc.
From the following reactions
\[ e^{-}+\mu^{+}\rightarrow\nu_e+\bar{\nu}_\mu,~~ e^{-}+p\rightarrow\nu_e+n,~~
\mu^{-}+p\rightarrow\nu_\mu+n \] we infer the equilibrium conditions
\begin{equation}
\mu_{e^{-}}-\mu_{\nu_e}=\mu_{\mu^{-}}-\mu_{\nu_\mu}=\mu_n-\mu_p.
\end{equation}
As independent chemical potentials we can thus choose
\begin{equation}
\fbox{$\displaystyle \mu_p,~\mu_{e^{-}},~\mu_{\nu_e},~\mu_{\nu_\mu},~\mu_{\nu_\tau}.$}
\end{equation}

Because of local electric charge neutrality, the charge number density $n_Q$ vanishes. From
observations (see subsection E) we also know that the baryon number density $n_b$ is much
smaller than the photon number density ($\sim$ entropy density $s_\gamma$). The ratio
$n_B/s_\gamma$ remains constant for adiabatic expansion (both decrease with $a^{-3}$; see
the next section). Moreover, the lepton number densities are
\begin{equation}
n_{L_e}=n_{e^{-}}+n_{\nu_e}-n_{e^{+}}-n_{\bar{\nu}_e},~~n_{L_\mu}=
n_{\mu^{-}}+n_{\nu_\mu}-n_{\mu^{+}}-n_{\bar{\nu}_\mu},~~etc.
\end{equation}
Since in the present Universe the number density of electrons is equal to that of the
protons (bound or free), we know that after the disappearance of the muons $n_{e^{-}}\simeq
n_{e^{+}}$ (recall $n_B\ll n_\gamma$), thus $\mu_{e^{-}}~(=-\mu_{e^{+}})\simeq0$. It is
conceivable that the chemical potentials of the neutrinos and antineutrinos can not be
neglected, i.e., that $n_{L_e}$ is not much smaller than the photon number density. In
analogy to what we know about the baryon density we make the reasonable \textit{asumption}
that the lepton number densities are also much smaller than $s_\gamma$. Then we can take
the chemical potentials of the neutrinos equal to zero ($|\mu_\nu|/kT\ll1$). With what we
said before, we can then put the five chemical potentials (75) equal to zero, because the
charge number densities are all odd in them. Of course, $n_B$ does not really vanish
(otherwise we would not be here), but for the thermal history in the era we are considering
they can be ignored.

------------

\textbf{Exercise}. Suppose we are living in a degenerate
$\bar{\nu}_e$-see. Use the current mass limit for the electron
neutrino mass coming from tritium decay to deduce a limit for the
magnitude of the chemical potential $\mu_{\nu_e}$.

------------

\subsection*{C. Constancy of entropy}

Let $\rho_{eq},p_{eq}$ denote (in this subsection only) the total energy density and
pressure of all particles in thermodynamic equilibrium. Since the chemical potentials of
the leptons vanish, these quantities are only functions of the temperature $T$. According
to the second law, the differential of the entropy $S(V,T)$ is given by
\begin{equation}
dS(V,T)=\frac{1}{T}[d(\rho_{eq}(T)V)+p_{eq}(T)dV].
\end{equation}
This implies
\begin{eqnarray*}
d(dS)=0&=&d\left(\frac{1}{T}\right)\wedge
d(\rho_{eq}(T)V)+d\left(\frac{p_{eq}(I)}{T}\right)\wedge dV\\
&=&-\frac{\rho_{eq}}{T^2}dT\wedge dV+\frac{d}{dT}\left(\frac{p_{eq}(T)}{T}\right)dT\wedge
dV,
\end{eqnarray*}
i.e., the Maxwell relation
\begin{equation}
\fbox{$\displaystyle \frac{dp_{eq}(T)}{dT}=\frac{1}{T}[\rho_{eq}(T)+p_{eq}(T)].$}
\end{equation}
If we use this in (77), we get
\[dS=d\left[\frac{V}{T}(\rho_{eq}+p_{eq})\right],\]
so the entropy density of the particles in equilibrium is
\begin{equation}
\fbox{$\displaystyle s=\frac{1}{T}[\rho_{eq}(T)+p_{eq}(T)].$}
\end{equation}
For an adiabatic expansion the entropy in a comoving volume remains constant:
\begin{equation}
S=a^3s =const.
\end{equation}
This constancy is equivalent to the energy equation (21) for the equilibrium part. Indeed ,
the latter can be written as
\[a^3\frac{dp_{eq}}{dt}=\frac{d}{dt}[a^3(\rho_{eq}+p_{eq})], \]
and by (79) this is equivalent to $dS/dt=0$.

In particular, we obtain for massless particles ($p=\rho/3$) from (78) again $\rho\propto
T^4$ and from (79) that $S=$ constant implies $T\propto a^{-1}$.

------------

\textbf{Exercise}. Assume that all components are in equilibrium and use the results of
this subsection to show that the temperature evolution is for $k=0$ given by
\[ \frac{dT}{dt}=-\sqrt{24\pi
G}\frac{\sqrt{\rho(T)}}{\frac{d}{dT}\left[\ln\frac{dp}{dT}\right]}.\]

------------

Once the electrons and positrons have annihilated below $T\sim m_e$, the equilibrium
components consist of photons, electrons, protons and -- after the big bang nucleosynthesis
-- of some light nuclei (mostly $He^4$). Since the charged particle number densities are
much smaller than the photon number density, the photon temperature $T_\gamma$ still
decreases as $a^{-1}$. Let us show this formally. For this we consider beside the photons
an ideal gas in thermodynamic equilibrium with the black body radiation. The total pressure
and energy density are then (we use units with $\hbar=c=k_B=1;~ n$ is the number density of
the non-relativistic gas particles with mass $m$):
\begin{equation}
p=nT +\frac{\pi^2}{45}T^4,~~\rho=nm+\frac{nT}{\gamma-1}+\frac{\pi^2}{15}T^4
\end{equation}
($\gamma=5/3$ for a monoatomic gas). The conservation of the gas particles, $na^3=const.$,
together with the energy equation (22) implies, if $\sigma:=s_\gamma/n,$
\[\frac{d\ln T}{d\ln a}=-\left[\frac{\sigma+1}{\sigma+1/3(\gamma-1)}\right].\]
For $\sigma\ll1$ this gives the well-known relation $T\propto a^{3(\gamma-1)}$ for an
adiabatic expansion of an ideal gas.

We are however dealing with the opposite situation $\sigma\gg1$, and then we obtain, as
expected, $a\cdot T=const$.

Let us look more closely at the famous ratio $n_B/s_\gamma$. We need
\begin{equation}
s_\gamma=\frac{4}{3T}\rho_\gamma=\frac{4\pi^2}{45}T^3=3.60n_\gamma,~~n_B=\rho_B/m_p=\Omega_B
\rho_{crit}/m_p.
\end{equation}
From the present value of $T_\gamma\simeq2.7~K$ and (50),
$\rho_{crit}=1.12\times10^{-5}~h_0^2 (m_p/cm^3)$, we obtain as a measure for the baryon
asymmetry of the Universe
\begin{equation}
\fbox{$\displaystyle \frac{n_B}{s_\gamma}=0.75\times10^{-8}(\Omega_Bh_0^2).$}
\end{equation}
It is one of the great challenges to explain this tiny number. So far, this has been
achieved at best qualitatively in the framework of grand unified theories (GUTs).

\subsection*{D. Neutrino temperature}

During the electron-positron annihilation below $T=m_e$ the $a$-dependence is complicated,
since the electrons can no more be treated as massless. We want to know at this point what
the ratio $T_\gamma/T_\nu$ is after the annihilation. This can easily be obtained by using
the constancy of comoving entropy for the photon-electron-positron system, which is
sufficiently strongly coupled to maintain thermodynamic equilibrium.

We need the entropy for the electrons and positrons at $T\gg m_e$, long before annihilation
begins. To compute this note the identity
\[ \int_0^\infty\frac{x^n}{e^x-1}dx-\int_0^\infty\frac{x^n}{e^x+1}dx=2\int_0^\infty\frac{x^n}{e^{2x}-1}dx
=\frac{1}{2^n}\int_0^\infty\frac{x^n}{e^x-1}dx, \] whence
\begin{equation}
\int_0^\infty\frac{x^n}{e^x+1}dx=(1-2^{-n})\int_0^\infty\frac{x^n}{e^x-1}dx.
\end{equation}
In particular, we obtain for the entropies $s_e,s_\gamma$ the following relation
\begin{equation}
s_e=\frac{7}{8}s_\gamma ~~~~(T\gg m_e).
\end{equation}
Equating the entropies for $T_\gamma\gg m_e$ and $T_\gamma\ll m_e$ gives
\[
\left.(T_\gamma a)^3\right|_{before}\left[1+2\times\frac{7}{8}\right]= \left.(T_\gamma
a)^3\right|_{after}\times1, \] because the neutrino entropy is conserved. Therefore, we
obtain
\begin{equation}
\left.(aT_\gamma)\right|_{after}=\left(\frac{11}{4}\right)^{1/3}\left.(aT_\gamma)\right|_{before}.
\end{equation}
But
$\left.(aT_\nu)\right|_{after}=\left.(aT_\nu)\right|_{before}=\left.(aT_\gamma)\right|_{before}$,
hence we obtain the important relation
\begin{equation}
\fbox{$\displaystyle
\left.\left(\frac{T_\gamma}{T_\nu}\right)\right|_{after}=\left(\frac{11}{4}\right)^{1/3}=1.401.$}
\end{equation}

\subsection*{E. Epoch of matter-radiation equality}

In the main parts of these lectures the epoch when radiation (photons and neutrinos) have
about the same energy density as non-relativistic matter (Dark Matter and baryons) plays a
very important role. Let us determine the redshift, $z_{eq}$, when there is equality.

For the three neutrino and antineutrino flavors the energy density is according to (84)
\begin{equation}
\rho_\nu=3\times\frac{7}{8}\times\left(\frac{4}{11}\right)^{4/3}\rho_\gamma.
\end{equation}
Using
\begin{equation}
\frac{\rho_\gamma}{\rho_{crit}}=2.47\times10^{-5}h_0^{-2}(1+z)^4,
\end{equation}
we obtain for the total radiation energy density, $\rho_r$,
\begin{equation}
\frac{\rho_r}{\rho_{crit}}=4.15\times10^{-5}h_0^{-2}(1+z)^4,
\end{equation}
Equating this to
\begin{equation}
\frac{\rho_M}{\rho_{crit}}=\Omega_M(1+z)^3
\end{equation}
we obtain
\begin{equation}
\fbox{$\displaystyle 1+z_{eq}=2.4\times10^4\Omega_Mh_0^2.$}
\end{equation}

Only a small fraction of $\Omega_M$ is baryonic. There are several methods to determine the
fraction $\Omega_B$ in baryons. A traditional one comes from the abundances of the light
elements. This is treated in most texts on cosmology. (German speaking readers find a
detailed discussion in my lecture notes \cite {NS4}, which are available in the internet.)
The comparison of the straightforward theory with observation gives a value in the range
$\Omega_Bh_0^2=0.021\pm0.002$. Other determinations are all compatible with this value. In
Part III we shall obtain $\Omega_B$ from the CMB anisotropies. The striking agreement of
different methods, sensitive to different physics, strongly supports our standard big bang
picture of the Universe.

\part{Cosmological Perturbation Theory}

\section*{Introduction} The astonishing isotropy of the cosmic microwave background
radiation provides direct evidence that the early universe can be described in a good first
approximation by a Friedmann model\footnote{For detailed treatments, see for instance the
recent textbooks on cosmology \cite{Cos1}, \cite{Cos2}, \cite{Cos3}, \cite{Cos4},
\cite{Cos5}. For GR I usually refer to \cite{NS1}.}. At the time of recombination
deviations from homogeneity and isotropy have been very small indeed ($\sim 10^{-5}$). Thus
there was a long period during which deviations from Friedmann models can be studied
perturbatively, i.e., by linearizing the Einstein and matter equations about solutions of
the idealized Friedmann-Lema\^{\i}tre models.

Cosmological perturbation theory is a very important tool that is by now well developed.
Among the various reviews I will often refer to \cite{KS84}, abbreviated as KS84. Other
works will be cited later, but the present notes should be self-contained. Almost always I
will provide detailed derivations. Some of the more lengthy calculations are deferred to
appendices.

The formalism, developed in this part, will later be applied to two main problems: (1) The
generation of primordial fluctuations during an inflationary era. (2) The evolution of
these perturbations during the linear regime. A main goal will be to determine the CMB
power spectrum as a function of certain cosmological parameters. Among these the fractions
of \textit{Dark Matter} and \textit{Dark Energy} are particularly interesting.


\chapter{Basic Equations}

In this chapter we develop the model independent parts of cosmological perturbation theory.
This forms the basis of all that follows.

\section{Generalities}

For the unperturbed Friedmann models the metric is denoted by $g^{(0)}$, and has the form
\begin{equation}
g^{(0)}= -dt^2 + a^2(t)\gamma = a^2(\eta)\left[-d\eta^2 + \gamma\right];
\end{equation}
$\gamma$ is the metric of a space with constant curvature $K$. In addition, we have matter
variables for the various components (radiation, neutrinos, baryons, cold dark matter
(CDM), etc). We shall linearize all basic equations about the unperturbed solutions.

\subsection{Decomposition into scalar, vector, and tensor contributions} We may regard the
various perturbation amplitudes as time dependent functions on a three-dimensional
Riemannian space $(\Sigma,\gamma)$ of constant curvature $K$. Since such a space is highly
symmetric, we can perform two types of decompositions.

Consider first the set $\mathcal{X}(\Sigma)$ of smooth vector fields on $\Sigma$. This
module can be decomposed into an orthogonal sum of `scalar' and `vector' contributions
\begin{equation}
\mathcal{X}(\Sigma) = \mathcal{X}^S \bigoplus \mathcal{X}^V~,
\end{equation}
where $\mathcal{X}^S$ consists of all gradients and $\mathcal{X}^V$ of all vector fields
with vanishing divergence.

More generally, we have for the $p$-forms $\bigwedge^p(\Sigma)$ on $\Sigma$ the orthogonal
decomposition\footnote{This is a consequence of the Hodge decomposition theorem. The scalar
product in $\bigwedge^p(\Sigma)$ is defined as
\[ (\alpha,\beta)=\int_\Sigma \alpha\wedge\star\beta;\]
see also Sect.13.9 of \cite{NS1}.}
\begin{equation}
\bigwedge\nolimits^p(\Sigma)= d\bigwedge \nolimits^{p-1}(\Sigma) \bigoplus ker\delta~,
\end{equation}
where the last summand denotes the kernel of the co-differential $\delta$ (restricted to
$\bigwedge^p(\Sigma)$).

Similarly, we can decompose a symmetric tensor $t\in\mathcal{S}(\Sigma)$ (= set of all
symmetric tensor fields on $\Sigma $) into `scalar', `vector', and `tensor' contributions:
\begin{equation}
t_{ij}=t^S_{ij} + t^V_{ij} + t^T_{ij} ~,
\end{equation}
where
\begin{eqnarray}
t_{ij}^S &= & Tr(t)\gamma_{ij} + (\nabla_i\nabla_j - \frac{1}{3}\gamma_{ij}\triangle)f~,\\
t_{ij}^V &= & \nabla_i\xi_j + \nabla_j\xi_i,\\
t_{ij}^T &: & Tr(t^T)=0;~~ \nabla\cdot t^T=0.
\end{eqnarray}
In these equations $f$ is a function on $\Sigma$ and $\xi^i$ a vector field  with vanishing
divergence. One can show that these decompositions are respected by the covariant
derivatives. For example, if $\xi\in\mathcal{X}(\Sigma),~ \xi=\xi_\ast +\nabla f,~
\nabla\cdot\xi_\ast=0$, then
\begin{equation}
\triangle\xi=\triangle\xi_\ast+\nabla\left[\triangle f +2Kf\right]
\end{equation}
(prove this as an exercise). Here, the first term on the right has a vanishing divergence
(show this), and the second (the gradient) involves only $f$. For other cases, see Appendix
B of \cite{KS84}. Is there a conceptual proof based on the isometry group of
$(\Sigma,\gamma)$?

\subsection{Decomposition into spherical harmonics}

In a second step we perform a harmonic decomposition. For $K=0$ this is just  Fourier
analysis. The spherical harmonics $\{Y\}$ of $(\Sigma,\gamma)$ are in this case the
functions $Y(\mathbf{x};\mathbf{k})=\exp(i\mathbf{k\cdot x})$ (for
$\gamma=\delta_{ij}dx^idx^j$). The \textit{scalar} parts of vector and symmetric tensor
fields can be expanded in terms of
\begin{eqnarray}
Y_i: & = & -k^{-1}\nabla_iY, \\
Y_{ij}: & = & k^{-2}\nabla_i\nabla_jY +\frac{1}{3}\gamma_{ij}Y,
\end{eqnarray}
and $\gamma_{ij}Y$.

There are corresponding complete sets of spherical harmonics for $K\neq 0$. They are
eigenfunctions of the Laplace-Beltrami operator on $(\Sigma,\gamma)$:
\begin{equation}
(\triangle+k^2)Y=0.
\end{equation}
Indices referring to the various modes are usually suppressed. By making use of the Riemann
tensor of $(\Sigma,\gamma)$ one can easily derive the following identities:

\begin{eqnarray}
\nabla_iY^i & = & k Y,\nonumber\\
\triangle Y_i & = & -(k^2-2K)Y_i,\nonumber\\
\nabla_jY_i & = & -k(Y_{ij}-\frac{1}{3}\gamma_{ij}Y),\nonumber\\
\nabla^jY_{ij} & = & \frac{2}{3}k^{-1}(k^2-3K)Y_i,\nonumber\\
\nabla_j\nabla^mY_{im} & = &
\frac{2}{3}(3K-k^2)(Y_{ij}-\frac{1}{3}\gamma_{ij}Y),\nonumber\\
\triangle Y_{ij} & = & -(k^2-6K)Y_{ij},\nonumber\\
\nabla_mY_{ij}-\nabla_jY_{im} & = &
\frac{k}{3}\left(1-\frac{3K}{k^2}\right)(\gamma_{im}Y_j-\gamma_{ij}Y_m).
\end{eqnarray}

------------------

\textbf{Exercise}. Verify some of the relations in (1.12).

------------------

The main point of the harmonic decomposition is, of course, that different modes in the
linearized approximation do not couple. Hence, it suffices to consider a generic mode.

For the time being, we consider only  scalar perturbations. Tensor perturbations (gravity
modes) will be studied later. For the harmonic analysis of vector and tensor perturbations
I refer again to \cite{KS84}.

\subsection{Gauge transformations, gauge invariant\\ amplitudes}

In GR the diffeomorphism group of spacetime is an invariance group. This means that we can
replace the metric $g$ and the matter fields by their pull-backs $\phi^\star(g)$, etc., for
any diffeomorphism $\phi$, without changing the physics. For small-amplitude departures in
\begin{equation}
g=g^{(0)} +\delta g, ~etc,
\end{equation}
we have, therefore, the \textit{gauge freedom}
\begin{equation}
\delta g\rightarrow \delta g + L_\xi g^{(0)},~etc.,
\end{equation}
where $\xi$ is any vector field and $L_\xi$ denotes its Lie derivative. (For further
explanations, see \cite{NS1}, Sect. 4.1). These transformations will induce changes in the
various perturbation amplitudes. It is clearly desirable to write all independent
perturbation equations in a manifestly \textit{gauge invariant} manner. In this way one
can, for instance, avoid misinterpretations of the growth of density fluctuations,
especially on superhorizon scales. Moreover, one gets rid of uninteresting gauge modes.

I find it astonishing that it took so long until the gauge invariant formalism was widely
used.

\subsection{Parametrization of the metric perturbations}

The most general \textit{scalar} perturbation of the metric can be parametrized as follows
\begin{equation}
\delta g = a^2(\eta)\left[-2Ad\eta^2 - 2B,_i dx^id\eta + (2D\gamma_{ij} + 2E_{\mid
ij})dx^idx^j\right].
\end{equation}
The functions $A(\eta,x^i),~B,~D,~E$ are the scalar perturbation amplitudes; $E_{\mid ij}$
denotes $\nabla_i\nabla_j E$ on $(\Sigma,\gamma)$. Thus the true metric is
\begin{equation}
\fbox{$\displaystyle g=a^2(\eta)\left\{-(1+2A)d\eta^2 - 2B,_idx^id\eta + [(1+2D)\gamma_{ij}
+ 2E_{\mid ij}]dx^idx^j\right\}.$}
\end{equation}

Let us work out how $A,B,D,E$ change under a gauge transformation (1.14), provided the
vector field is of the `scalar' type\footnote{It suffices to consider this type of vector
fields, since vector fields from $\mathcal{X}^V$ do not affect the scalar amplitudes; check
this.}:
\begin{equation}
\xi = \xi^0\partial_0 +\xi^i\partial_i,~~ \xi^i=\gamma^{ij}\xi_{\mid j}.
\end{equation}
(The index 0 refers to the conformal time $\eta$.) For this we need ('$\equiv d/d\eta $)
\[ L_\xi a^2(\eta)=2aa'\xi^0=2a^2\mathcal{H}\xi^0,~~ \mathcal{H}:=a'/a,\]
\[L_\xi d\eta=dL_\xi\eta=(\xi^0)'d\eta+\xi^0{}_{\mid i}dx^i,\]
\[L_\xi dx^i=dL_\xi x^i=d\xi^i=\xi^i,_jdx^j+(\xi^i)'d\eta=\xi^i,_jdx^j+\xi'^{\mid i}d\eta,\]
implying
\begin{eqnarray}
L_\xi\left(a^2(\eta)d\eta^2\right) & = &
2a^2\left\{(\mathcal{H}\xi^0+(\xi^0)')d\eta^2+\xi^0{}_{\mid i}dx^id\eta\right\},\nonumber\\
L_\xi\left(\gamma_{ij}dx^idx^j\right) & = & 2\xi_{\mid ij}dx^idx^j+2\xi'_{\mid
i}dx^id\eta.\nonumber
\end{eqnarray}
This gives the transformation laws:
\begin{equation}
A\rightarrow A+\mathcal{H}\xi^0+(\xi^0)',~~ B\rightarrow B +\xi^0 - \xi',~~ D\rightarrow
D+\mathcal{H}\xi^0,~~ E\rightarrow E+\xi.
\end{equation}
From this one concludes that the following \textit{Bardeen potentials}
\begin{eqnarray}
\Psi & = & A-\frac{1}{a}\left[a(B+E')\right]', \\
\Phi & = & D-\mathcal{H}(B+E'),
\end{eqnarray}
are gauge invariant.

Note that the transformations of $A$ and $D$ involve \textit{only} $\xi^0$. This is also
the case for the combinations
\begin{equation}
\chi:=a(B+E')\rightarrow \chi+a\xi^0
\end{equation}
and
\begin{eqnarray}
&&\kappa:=\frac{3}{a}(\mathcal{H}A-D')-\frac{1}{a^2}\triangle\chi  \rightarrow \\ &&
\kappa+\frac{3}{a}\left[\mathcal{H}(\mathcal{H}\xi^0+(\xi^0)')-(\mathcal{H}\xi^0)'\right]
-\frac{1}{a^2}\triangle\xi^0.
\end{eqnarray}
Therefore, it is good to work with $A,D,\chi,\kappa$. This was emphasized in \cite{JHw}.
Below we will show that $\chi$ and $\kappa$ have a simple geometrical meaning. Moreover, it
will turn out that the perturbation of the Einstein tensor can be expressed completely in
terms of the amplitudes $A,D,\chi,\kappa$.

------------------

\textbf{Exercise}. The most general vector perturbation of the metric is obviously of the
form
\[ \left(\delta g_{\mu\nu}\right)=a^2(\eta)\left(\begin{array}{cc}
0 & \beta_i\\
\beta_i & H_{i\mid j}+H_{j\mid i}\end{array} \right), \] with $B_i{}^{\mid i}=H_i{}^{\mid
i}=0$. Derive the gauge transformations for $\beta_i$ and $H_i$. Show that $H_i$ can be
gauged away. Compute $R^0{}_j$ in this gauge. Result:
\[ R^0{}_j=\frac{1}{2}\left(\triangle\beta_j+2K\beta_j\right).\]

------------------

\subsection{Geometrical interpretation}

Let us first compute the scalar curvature $R^{(3)}$ of the slices with constant time $\eta$
with the induced metric
\begin{equation}
g^{(3)}=a^2(\eta)\left[(1+2D)\gamma_{ij}+2E_{\mid ij}\right]dx^idx^j.
\end{equation}
If we drop the factor $a^2$, then the Ricci tensor does not change, but $R^{(3)}$ has to be
multiplied afterwards with $a^{-2}$.

For the metric $\gamma_{ij}+h_{ij}$ the \textit{Palatini identity} (eq. (4.20) in
\cite{NS1})
\begin{equation}
\delta R_{ij}=\frac{1}{2}\left[h^k{}_{i\mid jk}-h^k{}_{k\mid ij}+h^k{}_{j\mid ik}-\triangle
h_{ij}\right]
\end{equation}
gives
\[ \delta R^i{}_i=h^{ij}{}_{\mid ij}-\triangle h~~~(h:=h^i_i),~~ h_{ij}=2D\gamma_{ij}+2E_{\mid
ij}. \]
We also use
\begin{eqnarray}
h=6D+2\triangle E,~~ E^{\mid ij}{}_{\mid ij}
&=&\nabla_j(\triangle\nabla^jE)=\nabla_j(\nabla^j\triangle E-2K\nabla^j E)\nonumber\\&
=&\triangle^2E-2K\triangle E \nonumber
\end{eqnarray}
(we used ~ $(\nabla_i\triangle-\triangle\nabla_i)f=-R^{(0)}_{ij}\nabla^jf$, for a function
$f$). This implies
\begin{eqnarray}
h^{ij}{}_{\mid ij}&=&2\triangle D+2(\triangle^2E-2K\triangle E),\nonumber\\
\delta R^i{}_i &=&-4D-4K\triangle E),\nonumber
\end{eqnarray}
whence
\[ \delta R=\delta R^i{}_i+h^{ij}R^{(0)}_{ij}=-4\triangle D+12KD.\]
This shows that $D$ determines the scalar curvature perturbation
\begin{equation}
\fbox{$\displaystyle \delta R^{(3)}=\frac{1}{a^2}(-4\triangle D+12KD).$}
\end{equation}

Next, we compute the second fundamental form\footnote{This geometrical concept is
introduced in Appendix A of \cite{NS1}.} $K_{ij}$ for the time slices. We shall show that
\begin{equation}
\fbox{$\displaystyle \kappa=\delta K^i{}_i,$}
\end{equation}
and
\begin{equation}
K_{ij}-\frac{1}{3}g_{ij}K^l{}_l=-(\chi_{\mid ij}-\frac{1}{3}\gamma_{ij}\triangle\chi).
\end{equation}

\textit{Derivation}. In the following derivation we make use of Sect. 2.9 of \cite{NS1} on
the $3+1$ formalism. According to eq. (2.287) of this reference, the second fundamental
form is determined in terms of the lapse $\alpha$, the shift $\beta=\beta^i\partial_i$, and
the induced metric $\bar{g}$ as follows (dropping indices)
\begin{equation}
K=-\frac{1}{2\alpha}(\partial_t-L_\beta)\bar{g}.
\end{equation}
To first order this gives in our case
\begin{equation}
K_{ij}=-\frac{1}{2a(1+A)}\left[a^2(1+2D)\gamma_{ij}+2a^2E_{\mid ij}\right]'-aB_{\mid ij}.
\end{equation}
(Note that $\beta_i=-a^2B,_i,~\beta^i=-\gamma^{ij}B,_j$.)

In zeroth order this gives
\begin{equation}
K^{(0)}_{ij}=-\frac{1}{a}\mathcal{H}g^{(0)}_{ij}.
\end{equation}
Collecting the first order terms gives the claimed equations (1.27) and (1.28). (Note that
the trace-free part must be of first order, because the zeroth order vanishes according to
(1.31).)

\paragraph{Conformal gauge.}

According to (1.18) and (1.21) we can always chose the gauge such that $B=E=0$. This
so-called \textit{conformal Newtonian (or longitudinal) gauge} is often particularly
convenient to work with. Note that in this gauge
\[ \chi=0, ~~ A=\Psi, ~~ D=\Phi, ~~ \kappa=\frac{3}{a}(\mathcal{H}\Psi-\Phi'). \]

\subsection{Scalar perturbations of the energy-\\momentum tensor}

At this point we do not want to specify the matter model. For a convenient parametrization
of the scalar perturbations of the energy-momentum tensor
$T_{\mu\nu}=T^{(0)}_{\mu\nu}+\delta T_{\mu\nu}$, we define the four-velocity $u^\mu$ as a
normalized timelike eigenvector of $T^{\mu\nu}$:
\begin{eqnarray}
T^\mu{}_\nu u^\nu &=& -\rho u^\mu,\\
g_{\mu\nu}u^\mu u^\nu &=& -1.
\end{eqnarray}
The eigenvalue $\rho$ is the \textit{proper energy-mass density}.

For the unperturbed situation we have
\begin{equation}
u^{(0)0}=\frac{1}{a},~~ u^{(0)}_0=-a,~~ u^{(0)i}=0,~~ T^{(0)0}{}_0=-\rho^{(0)},~~
T^{(0)i}{}_j=p^{(0)}\delta^i{}_j,~~ T^{(0)0}{}_i=0.
\end{equation}
Setting $\rho=\rho^{(0)}+\delta\rho,~~ u^\mu=u^{(0)\mu}+\delta u^\mu$, etc, we obtain from
(1.33)
\begin{equation}
\delta u^0=-\frac{1}{a}A,~~ \delta u_0=-aA.
\end{equation}
The first order terms of (1.32) give, using (1.34),
\[ \delta
T^\mu{}_0u^{(0)0}+\delta^\mu{}_0u^{(0)0}\delta\rho+\left(T^{(0)\mu}{}_\nu+\rho^{(0)}\delta^\mu{}_\nu\right)\delta
u^\nu=0. \] For $\mu=0$ and $\mu=i$ this leads to
\begin{eqnarray}
\delta T^0{}_0 &=& -\delta\rho,\\
\delta T^i{}_0 &=& -a(\rho^{(0)}+p^{(0)})\delta u^i.
\end{eqnarray}
From this we can determine the components of $\delta T^0{}_j$ :
\begin{eqnarray}
\delta T^0{}_j &=& \delta\left[g^{0\mu}g_{j\nu}T^\nu{}_\mu\right]\nonumber\\
&=& \delta g^{0k}g^{(0)}_{ij}T^{(0)i}{}_k+g^{(0)00}\delta
g_{0j}T^{(0)0}{}_0+g^{(0)00}g^{(0)}_{ij}\delta T^i{}_0\nonumber\\
&=& \left(-\frac{1}{a^2}\gamma^{ki}B_{\mid i}\right)(a^2\gamma_{ij})p^{(0)}\delta^i{}_k
+\left(-\frac{1}{a^2}\right)(-a^2B_{\mid j})(-\rho^{(0)}) -\gamma_{ij}\delta
T^i{}_0\nonumber.
\end{eqnarray}
Collecting terms gives
\begin{equation}
\delta T^0{}_j=a(\rho^{(0)}+p^{(0)})\Bigl[\underbrace{\gamma_{ij}\delta
u^i-\frac{1}{a}B_{\mid j}}_{a^{-2}\delta u_j}\Bigr].
\end{equation}
Scalar perturbations of $\delta u^i $ can be represented as
\begin{equation}
\delta u^i=\frac{1}{a}\gamma^{ij}v_{\mid j}.
\end{equation}
Inserting this above gives
\begin{equation}
\delta T^0{}_j=(\rho^{(0)}+p^{(0)})(v-B)_{\mid j}.
\end{equation}

The scalar perturbations of the spatial components $ \delta T^i{}_j $ can be represented as
follows
\begin{equation}
\delta T^i{}_j=\delta^i{}_j~\delta p+p^{(0)}\left(\Pi^{\mid i}{}_{\mid
j}-\frac{1}{3}\delta^i{}_j~\triangle\Pi\right).
\end{equation}

Let us collect these formulae (dropping (0) for the unperturbed quantities $\rho^{(0)}$,
etc):
\begin{eqnarray}
\delta u^0 &=& -\frac{1}{a}A,~~ \delta u_0=-aA, ~~\delta u^i=\frac{1}{a}\gamma^{ij}v_{\mid
j}~~\Rightarrow
\delta u_i=a(v-B)_{\mid i};\nonumber \\
\delta T^0{}_0 &=& -\delta\rho,\nonumber\\
\delta T^0{}_i &=&(\rho+p)(v-B)_{\mid i},~~ \delta T^i{}_0=-(\rho+p)\gamma^{ij}v_{\mid
j},\nonumber\\
\delta T^i{}_j &=& \delta p~\delta^i{}_j+p\left(\Pi^{\mid i}{}_{\mid
j}-\frac{1}{3}\delta^i{}_j~\triangle\Pi\right).
\end{eqnarray}

Sometimes we shall also use the quantity
\[ \mathcal{Q}:= a(\rho+p)(v-B), \]
in terms of which the energy flux density can be written as
\begin{equation}
\delta T^0{}_i=\frac{1}{a}\mathcal{Q}_{,i}~,~~ (\Rightarrow T^t{}_i=\mathcal{Q}_{,i}).
\end{equation}
For fluids one often decomposes $\delta p $ as
\begin{equation}
p\pi_L :=\delta p=c^2_s\delta\rho+p\Gamma,
\end{equation}
where $c_s$ is the sound velocity
\begin{equation}
c^2_s=\dot{p}/\dot{\rho}.
\end{equation}
$\Gamma$ measures the deviation between $\delta p/\delta\rho$ and $\dot{p}/\dot{\rho}$.

As for the metric we have four perturbation amplitudes:
\begin{equation}
\fbox{$\displaystyle \delta :=\delta\rho/\rho~,~~v~,~~\Gamma~,~~\Pi~.$}
\end{equation}
Let us see how they change under gauge transformations:
\begin{equation}
\delta T^\mu{}_\nu \rightarrow \delta T^\mu{}_{\nu} +(L_\xi T^{(0)})^\mu{}_{\nu},~~(L_\xi
T^{(0)})^\mu{}_{\nu}=\xi^\lambda
T^{(0)\mu}{}_{\nu,\lambda}-T^{(0)\lambda}{}_{\nu}\xi^\mu{}_{,\lambda}+T^{(0)\mu}{}_\lambda\xi^{\lambda}{}_{,\nu}.
\end{equation}
Now,
\[(L_\xi T^{(0)})^0{}_0=\xi^0T^{(0)0}{}_{0,0}=\xi^0(-\rho)',\]
hence
\begin{equation}
\delta\rho\rightarrow\delta\rho+\rho'\xi^0~;~~
\delta\rightarrow\delta+\frac{\rho'}{\rho}\xi^0=\delta-3(1+w)\mathcal{H}\xi^0
\end{equation}
($w:=p/\rho$). Similarly ($\xi^i=\gamma^{ij}\xi_{\mid j}$):
\[(L_\xi T^{(0)})^0{}_i=0-T^{(0)j}{}_i\xi^0{}_{\mid
j}+T^{(0)0}{}_0\xi^0{}_{,i}=-\rho\xi^0{}_{\mid i}-p\xi^0{}_{\mid i};\] so
\begin{equation}
v-B\rightarrow (v-B)-\xi^0.
\end{equation}
Finally,
\[(L_\xi T^{(0)})^i{}_j=p'\delta^i{}_j\xi^0,\]
hence
\begin{eqnarray}
\delta p &\rightarrow & \delta p+p'\xi^0,\\
\Pi &\rightarrow & \Pi.
\end{eqnarray}
From (1.44), (1.48) and (1.50) we also obtain
\begin{equation}
\Gamma\rightarrow\Gamma.
\end{equation}
We see that $\Gamma,~\Pi$ are gauge invariant. Note that the transformation of $\delta$ and
$v-B$ involve only $\xi^0$, while $v$ transforms as
\[ v\rightarrow v-\xi'.\]
For $\mathcal{Q}$ we get
\begin{equation}
\mathcal{Q}\rightarrow \mathcal{Q}-a(\rho+p)\xi^0.
\end{equation}

 We can introduce various gauge invariant quantities. It is useful to
 adopt the following notation: For example, we use the symbol $\delta_\mathcal{Q}$ for that gauge
 invariant quantity which is equal to $\delta$ in the gauge where $\mathcal{Q}=0$, thus
 \begin{equation}
 \delta_{\mathcal{Q}}=\delta-\frac{3}{a\rho}\mathcal{H}\mathcal{Q}=\delta-3(1+w)\mathcal{H}(v-B).
 \end{equation}
 Similarly,
 \begin{eqnarray}
 \delta_\chi &=& \delta+3\frac{(1+w)\mathcal{H}}{a}\chi=\delta+3\mathcal{H}(1+w)(B+E');\\
 V:&=&(v-B)_\chi=v-B+a^{-1}\chi=v+E'=\frac{1}{a}\left(\chi+\frac{1}{\rho+p}\mathcal{Q}\right);\\
 \mathcal{Q}_\chi &=& \mathcal{Q}+(\rho+p)\chi=a(\rho+p)V.
 \end{eqnarray}
 Another important gauge invariant amplitude, often called the \textit{curvature
 perturbation} (see (1.26)), is
 \begin{equation}
 \mathcal{R}:=D_\mathcal{Q}=D+\mathcal{H}(v-B)=D_\chi+\mathcal{H}(v-B)_\chi=D_\chi+\mathcal{H}V.
 \end{equation}

\section{Explicit form of the energy-momentum conservation}

After these preparations we work out the consequences $\nabla\cdot T=0$ of Einstein's field
equations for the metric (1.16) and $T^\mu{}_\nu$ as given by (1.34) and (1.42). The
details of the calculations are presented in Appendix A of this chapter.

The energy equation reads (see (1.238)):
\begin{equation}
(\rho\delta)'+3\mathcal{H}\rho\delta+3\mathcal{H}p\pi_L+(\rho+p)\left[\triangle(v+E')+3D'\right]=0
\end{equation}
or, with $(\rho\delta)'/\rho=\delta'-3\mathcal{H}(1+w)\delta$ and (1.56),
\begin{equation}
\delta'+3\mathcal{H}(c^2_s-w)\delta+3\mathcal{H}w\Gamma=-(1+w)(\triangle V+3D').
\end{equation}
This gives, putting an index $\chi$, the gauge invariant equation
\begin{equation}
\delta'_\chi+3\mathcal{H}(c^2_s-w)\delta_\chi+3\mathcal{H}w\Gamma=-(1+w)(\triangle
V+3D'_\chi).
\end{equation}
Conversely, eq.(1.60) follows from (1.61): the $\chi$-terms cancel, as is easily verified
by using the zeroth order equation
\begin{equation}
 w'=-3(c^2_s-w)(1+w)\mathcal{H},
\end{equation}
that is easily derived from the Friedman equations in Sect. 0.1.3. From the definitions it
follows readily that the last factor in (1.60) is equal to
$-(a\kappa-3\mathcal{H}A-\triangle(v-B))$.

The momentum equation becomes (see (1.244)):
\begin{equation}
\left[(\rho+p)(v-B)\right]'+4\mathcal{H}(\rho+p)(v-B)+(\rho+p)A+p\pi_L+\frac{2}{3}(\triangle+3K)p\Pi=0.
\end{equation}
Using (1.44) in the form
\begin{equation}
 p\pi_L=\rho(c^2_s\delta+w\Gamma),
\end{equation}
and putting the index $\chi$ at the perturbation amplitudes gives the gauge invariant
equation
\begin{equation}
\left[(\rho+p)V\right]'+4\mathcal{H}(\rho+p)V+(\rho+p)A_\chi+\rho c^2_s\delta_\chi+\rho
w\Gamma+\frac{2}{3}(\triangle+3K)p\Pi=0
\end{equation}
or\footnote{Note that $h:=\rho+p$ satisfies $h'=-3\mathcal{H}(1+c^2_s)h$.}
\begin{equation}
V'+(1-3c^2_s)\mathcal{H}V+A_\chi+\frac{c^2_s}{1+w}\delta_\chi+\frac{w}{1+w}\Gamma+
\frac{2}{3}(\triangle+3K)\frac{w}{1+w}\Pi=0.
\end{equation}

For later use we write (1.63) also as
\begin{equation}
(v-B)'+(1-3c^2_s)\mathcal{H}(v-B)+A+\frac{c^2_s}{1+w}\delta+\frac{w}{1+w}\Gamma+
\frac{2}{3}(\triangle+3K)\frac{w}{1+w}\Pi=0
\end{equation}
(from which (1.66) follows immediately).

\section{Einstein equations}

A direct computation of the first order changes $\delta G^\mu{}_\nu$
of the Einstein tensor for (1.15) is complicated. It is much simpler
to proceed as follows: Compute first $\delta G^\mu{}_\nu$ in the
\textit{longitudinal gauge} $B=E=0$. (That these gauge conditions
can be imposed follows from (1.18).) Then we write the perturbed
Einstein equations in a gauge invariant form. It is then easy to
rewrite these equations without imposing any gauge conditions, thus
obtaining the equations one would get for the general form (1.15).

$\delta G^\mu{}_\nu$ is computed for the longitudinal gauge in Appendix B to this chapter.
Let us first consider the component $\mu=\nu=0$ (see eq. (1.256)):
\begin{eqnarray}
\delta G^0{}_0 &=& \frac{2}{a^2}\left[3\mathcal{H}(\mathcal{H}A-D')+(\triangle
+3K)D\right]\nonumber \\ & =& 2\left[3H(HA-\dot{D})+\frac{1}{a^2}(\triangle+3K)D\right].
\end{eqnarray}
Since $\delta T^0{}_0 = -\delta\rho$ (see (1.42)), we obtain the perturbed Einstein
equation in the longitudinal gauge
\begin{equation}
3H(HA-\dot{D})+\frac{1}{a^2}(\triangle+3K)D=-4\pi G\rho\delta.
\end{equation}
Since in the longitudinal gauge $\chi=0$ and
\begin{equation}
\kappa=3(HA-\dot{D}),
\end{equation}
we can write (1.69) as follows
\begin{equation}
\frac{1}{a^2}(\triangle+3K)D+H\kappa=-4\pi G\rho\delta.
\end{equation}
Obviously, the gauge invariant form of this equation is
\begin{equation}
\frac{1}{a^2}(\triangle+3K)D_\chi+H\kappa_\chi=-4\pi G\rho\delta_\chi,
\end{equation}
because it reduces to (1.71) for $\chi=0$. Recall in this connection the remark in
Sect.1.1.4 that the gauge transformations of the amplitudes $A,D,\chi,\kappa$ involve only
$\xi^0$. Therefore, $A_\chi,~D_\chi,~\kappa_\chi$ are \textit{uniquely} defined; the same
is true for $\delta_\chi$ (see (1.55)).

From (1.72) we can now obtain the generalization of (1.71) in \textit{any gauge}. First
note that as a consequence of
\begin{equation}
A_\chi=A-\dot{\chi},~~ D_\chi=D-H\chi
\end{equation}
(verify this), we have, using also (1.22),
\begin{eqnarray}
\kappa_\chi&=&3(HA_\chi-\dot{D}_\chi)=3(HA-\dot{D})+3\dot{H}\chi\nonumber\\
&=& \kappa+(3\dot{H}+\frac{1}{a^2}\triangle)\chi.
\end{eqnarray}
From this, (1.73) and (1.55) one readily sees that (1.72) is equivalent to
\begin{equation}
\fbox{$\displaystyle\frac{1}{a^2}(\triangle+3K)D+H\kappa=-4\pi G\rho\delta~~~\mathrm{(any~
gauge)},$}
\end{equation}
in \textit{any gauge}.

For the other components we proceed similarly. In the longitudinal gauge we have (see eqs.
(1.257) and (1.70)):
\begin{eqnarray}
\delta G^0{}_j &=&
-\frac{2}{a^2}(\mathcal{H}A-D')_{,j}=-\frac{2}{a}(H-\dot{D})_{,j}=-\frac{2}{3a}\kappa_{,j},\\
\delta T^0{}_j &=& (\rho+p)(v-B)_{,j}=\frac{1}{a}\mathcal{Q}_{,j}.
\end{eqnarray}
This gives, up to an (irrelevant) spatially homogeneous term,
\begin{equation}
\kappa=-12\pi G\mathcal{Q}~~~\mathrm{(long.~ gauge)}.
\end{equation}
The gauge invariant form of this is
\begin{equation}
\kappa_\chi=-12\pi G\mathcal{Q}_\chi.
\end{equation}
Inserting here (1.74), (1.57), and using the unperturbed equation
\begin{equation}
\dot{H}=\frac{K}{a^2}-4\pi G(\rho+p)
\end{equation}
(derive this), one obtains in any gauge
\begin{equation}
\fbox{$\displaystyle\kappa+\frac{1}{a^2}(\triangle+3K)\chi=-12\pi
G\mathcal{Q}~~~\mathrm{(any~ gauge)}.$}
\end{equation}

Next, we use (1.258):
\begin{eqnarray}
\frac{a^2}{2}\delta G^i{}_j=\delta^i{}_j\Bigl[(2\mathcal{H}'+
\mathcal{H}^2)A+\mathcal{H}A'-D''\nonumber \\
-2\mathcal{H}D'+K D+ \frac{1}{2}\triangle(A+D)\Bigr]
 -\frac{1}{2}(A+D)^{\mid i}{}_{\mid j}.
\end{eqnarray}
This implies
\begin{equation}
\frac{a^2}{2}(\delta G^i{}_j-\frac{1}{3}\delta^i{}_j~\delta
G^k{}_k)=-\frac{1}{2}\left[(A+D)^{\mid i}{}_{\mid j}-\frac{1}{3}\delta^i{}_j(A+D)^{\mid
k}{}_{\mid k}\right].
\end{equation}
Since
\[
\delta T^i{}_j-\frac{1}{3}\delta^i{}_j\delta T^k{}_k=p\left[ \Pi^{\mid i}{}_{\mid
j}-\frac{1}{3}\delta^i{}_j\triangle\Pi\right]\] we get following field equation for
$S:=A+D$
\[S^{\mid i}{}_{\mid
j}-\frac{1}{3}\delta^i{}_j\triangle S=-8\pi Ga^2p\left(\Pi^{\mid i}{}_{\mid
j}-\frac{1}{3}\delta^i{}_j\triangle\Pi\right).\] Modulo an irrelevant homogeneous term (use
the harmonic decomposition) this gives in the longitudinal gauge
\begin{equation}
A+D=-8\pi Ga^2p\Pi
\end{equation}

The gauge invariant form is
\begin{equation}
A_\chi+D_\chi=-8\pi Ga^2p\Pi,
\end{equation}
from which we obtain with (1.73) in any gauge
\begin{equation}
\fbox{$\displaystyle\dot{\chi}+H\chi-A-D=8\pi Ga^2p\Pi ~~~\mathrm{(any~ gauge)}.$}
\end{equation}

Finally, we consider the combination
\[
\frac{1}{2}(\delta G^i{}_i-\delta
G^0{}_0)=3\left\{2(\dot{H}+H^2)A+H\dot{A}-\ddot{D}-2H\dot{D}\right\}+\frac{1}{a^2}\triangle
A.\] Since
\[\frac{1}{2}(\delta T^i{}_i-\delta
T^0{}_0)=\frac{1}{2}\rho\Bigl[\underbrace{(1+3c_s^2)\delta+3w\Gamma}_{\delta+3w\pi_L}\Bigr]\]
we obtain in the longitudinal gauge the field equation
\begin{equation}
6\dot{H}A+6H^2A+3H\dot{A}-3\ddot{D}-6H\dot{D}=-\frac{1}{a^2}\triangle A+4\pi
G(1+3s_s^2)\rho\delta+12\pi Gp\Gamma .
\end{equation}
The gauge invariant form is obviously
\begin{equation}
6\dot{H}A_\chi+6H^2A_\chi+3H\dot{A}_\chi-3\ddot{D_\chi}-6H\dot{D_\chi}=-\frac{1}{a^2}\triangle
A_\chi+4\pi G(1+3s_s^2)\rho\delta_\chi+12\pi Gp\Gamma .
\end{equation}
or
\begin{eqnarray*}
\lefteqn{3(HA_\chi-\dot{D}_\chi)^{\cdot}+6H(HA_\chi-\dot{D}_\chi)=}
\\ &&-\left(\frac{1}{a^2}\triangle+3\dot{H}\right)A_\chi+4\pi G(1+3c_s^2)\rho\delta_\chi+12\pi
Gp\Gamma.
\end{eqnarray*}
With (1.74) we can write this as
\begin{equation}
\dot{\kappa}_\chi+2H\kappa_\chi=-\left(\frac{1}{a^2}\triangle+3\dot{H}\right)A_\chi+4\pi
G(1+3c_s^2)\rho\delta_\chi+12\pi Gp\Gamma.
\end{equation}
In an arbitrary gauge we obtain (the $\chi$-terms cancel)
\begin{equation}
\fbox{$\displaystyle\dot{\kappa}+2H\kappa=-\left(\frac{1}{a^2}\triangle+3\dot{H}\right)A\underbrace{+4\pi
G(1+3c_s^2)\rho\delta+12\pi Gp\Gamma}_{4\pi G\rho(\delta+3w\pi_L)}.$}
\end{equation}
\subsubsection{Intermediate summary}
This exhausts the field equations. For reference we summarize the results obtained so far.
First, we collect the equations that are valid in any gauge (indicating also their origin).
As perturbation amplitudes we use $A,D,\chi,\kappa$ (metric functions) and
$\delta,\mathcal{Q},\Pi,\Gamma$ (matter functions), because these are either gauge
invariant or their gauge transformations involve only the component $\xi^0$ of the vector
field $\xi^\mu$.

\begin{itemize}
\item definition of $\kappa$:
\begin{equation}
\kappa=3(HA-\dot{D})-\frac{1}{a^2}\triangle\chi;
\end{equation}
\item $\delta G^0{}_0$:
\begin{equation}
\frac{1}{a^2}(\triangle+3K)D+H\kappa=-4\pi G\rho\delta;
\end{equation}
\item $\delta G^0{}_j$:
\begin{equation}
\kappa+\frac{1}{a^2}(\triangle+3K)\chi=-12\pi G\mathcal{Q};
\end{equation}
\item $\delta G^i{}_j-\frac{1}{3}\delta^i{}_j~\delta G^k{}_k$:
\begin{equation}
\dot{\chi}+H\chi-A-D=8\pi Ga^2p\Pi;
\end{equation}
\item $\delta G^i{}_i-\delta G^0{}_0$:
\begin{equation}
\dot{\kappa}+2H\kappa=-\left(\frac{1}{a^2}\triangle+3\dot{H}\right)A\underbrace{+4\pi
G(1+3c_s^2)\rho\delta+12\pi Gp\Gamma}_{4\pi G\rho(\delta+3w\pi_L)};
\end{equation}
\item $T^{0\nu}{}_{;\nu}$ (eq. (1.60)):
\begin{equation}
\dot{\delta}+3H(c_s^2-w)\delta+3Hw\Gamma=(1+w)(\kappa-3HA)-\frac{1}{\rho a^2}\triangle
\mathcal{Q}
\end{equation}
or
\begin{equation}
(\rho\delta)^{\cdot}+3H\rho(\delta+\underbrace{w\pi_L}_{c_s^2\delta+w\Gamma})=
(\rho+p)(\kappa-3HA)-\frac{1}{a^2}\triangle \mathcal{Q};
\end{equation}
\item $T^{i\nu}{}_{;\nu}=0$ (eq. (1.63)):
\begin{equation}
\dot{\mathcal{Q}}+3H\mathcal{Q}=-(\rho+p)A-p\pi_L-\frac{2}{3}(\triangle+3K)p\Pi.
\end{equation}
\end{itemize}

These equations are, of course, not all independent. Putting an index $\chi$ or
$\mathcal{Q}$, etc at the perturbation amplitudes in any of them gives a gauge invariant
equation. We write these down for $A_\chi,D_\chi,\cdots$ (instead of $\mathcal{Q}_\chi$ we
use $V$; see also (1.61) and (1.66)):
\begin{equation}
\kappa_\chi =  3(HA_\chi-\dot{D}_\chi);
\end{equation}
\begin{equation}
\frac{1}{a^2}(\triangle+3K)D_\chi+H\kappa_\chi =  -4\pi G\rho\delta_\chi;
\end{equation}
\begin{equation}
\kappa_\chi = -12\pi G\mathcal{Q}_\chi;
\end{equation}
\begin{equation}
A_\chi+D_\chi = -8\pi Ga^2p\Pi;
\end{equation}
\begin{equation}
\dot{\kappa}_\chi+2H\kappa_\chi =
-\left(\frac{1}{a^2}\triangle+3\dot{H}\right)A_\chi+\underbrace{4\pi G
(1+3c_s^2)\rho\delta_\chi+12\pi Gp\Gamma}_{4\pi G\rho(\delta_\chi+3w\pi_L)};
\end{equation}
\begin{equation}
\dot{\delta}_\chi+3H(c_s^2-w)\delta_\chi+3Hw\Gamma=-3(1+w)\dot{D}_\chi-\frac{1+w}{a}\triangle
V;
\end{equation}
\begin{equation}
\dot{V}+(1-3c_s^2)HV=-\frac{1}{a}A_\chi-\frac{1}{a}\left[\frac{c_s^2}{1+w}\delta_\chi+\frac{w}{1+w}\Gamma
+\frac{2}{3}(\triangle+3K)\frac{w}{1+w}\Pi\right].
\end{equation}

\subsubsection{Harmonic decomposition}

We write these equations once more for the amplitudes of harmonic decompositions, adopting
the following conventions. For those amplitudes which enter in $g_{\mu\nu}$ and
$T_{\mu\nu}$ without spatial derivatives (i.e., $A,D,\delta,\Gamma$) we set
\begin{equation}
A=A_{(k)}Y_{(k)}~, etc~;
\end{equation}
those which appear only through their gradients ($B,v$) are decomposed as
\begin{equation}
B=-\frac{1}{k}B_{(k)}Y_{(k)}~,etc~,
\end{equation}
and, finally,, we set for $E$ and $\Pi$, entering only through second derivatives,
\begin{equation}
E=\frac{1}{k^2}E_{(k)}Y_{(k)}~~~(\Rightarrow \triangle E=-E_{(k)}Y_{(k)}).
\end{equation}
The reason for this is that we then have, using the definitions (1.9) and (1.10),
\begin{equation}
B_{\mid i}=B_{(k)}Y_{(k)i},~~~\Pi_{\mid
ij}-\frac{1}{3}\gamma_{ij}\triangle\Pi=\Pi_{(k)}Y_{(k)ij}.
\end{equation}

The spatial part of the metric in (1.16) then becomes
\begin{equation}
g_{ij}dx^idx^j=a^2(\eta)\left[\gamma_{ij}+2(D-\frac{1}{3}E)\gamma_{ij}Y+2EY_{ij}\right]dx^idx^j.
\end{equation}

The basic equations (1.91)-(1.98) imply for $A_{(k)}, B_{(k)}$, etc\footnote{We replace
$\chi$ by $\chi_{(k)}Y_{(k)}$, where according to (1.21) $\chi_{(k)}=-(a/k)(B-k^{-1}E')$;
eq. (1.111) is then just the translation of (1.22) to the Fourier amplitudes, with $\kappa
\rightarrow \kappa_{(k)}Y_{(k)}$. Similarly, $\mathcal{Q}\rightarrow
\mathcal{Q}_{(k)}Y_{(k)},~\mathcal{Q}_{(k)}=-(1/k)a(\rho+p)(v-B)_{(k)}$.}, dropping the
index $(k)$,
\begin{equation}
\kappa=3(HA-\dot{D})+\frac{k^2}{a^2}\chi,
\end{equation}
\begin{equation}
-\frac{k^2-3K}{a^2}D+H\kappa=-4\pi G\rho\delta,
\end{equation}
\begin{equation}
\kappa-\frac{k^2-3K}{a^2}\chi=-12\pi G\mathcal{Q},
\end{equation}
\begin{equation}
\dot{\chi}+H\chi-A-D=8\pi Ga^2p\Pi/k^2,
\end{equation}
\begin{equation}
\dot{\kappa}+2H\kappa=\left(\frac{k^2}{a^2}-3\dot{H}\right)A\underbrace{+4\pi
G(1+3c_s^2)\rho\delta+12\pi Gp\Gamma}_{4\pi G\rho(\delta+3w\pi_L)},
\end{equation}
\begin{equation}
(\rho\delta)^{\cdot}+3H\rho(\delta+\underbrace{w\pi_L}_{c_s^2\delta+w\Gamma})=
(\rho+p)(\kappa-3HA)+\frac{k^2}{a^2} \mathcal{Q},
\end{equation}
\begin{equation}
\dot{\mathcal{Q}}+3H\mathcal{Q}=-(\rho+p)A-p\pi_L+\frac{2}{3}\frac{k^2-3K}{k^2}p\Pi.
\end{equation}

For later use we also collect the gauge invariant eqs. (1.99)-(1.105) for the Fourier
amplitudes:
\begin{equation}
\kappa_\chi =  3(HA_\chi-\dot{D}_\chi),
\end{equation}
\begin{equation}
-\frac{k^2-3K}{a^2}D_\chi+H\kappa_\chi =  -4\pi G\rho\delta_\chi,
\end{equation}
\begin{equation}
\kappa_\chi = -12\pi
G\mathcal{Q}_\chi~~~~~\left(\mathcal{Q}_\chi=-\frac{a}{k}(\rho+p)V\right),
\end{equation}
\begin{equation}
k^2(A_\chi+D_\chi) = -8\pi Ga^2p\Pi,
\end{equation}
\begin{equation}
\dot{\kappa}_\chi+2H\kappa_\chi =
\left(\frac{k^2}{a^2}-3\dot{H}\right)A_\chi\underbrace{+4\pi
G(1+3c_s^2)\rho\delta_\chi+12\pi Gp\Gamma}_{4\pi G\rho(\delta_\chi+3w\pi_L)},
\end{equation}
\begin{equation}
\dot{\delta}_\chi+3H(c_s^2-w)\delta_\chi+3Hw\Gamma=-3(1+w)\dot{D}_\chi-(1+w)\frac{k}{a}V,
\end{equation}
\begin{equation}
\dot{V}+(1-3c_s^2)HV=\frac{k}{a}A_\chi+\frac{c_s^2}{1+w}\frac{k}{a}\delta_\chi+\frac{w}{1+w}\frac{k}{a}\Gamma
-\frac{2}{3}\frac{w}{1+w}\frac{k^2-3K}{k^2}\frac{k}{a}\Pi.
\end{equation}

\subsubsection{Alternative basic systems of equations}

From the basic equations (1.91)-(1.105) we now derive another set
which is sometimes useful, as we shall see. We want to work with
$\delta_\mathcal{Q},V$ and $D_\chi$.

The energy equation (1.96) with index $\mathcal{Q}$ gives
\begin{equation}
\dot{\delta}_\mathcal{Q}+3H(c_s^2-w)\delta_\mathcal{Q}+3Hw\Gamma=(1+w)(\kappa_\mathcal{Q}-3HA_\mathcal{Q}).
\end{equation}
Similarly, the momentum equation (1.98) implies
\begin{equation}
A_\mathcal{Q}=-\frac{1}{1+w}\left[c_s^2\delta_\mathcal{Q}+w\Gamma+\frac{2}{3}(\triangle+3K)w\Pi\right].
\end{equation}
From (1.93) we obtain
\begin{equation}
\kappa_\mathcal{Q}+\frac{1}{a^2}(\triangle+3K)\chi_\mathcal{Q}=0.
\end{equation}
But from (1.56) we see that
\begin{equation}
\chi_\mathcal{Q}=aV,
\end{equation}
hence
\begin{equation}
\kappa_\mathcal{Q}=-\frac{1}{a}(\triangle+3K)V.
\end{equation}
Now we insert (1.126) and (1.129) in (1.125) and obtain
\begin{equation}
\fbox{$\displaystyle\dot{\delta}_{\mathcal{Q}}-3Hw\delta_{\mathcal{Q}}=
-(1+w)\frac{1}{a}(\triangle+3K)V+2H(\triangle+3K)w\Pi.$}
\end{equation}

Next, we use (1.105) and the relation
\begin{equation}
\delta_\chi=\delta_\mathcal{Q}+3(1+w)HV,
\end{equation}
which follows from (1.54), to obtain
\begin{equation}
\dot{V}+HV=-\frac{1}{a}A_\chi-\frac{1}{a(1+w)}\left[c_s^2\delta_\mathcal{Q}+w\Gamma+\frac{2}{3}(\triangle+3K)w\Pi\right].
\end{equation}
Here we make use of (1.102), with the result
\begin{equation}
\fbox{$ \dot{V}+HV=\frac{1}{a}D_\chi-\frac{1}{a(1+w)}\left[c_s^2\delta_\mathcal{Q}+w\Gamma
-8\pi Ga^2(1+w)p\Pi +\frac{2}{3}(\triangle+3K)w\Pi\right]$}
\end{equation}

From (1.99), (1.101), (1.102) and (1.57) we find
\begin{equation}
\fbox{$\displaystyle\dot{D}_\chi+HD_\chi=4\pi Ga(\rho+p)V-8\pi Ga^2Hp\Pi.$}
\end{equation}

Finally, we replace in (1.100) $\delta_\chi$ by $\delta_\mathcal{Q}$ (making use of
(1.131)) and $\kappa_\chi$ by $V$ according to (1.101), giving the Poisson-like equation
\begin{equation}
\fbox{$\displaystyle \frac{1}{a^2}(\triangle+3K)D_\chi=-4\pi G\rho\delta_\mathcal{Q}.$}
\end{equation}

The system we were looking for consists of (1.130), (1.133), (1.134) and (1.135).

From these equations we now derive an interesting expression for $\dot{\mathcal{R}}$.
Recall (1.58):
\begin{equation}
\mathcal{R}=D_\mathcal{Q}=D_\chi+aHV=D_\chi+\dot{a}V.
\end{equation}
Thus
\[\dot{\mathcal{R}}=\dot{D}_\chi+\ddot{a}V+\dot{a}\dot{V}.\]
On the right of this equation we use for the first term (1.134), for the second the
following consequence of the Friedmann equations (17) and (23)
\begin{equation}
\ddot{a}=-\frac{1}{2}(1+3w)a\left(H^2+\frac{K}{a^2}\right),
\end{equation}
and for the last term we use (1.133). The result becomes relatively simple for $K=0$ (the
$V$-terms cancel):
\[\dot{\mathcal{R}}=-\frac{H}{1+w}\left[c_s^2\delta_\mathcal{Q}+w\Gamma+\frac{2}{3}w\triangle\Pi\right].\]
Using also (1.135) and the Friedmann equation (17) (for $K=0$) leads to
\begin{equation}
\dot{\mathcal{R}}=\frac{H}{1+w}\left[\frac{2}{3}c_s^2\frac{1}{(Ha)^2}\triangle
D_\chi-w\Gamma-\frac{2}{3}w\triangle\Pi\right].
\end{equation}
This is an important equation that will show, for instance, that $\mathcal{R}$ remains
constant on superhorizon scales, provided $\Gamma$ and $\Pi$ can be neglected.

As another important application, we can derive through elimination a second order equation
for $\delta_\mathcal{Q}$. For this we perform again a harmonic decomposition and rewrite
the basic equations (1.130), (1.133), (1.134) and (1.135) for the Fourier amplitudes:
\begin{equation}
\dot{\delta}_\mathcal{Q}-3Hw\delta_\mathcal{Q}=-(1+w)\frac{k}{a}\frac{k^2-3K}{k^2}V-2H\frac{k^2-3K}{k^2}w\Pi,
\end{equation}
\begin{equation}
\dot{V}+HV=-\frac{k}{a}D_\chi+\frac{1}{1+w}\frac{k}{a}\left[c_s^2\delta_\mathcal{Q}+w\Gamma
-8\pi G(1+w)\frac{a^2}{k^2}p\Pi -\frac{2}{3}\frac{k^2-3K}{k^2}w\Pi\right]
\end{equation}
\begin{equation}
\frac{k^2-3K}{a^2}D_\chi=4\pi G\rho\delta_\mathcal{Q},
\end{equation}
\begin{equation}
\dot{D}_\chi+HD_\chi=-4\pi G(\rho+p)\frac{a}{k}V-8\pi GH\frac{a^2}{k^2}p\Pi.
\end{equation}

Through elimination one can derive the following important second order equation for
$\delta_\mathcal{Q}$ (including the $\Lambda$ term)
\begin{eqnarray}
\ddot{\delta}_\mathcal{Q}+(2+3c_s^2-6w)H\dot{\delta}_\mathcal{Q}+\left[c_s^2\frac{k^2}{a^2}-4\pi
G\rho(1-6c_s^2+8w-3w^2)\right.\nonumber  \\
\left.+12(w-c_s^2)\frac{K}{a^2}+ (3c_s^2-5w)\Lambda \right]\delta_\mathcal{Q}=\mathcal{S},
\end{eqnarray}
where
\begin{eqnarray}
\lefteqn{\mathcal{S}=-\frac{k^2-3K}{a^2}w\Gamma-2\left(1-\frac{3K}{k^2}\right)Hw\dot{\Pi}}\nonumber
\\ & &-\left(1-\frac{3K}{k^2}\right)\left[-\frac{1}{3}\frac{k^2}{a^2}
+2\dot{H}+(5-3c_s^2/w)H^2\right]2w\Pi.
\end{eqnarray}
This is obtained by differentiating (1.139), and eliminating $V$ and $\dot{V}$ with the
help of (1.139) and (1.140). In addition one has to use several zeroth order equations. We
leave the details to the reader. Note that $\mathcal{S}=0$ for $\Gamma=\Pi=0$.

\section{Extension to multi-component systems}

The \textit{phenomenological} description of multi-component systems in this section
follows closely the treatment in \cite{KS84}.

Let $T^\mu_{(\alpha){}\nu}$ denote the energy-momentum tensor of species $(\alpha)$. The
total $T^\mu{}_\nu$ is assumed to be just the sum
\begin{equation}
T^\mu{}_\nu=\sum_{(\alpha)}T^\mu_{(\alpha){}\nu},
\end{equation}
and is, of course, `conserved'. For the unperturbed background we have, as in (1.34),
\begin{equation}
T^{(0)}_{(\alpha)\mu}{}^\nu=(\rho_\alpha^{(0)}+p_\alpha^{(0)})u_\mu^{(0)}
u^{(0)\nu}+p_\alpha^{(0)}\delta_\mu{}^\nu,
\end{equation}
with
\begin{equation}
\left(u^{(0)\mu}\right)=\left(\frac{1}{a},\mathbf{0}\right).
\end{equation}

The divergence of $T^\mu_{(\alpha){}\nu}$ does, in general, not vanish. We set
\begin{equation}
T^\nu_{(\alpha){}\mu;\nu}=Q_{(\alpha)\mu};~~~~ \sum_\alpha Q_{(\alpha)\mu}=0.
\end{equation}

The unperturbed $Q_{(\alpha)\mu}$ must be of the form
\begin{equation}
Q^{(0)}_{(\alpha)\mu}=\left(-aQ^{(0)}_\alpha,\mathbf{0}\right),
\end{equation}
and we obtain from (1.148) for the background
\begin{equation}
\dot{\rho}_\alpha^{(0)}=-3H(\rho_\alpha^{(0)}+p_\alpha^{(0)})
+Q^{(0)}_\alpha=-3H(1-q^{(0)}_\alpha)h_\alpha,
\end{equation}
where
\begin{equation}
h_\alpha=\rho_\alpha^{(0)}+p_\alpha^{(0)},~~~~q^{(0)}_\alpha:=Q^{(0)}_\alpha/(3Hh_\alpha).
\end{equation}
Clearly,
\begin{equation}
\rho^{(0)}=\sum_\alpha \rho_\alpha^{(0)},~~ p^{(0)}=\sum_\alpha p_\alpha^{(0)},~~
h:=\rho^{(0)}+p^{(0)}=\sum_\alpha h_\alpha,
\end{equation}
and (1.148) implies
\begin{equation}
\sum_\alpha Q^{(0)}_\alpha=0~~~~~\Leftrightarrow ~~~~~\sum_\alpha h_\alpha
q^{(0)}_\alpha=0.
\end{equation}

We again consider only \textit{scalar perturbations}, and proceed with each component as in
Sect.1.1.6. In particular, eqs. (1.32), (1.33), (1.42) and (1.44) become
\begin{equation}
T^\mu_{(\alpha)\nu} u_{(\alpha)}^\nu=-\rho_{(\alpha)}u_{(\alpha)}^\mu ,
\end{equation}
\begin{equation}
g_{\mu\nu}u_{(\alpha)}^\mu u_{(\alpha)}^\nu=-1,
\end{equation}
\begin{eqnarray}
\delta u_{(\alpha)}^0 &=&-\frac{1}{a}A,~~~\delta
u_{(\alpha)}^i=\frac{1}{a}\gamma^{ij}v_{\alpha\mid j}~~~\Rightarrow \delta
u_{(\alpha)i}=a(v_\alpha-B)_{\mid i}, \nonumber \\
\delta T^0_{(\alpha)0} &=& -\delta\rho_\alpha ,\nonumber \\
\delta T^0_{(\alpha)j} &=& h_\alpha(v_\alpha-B)_{\mid j},~~~T^i_{(\alpha)0}=
-h_\alpha\gamma^{ij}v_{\alpha\mid j}, \nonumber \\
\delta T^i_{(\alpha)j} &=& \delta p_\alpha\delta^i{}_j+p_\alpha\left(\Pi^{\mid
i}_{\alpha\mid
j}-\frac{1}{3}\delta^i{}_j\triangle\Pi_\alpha\right), \nonumber \\
\delta p_\alpha  &=& c_\alpha^2\delta\rho_\alpha+p_\alpha\Gamma_\alpha \equiv
p_\alpha\pi_{L\alpha},~~~c_\alpha^2:=\dot{p}_\alpha/\dot{\rho}_\alpha.
\end{eqnarray}
In (1.156) and in what follows the index $(0)$ is dropped.

Summation of these equations give ($\delta_\alpha:=\delta\rho_\alpha/\rho_\alpha$):
\begin{eqnarray}
\rho\delta &=&\sum_\alpha\rho_\alpha\delta_\alpha ,\\
h v &=&\sum_\alpha v_\alpha, \\
p\pi_L &= &\sum_\alpha\pi_{L\alpha}, \\
p\Pi &=&\sum_\alpha p\Pi_\alpha.
\end{eqnarray}

The only new aspect is the appearance of the perturbations $\delta Q_{(\alpha)\mu}$. We
decompose $Q_{(\alpha)\mu}$ into energy and momentum transfer rates:
\begin{equation}
Q_{(\alpha)\mu}=Q_\alpha u_\mu + f_{(\alpha)\mu},~~~u^\mu f_{(\alpha)\mu}=0.
\end{equation}
Since $u_i$ and $f_{(\alpha)i}$ are of first order, the orthogonality condition in (1.161)
implies
\begin{equation}
f_{(\alpha)0}=0.
\end{equation}
We set (for scalar perturbations)
\begin{eqnarray}
\delta Q_{(\alpha)} &=&  Q^{(0)}_\alpha \varepsilon_\alpha,\\
f_{(\alpha)j} &=& \mathcal{H}h_\alpha f_{\alpha\mid j},
\end{eqnarray}
with two perturbation functions $\varepsilon_\alpha,~f_\alpha$ for each component. Now,
recall from (1.42) that
\[\delta u_0=-aA,~~~\delta u_i=a(v-B)_{\mid i}.\]
Using all this in (1.161) we obtain
\begin{eqnarray}
\delta Q_{(\alpha)0} &=& -aQ^{(0)}_\alpha(\varepsilon_\alpha+A),\\
\delta Q_{(\alpha)j} &=& a\left[Q^{(0)}_\alpha(v-B)+Hh_\alpha f_\alpha\right]_{\mid i}.
\end{eqnarray}
The constraint in (1.148) can now be expressed as
\begin{equation}
\sum_\alpha Q^{(0)}_\alpha\varepsilon_\alpha=0,~~~\sum_\alpha h_\alpha f_\alpha=0
\end{equation}
(we have, of course, made use of (1.153)).

From now on we drop the index $(0)$.

We turn to the gauge transformation properties. As long as we do not use the zeroth-order
energy equation (1.150), the transformation laws for
$\delta_\alpha,v_\alpha,\pi_{L\alpha},\Pi_\alpha$ remain the same as those in Sect.1.1.6
for $\delta,v,\pi_L,$ and $\Pi$. Thus, using (1.150) and the notation
$w_\alpha=p_\alpha/\rho_\alpha$, we have
\begin{eqnarray}
\delta_\alpha & \rightarrow &
\delta_\alpha+\frac{\rho'}{\rho}\xi^0=\delta_\alpha-3(1+w_\alpha)\mathcal{H}
(1-q_\alpha)\xi^0,\nonumber \\
v_\alpha-B & \rightarrow & (v_\alpha-B)-\xi^0, \nonumber \\
\delta p_\alpha & \rightarrow & \delta p_\alpha+p'_\alpha\xi^0, \nonumber \\
\Pi_\alpha &\rightarrow & \Pi_\alpha, \nonumber \\
\Gamma_\alpha &\rightarrow & \Gamma_\alpha.
\end{eqnarray}

The quantity $\mathcal{Q}$, introduced below (1.42), will also be used for each component:
\begin{equation}
\delta T^0_{(\alpha)i}=: \frac{1}{a}\mathcal{Q}_{\alpha\mid i}, ~~~
\Rightarrow~~~\mathcal{Q}=\sum_{\alpha} \mathcal{Q}_{\alpha\mid i} .
\end{equation}
The transformation law of $\mathcal{Q}_{\alpha}$ is
\begin{equation}
\mathcal{Q}_\alpha \rightarrow \mathcal{Q}_\alpha -ah_\alpha \xi^0.
\end{equation}

For each $\alpha$ we define gauge invariant density perturbations
$\left(\delta_{\alpha}\right)_{\mathcal{Q}_\alpha}, (\delta_{\alpha})_{\chi}$ and
velocities $V_\alpha=(v_\alpha-B)_{\chi}$. Because of the modification in the first of eq.
(1.168), we have instead of (1.54)
\begin{equation}
\Delta_\alpha:=\left(\delta_{\alpha}\right)_{\mathcal{Q}_\alpha}=
\delta_\alpha-3\mathcal{H}(1+w_\alpha)(1-q_\alpha)(v_\alpha-B).
\end{equation}
Similarly, adopting the notation of \cite{KS84}, eq. (1.55) generalizes to
\begin{equation}
\Delta_{s\alpha}:= (\delta_{\alpha})_{\chi}=\delta_\alpha+3(1+w_\alpha)(1-q_\alpha)H\chi.
\end{equation}
If we replace in (1.171) $v_\alpha-B$ by $v-B$ we obtain another gauge invariant density
perturbation
\begin{equation}
\Delta_{c\alpha}:=
\left(\delta_{\alpha}\right)_{\mathcal{Q}}=\delta_\alpha-3\mathcal{H}(1+w_\alpha)(1-q_\alpha)(v-B),
\end{equation}
which reduces to $\delta_\alpha$ for the \textit{comoving gauge}: $v=B$.

The following relations between the three gauge invariant density perturbations are useful.
Putting an index $\chi$ on the right of (1.171) gives
\begin{equation}
\Delta_\alpha=\Delta_{s\alpha}-3\mathcal{H}(1+w_\alpha)(1-q_\alpha)V_\alpha.
\end{equation}
Similarly, putting $\chi$ as an index on the right of (1.173) implies
\begin{equation}
\Delta_{cs}=\Delta_{s\alpha}-3\mathcal{H}(1+w_\alpha)(1-q_\alpha)V.
\end{equation}

For $V_\alpha$ we have, as in (1.56),
\begin{equation}
V_\alpha=v_\alpha+E'.
\end{equation}

From now on we use similar notations for the total density perturbations:
\begin{equation}
\Delta:=\delta_{\mathcal{Q}},~~~\Delta_s:=\delta_\chi~~~(\Delta\equiv\Delta_c).
\end{equation}

Let us translate the identities (1.157)-(1.160). For instance,
\[\sum_\alpha\rho_\alpha\Delta_{c\alpha}=\sum\alpha\rho_\alpha\delta_\alpha+3\mathcal{H}(v-B)\sum_\alpha
h_\alpha(1-q_\alpha)=\rho\delta+3\mathcal{H}(v-B)h=\rho\Delta.\] We collect this and
related identities:
\begin{eqnarray}
\rho\Delta &=& \sum_\alpha\rho_\alpha\Delta_{c\alpha} \\
&=& \sum_\alpha\rho_\alpha\Delta_\alpha-a\sum_\alpha Q_\alpha V_\alpha, \\
\rho\Delta_s &=& \sum_\alpha\rho_\alpha\Delta_{s\alpha}, \\
hV &=& \sum_\alpha h_\alpha V_\alpha, \\
p\Pi &=& \sum_\alpha p_\alpha\Pi_\alpha.
\end{eqnarray}

We would like to write also $p\Gamma$ in a manifestly gauge invariant form. From (using
(1.157), (1.159) and (1.156))
\[p\Gamma=p\pi_L-c_s^2\rho\delta=\sum_\alpha\underbrace{
p_\alpha\pi_{L\alpha}}_{c_\alpha^2\rho_\alpha\delta_\alpha+
p_\alpha\Gamma_\alpha}-c_s^2\sum_\alpha\rho_\alpha\delta_\alpha
=\sum_\alpha p_\alpha\Gamma_\alpha+\sum_\alpha(c_\alpha^2-c_s^2)\rho_\alpha\delta_\alpha
\]we get
\begin{equation}
p\Gamma=p\Gamma_{int}+p\Gamma_{rel},
\end{equation}
with
\begin{equation}
p\Gamma_{int}=\sum_\alpha p_\alpha\Gamma_\alpha
\end{equation}
and
\begin{equation}
p\Gamma_{rel}=\sum_\alpha(c_\alpha^2-c_s^2)\rho_\alpha\delta_\alpha.
\end{equation}
Since $p\Gamma_{int}$ is obviously gauge invariant, this must also be the case for
$p\Gamma_{rel}$. We want to exhibit this explicitly. First note, using (1.152) and (1.150),
that
\begin{equation}
c_s^2=\frac{p'}{\rho'}=\sum_\alpha\frac{p'_\alpha}{\rho'}=\sum_\alpha
c_\alpha^2\frac{\rho'_\alpha}{\rho'}=\sum_\alpha c_\alpha^2\frac{h_\alpha}{h}(1-q_\alpha),
\end{equation}
i.e.,
\begin{equation}
c_s^2=\bar{c}_s^2-\sum_\alpha\frac{h_\alpha}{h}q_\alpha c_\alpha^2,
\end{equation}
where
\begin{equation}
\bar{c}_s^2=\sum_\alpha\frac{h_\alpha}{h}c_\alpha^2.
\end{equation}
Now we replace $\delta_\alpha$ in (1.185) with the help of (1.173) and use (1.186), with
the result
\begin{equation}
p\Gamma_{rel}=\sum_\alpha(c_\alpha^2-c_s^2)\rho_\alpha\Delta_{c\alpha}.
\end{equation}
One can write this in a physically more transparent fashion by using once more (1.186), as
well as (1.152) and (1.153),
\[p\Gamma_{rel}=\sum_{\alpha,\beta}(c_\alpha^2-c_\beta^2)\frac{h_\beta}{h}(1-q_\beta)\rho_\alpha\Delta_{c\alpha},\]
or
\begin{eqnarray}
\lefteqn{p\Gamma_{rel}=\frac{1}{2}\sum_{\alpha,\beta}(c_\alpha^2-c_\beta^2)\frac{h_\alpha
h_\beta}{h}(1-q_\alpha)(1-q_\beta)} \nonumber
\\ & & \cdot\left[\frac{\Delta_{c\alpha}}{(1+w_\alpha)(1-q_\alpha)}-
\frac{\Delta_{c\beta}}{(1+w_\beta)(1-q_\beta)}\right].
\end{eqnarray}

For the special case $q_\alpha=0$, for all $\alpha$, we obtain
\begin{eqnarray}
p\Gamma_{rel} &=& \frac{1}{2}\sum_{\alpha,\beta}(c_\alpha^2-c_\beta^2)\frac{h_\alpha
h_\beta}{h}S_{\alpha\beta}~;\\
S_{\alpha\beta}: &=& \frac{\Delta_{c\alpha}}{1+w_\alpha}-
\frac{\Delta_{c\beta}}{1+w_\beta}.
\end{eqnarray}

The gauge transformation properties of $\varepsilon_\alpha,f_\alpha$ are obtained from
\begin{equation}
\delta Q_{(\alpha)\mu} \rightarrow \delta Q_{(\alpha\mu)}+\xi^\lambda
Q_{(\alpha)\mu,\lambda}+ Q_{(\alpha)\lambda}\xi^\lambda{}_{,\mu}.
\end{equation}
For $\mu=0$ this gives, making use of (1.149) and (1.165),
\[ \varepsilon_\alpha+A \rightarrow
\varepsilon_\alpha+A+\xi^0\frac{(aQ_\alpha)'}{aQ_\alpha}+(\xi^0)'~.\] Recalling (1.18), we
obtain
\begin{equation}
\varepsilon_\alpha \rightarrow \varepsilon_\alpha +\frac{(Q_\alpha)'}{Q_\alpha}\xi^0.
\end{equation}
For $\mu=i$ we get
\[\delta Q_{(\alpha)i}\rightarrow \delta Q_{(\alpha)i}+Q_{(\alpha)0}\xi^0{}_i ,\]
thus \[v-B+Hh_\alpha f_\alpha \rightarrow v-B+Hh_\alpha f_\alpha-\xi^0.\] But according to
(1.49) $v-B$ transforms the same way, whence
\begin{equation}
\fbox{$\displaystyle f_\alpha \rightarrow f_\alpha .$}
\end{equation}
We see that the following quantity is a gauge invariant version of $\varepsilon_\alpha$
\begin{equation}
E_{c\alpha}:=\left(\varepsilon_\alpha\right)_{\mathcal{Q}}=
\varepsilon_\alpha+\frac{(Q_\alpha)'}{Q_\alpha}(v-B).
\end{equation}
We shall also use
\begin{equation}
E_\alpha:=\left(\varepsilon_\alpha\right)_{\mathcal{Q}_\alpha}=\varepsilon_\alpha+
\frac{(Q_\alpha)'}{Q_\alpha}(v_\alpha-B)=
E_{c\alpha}+\frac{(Q_\alpha)'}{Q_\alpha}(V_\alpha-V)
\end{equation}
and
\begin{equation}
E_{s\alpha}:=\left(\varepsilon_\alpha\right)_\chi=
\varepsilon_\alpha-\frac{\dot{Q}_\alpha}{Q_\alpha}\chi.
\end{equation}
Beside
\begin{equation}
F_{c\alpha}:=f_\alpha
\end{equation}
we also make use of
\begin{equation}
F_\alpha:=F_{c\alpha}-3q_\alpha(V_\alpha-V).
\end{equation}

In terms of these gauge invariant amplitudes the constraints (1.167) can be written as
(using (1.153))
\begin{eqnarray}
\sum_\alpha Q_\alpha E_{c\alpha} &=& 0 ,\\
\sum_\alpha Q_\alpha E_\alpha &=& \sum_\alpha (Q_\alpha)'V_\alpha ,\\
\sum_\alpha h_\alpha F_{c\alpha} &=& 0.
\end{eqnarray}

\subsection*{Dynamical equations}

We now turn to the dynamical equations that follow from
\begin{equation}
\delta T^\nu_{(\alpha){}\mu;\nu}=\delta Q_{(\alpha)\mu},
\end{equation}
and the expressions for $\delta T^\nu_{(\alpha){}\mu;\nu}$ and $\delta Q_{(\alpha)\mu}$
given in (1.156), (1.165) and (1.166). Below we write these in a harmonic decomposition,
making use of the formulae in Appendix A for $\delta T^\nu_{(\alpha){}\mu;\nu}$ (see
(1.235) and (1.243)). In the harmonic decomposition eqs. (1.165) and (1.166) become
\begin{eqnarray}
\delta Q_{(\alpha)0} &=& -aQ_\alpha(\varepsilon_\alpha+A)Y,\\
\delta Q_{(\alpha)j} &=& a\left[Q_\alpha(v-B)+Hh_\alpha f_\alpha\right]Y_j.
\end{eqnarray}

From (1.235) we obtain, following the conventions adopted in the harmonic decompositions
and using the last line in (1.156),
\begin{equation}
(\rho_\alpha\delta_\alpha)'+3\frac{a'}{a}\rho_\alpha\delta_\alpha+3\frac{a'}{a}p_\alpha\pi_{L\alpha}
+h_\alpha(kv_\alpha+3D'-E')=aQ_\alpha(A+\varepsilon_\alpha).
\end{equation}
In the longitudinal gauge we have $\Delta_{s\alpha}=\delta_\alpha, V_\alpha=v_\alpha,
E_{s\alpha}=\varepsilon_\alpha, E=0$, and (see (1.73) $A=A_\chi, D=D_\chi$. We also note
that, according to the definitions (1.19), (1.20), the Bardeen potentials can be expressed
as
\begin{equation}
\fbox{$\displaystyle A_\chi=\Psi ,~~~ D_\chi=\Phi.$}
\end{equation}
Eq. (1.207) can thus be written in the following gauge invariant form
\begin{equation}
(\rho_\alpha\Delta_{s\alpha})'+3\frac{a'}{a}\rho_\alpha\Delta_{s\alpha}
+3\frac{a'}{a}p_\alpha\left(\frac{c_\alpha^2}{w_\alpha}\Delta_{s\alpha}+\Gamma_\alpha\right)
+h_\alpha(kV_\alpha+3\Phi')=aQ_\alpha(\Psi+E_{s\alpha}).
\end{equation}

Similarly, we obtain from (1.243) the momentum equation
\begin{eqnarray}
\lefteqn{\left[h_\alpha(v_\alpha-B)\right]'+4\frac{a'}{a}h_\alpha(v_\alpha-B)-kh_\alpha A
-kp_\alpha\pi_{L\alpha}}\nonumber \\ &  &+\frac{2}{3}\frac{k^2-3K}{k}p_\alpha\Pi_\alpha=
a[Q_\alpha(v-B)+\frac{\dot{a}}{a}h_\alpha f_\alpha].
\end{eqnarray}
The gauge invariant form of this is (remember that $f_\alpha$ is gauge invariant)
\begin{eqnarray}
\lefteqn{(h_\alpha V_\alpha)'+4\frac{a'}{a}h_\alpha
V_\alpha-kp_\alpha\left(\frac{c_\alpha^2}{w_\alpha}\Delta_{s\alpha}+\Gamma_\alpha\right)}
\nonumber \\ &  &  - kh_\alpha\Psi+\frac{2}{3}\frac{k^2-3K}{k}p_\alpha\Pi_\alpha=a[Q_\alpha
V+\frac{\dot{a}}{a}h_\alpha f_\alpha].
\end{eqnarray}
Eqs. (1.209) and (1.211) constitute our basic system describing the dynamics of matter. It
will be useful to rewrite the momentum equation by using \[ (h_\alpha V_\alpha)'=h_\alpha
V'_\alpha+ V_\alpha
h'_\alpha,~~~h'_\alpha=\rho'_\alpha(1+c_s^2)=-3\frac{a'}{a}(1-q_\alpha)(1+c_\alpha^2)h_\alpha.\]
Together with (1.151) and (1.200) we obtain
\begin{eqnarray}
\lefteqn{V'_\alpha-3\frac{a'}{a}(1-q_\alpha)(1+c_\alpha^2)V_\alpha+4\frac{a'}{a}V_\alpha
-k\frac{p_\alpha}{h_\alpha}\left(\frac{c_\alpha^2}{w_\alpha}\Delta_{s\alpha}+\Gamma_\alpha\right)}\nonumber
\\ &  & -k\Psi+\frac{2}{3}\frac{k^2-3K}{k}\frac{p_\alpha}{h_\alpha}\Pi_\alpha=a[\frac{Q_\alpha}{h_\alpha}
V+\frac{\dot{a}}{a} f_\alpha]=\frac{a'}{a}(F_\alpha+3q_\alpha V_\alpha)\nonumber
\end{eqnarray}
or
\begin{eqnarray}
\lefteqn{V'_\alpha+\frac{a'}{a}V_\alpha=k\Psi +\frac{a'}{a}F_\alpha
+3\frac{a'}{a}(1-q_\alpha)c_\alpha^2V_\alpha }\nonumber
\\& &
+k\left[\frac{c_\alpha^2}{1+w_\alpha}\Delta_{s\alpha}+\frac{w_\alpha}{1+w_\alpha}\Gamma_\alpha\right]
-\frac{2}{3}\frac{k^2-3K}{k}\frac{w_\alpha}{1+w_\alpha}\Pi_\alpha.
\end{eqnarray}
Here we use (1.174) in the harmonic decomposition, i.e.,
\begin{equation}
\Delta_\alpha=\Delta_{s\alpha}+3(1+w_\alpha)(1-q_\alpha)\frac{a'}{a}\frac{1}{k}V_\alpha ,
\end{equation}
and finally get
\begin{eqnarray}
\lefteqn{V'_\alpha+\frac{a'}{a}V_\alpha=k\Psi +\frac{a'}{a}F_\alpha }\nonumber
\\& &
+k\left[\frac{c_\alpha^2}{1+w_\alpha}\Delta_{\alpha}+\frac{w_\alpha}{1+w_\alpha}\Gamma_\alpha\right]
-\frac{2}{3}\frac{k^2-3K}{k}\frac{w_\alpha}{1+w_\alpha}\Pi_\alpha.
\end{eqnarray}

In applications it is useful to have an equation for $V_{\alpha\beta}:=V_\alpha-V_\beta$.
We derive this for $q_\alpha=\Gamma_\alpha=0~~(\Rightarrow
\Gamma_{int}=0,~F_\alpha=F_{c\alpha}=f_\alpha)$. From (1.214) we get
\begin{eqnarray}
\lefteqn{V'_{\alpha\beta}+\frac{a'}{a}V_{\alpha\beta}=\frac{a'}{a}F_{\alpha\beta}
}\nonumber
\\& &
+k\left[\frac{c_\alpha^2}{1+w_\alpha}\Delta_{\alpha}-\frac{c_\beta^2}{1+w_\beta}\Delta_{\beta}\right]
-\frac{2}{3}\frac{k^2-3K}{k}\Pi_{\alpha\beta},
\end{eqnarray}
where
\begin{equation}
\Pi_{\alpha\beta}=\frac{w_\alpha}{1+w_\alpha}\Pi_\alpha-\frac{w_\beta}{1+w_\beta}\Pi_\beta.
\end{equation}
Beside (1.213) we also use (1.175) in the harmonic decomposition,
\begin{equation}
\Delta_{c\alpha}=\Delta_{s\alpha}+3(1+w_\alpha)(1-q_\alpha)\frac{a'}{a}\frac{1}{k}V ,
\end{equation}
to get
\begin{equation}
\Delta_\alpha=\Delta_{c\alpha}+3(1+w_\alpha)(1-q_\alpha)\frac{a'}{a}\frac{1}{k}(V_\alpha
-V).
\end{equation}

From now on we consider only a two-component system $\alpha,\beta$. (The generalization is
easy; see \cite{KS84}.) Then $V_\alpha-V=(h_\beta/h)V_{\alpha\beta}$, and therefore the
second term on the right of (1.215) is (remember that we assume $q_\alpha=0$)
\begin{eqnarray}
\lefteqn{k\left[\frac{c_\alpha^2}{1+w_\alpha}\Delta_{\alpha}-\frac{c_\beta^2}{1+w_\beta}\Delta_{\beta}\right]=
}\nonumber
\\ & &  k\left[\frac{c_\alpha^2}{1+w_\alpha}\Delta_{c\alpha}-\frac{c_\beta^2}{1+w_\beta}\Delta_{c\beta}\right]
+3\frac{a'}{a}\left(c_\alpha^2 V_{\alpha\beta}\frac{h_\beta}{h}+ c_\beta^2
V_{\alpha\beta}\frac{h_\alpha}{h}\right)
\end{eqnarray}
At this point we use the identity\footnote{From (1.192) we obtain for an arbitrary number
of components (making use of (1.178))
\[\sum_\beta\frac{h_\beta}{h}S_{\alpha\beta}=\frac{\Delta_{c\alpha}}{1+w_\alpha}-\sum_\beta\underbrace{\frac{h_\beta}{h}
\frac{1}{1+w_\beta}}_{\rho_\beta/h}\Delta_{c\beta}=\frac{\Delta_{c\alpha}}{1+w_\alpha}-
\frac{\rho}{h}\Delta=\frac{\Delta_{c\alpha}}{1+w_\alpha}-\frac{\Delta}{1+w}.\]}
\begin{equation}
\frac{\Delta_{c\alpha}}{1+w_\alpha}=\frac{\Delta}{1+w}+\frac{h_\beta}{h}S_{\alpha\beta}.
\end{equation}
Introducing also the abbreviation
\begin{equation}
c_z^2:=c_\alpha^2\frac{h_\beta}{h}+c_\beta^2\frac{h_\alpha}{h}
\end{equation}
the right hand side of (1.219) becomes
$k(c_\alpha^2-c_\beta^2)\frac{\Delta}{1+w}+kc_z^2S_{\alpha\beta}+3\frac{a'}{a}c_z^2V_{\alpha\beta}$.
So finally we arrive at
\begin{eqnarray}
\lefteqn{V'_{\alpha\beta}+\frac{a'}{a}(1-3c_z^2)V_{\alpha\beta}}\nonumber \\
& &=k(c_\alpha^2-c_\beta^2)\frac{\Delta}{1+w}+kc_z^2S_{\alpha\beta}+
\frac{a'}{a}F_{\alpha\beta}-\frac{2}{3}\frac{k^2-3K}{k}\Pi_{\alpha\beta}.
\end{eqnarray}
For the generalization of this equation, without the simplifying assumptions, see (II.5.27)
in \cite{KS84}.

Under the same assumptions we can simplify the energy equation (1.209). Using
\[\left(\frac{\rho_\alpha\Delta_{s\alpha}}{h_\alpha}\right)'=\frac{1}{h_\alpha}(\rho_\alpha\Delta_{s\alpha})'
-\frac{h'_\alpha}{h_\alpha}\frac{\rho_\alpha}{h_\alpha}\Delta_{s\alpha},~~~
\frac{h'_\alpha}{h_\alpha}\frac{\rho_\alpha}{h_\alpha}=-3\frac{a'}{a}(1+c_\alpha^2)\frac{1}{1+w_\alpha}\]
in (1.209) yields
\begin{equation}
\fbox{$\displaystyle\left(\frac{\Delta_{s\alpha}}{1+w_\alpha}\right)'=-kV_\alpha-3\Phi'.$}
\end{equation}
From this, (1.217) and the defining equation (1.192) of $S_{\alpha\beta}$ we obtain the
useful equation
\begin{equation}
\fbox{$\displaystyle S'_{\alpha\beta}=-kV_{\alpha\beta}.$}
\end{equation}

It is sometimes useful to have an equation for
$(\Delta_{c\alpha}/(1+w_\alpha))'$. From (1.217) and (1.223) (for
$q_\alpha=0$) we get
\[\left(\frac{\Delta_{c\alpha}}{1+w_\alpha}\right)'=-kV_\alpha-3\Phi'+3\left(\frac{a'}{a}\frac{1}{k}V\right)'.\]
For the last term make use of (1.137), (1.140) and (1.121). If one uses also the following
consequence of (1.118) and (1.120)
\begin{equation}
\frac{a'}{a}\Psi-\Phi'=4\pi G\rho
a^2(1+w)k^{-1}V=\frac{3}{2}\left[\left(\frac{a'}{a}\right)^2+K\right](1+w)k^{-1}V
\end{equation}
one obtains after some manipulations
\begin{eqnarray}
\left(\frac{\Delta_{c\alpha}}{1+w_\alpha}\right)' &=& -kV_\alpha +3\frac{K}{k}V
+3\frac{a'}{a}c_s^2\frac{\Delta}{1+w}+3\frac{a'}{a}\frac{w}{1+w}\Gamma\nonumber \\
&-& 3\frac{a'}{a}\frac{w}{1+w} \frac{2}{3}\left(1-\frac{3K}{k^2}\right)\Pi.
\end{eqnarray}

\section{Appendix to Chapter 1}

In this Appendix we give derivations of some results that were used in previous sections.

\subsection*{A. Energy-momentum equations}

In what follows we derive the explicit form of the perturbation equations $\delta
T^\mu{}_{\nu;\mu}=0$ for scalar perturbations, i.e., for the metric (1.16) and the
energy-momentum tensor given by (1.34) and (1.42).

\subsubsection*{Energy equation}

From
\begin{equation}
T^\mu{}_{\nu;\mu}=T^\mu{}_{\nu,\mu}+\Gamma^\mu{}_{\mu\lambda}T^\lambda{}_\nu
-\Gamma^\lambda{}_{\mu\nu}T^\mu{}_\lambda
\end{equation}
we get for $\nu=0$:
\begin{equation}
\delta(T^\mu{}_{0;\mu})=\delta
T^\mu{}_{0,\mu}+\delta\Gamma^\mu{}_{\mu\lambda}T^\lambda{}_0+\Gamma^\mu{}_{\mu\lambda}
\delta
T^\lambda{}_0-\delta\Gamma^\lambda{}_{\mu0}T^\mu{}_\lambda-\Gamma^\lambda{}_{\mu0}\delta
T^\mu{}_\lambda
\end{equation}
(quantities without a $\delta$ in front are from now on the zeroth order contributions). On
the right we have more explicitly for the first three terms
\begin{eqnarray*}
\delta T^\mu{}_{0,\mu} &=& \delta T^i{}_{0,i}+\delta T^0{}_{0,0},  \\
\delta\Gamma^\mu{}_{\mu\lambda}T^\lambda{}_0 &=&
\delta\Gamma^\mu{}_{\mu0}T^0{}_0=\delta\Gamma^i{}_{i0}T^0{}_0+\delta\Gamma^0{}_{00}T^0{}_0,
 \\
\Gamma^\mu{}_{\mu\lambda} \delta T^\lambda{}_0 &=& \Gamma^\mu{}_{\mu0} \delta T^0{}_0+
\Gamma^\mu{}_{\mu i} \delta T^i{}_0=4\mathcal{H}\delta T^0{}_0+\Gamma^j{}_{ji} \delta
T^i{}_0;  \\
\end{eqnarray*}
we used some of the unperturbed Christoffel symbols:
\begin{equation}
\Gamma^0{}_{00}=\mathcal{H},~~ \Gamma^0{}_{0i}=\Gamma^i{}_{00}=0,~~ \Gamma^0{}_{ij}=
\mathcal{H}\gamma_{ij},~~ \Gamma^i{}_{0j}=\mathcal{H}\delta^i{}_j,~~
\Gamma^i{}_{jk}=\bar{\Gamma}^i{}_{jk},
\end{equation}
where $\bar{\Gamma}^i{}_{jk}$ are the Christoffel symbols for the metric $\gamma_{ij}$.
With these the other terms become
\begin{eqnarray*}
-\delta\Gamma^\lambda{}_{\mu0}T^\mu{}_\lambda &=& -\delta\Gamma^0{}_{\mu0}T^\mu{}_0
-\delta\Gamma^i{}_{\mu0}T^\mu{}_i=-\delta\Gamma^0{}_{00}T^0{}_0-\delta\Gamma^i{}_{j0}T^j{}_i,\\
-\Gamma^\lambda{}_{\mu0}\delta T^\mu{}_\lambda &=& -\Gamma^0{}_{\mu0}\delta
T^\mu{}_0-\Gamma^i{}_{\mu0}\delta T^\mu{}_i=-\mathcal{H}\delta T^0{}_0-\mathcal{H}\delta
T^i{}_i.
\end{eqnarray*}
Collecting terms gives
\begin{equation}
\delta(T^\mu{}_{0;\mu})=(\delta T^i{}_0)_{\mid i}+\delta T^0{}_{0,0}-\mathcal{H}\delta
T^i{}_i+3\mathcal{H}\delta T^0{}_0-(\rho+p)\delta\Gamma^i{}_{i0}.
\end{equation}
We recall part of (1.42)
\begin{equation}
\delta T^0{}_0=-\delta\rho,~~\delta T^i{}_0=-(\rho+p)v^{\mid i},~~\delta T^i{}_j= \delta
p\delta^i{}_j+p\Pi^i{}_j,
\end{equation}
where
\begin{equation}
\Pi^i{}_j:=\Pi^{\mid i}{}_{\mid j}-\frac{1}{3}\delta^i{}_j~\triangle\Pi.
\end{equation}
Inserting this gives
\begin{equation}
\delta(T^\mu{}_{0;\mu})=-\delta\rho_{,0}-(\rho+p)\triangle v-3\mathcal{H}(\delta\rho+\delta
p)-(\rho+p)\delta\Gamma^i{}_{i0}.
\end{equation}
We need $\delta\Gamma^i{}_{i0}$. In a first step we have
\[\delta\Gamma^i{}_{i0}=\frac{1}{2}g^{ij}(\delta g_{ij,0}+\delta g_{j0,i}-\delta g_{i0,j})+
\frac{1}{2}\delta g^{i\nu}(g_{\nu i,0}+g_{\nu0,i}-\delta g_{i0,\nu}),\] so
\[\delta\Gamma^i{}_{i0}=\frac{1}{2}\left(\frac{1}{a^2}\gamma^{ij}\delta g_{ij,0}+\delta
g^{ij}(a^2)_{,0}\gamma_{ij}\right). \] Inserting here (1.16), i.e.,
\[\delta g_{ij}=2a^2(D\gamma_{ij}+E_{\mid ij}),~~\delta g^{ij}=-2a^2(D\gamma^{ij}+E^{\mid
ij}),\] gives
\begin{equation}
\delta\Gamma^i{}_{i0}=(3D+\triangle E)'.
\end{equation}
Hence (1.233) becomes
\begin{equation}
-\delta(T^\mu{}_{0;\mu})=(\delta\rho)'+3\mathcal{H}(\delta\rho+\delta
p)+(\rho+p)[\triangle(v+E')+3D'],
\end{equation}
giving the \textit{energy equation}:
\begin{equation}
\fbox{$\displaystyle(\delta\rho)'+3\mathcal{H}(\delta\rho+\delta
p)+(\rho+p)[\triangle(v+E')+3D']=0$}
\end{equation}
or
\begin{equation}
(\delta\rho)^{\cdot}+3H(\delta\rho+\delta p)+(\rho+p)[\triangle(v+\dot{E})+3\dot{D}]=0.
\end{equation}

We rewrite (1.236) in terms of $\delta:=\delta\rho/\rho$, using also (1.44) and (1.56),
\begin{equation}
(\rho\delta)'+3\mathcal{H}\rho\delta+3\mathcal{H}p\pi_L+(\rho+p)[\triangle V+3D']=0.
\end{equation}

\subsubsection*{Momentum equation}

For $\nu=i$ eq. (1.227) gives
\begin{equation}
\delta(T^\mu{}_{i;\mu})=\delta
T^\mu{}_{i,\mu}+\delta\Gamma^\mu{}_{\mu\lambda}T^\lambda{}_i+\Gamma^\mu{}_{\mu\lambda}
\delta T^\lambda{}_i-\delta\Gamma^\lambda{}_{\mu i}T^\mu{}_\lambda -\Gamma^\lambda{}_{\mu
i}\delta T^\mu{}_\lambda.
\end{equation}
On the right we have more explicitly, again using (1.229),
\begin{eqnarray*}
\delta T^\mu{}_{i,\mu} &=&\delta T^j{}_{i,j}+\delta T^0{}_{i,0},\\
\delta\Gamma^\mu{}_{\mu j}T^\lambda{}_i &=& \delta\Gamma^\mu{}_{\mu
j}T^j{}_i=\delta\Gamma^0{}_{0j}T^j{}_i+\delta\Gamma^k{}_{kj}T^j{}_i,\\
\Gamma^\mu{}_{\mu\lambda} \delta T^\lambda{}_i &=& \Gamma^\mu{}_{\mu0} \delta
T^0{}_i+\Gamma^\mu{}_{\mu j} \delta T^j{}_i=4\mathcal{H}\delta T^0{}_i+\Gamma^k{}_{k j}
\delta T^j{}_i,\\
-\delta\Gamma^\lambda{}_{\mu i}T^\mu{}_\lambda &=& -\delta\Gamma^0{}_{\mu i}T^\mu{}_0
-\delta\Gamma^j{}_{\mu
i}T^\mu{}_j=-\delta\Gamma^0{}_{0i}T^0{}_0-\delta\Gamma^j{}_{ki}T^k{}_j,\\
-\Gamma^\lambda{}_{\mu i}\delta T^\mu{}_\lambda &=& -\Gamma^0{}_{\mu i}\delta T^\mu{}_0
-\Gamma^j{}_{\mu i}\delta T^\mu{}_j =-\mathcal{H}\gamma_{ij}\delta
T^j{}_0-\mathcal{H}\delta T^0{}_i-\Gamma^j{}_{ki}\delta T^k{}_j.
\end{eqnarray*}
Collecting terms gives
\begin{equation}
\delta(T^\mu{}_{i;\mu})=(\delta T^j{}_i)_{\mid j}+\delta T^0{}_{i,0}+3\mathcal{H}\delta
T^0{}_i-\mathcal{H}\gamma_{ij}\delta T^j{}_0+(\rho+p)\delta\Gamma^0{}_{0i}.
\end{equation}
One readily finds
\begin{equation}
\delta\Gamma^0{}_{0i}=(A-\mathcal{H}B)_{\mid i}
\end{equation}
We insert this and (1.231) into the last equation and obtain
\begin{eqnarray*}
\delta(T^\mu{}_{i;\mu})=\left\{\delta p+(\rho+p)'(v-B) + (\rho+p)\right.\\
\left.\cdot[(v-B)'+4\mathcal{H}(v-B)+A]\right\}_{\mid i}+p\Pi^j{}_{i\mid j}.
\end{eqnarray*}
From (1.232) we obtain ($R(\gamma)_{ij}$ denotes the Ricci tensor for the metric
$\gamma_{ij}$)
\begin{equation}
\Pi^j{}_{i \mid j}=\Pi^{\mid j}{}_{\mid ij}-\frac{1}{3}\Pi_{\mid i}=\Pi^{\mid j}{}_{\mid
ji}+R(\gamma)_{ij}\Pi^{\mid j}-\frac{1}{3}\Pi_{\mid
i}=\left[\frac{2}{3}(\triangle+3K)\Pi\right]_{\mid i}.
\end{equation}
As a result we see that $\delta(T^\mu{}_{i;\mu})$ is equal to $\partial_i$ of the function
\begin{equation}
[(\rho+p)(v-B)]'+4\mathcal{H}(\rho+p)(v-B)+(\rho+p)A+p\pi_L+\frac{2}{3}(\triangle+3K)p\Pi,
\end{equation}
and the momentum equation becomes explicitly ($h=\rho+p$)
\begin{equation}
\fbox{$\displaystyle[h(v-B)]'+4\mathcal{H}h(v-B)+hA+p\pi_L+\frac{2}{3}(\triangle+3K)p\Pi=0.$}
\end{equation}

\subsection*{B. Calculation of the Einstein tensor \\for the longitudinal gauge}

In the longitudinal gauge the metric is equal to $g_{\mu\nu}+\delta g_{\mu\nu}$, with
\begin{equation}
g_{00}=-a^2,~~ g_{0i}=0,~~ g_{ij}=a^2\gamma_{ij},~~ g^{00}=-a^{-2},~~ g^{0i}=0,~~
g^{ij}=a^{-2}\gamma^{ij};
\end{equation}
\begin{eqnarray}
\delta g_{00} &=& -2a^2A,~~ \delta g_{0i}=0,~~ \delta g_{ij}=2a^2D\gamma_{ij},~~ \nonumber
\\ \delta g^{00} &=& 2a^{-2}A,~~ \delta g^{0i}=0,~~ \delta g^{ij}=-2a^{-2}D\gamma^{ij}.
\end{eqnarray}
The unperturbed Christoffel symbols have been given before in (1.229).

Next we need
\begin{equation}
\delta\Gamma^\mu{}_{\alpha\beta}=\frac{1}{2}\delta
g^{\mu\nu}(g_{\nu\alpha,\beta}+g_{\nu\beta,\alpha}-g_{\alpha\beta,\nu})+
\frac{1}{2}g^{\mu\nu}(\delta g_{\nu\alpha,\beta}+\delta g_{\nu\beta,\alpha}-\delta
g_{\alpha\beta,\nu}).
\end{equation}
For example, we have
\[\delta\Gamma^0{}_{00}=\frac{1}{2}2a^{-2}A(-a^2)'+\frac{1}{2}(-a^2)(-2a^2A)'=A'.\]
Some of the other components have already been determined in Sect.A. We list, for further
use, all $\delta\Gamma^\mu{}_{\alpha\beta}$:

\begin{eqnarray}
\delta\Gamma^0{}_{00}&=&A',~~
\delta\Gamma^0{}_{0i}=A_{,i},~~\delta\Gamma^0{}_{ij}=\left[2\mathcal{H}(D-A)+D'\right]\gamma_{ij},~~\nonumber \\
\delta\Gamma^i{}_{00} &=& A^{,i},~~\delta\Gamma^i{}_{0j}=D'\delta^i{}_j,~~
\delta\Gamma^i{}_{jk} = D_{,k}\delta^i{}_j+D_{,j}\delta^i{}_k-D^{,i}\delta_{jk}\nonumber \\
\end{eqnarray}
(indices are raised with $\gamma^{ij})$.

For $\delta R_{\mu\nu}$ we have the general formula
\begin{equation}
\delta R_{\mu\nu}=\partial_\lambda\delta\Gamma^\lambda{}_{\nu\mu}-
\partial_\nu\delta\Gamma^\lambda{}_{\lambda\mu}
+\delta\Gamma^\sigma{}_{\nu\mu}\Gamma^\lambda{}_{\lambda\sigma}+\Gamma^\sigma{}_{\nu\mu}\delta
\Gamma^\lambda{}_{\lambda\sigma}
-\delta\Gamma^\sigma{}_{\lambda\mu}\Gamma^\lambda{}_{\nu\sigma}-
\Gamma^\sigma{}_{\lambda\mu}\delta\Gamma^\lambda{}_{\nu\sigma}.
\end{equation}
We give the details for $\delta R_{00}$,
\begin{equation}
\delta R_{00}=\partial_\lambda\delta\Gamma^\lambda{}_{00}-
\partial_0\delta\Gamma^\lambda{}_{\lambda 0}
+\delta\Gamma^\sigma{}_{00}\Gamma^\lambda{}_{\lambda\sigma}+\Gamma^\sigma{}_{00}\delta
\Gamma^\lambda{}_{\lambda\sigma} -\delta\Gamma^\sigma{}_{\lambda
0}\Gamma^\lambda{}_{0\sigma}- \Gamma^\sigma{}_{\lambda 0}\delta\Gamma^\lambda{}_{0\sigma}.
\end{equation}
The individual terms on the right are:
\begin{eqnarray*}
\partial_\lambda\delta\Gamma^\lambda{}_{00} &=&
(\delta\Gamma^0{}_{00})'+(\delta\Gamma^i{}_{00})_{,i}=A''+A^{,i}{}_{,i}, \\
-\partial_0\delta\Gamma^\lambda{}_{\lambda 0} &=& -A''-3D'',\\
\delta\Gamma^\sigma{}_{00}\Gamma^\lambda{}_{\lambda\sigma} &=&
\delta\Gamma^0{}_{00}\Gamma^\lambda{}_{\lambda0}+\delta\Gamma^i{}_{00}\Gamma^\lambda{}_{\lambda
i}=4\mathcal{H}A'+\bar{\Gamma}^l{}_{li}A^{,i},\\
\Gamma^\sigma{}_{00}\delta \Gamma^\lambda{}_{\lambda\sigma} &=& \Gamma^0{}_{00}\delta
\Gamma^\lambda{}_{\lambda 0}+\Gamma^i{}_{00}\delta \Gamma^\lambda{}_{\lambda i}=\mathcal{H}(A'+3D'),\\
-\delta\Gamma^\sigma{}_{\lambda 0}\Gamma^\lambda{}_{0\sigma} &=& -\delta\Gamma^0{}_{\lambda
0}\Gamma^\lambda{}_{00}-\delta\Gamma^i{}_{\lambda
0}\Gamma^\lambda{}_{0i}=-\mathcal{H}(A'+3D'),\\
- \Gamma^\sigma{}_{\lambda 0}\delta\Gamma^\lambda{}_{0\sigma} &=& - \Gamma^0{}_{\lambda
0}\delta\Gamma^\lambda{}_{00}- \Gamma^i{}_{\lambda
0}\delta\Gamma^\lambda{}_{0i}=-\mathcal{H}(A'+3D').
\end{eqnarray*}
Summing up gives the desired result
\begin{equation}
\fbox{$\displaystyle\delta R_{00}=\triangle A+3\mathcal{H}A'-3D''-3\mathcal{H}D'.$}
\end{equation}

Similarly one finds (unpleasant exercise)
\begin{equation}
\delta R_{0j} = 2(\mathcal{H}A-D')_{,j}~,
\end{equation}
\begin{eqnarray}
\delta R_{ij} = -(A+D)_{\mid ij}+\left[-\triangle
D-(4\mathcal{H}^2+2\mathcal{H}')A-\mathcal{H}A'\right.\\
\left.+(4\mathcal{H}^2+2\mathcal{H}')D-5\mathcal{H}D'+D''\right]\gamma_{ij}.
\end{eqnarray}

Using also the zeroth order expressions for the Ricci tensor
\begin{equation}
R_{00}=-3\mathcal{H}',~~ R_{ij}=[\mathcal{H}'+2\mathcal{H}^2+2K]\gamma_{ij},~~ R_{0i}=0,
\end{equation}
one finds for the Einstein tensor\footnote{Note that $\delta R^\mu{}_\nu=\delta
g^{\mu\lambda}R_{\lambda\nu}+g^{\mu\lambda}\delta R_{\lambda\nu}$.}
\begin{equation}
\delta G^0{}_0=\frac{2}{a^2}[3\mathcal{H}(\mathcal{H}A-D')+\triangle D+3KD],
\end{equation}
\begin{equation}
\delta G^0{}_j=-\frac{2}{a^2}[\mathcal{H}A-D']_{,j}~,
\end{equation}
\begin{eqnarray}
\delta G^i{}_j=\frac{2}{a^2}\Bigl\{(2\mathcal{H}'+\mathcal{H}^2)A+\mathcal{H}A'-D'' \nonumber\\
-2\mathcal{H}D'+K
D+\frac{1}{2}\triangle(A+D)\Bigr\}\delta^i{}_j-\frac{1}{a^2}(A+D)^{\mid
i}{}_{\mid j}.
\end{eqnarray}

These results can be derived less tediously with the help of the '3+1 formalism',
developed, for instance, in Sect. 2.9 of \cite{NS1}. This was sketched in \cite{DS}.

\subsection*{C. Summary of notation and basic equations}

Notation in cosmological perturbation theory is a nightmare.
Unfortunately, we had to introduce lots of symbols and many
equations. For convenience, we summarize the adopted notation and
indicate the location of the most important formulae. Some of them
are repeated for further reference.

\subsubsection{Recapitulation of the basic perturbation equations}

For \textit{scalar} perturbations we use the following gauge invariant amplitudes:

\textit{metric}: $\Psi,~\Phi$ (Bardeen potentials)
\begin{equation}
\Psi\equiv A_\chi,~~ \Phi\equiv D_\chi;
\end{equation}

\textit{total energy-momentum tensor} $T^{\mu\nu}:~ \Delta,~V$; instead of $\Delta$ we also
use
\begin{equation}
\Delta_s=\Delta-3(1+w)H\frac{a}{k}V.
\end{equation}

The basic equations, derived from Einstein's field equations, and some of the consequences,
can be summarized in the harmonic decomposition as follows: \\$\bullet$ constraint
equations:
\begin{equation}
(k^2-3K)\Phi=4\pi G\rho a^2\Delta,
\end{equation}
\begin{equation}
\dot{\Phi}-H\Psi=-4\pi G(\rho+p)\frac{a}{k}V;
\end{equation}
$\bullet$ relevant dynamical equation:
\begin{equation}
\Phi+\Psi = -8\pi G\frac{a^2}{k^2}p\Pi;
\end{equation}
$\bullet$ energy equation:
\begin{equation}
\dot{\Delta}-3Hw\Delta=-\left(1-\frac{3K}{k^2}\right)\left[(1+w)\frac{k}{a}V+2Hw\Pi\right];
\end{equation}
$\bullet$ momentum equation:
\begin{equation}
\dot{V}+HV=\frac{k}{a}\Psi+\frac{1}{1+w}\frac{k}{a}\left[c_s^2\Delta+w\Gamma
-\frac{2}{3}\frac{k^2-3K}{k^2}w\Pi\right].
\end{equation}

If $\Delta$ is replaced in (1.264) and (1.265) by $\Delta_s$ these equations become
\begin{equation}
\dot{\Delta}_s+3H(c_s^2-w)\Delta_s=-3(1+w)\dot{\Phi}-(1+w)\frac{k}{a}V-3Hw\Gamma,
\end{equation}
\begin{equation}
\dot{V}+(1-3c_s^2)HV=\frac{k}{a}\Psi+\frac{c_s^2}{1+w}\frac{k}{a}\Delta_s+\frac{w}{1+w}\frac{k}{a}\Gamma
-\frac{2}{3}\frac{w}{1+w}\frac{k^2-3K}{k^2}\frac{k}{a}\Pi.
\end{equation}

\textit{multi-component systems}:
\begin{equation}
T^\mu{}_\nu=\sum_{(\alpha)}T^\mu_{(\alpha){}\nu},~~~~T^\nu_{(\alpha){}\mu;\nu}=Q_{(\alpha)\mu},~~~~
\sum_\alpha Q_{(\alpha)\mu}=0.
\end{equation}
$\bullet$ additional \textit{unperturbed} quantities, beside
$\rho_\alpha,,p_\alpha,h_\alpha,c_\alpha$, : $Q_\alpha,q_\alpha$, satisfy:
\begin{eqnarray}
\rho &=& \sum_\alpha \rho_\alpha,~~ p=\sum_\alpha p_\alpha,~~h:=\rho+p=\sum_\alpha
h_\alpha, \\
Q_\alpha &=& 3Hh_\alpha q_\alpha,~~\sum_\alpha Q_\alpha =0,~~ \sum_\alpha h_\alpha
q_\alpha=0, \\
\dot{\rho}_\alpha &=&-3H(1-q_\alpha)h_\alpha.
\end{eqnarray}
$\bullet$ \textit{perturbations}: gauge invariant amplitudes:
$\Delta_\alpha,\Delta_{s\alpha},\Delta_{c\alpha},\Pi_\alpha,\Gamma_\alpha$,
\begin{eqnarray}
\rho\Delta &=& \sum_\alpha\rho_\alpha\Delta_{c\alpha} \\
&=& \sum_\alpha\rho_\alpha\Delta_\alpha-a\sum_\alpha Q_\alpha V_\alpha, \\
\rho\Delta_s &=& \sum_\alpha\rho_\alpha\Delta_{s\alpha}, \\
hV &=& \sum_\alpha h_\alpha V_\alpha, \\
p\Pi &=& \sum_\alpha p_\alpha\Pi_\alpha,\\
 p\Gamma &=&p\Gamma_{int}+p\Gamma_{rel},\\
p\Gamma_{int} &=&\sum_\alpha p_\alpha\Gamma_\alpha,\\
p\Gamma_{rel} &=& \sum_\alpha(c_\alpha^2-c_s^2)\rho_\alpha\Delta_{c\alpha}
\end{eqnarray}
or
\begin{eqnarray}
\lefteqn{p\Gamma_{rel}=\frac{1}{2}\sum_{\alpha,\beta}(c_\alpha^2-c_\beta^2)\frac{h_\alpha
h_\beta}{h}(1-q_\alpha)(1-q_\beta)} \nonumber
\\ & & \cdot\left[\frac{\Delta_{c\alpha}}{(1+w_\alpha)(1-q_\alpha)}-
\frac{\Delta_{c\beta}}{(1+w_\beta)(1-q_\beta)}\right];
\end{eqnarray}
for the special case $q_\alpha=0$, for all $\alpha$:
\begin{eqnarray}
p\Gamma_{rel} &=& \frac{1}{2}\sum_{\alpha,\beta}(c_\alpha^2-c_\beta^2)\frac{h_\alpha
h_\beta}{h}S_{\alpha\beta}~;\\
S_{\alpha\beta}: &=& \frac{\Delta_{c\alpha}}{1+w_\alpha}-
\frac{\Delta_{c\beta}}{1+w_\beta}.
\end{eqnarray}
$\bullet$ additional gauge invariant perturbations from $\delta Q_{(\alpha)\mu}$: \\energy:
$E_\alpha,E_{c\alpha},E_{s\alpha}$;~~ momentum: $F_\alpha,F_{c\alpha}$; ~~ constraints:
\begin{eqnarray}
\sum_\alpha Q_\alpha E_{c\alpha} &=& 0 ,\\
\sum_\alpha Q_\alpha E_\alpha &=& \sum_\alpha (Q_\alpha)'V_\alpha ,\\
\sum_\alpha h_\alpha F_{c\alpha} &=& 0.
\end{eqnarray}
$\bullet$ \textit{dynamical equations} for $q_\alpha=\Gamma_\alpha=0~(\Rightarrow
\Gamma_{int}=0)$; some of the equations below hold only for two-component systems;
\begin{equation}
\left(\frac{\Delta_{s\alpha}}{1+w_\alpha}\right)'=-kV_\alpha-3\Phi';
\end{equation}
eq. (1.226) for $K=0$:
\begin{equation}
\left(\frac{\Delta_{c\alpha}}{1+w_\alpha}\right)'=-kV_\alpha
+3\frac{a'}{a}c_s^2\frac{\Delta}{1+w}+3\frac{a'}{a}\frac{w}{1+w}\Gamma-3\frac{a'}{a}\frac{w}{1+w}
\frac{2}{3}\Pi;
\end{equation}
\begin{equation}
V'_\alpha+\frac{a'}{a}V_\alpha=k\Psi +\frac{a'}{a}F_\alpha
+k\frac{c_\alpha^2}{1+w_\alpha}\Delta_{\alpha}-\frac{2}{3}\frac{k^2-3K}{k}\Pi_\alpha;
\end{equation}
for $V_{\alpha\beta}:=V_\alpha-V_\beta$:
\begin{eqnarray}
\lefteqn{V'_{\alpha\beta}+\frac{a'}{a}(1-3c_z^2)V_{\alpha\beta}}\nonumber \\
& &=k(c_\alpha^2-c_\beta^2)\frac{\Delta}{1+w}+kc_z^2S_{\alpha\beta}+
\frac{a'}{a}F_{\alpha\beta}-\frac{2}{3}\frac{k^2-3K}{k}\Pi_{\alpha\beta},
\end{eqnarray}
relation between $S_{\alpha\beta}$ and $V_{\alpha\beta}$:
\begin{equation}
 S'_{\alpha\beta}=-kV_{\alpha\beta}.
\end{equation}
When working with $\Delta_{s\alpha}$ it is natural to substitute in (1.288) $\Delta_\alpha$
with the help of (1.174) in terms of $\Delta_{s\alpha}$:
\begin{equation}
V'_\alpha+\frac{a'}{a}(1-3c^2_\alpha)V_\alpha=k\Psi +\frac{a'}{a}F_\alpha
+k\frac{c_\alpha^2}{1+w_\alpha}\Delta_{s\alpha}-\frac{2}{3}\frac{k^2-3K}{k}\frac{w_\alpha}{1+w_\alpha}\Pi_\alpha,
\end{equation}


\chapter{Some Applications of Cosmological Perturbation Theory}

In this Chapter we discuss some applications of the general formalism. More relevant
applications will follow in later Parts.

Before studying realistic multi-component fluids, we consider first the simplest case when
one component, for instance CDM, dominates. First, we study, however, a general problem

Let us write down the basic equations (1.139)-(1.142) in the notation adopted later
($A_\chi=\Psi, D_\chi=\Phi, \delta_{\mathcal{Q}}=\Delta$):
\begin{equation}
\dot{\Delta}-3Hw\Delta=-(1+w)\frac{k}{a}\frac{k^2-3K}{k^2}V-2H\frac{k^2-3K}{k^2}w\Pi,
\end{equation}
\begin{equation}
\dot{V}+HV=-\frac{k}{a}\Phi+\frac{1}{1+w}\frac{k}{a}\left[c_s^2\Delta+w\Gamma -8\pi
G(1+w)\frac{a^2}{k^2}p\Pi -\frac{2}{3}\frac{k^2-3K}{k^2}w\Pi\right],
\end{equation}
\begin{equation}
\frac{k^2-3K}{a^2}\Phi=4\pi G\rho\Delta,
\end{equation}
\begin{equation}
\dot{\Phi}+H\Phi=-4\pi G(\rho+p)\frac{a}{k}V-8\pi GH\frac{a^2}{k^2}p\Pi.
\end{equation}
Recall also (1.121):
\begin{equation}
\Phi+\Psi = -8\pi G\frac{a^2}{k^2}p\Pi.
\end{equation}
Note that $\Phi=-\Psi$ for $\Pi=0$.

From these perturbation equations we derived through elimination the second order equation
(1.143) for $\Delta$, which we repeat for $\Pi=0$ (vanishing anisotropic stresses) and
$\Gamma=0$ (vanishing entropy production):
\begin{eqnarray}
\ddot{\Delta}+(2+3c_s^2-6w)H\dot{\Delta}+\left[c_s^2\frac{k^2}{a^2}-4\pi
G\rho(1-6c_s^2+8w-3w^2)\right.\nonumber  \\
\left.+12(w-c_s^2)\frac{K}{a^2}+ (3c_s^2-5w)\Lambda \right]\Delta=0.
\end{eqnarray}
Sometimes it is convenient to write this in terms of the conformal time for the quantity
$\rho a^3\Delta$. Making use of $(\rho a^3)^{\cdot}=-3Hw(\rho a^3)$ (see (0.22)) one finds
\begin{equation}
(\rho a^3\Delta)''+(1+3c_s^2)\mathcal{H}(\rho a^3\Delta)'+\left[(k^2-3K)c_s^2-4\pi
G(\rho+p)a^2\right](\rho a^3\Delta)=0.
\end{equation}
Similarly, one can derive a second order equation for $\Phi$:
\begin{equation}
\ddot{\Phi}+(4+3c_s^2)H\dot{\Phi}+\left[c_s^2\frac{k^2}{a^2}+8\pi
G\rho(c_s^2-w)-2(1+3c_s^2)\frac{K}{a^2}+(1+c_s^2)\Lambda\right]\Phi=0.
\end{equation}
Remarkably, for $p=p(\rho)$ this can be written as \cite{JHN}
\begin{equation}
\frac{\rho+p}{H}\left[\frac{H^2}{a(\rho+p)}\left(\frac{a}{H}\Phi\right)^{\cdot}\right]^{\cdot}
+c_s^2\frac{k^2}{a^2}\Phi=0
\end{equation}
(Exercise).

\section{Non-relativistic limit}

It is instructive to first consider a one-component non-relativistic fluid. The
non-relativistic limit of the second order equation (2.6) is
\begin{equation}
\ddot{\Delta}+2H\dot{\Delta}=4\pi G\rho\Delta-c_s^2\left(\frac{k}{a}\right)^2\Delta.
\end{equation}
From this basic equation one can draw various conclusions.

\subsection*{The Jeans criterion}

One sees from (2.10) that gravity wins over the pressure term $\propto c_s^2$ for $k<k_J$,
where
\begin{equation}
k_J^2\left(\frac{c_s}{a}\right)^2=4\pi G\rho
\end{equation}
defines the \textit{comoving Jeans wave number}. The corresponding \textit{Jeans length}
(wave length) is
\begin{equation} \lambda_J=\frac{2\pi}{k_J}a=\left(\frac{\pi
c_s^2}{G\rho}\right)^{1/2}, ~~~\frac{\lambda_J}{2\pi}\simeq \frac{c_s}{H}.
\end{equation}
For $\lambda<\lambda_J$ we expect that the fluid oscillates, while for
$\lambda\gg\lambda_J$ an over-density will increase.

Let us illustrate this for a polytropic equation of state $p=const~\rho^\gamma$. We
consider, as a simple example, a matter dominated Einstein-de Sitter model ($K=0$), for
which $a(t)\propto t^{2/3},~H=2/(3t)$. Eq. (2.10) then becomes (taking $\rho$ from the
Friedmann equation, $\rho=1/(6\pi Gt^2)$)
\begin{equation}
\ddot{\Delta}+\frac{4}{3t}\dot{\Delta}+\left(\frac{L^2}{t^{2\gamma-2/3}}-\frac{2}{3t^2}\right)\Delta=0,
\end{equation}
where $L^2$ is the constant
\begin{equation}
L^2=\frac{t^{2\gamma-2/3}c_s^2k^2}{a^2}.
\end{equation}
The solutions of (2.27)are
\begin{equation}
\Delta_{\pm}(t)\propto t^{-1/6}J_{\mp
5/6\nu}\left(\frac{Lt^{-\nu}}{\nu}\right),~~~\nu:=\gamma-\frac{4}{3}>0.
\end{equation}
The Bessel functions $J$ oscillate for $t\ll L^{1/\nu}$, whereas for $t\gg L^{1/\nu}$ the
solutions behave like
\begin{equation}
\Delta_{\pm}(t)\propto t^{-\frac{1}{6}\pm\frac{5}{6}}.
\end{equation}
Now, $t>L^{1/\nu}$ signifies $c_s^2k^2/a^2< 6\pi G\rho$. This is essentially again the
Jeans criterion $k<k_J$. At the same time we see that
\begin{eqnarray}
\Delta_+ &\propto& t^{2/3}\propto a,\\
\Delta_- &\propto& t^{-1}.
\end{eqnarray}
Thus the \textit{growing mode increases like the scale factor}.

\section{Large scale solutions}

Consider, as an important application, wavelengths \textit{larger than the Jeans length},
i.e., $c_s(k/aH)\ll 1$. Then we can drop the last term in equation (2.9) and solve for
$\Phi$ in terms of quadratures:
\begin{equation}
\Phi(t,\mathbf{k})=C(\mathbf{k})\frac{H}{a}\int_{0}^t\frac{a(\rho+p)}{H^2}dt+\frac{H}{a}d(\mathbf{k}).
\end{equation}
We write this differently by using in the integrand the following background equation (
implied by (1.80))
\[\frac{a(\rho+p)}{H^2}=\left(\frac{a}{H}\right)^{\cdot}-a\left(1-\frac{K}{\dot{a}^2}\right).\]
With this we obtain
\begin{equation}
\Phi(t,\mathbf{k})=C(\mathbf{k})\left[1-\frac{H}{a}\int_0^t
a\left(1-\frac{K}{\dot{a}^2}\right)dt\right]+\frac{H}{a}d(\mathbf{k}).
\end{equation}

Let us work this out for a mixture of dust ($p=0$) and radiation
($p=\frac{1}{3}\rho$). We use the `normalized' scale factor
$\zeta:=a/a_{eq}$, where $a_{eq}$ is the value of $a$ when the
energy densities of dust (CDM) and radiation are equal. Then (see
Sect. 0.1.3)
\begin{equation}
\rho=\frac{1}{2}\zeta^{-4}+\frac{1}{2}\zeta^{-3},~~ p=\frac{1}{6}\zeta^{-4}.
\end{equation}
Note that
\begin{equation}
\zeta'=kx\zeta,~~~x:=\frac{Ha}{k}.
\end{equation}

From now on we assume $K=0,~\Lambda=0$. Then the Friedmann equation gives
\begin{equation}
H^2=H^2_{eq}\frac{\zeta+1}{2}\zeta^{-4},
\end{equation}
thus
\begin{equation}
x^2=\frac{\zeta+1}{2\zeta^2}\frac{1}{\omega^2},~~~\omega:=\frac{1}{x_{eq}}=\left(\frac{Ha}{k}\right)_{eq}.
\end{equation}
In (2.11) we need the integral
\[ \frac{H}{a}\int_0^t adt=Ha_{eq}\frac{1}{\zeta}\int_0^\eta\zeta^2d\eta
=\frac{\sqrt{\zeta+1}}{\zeta^3}\int_0^\zeta\frac{\zeta^2}{\sqrt{\zeta+1}}d\zeta.\] As a
result we get for the growing mode
\begin{equation}
\Phi(\zeta,\mathbf{k})=C(\mathbf{k})\left[1-\frac{\sqrt{\zeta+1}}
{\zeta^3}\int_0^\zeta\frac{\zeta^2}{\sqrt{\zeta+1}}d\zeta\right].
\end{equation}
From (2.3) and the definition of $x$ we obtain
\begin{equation}
\Phi=\frac{3}{2}x^2\Delta,
\end{equation}
hence with (2.15)
\begin{equation}
\Delta=\frac{4}{3}\omega^2C(\mathbf{k})\frac{\zeta^2}{\zeta+1}\left[1-\frac{\sqrt{\zeta+1}}
{\zeta^3}\int_0^\zeta\frac{\zeta^2}{\sqrt{\zeta+1}}d\zeta\right].
\end{equation}
The integral is elementary. One finds that $\Delta$ is proportional to
\begin{equation}
U_g=\frac{1}{\zeta(\zeta+1)}\left[\zeta^3+\frac{2}{9}\zeta^2-\frac{8}{9}\zeta-\frac{16}{9}+\frac{16}{9}
\sqrt{\zeta+1}\right].
\end{equation}
This is a well-known result.

The decaying mode corresponds to the second term in (2.11), and is thus proportional to
\begin{equation}
U_d=\frac{1}{\zeta\sqrt{\zeta+1}}.
\end{equation}

Limiting approximations of (2.19) are
\begin{equation}
U_g = \left\{\begin{array}{r@{\quad:\quad}l} \frac{10}{9}\zeta^2 & \zeta\ll 1\\ \zeta &
\zeta\gg 1 .\end{array} \right.
\end{equation}
For the potential $\Phi\propto x^2\Delta$ the growing mode is given by
\begin{equation}
\Phi(\zeta)=\Phi(0)\frac{9}{10}\frac{\zeta+1}{\zeta^2}U_g.
\end{equation}
Thus
\begin{equation}
\Phi(\zeta)=\Phi(0)\left\{\begin{array}{r@{\quad:\quad}l} 1 & \zeta\ll 1\\
\frac{9}{10} & \zeta\gg 1 .\end{array} \right.
\end{equation}
In particular, $\Phi$ stays \textit{constant both in the radiation
and in the matter dominated eras}. Recall that this holds only for
$c_s(k/aH)\ll 1$. We shall later study eq. (2.9) for arbitrary
scales.

\section{Solution of (2.6) for dust}

Using the Poisson equation (2.3) we can write (2.9) in terms of $\Delta$
\begin{equation}
\frac{1+w}{a^2H}\left[\frac{H^2}{a(\rho+p)}\left(\frac{a^3\rho}{H}\Delta\right)^{\cdot}\right]^{\cdot}
+c_s^2\frac{k^2}{a^2}\Delta=0.
\end{equation}
For dust this reduces to (using $\rho a^3=const$)
\begin{equation}
\left[a^2H^2\left(\frac{\Delta}{H}\right)^{\cdot}\right]^{\cdot}=0.
\end{equation}
The general solution of this equation is
\begin{equation}
\Delta(t,\mathbf{k})=C(\mathbf{k})H(t)\int_0^t\frac{dt'}{a^2(t')H^2(t')} +d(k)H(t).
\end{equation}
This result can also be obtained in Newtonian perturbation theory. The first term gives the
growing mode and the second the decaying one.

Let us rewrite (2.35) in terms of the redshift $z$. From $1+z=a_0/a$ we get $dz=-(1+z)Hdt$,
so by (0.52)
\begin{equation}
\frac{dt}{dz}=-\frac{1}{H_0(1+z)E(z)},~~~H(z)=H_0 E(z).
\end{equation}
Therefore, the growing mode $D_g(z)$ can be written in the form
\begin{equation}
D_g(z)=\frac{5}{2}\Omega_ME(z)\int_z^\infty\frac{1+z'}{E^3(z')}dz'.
\end{equation}
Here the normalization is chosen such that $D_g(z)=(1+z)^{-1}=a/a_0$ for
$\Omega_M=1,~\Omega_\Lambda=0$. This growth function is plotted in Fig. 7.12 of
\cite{Cos4}.

\section{A simple relativistic example}

As an additional illustration we now solve (2.7) for a single perfect fluid with
$w=const,~K=\Lambda=0$. For a flat universe the background equations are then
\[\rho'+3\frac{a'}{a}(1+w)\rho=0,~~~\left(\frac{a'}{a}\right)^2=\frac{8\pi G}{3}a^2\rho.\]
Inserting the ansatz
\[\rho a^2=A\eta^{-\nu},~~~a=a_0(\eta/\eta_0)^\beta \]
we get
\[ \frac{\beta}{\eta^2}=\frac{8\pi G}{3}A\eta^{-\nu}~~~\Rightarrow \nu=2,~A=\frac{3}{8\pi
G}\beta^2.\] The energy equation then gives $\beta=2/(1+3w)~(=1$ if radiation dominates).
Let $x:=k\eta$ and
\[f:=x^{\beta-2}\Delta\propto\rho a^3\Delta.\] Also note that $k/(aH)=x/\beta$. With all this
we obtain from (2.7) for $f$
\begin{equation}
\left[\frac{d^2}{dx^2}+\frac{2}{x}\frac{d}{dx}+c_s^2-\frac{\beta(\beta+1)}{x^2}\right]f=0.
\end{equation}
The solutions are given in terms of Bessel functions:
\begin{equation}
f(x)=C_0j_\beta(c_s x)+ D_0n_\beta(c_sx).
\end{equation}

This implies acoustic oscillations for $c_sx\gg 1$, i.e., for $\beta(k/aH)\gg1$ (subhorizon
scales). In particular, if the radiation dominates ($\beta=1$)
\begin{equation}
\Delta\propto x[C_0j_1(c_sx)+D_0n_1(c_sx)],
\end{equation}
and the growing mode is soon proportional to $x\cos(c_sx)$, while the term going with $\sin
(c_sx)$ dies out.

On the other hand, on superhorizon scales ($c_sx\ll 1$) one obtains
\[ f\simeq Cx^\beta+Dx^{-(\beta+1)},\]
and thus
\begin{eqnarray}
\Delta &\simeq & Cx^2+Dx^{-(2\beta-1)},\nonumber \\
\Phi   &\simeq & \frac{3}{2}\beta^2(C+Dx^{-(2\beta+1)}, \nonumber \\
V      &\simeq & \frac{3}{2}\beta\left(-\frac{1}{\beta+1}Cx+Dx^{-2\beta}\right).
\end{eqnarray}
We see that the growing mode behaves as $\Delta\propto a^2$ in the radiation dominated
phase and $\Delta\propto a$ in the matter dominated era.

The characteristic Jeans wave number is obtained when the square bracket in (2.7) vanishes.
This gives
\begin{equation}
\lambda_J=\left(\frac{\pi c_s^2}{Gh}\right)^{1/2},~~~h=\rho+p.
\end{equation}

-----------------

\textbf{Exercise}. Derive the exact expression for $V$.

-----------------

In Part III we shall study more complicated coupled fluid models that are important for the
evolution of perturbations before recombination. In the next part the general theory will
be applied in attempts to understand the generation of primordial perturbations from
original quantum fluctuations.

\part{Inflation and Generation of Fluctuations}

\chapter{Inflationary Scenario}

\section{Introduction}

The horizon and flatness problems of standard big bang cosmology are so serious that the
proposal of a very early accelerated expansion, preceding the hot era dominated by
relativistic fluids, appears quite plausible. This general qualitative aspect of
`inflation' is now widely accepted. However, when it comes to concrete model building the
situation is not satisfactory. Since we do not know the fundamental physics at superhigh
energies not too far from the Planck scale, models of inflation are usually of a
phenomenological nature. Most models consist of a number of scalar fields, including a
suitable form for their potential. Usually there is no direct link to fundamental theories,
like supergravity, however, there have been many attempts in this direction. For the time
being, inflationary cosmology should be regarded as an attractive scenario, and not yet as
a theory.

The most important aspect of inflationary cosmology is that \textit{the generation of
perturbations on large scales from initial quantum fluctuations is unavoidable and
predictable}. For a given model these fluctuations can be calculated accurately, because
they are tiny and cosmological perturbation theory can be applied. And, most importantly,
these predictions can be \textit{confronted with the cosmic microwave anisotropy
measurements}. We are in the fortunate position to witness rapid progress in this field.
The results from various experiments, most recently from WMAP, give already strong support
of the basic predictions of inflation. Further experimental progress can be expected in the
coming years.

In what follows I shall mainly concentrate on this aspect. It is, I think, important to
understand in sufficient detail how the involved calculations are done, and which aspects
are the most generic ones for inflationary models. We shall learn a lot in the coming
years, thanks to the confrontation of the theory with precise observations.

\section{The horizon problem and the general idea of inflation}

I begin by describing the famous horizon puzzle (topic belonging to Chap. 0), which is a
very serious causality problem of standard big bang cosmology.

\subsubsection*{Past and future light cone distances}

Consider our past light cone for a Friedmann spacetime model (Fig. 3.1). For a radial light
ray the differential relation $dt=a(t)dr/(1-kr^2)^{1/2}$ holds for the coordinates $(t,r)$
of the metric (0.40). The proper radius of the past light sphere at time $t$ (cross section
of the light cone with the hypersurface $\{t=const\}$) is
\begin{equation}
l_p(t)=a(t)\int_0^{r(t)}\frac{dr}{\sqrt{1-kr^2}},
\end{equation}
where the coordinate radius is determined by
\begin{equation}
\int_0^{r(t)}\frac{dr}{\sqrt{1-kr^2}}=\int_t^{t_0}\frac{dt'}{a(t')}.
\end{equation}
Hence,
\begin{equation}
l_p(t)=a(t)\int_t^{t_0}\frac{dt'}{a(t')}.
\end{equation}
We rewrite this in terms of the redhift variable, using (2.36),
\begin{equation}
l_p(z)=\frac{1}{H_0(1+z)}\int_0^z\frac{dz'}{E(z')}.
\end{equation}

\begin{figure}
\begin{center}
\includegraphics[height=0.3\textheight]{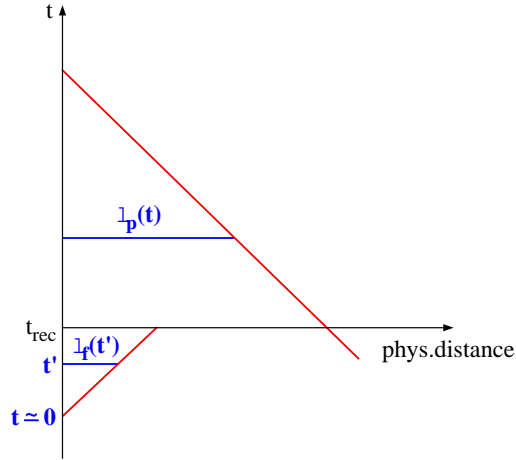}
\caption{Spacetime diagram illustrating the horizon problem.} \label{Fig-6}
\end{center}
\end{figure}

Similarly, the extension $l_f(t)$ of the forward light cone at time $t$ of a very early
event ($t\simeq0,~z\simeq\infty$) is
\begin{equation}
l_f(t)=a(t)\int_0^{t}\frac{dt'}{a(t')}=\frac{1}{H_0(1+z)}\int_z^\infty\frac{dz'}{E(z')}.
\end{equation}
For the present Universe ($t_0$) this becomes what is called the \textit{particle horizon
distance}
\begin{equation}
D_{hor}=H_0^{-1}\int_0^\infty\frac{dz'}{E(z')},
\end{equation}
and gives the size of the \textit{observable Universe}.

Analytical expressions for these distances are only available in special cases. For
orientation we consider first the Einstein-de Sitter model
($K=0,~\Omega_\Lambda=0,~\Omega_M=1$), for which $a(t)=a_0(t/t_0)^{2/3}$ and thus
\begin{equation}
D_{hor}=3t_0=2H_0^{-1},~~ l_f(t)=3t,~~
\frac{l_p}{l_f}=\left(\frac{t_0}{t}\right)^{1/3}-1=\sqrt{1+z}-1.
\end{equation}
For a flat Universe a good fitting formula for cases of interest is (Hu and White)
\begin{equation}
D_{hor}\simeq 2H_0^{-1}\frac{1+0.084\ln\Omega_M}{\sqrt{\Omega_M}}.
\end{equation}

It is often convenient to work with `comoving distances', by rescaling distances referring
to time $t$ (like $l_p(t),l_f(t)$) with the factor $a(t_0)/a(t)=1+z$ to the present. We
indicate this by the superscript $c$. For instance,
\begin{equation}
l_p^c(z)=\frac{1}{H_0}\int_0^z\frac{dz'}{E(z')}.
\end{equation}
This distance is plotted in Fig. 3 of Chap. 0 as $D_{com}(z)$. Note
that for $a_0=1:~l_f^c(\eta)=\eta,~l_p^c(\eta)=\eta_0-\eta$. Hence
(3.5) gives the following relation between $\eta$ and $z$:
\[ \eta=\frac{1}{H_0}\int_z^\infty\frac{dz'}{E(z')}.\]

\subsubsection*{The number of causality distances on the cosmic photosphere}

The number of causality distances at redshift $z$ between two antipodal emission points is
equal to $l_p(z)/l_f(z)$, and thus the ratio of the two integrals on the right of (3.4) and
(3.5). We are particularly interested in this ratio at the time of last scattering with
$z_{rec}\simeq1100$. Then we can use for the numerator a flat Universe with
non-relativistic matter, while for the denominator we can neglect in the standard hot big
bang model $\Omega_K$ and $\Omega_\Lambda$. A reasonable estimate is already obtained by
using the simple expression in (3.7), i.e., $z_{rec}^{1/2}\approx 30$. A more accurate
evaluation would increase this number to about 40. The length $l_f(z_{rec})$ subtends an
angle of about 1 degree (exercise). How can it be that there is such a large number of
causally disconnected regions we see on the microwave sky all having the same temperature?
This is what is meant by the \textit{horizon problem} and was a troublesome mystery before
the invention of inflation.

\subsubsection*{Vacuum-like energy and exponential expansion}

This causality problem is potentially avoided, if $l_f(t)$ would be increased in the very
early Universe as a result of different physics. If a vacuum-like energy density would
dominate, the Universe would undergo an \textit{exponential expansion}. Indeed, in this
case the Friedmann equation is
\begin{equation}
\left(\frac{\dot{a}}{a}\right)^2+ \frac{k}{a^2}=\frac{8\pi
G}{3}\rho_{vac},~~~\rho_{vac}\simeq const,
\end{equation}
and has the solutions
\begin{equation}
a(t) \propto \left\{\begin{array}{r@{\quad:\quad}l} \cosh~H_{vac}t & k=1\\ e^{H_{vac}t} & k=0\\
\sinh~H_{vac}t & k=1 ,\end{array} \right.
\end{equation}
with
\begin{equation}
H_{vac}=\sqrt{\frac{8\pi G}{3}\rho_{vac}}~.
\end{equation}

Assume that such an exponential expansion starts for some reason at time $t_i$ and ends at
the \textit{reheating time} $t_e$, after which standard expansion takes over. From
\begin{equation}
a(t)=a(t_i)e^{H_{vac}(t-t_i)}~~~(t_i<t<t_e),
\end{equation}
for $k=0$ we get
\[ l_f^c(t_e)\simeq a_0\int_{t_i}^{t_e}\frac{dt}{a(t)}=\frac{a_0}{H_{vac}a(t_i)}\left(1-e^{-H_{vac}\Delta
t}\right)\simeq \frac{a_0}{H_{vac}a(t_i)} ,\] where $\Delta t:=t_e-t_i$. We want to satisfy
the condition $l_f^c(t_e)\gg l_p^c(t_e)\simeq H_0^{-1}$ (see (3.8), i.e.,
\begin{equation}
a_iH_{vac}\ll a_0H_0~~~\Leftrightarrow \frac{a_i}{a_e}\ll\frac{a_0H_0}{a_eH_{vac}}
\end{equation}
or
\[e^{H_{vac}\Delta
t}\gg\frac{a_eH_{vac}}{a_0H_0}=\frac{H_{eq}a_{eq}}{H_0a_0}\frac{H_{vac}a_e}{H_{eq}a_{eq}}.\]
Here, $eq$ indicates the values at the time $t_{eq}$ when the energy densities of
non-relativistic and relativistic matter were equal. We now use the Friedmann equation for
$k=0$ and $w:=p/\rho=const.$ From (0.46) it follows that in this case
\[ Ha\propto a^{-(1+3w)/2},\]
and hence we arrive at
\begin{equation}
e^{H_{vac}\Delta
t}\gg\left(\frac{a_0}{a_{eq}}\right)^{1/2}\left(\frac{a_{eq}}{a_{e}}\right)=
(1+z_{eq})^{1/2}\left(\frac{T_e}{T_{eq}}\right)=(1+z_{eq})^{-1/2}\frac{T_{Pl}}{T_0}\frac{T_e}{T_{Pl}},
\end{equation}
where we used $aT=const.$ So the number of e-folding periods during the inflationary
period, $\mathcal{N}=H_{vac}\Delta t$, should satisfy
\begin{equation}
\mathcal{N}\gg\ln\left(\frac{T_{Pl}}{T_0}\right)-\frac{1}{2}\ln
z_{eq}+\ln\left(\frac{T_e}{T_{Pl}}\right)\simeq 70+\ln\left(\frac{T_e}{T_{Pl}}\right).
\end{equation}
For a typical GUT scale, $T_e\sim10^{14}~GeV$, we arrive at the condition
$\mathcal{N}\gg60$.

Such an exponential expansion would also solve the \textit{flatness problem}, another worry
of standard big bang cosmology. Let me recall how this problem arises.

The Friedmann equation (0.17) can be written as
\[ (\Omega^{-1}-1)\rho a^2=-\frac{3k}{8\pi G}=const.,\]
where
\begin{equation}
\Omega(t):=\frac{\rho(t)}{3H^2/8\pi G}
\end{equation}
($\rho$ includes vacuum energy contributions). Thus
\begin{equation}
\Omega^{-1}-1=(\Omega_0^{-1}-1)\frac{\rho_0a_0^2}{\rho a^2}.
\end{equation}
Without inflation we have
\begin{eqnarray}
\rho &=& \rho_{eq}\left(\frac{a_{eq}}{a}\right)^4~~~(z>z_{eq}), \\
\rho &=& \rho_0\left(\frac{a_0}{a}\right)^3~~~~~(z<z_{eq}).
\end{eqnarray}
According to (0.47) $z_{eq}$ is given by
\begin{equation}
1+z_{eq}=\frac{\Omega_M}{\Omega_R}\simeq 10^4~\Omega_0h_0^2.
\end{equation}

-----------------

\textbf{Exercise}: Derive the estimate on the right of (3.21).

-----------------

For $z>z_{eq}$ we obtain from (3.18) and (3.19)
\begin{equation}
\Omega^{-1}-1=(\Omega_0^{-1}-1)\frac{\rho_0a_0^2}{\rho_{eq}
a_{eq}^2}\frac{\rho_{eq}a_{eq}^2}{\rho
a^2}=(\Omega_0^{-1}-1)(1+z_{eq})^{-1}\left(\frac{a}{a_{eq}}\right)^2
\end{equation}
or
\begin{equation}
\Omega^{-1}-1=(\Omega_0^{-1}-1)(1+z_{eq})^{-1}\left(\frac{T_{eq}}{T}\right)^2\simeq
10^{-60}(\Omega_0^{-1}-1)\left(\frac{T_{Pl}}{T}\right)^2.
\end{equation}

Let us apply this equation for $T=1MeV,~\Omega_0\simeq 0.2-0.3$. Then $\mid\Omega-1\mid\leq
10^{-15}$, thus the Universe was already incredibly flat at modest temperatures, not much
higher than at the time of nucleosynthesis.

Such a fine tuning must have a physical reason. This is naturally provided by inflation,
because our observable Universe could originate from a small patch at $t_e$. (A tiny part
of the Earth surface is also practically flat.)

Beside the horizon scale $l_f(t)$, the \textit{Hubble length} $H^{-1}(t)=a(t)/\dot{a}(t)$
plays also an important role. One might call this the ``microphysics horizon'', because
this is the maximal distance microphysics can operate coherently in one expansion time. It
is this length scale which enters in basic evolution equations, such as the equation of
motion for a scalar field (see eq. (3.30) below).

We sketch in Figs. 3.2 -- 3.4  the various length scales in
inflationary models, that is for models with a period of accelerated
(e.g., exponential) expansion. From these it is obvious that there
can be -- at least in principle -- a \textit{causal generation
mechanism for perturbations}. This topic will be discussed in great
detail in later parts of these lectures.

Exponential inflation is just an example. What we really need is an early phase during
which the \textit{comoving Hubble length decreases} (Fig. 3.4). This means that (for
Friedmann spacetimes)
\begin{equation}
\fbox{$\displaystyle\left( H^{-1}(t)/a\right)^{\cdot}<0.$}
\end{equation}
This is the \textit{general definition of inflation}; equivalently, $\ddot{a}>0$
(accelerated expansion). For a Friedmann model eq. (0.23) tells us that
\begin{equation}
\ddot{a}>0 \Leftrightarrow p<-\rho/3.
\end{equation}
This is, of course, not satisfied for `ordinary' fluids.

Assume, as another example, \textit{power-law inflation}: $a\propto t^p$. Then $\ddot{a}>0
\Leftrightarrow p>1$.

\begin{figure}
\begin{center}
\includegraphics[height=0.3\textheight]{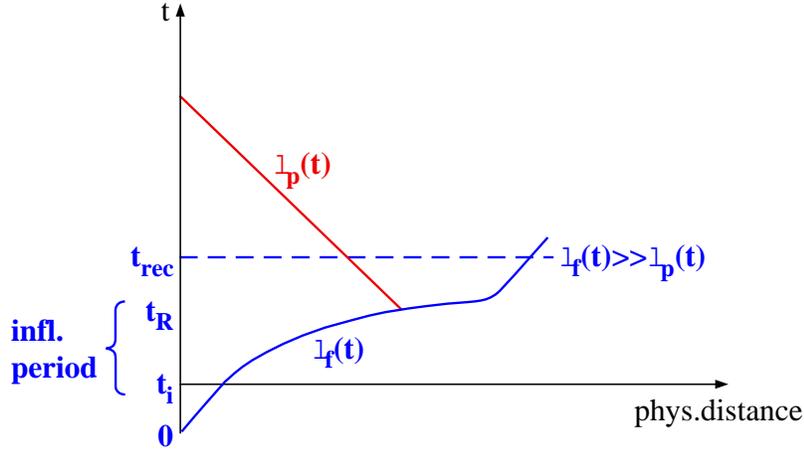}
\caption{Past and future light cones in models with an inflationary period.} \label{Fig-7}
\end{center}
\end{figure}

\begin{figure}
\begin{center}
\includegraphics[height=0.3\textheight]{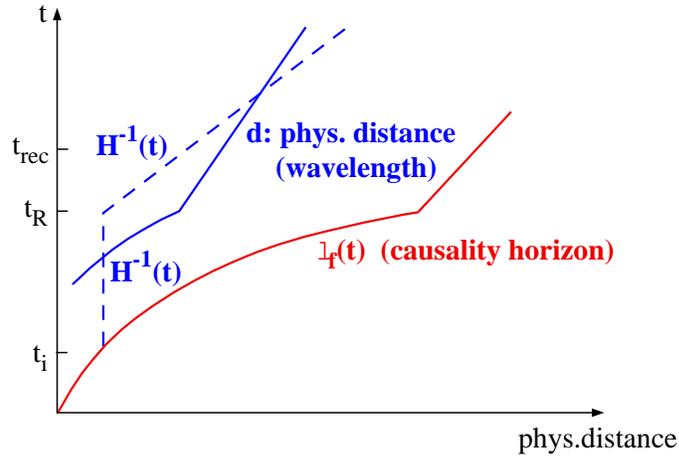}
\caption{Physical distance (e.g. between clusters of galaxies) and Hubble distance,
 and causality horizon in inflationary models.} \label{Fig-8}
\end{center}
\end{figure}

\begin{figure}
\begin{center}
\includegraphics[height=0.2\textheight]{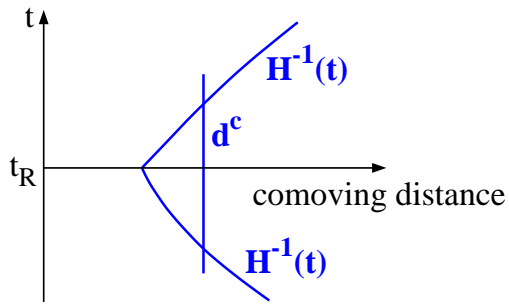}
\caption{Part of Fig. 3.3 expressed in terms of comoving distances.}
\label{Fig-9}
\end{center}
\end{figure}

\section{Scalar field models}

Models with $p<-\rho/3$ are naturally obtained in scalar field theories. Most of the time
we shall consider the simplest case of \textit{one} neutral scalar field $\varphi$
minimally coupled to gravity. Thus the Lagrangian density is assumed to be
\begin{equation}
\mathcal{L}=\frac{M_{pl}^2}{16\pi}R[g]-\frac{1}{2}\nabla_\mu\varphi\nabla^\mu\varphi-V(\varphi),
\end{equation}
where $R[g]$ is the Ricci scalar for the metric $g$. The scalar field equation is
\begin{equation}
\Box\varphi=V_{,\varphi},
\end{equation}
and the energy-momentum tensor in the Einstein equation
\begin{equation}
G_{\mu\nu}=\frac{8\pi}{M_{Pl}^2}T_{\mu\nu}
\end{equation}
is
\begin{equation}
T_{\mu\nu}=\nabla_\mu\varphi\nabla_\nu\varphi+g_{\mu\nu}\mathcal{L}_\varphi
\end{equation}
($\mathcal{L}_\varphi$ is the scalar field part of (3.26)).

We consider first Friedmann spacetimes. Using previous notation, we obtain from (0.1)
\[\sqrt{-g}=a^3\sqrt{\gamma},~~
\Box\varphi=\frac{1}{\sqrt{-g}}\partial_\mu(\sqrt{-g}g^{\mu\nu}\partial_\nu\varphi)=-
\frac{1}{a^3}(a^3\dot{\varphi})^{\cdot}+ \frac{1}{a^2}\triangle_\gamma\varphi. \] The field
equation (3.27) becomes
\begin{equation}
\fbox{$\displaystyle\ddot{\varphi}+3H\dot{\varphi}-\frac{1}{a^2}\triangle_\gamma\varphi=-V_{,\varphi}(\varphi).$}
\end{equation}
Note that the expansion of the Universe induces a `friction' term. In this basic equation
one also sees the appearance of the Hubble length. From (3.29) we obtain for the energy
density and the pressure of the scalar field
\begin{eqnarray}
\rho_\varphi &=& T_{00}=\frac{1}{2}\dot{\varphi}^2+V+\frac{1}{2a^2}(\nabla\varphi)^2,\\
p_\varphi &=&
\frac{1}{3}T^i{}_i=\frac{1}{2}\dot{\varphi}^2-V-\frac{1}{6a^2}(\nabla\varphi)^2.
\end{eqnarray}
(Here, $(\nabla\varphi)^2$ denotes the squared gradient on the 3-space $(\Sigma,\gamma)$.)

Suppose the gradient terms can be neglected, and that $\varphi$ is during a certain phase
slowly varying in time, then we get
\begin{equation}
\rho_\varphi\approx V,~~~p_\varphi\approx -V.
\end{equation}
Thus $p_\varphi\approx-\rho_\varphi$, as for a cosmological term.

Let us ignore for the time being the spatial inhomogeneities in the previous equations.
Then these reduce to
\begin{equation}
\ddot{\varphi}+3H\dot{\varphi}+V_{,\varphi}(\varphi)=0;
\end{equation}
\begin{equation}
\rho_\varphi=\frac{1}{2}\dot{\varphi}^2+V,~~~p_\varphi=\frac{1}{2}\dot{\varphi}^2-V.
\end{equation}

Beside (3.34) the other dynamical equation is the Friedmann equation
\begin{equation}
\fbox{$\displaystyle
H^2+\frac{K}{a^2}=\frac{8\pi}{3M_{Pl}^2}\left[\frac{1}{2}\dot{\varphi}^2+V(\varphi)\right].$}
\end{equation}
Eqs. (3.34) and (3.36) define a nonlinear dynamical system for the dynamical variables
$a(t),\varphi(t)$, which can be studied in detail (see, e.g., \cite{Bel}).

Let us ignore the curvature term $K/a^2$ in (3.36). Differentiating this equation and using
(3.34) shows that
\begin{equation}
\dot{H}=-\frac{4\pi}{M_{Pl}^2}\dot{\varphi}^2.
\end{equation}
Regard $H$ as a function of $\varphi$, then
\begin{equation}
\frac{dH}{d\varphi}=-\frac{4\pi}{M_{Pl}^2}\dot{\varphi}.
\end{equation}
This allows us to write the Friedmann equation as
\begin{equation}
\left(\frac{dH}{d\varphi}\right)^2-\frac{12\pi}{M_{Pl}^2}H^2(\varphi)=-\frac{32\pi^2}{M_{Pl}^4}V(\varphi).
\end{equation}
For a given potential $V(\varphi)$ this is a differential equation for $H(\varphi)$. Once
this function is known, we obtain $\varphi(t)$ from (3.38) and $a(t)$ from (3.37).

\subsection{Power-law inflation}

We now proceed in the reverse order, assuming that $a(t)$ follows a power law
\begin{equation}
a(t)=const.~t^p.
\end{equation}
Then $H=p/t$, so by (3.37)
\[
\dot{\varphi}=\sqrt{\frac{p}{4\pi}}M_{Pl}\frac{1}{t},~~\varphi(t)=\sqrt{\frac{p}{4\pi}}M_{Pl}\ln(t)+const.,\]
hence
\begin{equation}
H\propto\exp\left(-\sqrt{\frac{4\pi}{p}}\frac{\varphi}{M_{Pl}}\right).
\end{equation}
Using this in (3.39) leads to an exponential potential
\begin{equation}
V(\varphi)=V_0\exp\left(-4\sqrt{\frac{\pi}{p}}\frac{\varphi}{M_{Pl}}\right).
\end{equation}

\subsection{Slow-roll approximation}

An important class of solutions is obtained in the slow-roll approximation (SLA), in which
the basic eqs. (3.34) and (3.36) can be replaced by
\begin{eqnarray}
H^2 &=& \frac{8\pi}{3M_{Pl}^2}V(\varphi), \\
3H\dot{\varphi} &=& -V_{,\varphi}.
\end{eqnarray}
A necessary condition for their validity is that the \textit{slow-roll parameters}
\begin{eqnarray}
\varepsilon_V(\varphi): &=& \frac{M_{Pl}^2}{16\pi}\left(\frac{V_{,\varphi}}{V}\right)^2,\\
\eta_V(\varphi): &=& \frac{M_{Pl}^2}{8\pi}\frac{V_{,\varphi\varphi}}{V}
\end{eqnarray}
are small:
\begin{equation}
\varepsilon_V\ll 1,~~~\mid\eta_V\mid\ll1.
\end{equation}
These conditions, which guarantee that the potential is flat, are, however, not sufficient
(for details, see Sect. 5.1.2).

The simplified system (3.43) and (3.44) implies
\begin{equation}
\fbox{$\displaystyle
\dot{\varphi}^2=\frac{M_{Pl}^2}{24\pi}\frac{1}{V}\left(V_{,\varphi}\right)^2.$}
\end{equation}
This is a differential equation for $\varphi(t)$.

Let us  consider potentials of the form
\begin{equation}
V(\varphi)=\frac{\lambda}{n}\varphi^n.
\end{equation}
Then eq. (3.48) becomes
\begin{equation}
\fbox{$\displaystyle \dot{\varphi}^2=\frac{n^2M_{Pl}^2}{24\pi}\frac{1}{\varphi^2}V.$}
\end{equation}
Hence, (3.43) implies
\[\frac{\dot{a}}{a}=-\frac{4\pi}{nM_{Pl}^2}(\varphi^2)^{\cdot},\]
and so
\begin{equation}
\fbox{$\displaystyle
a(t)=a_0\exp\left[\frac{4\pi}{nM_{Pl}^2}(\varphi_0^2-\varphi^2(t))\right].$}
\end{equation}

We see from (3.50) that $\frac{1}{2}\dot{\varphi}^2\ll V(\varphi)$ for
\begin{equation}
\varphi\gg\frac{n}{4\sqrt{3\pi}}M_{Pl}.
\end{equation}

Consider first the example $n=4$. Then (3.50) implies
\begin{equation}
\frac{\dot{\varphi}}{\varphi}=\sqrt{\frac{\lambda}{6\pi}}M_{Pl}~\Rightarrow
\varphi(t)=\varphi_0\exp\left(-\sqrt{\frac{\lambda}{6\pi}}M_{Pl}~t\right).
\end{equation}
For $n\neq4$:
\begin{equation}
\varphi(t)^{2-n/2}=\varphi_0^{2-n/2}+t\left(2-\frac{n}{2}\right)\sqrt{\frac{n\lambda}{24\pi}}M_{Pl}^{3-n/2}.
\end{equation}
For the special case $n=2$ this gives, using the notation $V=\frac{1}{2}m^2\varphi^2$, the
simple result
\begin{equation}
\varphi(t)=\varphi_0-\frac{mM_{Pl}}{2\sqrt{3\pi}}t.
\end{equation}
Inserting this into (3.51) provides the time dependence of $a(t)$.

\section{Why did inflation start?}

Attempts to answer this and related questions are \textit{very speculative} indeed. A
reasonable direction is to imagine random initial conditions and try to understand how
inflation can emerge, perhaps generically, from such a state of matter. A.Linde first
discussed a scenario along these lines which he called \textit{chaotic inflation}. In the
context of a single scalar field model he argued that typical initial conditions correspond
to $\frac{1}{2}\dot{\varphi}^2\sim\frac{1}{2}(\partial_i\varphi)^2\sim V(\varphi)\sim1$ (in
Planckian units). The chance that the potential energy dominates in some domain of size
$>\mathcal{O}(1)$ is presumably not very small. In this situation inflation could begin and
$V(\varphi)$ would rapidly become even more dominant, which ensures continuation of
inflation. Linde concluded from such considerations that chaotic inflation occurs under
rather natural initial conditions. For this to happen, the form of the potential
$V(\varphi)$ can even be a simple power law of the form (3.49). Many questions remain,
however, open.

The chaotic inflationary Universe will look on very large scales -- much larger than the
present Hubble radius -- extremely inhomogeneous. For a review of this scenario I refer to
\cite{Lin}. A much more extended discussion of inflationary models, including references,
can be found in \cite{Cos3}.

\chapter{Cosmological Perturbation Theory for Scalar Field Models}

To keep this Chapter independent of the previous one, let us begin
by repeating the set up of Sect. 3.3.

We consider a minimally coupled scalar field $\varphi$, with Lagrangian density
\begin{equation}
\mathcal{L}=-\frac{1}{2}g^{\mu\nu}\partial_\mu\varphi\partial_\nu\varphi-U(\varphi)
\end{equation}
and corresponding field equation
\begin{equation}
\Box\varphi=U_{,\varphi}.
\end{equation}
As a result of this the energy-momentum tensor
\begin{equation}
T^\mu{}_\nu=\partial^\mu\varphi\partial_\nu\varphi-\delta^\mu{}_\nu\left(\frac{1}{2}
\partial^\lambda\varphi\partial_\lambda\varphi+U(\varphi)\right)
\end{equation}
is covariantly conserved. In the general multi-component formalism
(Sect. 1.4) we have, therefore, $Q_\varphi=0$.

The unperturbed quantities $\rho_\varphi$, etc, are
\begin{eqnarray}
\rho_\varphi &=& -T^0{}_0=\frac{1}{2a^2}(\varphi')^2+U(\varphi), \\
p_\varphi &=& \frac{1}{3}T^i{}_i=\frac{1}{2a^2}(\varphi')^2-U(\varphi), \\
h_\varphi &=& \rho_\varphi+p_\varphi=\frac{1}{a^2}(\varphi')^2.
\end{eqnarray}
Furthermore,
\begin{equation}
\rho'_\varphi=-3\frac{a'}{a}h_\varphi.
\end{equation}
It is not very sensible to introduce a ``velocity of sound'' $c_\varphi$.

\section{Basic perturbation equations}

Now we consider small deviations from the ideal Friedmann behavior:
\begin{equation}
\varphi\rightarrow \varphi_0+\delta\varphi,~~\rho_\varphi\rightarrow
\rho_\varphi+\delta\rho,~~ etc.
\end{equation}
(The index $0$ is only used for the unperturbed field $\varphi$.) Since
$L_\xi\varphi_0=\xi^0\varphi'_0$ the gauge transformation of $\delta\varphi$ is
\begin{equation}
\delta\varphi\rightarrow\delta\varphi+\xi^0\varphi'_0.
\end{equation}
Therefore, \begin{equation}
\delta\varphi_\chi=\delta\varphi-\frac{1}{a}\varphi'_0\chi=\delta\varphi-\varphi'_0(B+E')
\end{equation}
is gauge invariant (see (1.21)). Further perturbations are
\begin{eqnarray}
\delta T^0{}_0 &=&
-\frac{1}{a^2}\left[-\varphi_0^{'2} A+\varphi'_0\delta\varphi'+U_{,\varphi}a^2\delta\varphi\right],\\
\delta T^0{}_i &=& -\frac{1}{a^2}\varphi'_0\delta\varphi_{,i},\\
\delta T^i{}_j &=& -\frac{1}{a^2}[\varphi_0^{'2}A-\varphi'_0\delta\varphi'
+U_{,\varphi}a^2\delta\varphi ] \delta^i{}_j.
\end{eqnarray}
This gives (recall (1.43))
\begin{eqnarray}
\delta\rho &=& \frac{1}{a^2}[-\varphi_0^{'2}A+\varphi'_0\delta\varphi'+a^2U_{,\varphi}\delta\varphi],\\
\delta p &=& p\pi_L
=\frac{1}{a^2}[\varphi'_0\delta\varphi'-\varphi_0^{'2}A-a^2U_{,\varphi}\delta\varphi],\\
\Pi &=& 0,~~~\mathcal{Q}=-\dot{\varphi}_0\delta\varphi.
\end{eqnarray}

\subsection*{Einstein equations}

We insert these expressions into the general perturbation equations
(1.91)- (1.98) and obtain
\begin{equation}
\kappa=3(HA-\dot{D})-\frac{1}{a^2}\triangle\chi,
\end{equation}
\begin{equation}
\frac{1}{a^2}(\triangle+3K)D+H\kappa=-4\pi
G[\dot{\varphi}_0\delta\dot{\varphi}-\dot{\varphi}_0^2A+U_{,\varphi}\delta\varphi],
\end{equation}
\begin{equation}
\kappa+\frac{1}{a^2}(\triangle+3K)\chi=12\pi G\dot{\varphi}_0\delta\varphi,
\end{equation}
\begin{equation}
A+D=\dot{\chi}+H\chi
\end{equation}
Equation (1.95) is in the present notation
\[ \dot{\kappa}+2H\kappa=-\left(\frac{1}{a^2}\triangle+3\dot{H}\right)A+4\pi G[\delta\rho+3\delta p] ,\] with
\[ \delta\rho+3\delta
p=2(-2\dot{\varphi}_0^2A+2\dot{\varphi}_0\delta\dot{\varphi}-U_{,\varphi}\delta\varphi).\]
If we also use (recall (1.80))
\[ \dot{H}=-4\pi G\dot{\varphi}_0^2+\frac{K}{a^2}\]
we obtain
\begin{equation}
\dot{\kappa}+2H\kappa=-\left(\frac{\triangle+3K}{a^2}+4\pi G\dot{\varphi}_0^2\right)A+8\pi
G(2\dot{\varphi}_0\delta\dot{\varphi}-U_{,\varphi}\delta\varphi).
\end{equation}
The two remaining equations (1.97) and (1.98) are:
\begin{equation}
(\delta\rho)^{\cdot}+3H(\delta\rho+\delta p)= (\rho+p)(\kappa-3HA)-\frac{1}{a^2}\triangle
\mathcal{Q},
\end{equation}
and
\begin{equation}
\dot{\mathcal{Q}}+3H\mathcal{Q}=-(\rho+p)A-\delta p,
\end{equation}
with the expressions (4.14)-(4.16). Since these last two equations
express energy-momentum `conservation', they are not independent of
the others if we add the field equation for $\varphi$; we shall not
make use of them below.

Eqs. (4.17) -- (4.21) can immediately be written in a gauge
invariant form:
\begin{equation}
\kappa_\chi=3(HA_\chi-\dot{D}_\chi),
\end{equation}
\begin{equation}
\frac{1}{a^2}(\triangle+3K)D_\chi+H\kappa_\chi=-4\pi
G[\dot{\varphi}_0\delta\dot{\varphi}_\chi-\dot{\varphi}_0^2A_\chi+U_{,\varphi}\delta\varphi_\chi],
\end{equation}
\begin{equation}
\kappa_\chi=12\pi G\dot{\varphi}_0\delta\varphi_\chi,
\end{equation}
\begin{equation}
A_\chi+D_\chi=0
\end{equation}
\begin{equation}
\dot{\kappa}_\chi+2H\kappa_\chi=-\left(\frac{\triangle+3K}{a^2}+4\pi
G\dot{\varphi}_0^2\right)A_\chi+8\pi
G(2\dot{\varphi}_0\delta\dot{\varphi}_\chi-U_{,\varphi}\delta\varphi_\chi).
\end{equation}

From now on we set $\mathbf{K=0}$. Use of (4.27) then gives us the following four basic
equations:
\begin{equation}
\kappa_\chi=3(\dot{A}_\chi+ HA_\chi),
\end{equation}
\begin{equation}
\frac{1}{a^2}\triangle A_\chi-H\kappa_\chi=4\pi
G[\dot{\varphi}_0\delta\dot{\varphi}_\chi-\dot{\varphi}_0^2A_\chi+U_{,\varphi}\delta\varphi_\chi],
\end{equation}
\begin{equation}
\kappa_\chi=12\pi G\dot{\varphi}_0\delta\varphi_\chi,
\end{equation}
\begin{equation}
\dot{\kappa}_\chi+2H\kappa_\chi=-\frac{1}{a^2}\triangle A_\chi-4\pi
G\dot{\varphi}_0^2A_\chi+8\pi
G(2\dot{\varphi}_0\delta\dot{\varphi}_\chi-U_{,\varphi}\delta\varphi_\chi).
\end{equation}
Recall also
\begin{equation}
4\pi G\dot{\varphi}_0^2=-\dot{H}.
\end{equation}

From (4.29) and (4.31) we get
\begin{equation}
\fbox{$\displaystyle \dot{A}_\chi+ HA_\chi=4\pi G\dot{\varphi}_0\delta\varphi_\chi.$}
\end{equation}
The difference of (4.32) and (4.30) gives (using also (4.29))
\[(\dot{A}_\chi+ HA_\chi)^{\cdot}+3H(\dot{A}_\chi+ HA_\chi)=4\pi G(\dot{\varphi}_0\delta\dot{\varphi}_\chi-
U_{,\varphi}\delta\varphi_\chi)\] i.e.,
\begin{equation}
\fbox{$\displaystyle\ddot{A}_\chi+4H\dot{A}_\chi+(\dot{H}+3H^2)A_\chi=4\pi
G(\dot{\varphi}_0\delta\dot{\varphi}_\chi- U_{,\varphi}\delta\varphi_\chi).$}
\end{equation}
Beside (4.34) and (4.35) we keep (4.30) in the form (making use of (4.33))
\begin{equation}
\fbox{$\displaystyle\frac{1}{a^2}\triangle A_\chi-3H\dot{A}_\chi-(\dot{H}+3H^2)A_\chi=4\pi
G(\dot{\varphi}_0\delta\dot{\varphi}_\chi+ U_{,\varphi}\delta\varphi_\chi).$}
\end{equation}

\subsection*{Scalar field equation}

We now turn to the $\varphi$ equation (4.2). Recall (the index 0 denotes in this subsection
the $t$-coordinate)
\begin{eqnarray*}
g_{00} &=& -(1+2A),~~ g_{0j}=-aB_{,j},~~ g_{ij}=a^2[\gamma_{ij}+2D\gamma_{ij}+2E_{\mid
ij}];\\
g^{00} &=& -(1-2A),~~ g^{0j}=-\frac{1}{a}B^{,j},~~
g^{ij}=\frac{1}{a^2}[\gamma^{ij}-2D\gamma^{ij}-2E^{\mid
ij}];\\
\sqrt{-g} &=& a^3\sqrt{\gamma}(1+A+3D+\triangle E.
\end{eqnarray*}
Up to first order we have (note that $\partial_j\varphi$ and $g^{0j}$ are of first order)
\[ \Box\varphi=\frac{1}{\sqrt{-g}}\partial_\mu(\sqrt{-g}g^{\mu\nu}\partial_\nu\varphi)=
\frac{1}{\sqrt{-g}}(\sqrt{-g}g^{00}\dot{\varphi})^{\cdot}+\frac{1}{a^2}\triangle\delta\varphi-\frac{1}{a}
\dot{\varphi}_0\triangle B.\] Using the zeroth order field equation (3.34), we readily find
\begin{eqnarray*}
\lefteqn{
\delta\ddot{\varphi}+3H\delta\dot{\varphi}+\left(-\frac{1}{a^2}\triangle+
U_{,\varphi\varphi}\right)\delta\varphi =} \\
& &(\dot{A}-3\dot{D}-\triangle\dot{E}+3HA-\frac{1}{a}\triangle B)\dot{\varphi}_0
-(3H\dot{\varphi}_0+2U_{,\varphi})A.
\end{eqnarray*}
Recalling the definition of $\kappa$,
\[\kappa=3(HA-\dot{D})-\frac{1}{a}\triangle(B+a\dot{E}),\]
we finally obtain the perturbed field equation in the form
\begin{equation}
\delta\ddot{\varphi}+3H\delta\dot{\varphi}+\left(-\frac{1}{a^2}\triangle+
U_{,\varphi\varphi}\right)\delta\varphi
=(\kappa+\dot{A})\dot{\varphi}_0-(3H\dot{\varphi}_0+2U_{,\varphi})A.
\end{equation}
By putting the index $\chi$ at all perturbation amplitudes one obtains a gauge invariant
equation. Using also (4.29) one arrives at
\begin{equation}
\fbox{$\displaystyle\delta\ddot{\varphi}_\chi+3H\delta\dot{\varphi}_\chi+\left(-\frac{1}{a^2}\triangle+
U_{,\varphi\varphi}\right)\delta\varphi_\chi
=4\dot{\varphi}_0\dot{A}_\chi-2U_{,\varphi}A_\chi.$}
\end{equation}

Our basic -- but not independent -- equations are (4.34), (4.35), (4.36) and (4.38).

\section{Consequences and reformulations}

In (1.58) we have introduced the curvature perturbation (recall also (4.16))
\begin{equation}
\mathcal{R}:=D_\mathcal{Q}=D_\chi-\frac{H}{\dot{\varphi}_0}\delta\varphi_\chi
=D-\frac{H}{\dot{\varphi}_0}\delta\varphi.
\end{equation}
It will turn out to be convenient to work also with
\begin{equation}
u=-z\mathcal{R},~~~z:=\frac{a\dot{\varphi}_0}{H},
\end{equation}
thus
\begin{equation}
u=a\left[\delta\varphi_\chi-\frac{\dot{\varphi}_0}{H}D_\chi\right]
=a\left[\delta\varphi-\frac{\dot{\varphi}_0}{H}D\right].
\end{equation}
This amplitude will play an important role, because we shall obtain from the previous
formulae the simple equation
\begin{equation}
\fbox{$\displaystyle u''-\triangle u-\frac{z''}{z}u=0.$}
\end{equation}
This is a Klein-Gordon equation with a time-dependent mass.

We next rewrite the basic equations in terms of the conformal time:
\begin{equation}
\triangle A_\chi-3\mathcal{H}A'_\chi-(\mathcal{H}'+3\mathcal{H}^2)A_\chi=4\pi
G(\varphi'_0\delta\varphi'_\chi+ U_{,\varphi}a^2\delta\varphi_\chi),
\end{equation}
\begin{equation}
A'_\chi+ \mathcal{H}A_\chi=4\pi G\varphi'_0\delta\varphi_\chi,
\end{equation}
\begin{equation}
A''_\chi+3\mathcal{H}A'_\chi+(\mathcal{H}'+2\mathcal{H}^2)A_\chi=4\pi
G(\varphi'_0\delta\varphi'_\chi- U_{,\varphi}a^2\delta\varphi_\chi),
\end{equation}
\begin{equation}
\delta\varphi''_\chi+2\mathcal{H}\delta\varphi'_\chi-\triangle\delta\varphi_\chi+
U_{,\varphi\varphi}a^2\delta\varphi_\chi =4\varphi'_0 A'_\chi-2U_{,\varphi}a^2A_\chi.
\end{equation}

Let us first express $u$ (or $\mathcal{R}$) in terms of $A_\chi$. From (4.40), (4.39) we
obtain in a first step
\[ 4\pi Gzu=4\pi Gz^2A_\chi+4\pi G\frac{z^2\mathcal{H}}{\varphi'_0}\delta\varphi_\chi.\]
For the first term on the right we use the unperturbed equation (see (4.33))
\begin{equation}
4\pi G\varphi^{'2}_0=\mathcal{H}^2-\mathcal{H}',
\end{equation}
and in the second term we make use of (4.44). Collecting terms gives
\begin{equation}
\fbox{$\displaystyle 4\pi Gzu=\left(\frac{a^2A_\chi}{\mathcal{H}}\right)'.$}
\end{equation}

Next, we derive an equation for $A_\chi$ alone. For this we subtract (4.43) from (4.45) and
use (4.44) to express $\delta\varphi_\chi$ in terms of $A_\chi$ and $A'_\chi$. Moreover we
make use of (4.47) and the unperturbed equation (3.34),
\begin{equation}
\varphi''_0+2\mathcal{H}\varphi'_0+U_{,\varphi}(\varphi_0)a^2=0.
\end{equation}

\paragraph{Detailed derivation:}

The quoted equations give
\begin{eqnarray*}
\lefteqn{A''_\chi+6\mathcal{H}A'_\chi -\triangle
A_\chi+2(\mathcal{H}'+2\mathcal{H}^2)A_\chi=}\\
& & -8\pi GU_{,\varphi}a^2\delta\varphi_\chi=
\frac{2}{\varphi'_0}(\varphi''_0+2\mathcal{H}\varphi'_0) (A'_\chi+ \mathcal{H}A_\chi),
\end{eqnarray*}
thus
\[ A''_\chi+2(\mathcal{H}-\varphi''_0/\varphi'_0)A'_\chi-\triangle A_\chi+2(\mathcal{H}'-\mathcal{H}
\varphi''_0/\varphi'_0)A_\chi=0.\] Rewriting the coefficients of $A_\chi,A'_\chi$ slightly,
we obtain the important equation:
\begin{equation}
\fbox{$\displaystyle A''_\chi+2\frac{(a/\varphi'_0)'}{a/\varphi'_0}A'_\chi-\triangle
A_\chi+2\varphi'_0(\mathcal{H}/\varphi'_0)'A_\chi=0.$}
\end{equation}

Now we return to (4.48) and write this, using (4.47), as follows:
\begin{equation}
\frac{u}{z}=A_\chi+\frac{A'_\chi+\mathcal{H}A_\chi}{L},
\end{equation}
where
\begin{equation}
L=4\pi G\frac{z^2\mathcal{H}}{a^2}=4\pi
G(\varphi'_0)^2/\mathcal{H}=\mathcal{H}-\mathcal{H}'/\mathcal{H}.
\end{equation}
Differentiating (4.51) implies
\[\left(\frac{u}{z}\right)'=A'_\chi+\frac{A''_\chi+(\mathcal{H}A_\chi)'}{L}-\frac{A'_\chi+\mathcal{H}A_\chi}{L^2}L'\]
or, making use of (4.52) and (4.50),
\begin{eqnarray*}
\lefteqn{L\left(\frac{u}{z}\right)'=(\mathcal{H}-\mathcal{H}'/\mathcal{H})A'_\chi-
2\frac{(a/\varphi'_0)'}{a/\varphi'_0}A'_\chi+\triangle A_\chi }\\
& & -2\varphi'_0(\mathcal{H}/\varphi'_0)'A_\chi+(\mathcal{H}A_\chi)'-
(A'_\chi+\mathcal{H}A_\chi)\frac{(\varphi^{'2}_0/\mathcal{H})'}{\varphi^{'2}_0/\mathcal{H}}.
\end{eqnarray*}
From this one easily finds the simple equation
\begin{equation}
\fbox{$\displaystyle 4\pi G\frac{\mathcal{H}z^2}{a^2}\left(\frac{u}{z}\right)'=\triangle
A_\chi.$}
\end{equation}

Finally, we derive the announced eq. (4.42). To this end we rewrite the last equation as
\[\triangle A_\chi=4\pi G\frac{\mathcal{H}}{a^2}(zu'-z'u), \]
from which we get \[\triangle A'_\chi=4\pi G\left(\frac{\mathcal{H}}{a^2}\right)'(zu'-z'u)
+4\pi G\frac{\mathcal{H}}{a^2}(zu''-z''u).\] Taking the Laplacian of (4.51) gives
\[ 4\pi G\frac{\mathcal{H}}{a^2}z\triangle
u=L\triangle A_\chi+\triangle A'_\chi+\mathcal{H}\triangle A_\chi.\] Combining the last two
equations and making use of (4.52) shows that indeed (4.42) holds.

Summarizing, we have the basic equations
\begin{equation}
 u''-\triangle u-\frac{z''}{z}u=0,
\end{equation}
\begin{equation}
\triangle A_\chi=4\pi G\frac{\mathcal{H}}{a^2}(zu'-z'u),
\end{equation}
\begin{equation}
\left(\frac{a^2A_\chi}{\mathcal{H}}\right)'=4\pi Gzu.
\end{equation}

We now discuss some important consequences of these equations. The first concerns the
curvature perturbation $\mathcal{R}=-u/z$ (original definition in (4.39)). In terms of this
quantity eq. (4.55) can be written as
\begin{equation}
\frac{\dot{\mathcal{R}}}{H}=\frac{1}{1-\mathcal{H}'/\mathcal{H}^2}\frac{1}{(aH)^2}(-\triangle
A_\chi).
\end{equation}
The right-hand side is of order $(k/aH)^2$, hence very small on scales much larger than the
Hubble radius. It is common practice to use the terms ``Hubble length'' and ``horizon''
interchangeably, and to call length scales satisfying $k/aH\ll1$ to be
\textit{super-horizon}. (This can cause confusion; `super-Hubble' might be a better term,
but the jargon can probably not be changed anymore.)

We have studied $\dot{\mathcal{R}}$ already at the end of Sect.1.3. I recall (1.138):
\begin{equation}
\dot{\mathcal{R}}=\frac{H}{1+w}\left[\frac{2}{3}c_s^2\frac{1}{(Ha)^2}\triangle
D_\chi-w\Gamma-\frac{2}{3}w\triangle\Pi\right].
\end{equation}
This general equation also holds for our scalar field model, for which
$\Pi=0,~D_\chi=-A_\chi$. The first term on the right in (4.58) is again small on
super-horizon scales. So the non-adiabatic piece $p\Gamma=\delta p-c_s^2\delta\rho$ must
also be small on large scales. This means that the perturbations are \textbf{adiabatic}. We
shall show this more directly further below, by deriving the following expression for
$\Gamma$:
\begin{equation}
\fbox{$\displaystyle p\Gamma=-\frac{U_{,\varphi}}{6\pi
GH\dot{\varphi}}\frac{1}{a^2}\triangle A_\chi.$}
\end{equation}

After inflation, when relativistic fluids dominate the matter content, eq. (4.58) still
holds. The first term on the right is small on scales larger than the \textit{sound
horizon}. Since $\Gamma$ and $\Pi$ are then not important, we see that for super-horizon
scales $\mathcal{R}$ \textit{remains constant also after inflation}. This will become
important in the study of CMB anisotropies.

Later, it will be useful to have a handy expression of $A_\chi$ in terms of $\mathcal{R}$.
According to (1.58) and (1.57) we have
\begin{equation}
\mathcal{R}=D_\chi+\frac{\mathcal{H}}{a(\rho+p)}\mathcal{Q}.
\end{equation}
We rewrite this by combining (1.99) and (1.101)
\begin{equation}
\mathcal{R}=D_\chi-\frac{\mathcal{H}}{4\pi Ga^2(\rho+p)}(\mathcal{H}A_\chi-D'_\chi).
\end{equation}
At this point we specialize again to $K=0$, and use (1.80) in the form
\[4\pi Ga^2(\rho+p)=\mathcal{H}^2(1-\mathcal{H}'/\mathcal{H}^2)\]
and obtain
\begin{equation}
\fbox{$\displaystyle
\mathcal{R}=D_\chi-\frac{1}{\varepsilon\mathcal{H}}(\mathcal{H}A_\chi-D'_\chi),$}
\end{equation}
where
\begin{equation}
\varepsilon:=1-\mathcal{H}'/\mathcal{H}^2.
\end{equation}

If $\Pi=0$ then $D_\chi=-A_\chi$, so
\begin{equation}
\fbox{$\displaystyle
-\mathcal{R}=A_\chi+\frac{1}{\varepsilon\mathcal{H}}(\mathcal{H}A_\chi+A'_\chi),$}
\end{equation}
I claim that for a constant $\mathcal{R}$
\begin{equation}
A_\chi=-\left(1-\frac{\mathcal{H}}{a^2}\int a^2 d\eta\right)\mathcal{R}.
\end{equation}
We prove this by showing that (4.65) satisfies (4.64). Differentiating the last equation
gives by the same equation and (4.63) our claim.

As a special case we consider (always for $K=0$) $w=const$. Then, as shown in Sect.2.4,
\begin{equation}
a=a_0(\eta/\eta_0)^\beta,~~ \beta=\frac{2}{3w+1}.
\end{equation}
Thus
\[\frac{\mathcal{H}}{a^2}\int a^2 d\eta=\frac{\beta}{2\beta+1},\]
hence
\begin{equation}
\fbox{$\displaystyle A_\chi=-\frac{3(w+1)}{3w+5}\mathcal{R}.$}
\end{equation}
This will be important later.

\textit{Derivation of (4.59)}: By definition
\begin{equation}
p\Gamma=\delta p-c_s^2\delta\rho,~~ c_s^2=\dot{p}/\dot{\rho} \Rightarrow
p\Gamma=\frac{\dot{\rho}\delta p-\dot{p}\delta\rho}{\dot{\rho}}.
\end{equation}
Now, by (4.7) and (4.5)
\[\dot{\rho}=-3H\dot{\varphi}^2,~~
\dot{p}=\dot{\varphi}(\ddot{\varphi}-U_{,\varphi})=-\dot{\varphi}(3H\dot{\varphi}+2U_{,\varphi}),\]
and by (4.14) and (4.15)
\[ \delta\rho=-\dot{\varphi}^2A+\dot{\varphi}\delta\dot{\varphi}+U_{,\varphi}\delta\varphi~,~~ \delta p=
\dot{\varphi}\delta\dot{\varphi}-\dot{\varphi}^2A-U_{,\varphi}\delta\varphi.\] With these
expressions one readily finds
\begin{equation}
p\Gamma=-\frac{2}{3}\frac{U_{,\varphi}}{H\dot{\varphi}}[-\ddot{\varphi}\delta\varphi+
\dot{\varphi}(\delta\dot{\varphi}-\dot{\varphi}A)].
\end{equation}
Up to now we have not used the perturbed field equations. The square bracket on the right
of the last equation appears in the combination (4.18)-$H\cdot$ (4.19) for the right hand
sides. Since the right hand side of (4.69) must be gauge invariant, we can work in the
gauge $\chi=0$, and obtain (for $K=0$) from (4.18),(4.19)
\[ \frac{1}{a^2}\triangle A=4\pi G[-\ddot{\varphi}\delta\varphi+
\dot{\varphi}(\delta\dot{\varphi}-\dot{\varphi}A)],\] thus (4.59) since in the longitudinal
gauge $A=A_\chi$.

\textit{Application}. We return to eq. (4.57) and use there (4.59) to obtain
\begin{equation}
\fbox{$\displaystyle\dot{\mathcal{R}}=4\pi G\frac{\rho p}{\dot{U}}\Gamma.$}
\end{equation}

As a result of (4.59) $\Gamma$ is small on super-horizon scales, and hence (4.70) tells us
that $\mathcal{R}$ is almost constant (as we knew before).

The crucial conclusion is that the perturbations are \textbf{adiabatic}, which is not
obvious (I think). For multi-field inflation this is, in general, not the case (see, e.g.,
\cite{Sta}).

\chapter{Quantization, Primordial Power Spectra}

The main goal of this Chapter is to derive the primordial power spectra that are generated
as a result of quantum fluctuations during an inflationary period.

\section{Power spectrum of the inflaton field}

For the quantization of the scalar field that drives the inflation we note that the
equation of motion (4.42) for the scalar perturbation (4.41),
\begin{equation}
u=a\left[\delta\varphi_\chi-\frac{\dot{\varphi}_0}{H}D_\chi\right]=
a\left[\delta\varphi_\chi+\frac{\varphi'_0}{\mathcal{H}}A_\chi\right],
\end{equation}
is the Euler-Lagrange equation for the effective action
\begin{equation}
S_{eff}=\frac{1}{2}\int~ d^3x d\eta \left[(u')^2-(\nabla u)^2+\frac{z''}{z}u^2\right].
\end{equation}
The normalization is chosen such that $S_{eff}$ reduces to the correct action when gravity
is switched off. (In \cite{MFB} this action is obtained by considering the quadratic piece
of the full action with Lagrange density (3.26), but this calculation is extremely
tedious.)

The effective Lagrangian of (5.1) is
\begin{equation}
\mathcal{L}=\frac{1}{2}\left[(u')^2-(\nabla u)^2+\frac{z''}{z}u^2\right].
\end{equation}
This is just a free theory with a time-dependent mass $m^2=-z''/z$. Therefore the
quantization is straightforward. Once $u$ is quantized the quantization of $\Psi=A_\chi$ is
then also fixed (see eq. (4.55)).

The canonical momentum is
\begin{equation}
\pi=\frac{\partial\mathcal{L}}{\partial u'}=u',
\end{equation}
and the canonical commutation relations are the usual ones:
\begin{equation}
\left[\hat{u}(\eta,\mathbf{x}),\hat{u}(\eta,\mathbf{x}')\right]=
\left[\hat{\pi}(\eta,\mathbf{x}),\hat{\pi}(\eta,\mathbf{x}')\right]=0,~~
\left[\hat{u}(\eta,\mathbf{x}),\hat{\pi}(\eta,\mathbf{x}')\right]=
i\delta^{(3)}(\mathbf{x}-\mathbf{x}').
\end{equation}

Let us expand the field operator $\hat{u}(\eta,\mathbf{x})$ in terms of eigenmodes
$u_k(\eta)e^{i\mathbf{k}\cdot\mathbf{x}}$ of eq. (4.42), for which
\begin{equation}
u''_k+\left(k^2-\frac{z''}{z}\right)u_k=0.
\end{equation}
The time-independent normalization is chosen to be
\begin{equation}
u^\ast_ku'_k-u_ku'^{\ast}_k=-i.
\end{equation}
In the decomposition
\begin{equation}
\hat{u}(\eta,\mathbf{x})=(2\pi)^{-3/2}\int
d^3k\left[u_k(\eta)\hat{a}_{\mathbf{k}}e^{i\mathbf{k}\cdot\mathbf{x}}
+u^{\ast}_k(\eta)\hat{a}^{\dagger}_{\mathbf{k}}e^{-i\mathbf{k}\cdot\mathbf{x}}\right]
\end{equation}
the coefficients $\hat{a}_{\mathbf{k}},\hat{a}^{\dagger}_{\mathbf{k}}$ are annihilation and
creation operators with the usual commutation relations:
\begin{equation}
[\hat{a}_{\mathbf{k}},\hat{a}_{\mathbf{k}'}]=[\hat{a}^{\dagger}_{\mathbf{k}},\hat{a}^{\dagger}_{\mathbf{k}'}
]=0,~~
[\hat{a}_{\mathbf{k}},\hat{a}^{\dagger}_{\mathbf{k}'}]=\delta^{(3)}(\mathbf{k}-\mathbf{k}').
\end{equation}
With the normalization (5.7) these imply indeed the commutation relations (5.5). (Translate
(5.8) with the help of (4.55) into a similar expansion of $\Psi$, whose mode functions are
determined by $u_k(\eta)$.)

The modes $u_k(\eta)$ are chosen such that at very short distances
($k/aH\rightarrow\infty$) they approach the plane waves of the gravity free case with
positive frequences
\begin{equation}
u_k(\eta) \sim\frac{1}{\sqrt{2k}}e^{-ik\eta}~~~~~~(k/aH\gg 1).
\end{equation}
In the opposite long-wave regime, where $k$ can be neglected in (5.6), we see that the
\textit{growing mode} solution is
\begin{equation}
u_k\propto z ~~~~~~~~(k/aH\ll 1),
\end{equation}
i.e., $u_k/z$ and thus $\mathcal{R}$ is constant on super-horizon scales. This has to be so
on the basis of what we saw in Sect. 4.2. The power spectrum is conveniently defined in
terms of $\mathcal{R}$. We have (we do not put a hat on $\mathcal{R}$)
\begin{equation}
\mathcal{R}(\eta,\mathbf{x})=(2\pi)^{-3/2}\int\mathcal{R}_{\mathbf{k}}(\eta)e^{i\mathbf{k}\cdot\mathbf{x}}d^3k,
\end{equation}
with
\begin{equation}
\mathcal{R}_{\mathbf{k}}(\eta)=\left[\frac{u_k(\eta)}{z}\hat{a}_{\mathbf{k}}
+\frac{u^{\ast}_k(\eta)}{z}\hat{a}^{\dagger}_{-\mathbf{k}}\right].
\end{equation}
The \textit{power spectrum} is defined by (see also Appendix A)
\begin{equation}
\langle0|\mathcal{R}_{\mathbf{k}}\mathcal{R}^{\dagger}_{\mathbf{k}'}|0\rangle=
:\frac{2\pi^2}{k^3}P_{\mathcal{R}}(k)\delta^{(3)}(\mathbf{k}-\mathbf{k}').
\end{equation}
From (5.13) we obtain
\begin{equation}
\fbox{$\displaystyle P_{\mathcal{R}}(k)=\frac{k^3}{2\pi^2}\frac{|u_k(\eta)|^2}{z^2}.$}
\end{equation}

Below we shall work this out for the inflationary models considered in Chap. 4. Before, we
should address the question why we considered the two-point correlation for the Fock vacuum
relative to our choice of modes $u_k(\eta)$. A priori, the initial state could contain all
kinds of excitations. These would, however, be redshifted away by the enormous inflationary
expansion, and the final power spectrum on interesting scales, much larger than the Hubble
length,  should be largely independent of possible initial excitations. (This point should,
perhaps, be studied in more detail.)

\subsection{Power spectrum for power law inflation}

For power law inflation one can derive an exact expression for (5.15). For the mode
equation (5.6) we need $z''/z$. To compute this we insert in the definition (4.40) of $z$
the results of Sect. 3.3.1, giving immediately $z\propto a(t)\propto t^p$. In addition
(3.40) implies $t\propto\eta^{1/1-p}$, so $a(\eta)\propto\eta^{p/1-p}$. Hence,
\begin{equation}
\fbox{$\displaystyle \frac{z''}{z}=\left(\nu^2-\frac{1}{4}\right)\frac{1}{\eta^2}$}~,
\end{equation}
where
\begin{equation}
\nu^2-\frac{1}{4}=\frac{p(2p-1)}{(p-1)^2}.
\end{equation}
Using this in  (5.6) gives the mode equation
\begin{equation}
\fbox{$\displaystyle u''_k+\left(k^2-\frac{\nu^2-1/4}{\eta^2}\right)u_k=0.$}
\end{equation}
This can be solved in terms of Bessel functions. Before proceeding with this we note two
further relations that will be needed later. First, from $H=p/t$ and $a(t)=a_0t^p$ we get
\begin{equation}
\eta=-\frac{1}{aH}\frac{1}{1-1/p}.
\end{equation}
In addition, \[
\frac{z}{a}=\frac{\dot{\varphi}}{H}=\sqrt{\frac{p}{4\pi}}\frac{M_{Pl}/t}{(p/t)}=\frac{1}{\sqrt{4\pi
p}}M_{Pl},\] so
\begin{equation}
\fbox{$\displaystyle
\varepsilon:=-\frac{\dot{H}}{H^2}=\frac{1}{p}=\frac{4\pi}{M_{Pl}^2}\frac{z^2}{a^2}.$}
\end{equation}

Let us now turn to the mode equation (5.18). According to \cite{Abr}, 9.1.49, the functions
$w(z)=z^{1/2}\mathcal{C}_\nu(\lambda z),~ \mathcal{C}_\nu\propto
H^{(1)}_\nu,H^{(2)}_\nu,...$ satisfy the differential equation
\begin{equation}
w''+\left(\lambda^2-\frac{\nu^2-1/4}{z^2}\right)w=0.
\end{equation}
From the asymptotic formula for large $z$ (\cite{Abr}, 9.2.3),
\begin{equation}
H^{(1)}_\nu\sim\sqrt{\frac{2}{\pi z}}e^{i(z-\frac{1}{2}\nu\pi-\frac{1}{4}\pi)}~~~~
(-\pi<\arg z <\pi),
\end{equation}
we see that the correct solutions are
\begin{equation}
u_k(\eta)=\frac{\sqrt{\pi}}{2}e^{i(\nu+\frac{1}{2})\frac{\pi}{2}}(-\eta)^{1/2}H^{(1)}_\nu(-k\eta).
\end{equation}
Indeed, since $-k\eta=(k/aH)(1-1/p)^{-1},~ k/aH\gg 1$ means large $-k\eta$, hence (5.23)
satisfies (5.10). Moreover, the Wronskian is normalized according to (5.7) (use 9.1.9 in
\cite{Abr}).

In what follows we are interested in modes which are well outside the horizon: $(k/aH)\ll
1$. In this limit we can use (9.1.9 in \cite{Abr})
\begin{equation}
iH^{(1)}_\nu(z)\sim \frac{1}{\pi}\Gamma(\nu)\left(\frac{1}{2}z\right)^{-\nu} ~~~~
(z\rightarrow 0)
\end{equation}
to find
\begin{equation}
u_k(\eta)\simeq
2^{\nu-3/2}e^{i(\nu-1/2)\pi/2}\frac{\Gamma(\nu)}{\Gamma(3/2)}\frac{1}{\sqrt{2k}}(-k\eta)^{-\nu+1/2}.
\end{equation}
Therefore, by (5.19) and (5.20)
\begin{equation}
|u_k|=2^{\nu-3/2}\frac{\Gamma(\nu)}{\Gamma(3/2)}(1-\varepsilon)^{\nu-1/2}\frac{1}{\sqrt{2k}}
\left(\frac{k}{aH}\right)^{-\nu+1/2}.
\end{equation}
The form (5.26) will turn out to hold also in more general situations studied below,
however, with a different $\varepsilon$. We write (5.26) as
\begin{equation}
|u_k|=C(\nu)\frac{1}{\sqrt{2k}} \left(\frac{k}{aH}\right)^{-\nu+1/2},
\end{equation}
with
\begin{equation}
C(\nu)=2^{\nu-3/2}\frac{\Gamma(\nu)}{\Gamma(3/2)}(1-\varepsilon)^{\nu-1/2}
\end{equation}
(recall $\nu=\frac{3}{2}+\frac{1}{p-1}$).

The power spectrum is thus
\begin{equation}
P_{\mathcal{R}}(k)=\frac{k^3}{2\pi^2}\left|\frac{u_k(\eta)}{z^2}\right|^2
=\frac{k^3}{2\pi^2}\frac{1}{z^2}C^2(\nu)\frac{1}{2k}\left(\frac{k}{aH}\right)^{1-2\nu}.
\end{equation}
For $z$ we could use (5.20). There is, however, a formula which holds more generally: From
the definition (4.40) of $z$ and (3.38) we get
\begin{equation}
\fbox{$\displaystyle z=-\frac{M^2_{Pl}}{4\pi}\frac{a}{H}\frac{dH}{d\varphi}.$}
\end{equation}
Inserting this in the previous equation we obtain for the power spectrum on super-horizon
scales
\begin{equation}
 P_{\mathcal{R}}(k)=C^2(\nu)\frac{4}{M^4_{Pl}}\frac{H^4}{(dH/d\varphi)^2}
\left(\frac{k}{aH}\right)^{3-2\nu}.
\end{equation}
For power-law inflation a comparison of (5.20) and (5.30) shows that
\begin{equation}
\frac{M^2_{Pl}}{4\pi}\frac{(dH/d\varphi)^2}{H^2}=\frac{1}{p}=\varepsilon.
\end{equation}

The asymptotic expression (5.31), valid for $k/aH\ll 1$, remains, as we know, constant in
time\footnote{Let us check this explicitly. Using (5.32) we can write (5.31) as
\[P_{\mathcal{R}}(k)=C^2(\nu)\frac{1}{\pi M^2_{Pl}}\frac{H^2}{\varepsilon}
\left(\frac{k}{aH}\right)^{3-2\nu},\] and we thus have to show that $H^2(aH)^{2\nu-3}$ is
time independent. This is indeed the case since
$aH\propto1/\eta,~H=p/t,~t\propto\eta^{1/(1-p)}\Rightarrow~H\propto\eta^{-1/(1-p)}$.}.
Therefore, we can evaluate it at \textit{horizon crossing} $k=aH$:
\begin{equation}
\fbox{$\displaystyle
P_{\mathcal{R}}(k)=C^2(\nu)\frac{4}{M^4_{Pl}}\left.\frac{H^4}{(dH/d\varphi)^2}\right|_{k=aH}
.$}
\end{equation}
We emphasize that this is \textit{not} the value of the spectrum at the moment when the
scale crosses outside the Hubble radius. We have just rewritten the asymptotic value for
$k/aH\ll 1$ in terms of quantities at horizon crossing.

Note also that $C(\nu)\simeq 1$. The result (5.33) holds, as we shall see below, also in
the slow-roll approximation.

\subsection{Power spectrum in the slow-roll approximation}

We now define two slow-roll parameters and rewrite them with the help of (3.37) and (3.38):
\begin{eqnarray}
\varepsilon &=&
-\frac{\dot{H}}{H^2}=\frac{4\pi}{M^2_{Pl}}\frac{\dot{\varphi}^2}{H^2}=\frac{M^2_{Pl}}{4\pi}
\left(\frac{dH/d\varphi}{H(\varphi)}\right)^2,\\
\delta &=&
-\frac{\ddot{\varphi}}{H\dot{\varphi}}=\frac{M^2_{Pl}}{4\pi}\frac{d^2H/d\varphi^2}{H}
\end{eqnarray}
($\mid\varepsilon\mid,\mid\delta\mid\ll 1$ in the slow-roll approximation). These
parameters are approximately related to $\varepsilon_U,\eta_U$ introduced in (3.45) and
(3.46), as we now show. From (3.36) for $K=0$ and (3.37) we obtain
\begin{equation}
H^2(1-\frac{\varepsilon}{3})=\frac{8\pi}{3M^2_{Pl}}U(\varphi).
\end{equation}
For small $\mid\varepsilon\mid$ we obtain from this the following approximate expressions
for the slow-roll parameters:
\begin{eqnarray}
\varepsilon &\simeq& \frac{M^2_{Pl}}{16\pi}\left(\frac{U_{,\varphi}}{U}\right)^2 = \varepsilon_U,\\
\delta &\simeq&
\frac{M^2_{Pl}}{8\pi}\frac{U_{,\varphi\varphi}}{U}-\frac{M^2_{Pl}}{16\pi}\left(\frac{U_{,\varphi}}{U}\right)^2
=\eta_U-\varepsilon_U.
\end{eqnarray}
(In the literature the letter $\eta$ is often used instead of $\delta$, but $\eta$ is
already occupied for the conformal time.)

We use these small parameters to approximate various quantities, such as the effective mass
$z''/z$.

First, we note that (5.34) and (5.30) imply the relations\footnote{Note also that
\[\frac{\ddot{a}}{a}\equiv \dot{H}+H^2=(1-\varepsilon)H^2,\] so $\ddot{a}>0$ for
$\varepsilon<1$.}
\begin{equation}
\varepsilon=1- \frac{\mathcal{H}'}{\mathcal{H}^2}=\frac{4\pi}{M^2_{Pl}}\frac{z^2}{a^2}.
\end{equation}
According to (5.35) we have $\delta=1-\varphi''/\varphi'\mathcal{H}$. For the last term we
obtain from the definition $z=a\varphi'/\mathcal{H}$
\[\frac{\varphi''}{\varphi'\mathcal{H}}=\frac{z'}{z\mathcal{H}}-(1-\mathcal{H}'/\mathcal{H}^2).\]
Hence
\begin{equation}
\delta=1+\varepsilon-\frac{z'}{z\mathcal{H}}.
\end{equation}

Next, we look for a convenient expression for the conformal time. From (5.39) we get
\[\frac{\varepsilon}{a\mathcal{H}}da=\varepsilon d\eta=d\eta-(\mathcal{H}'/\mathcal{H}^2)d\eta
=d\eta+d\left(\frac{1}{\mathcal{H}}\right),\] so
\begin{equation}
\eta=-\frac{1}{\mathcal{H}}+\int\frac{\varepsilon}{a\mathcal{H}}da.
\end{equation}

Now we determine $z''/z$ to first order in $\varepsilon$ and $\delta$. From (5.40), i.e.,
$z'/z=\mathcal{H}(1+\varepsilon-\delta)$, we get
\[
\frac{z''}{z}-\left(\frac{z'}{z}\right)^2=(\varepsilon'-\delta')\mathcal{H}+(1+\varepsilon-\delta)\mathcal{H}',\]
hence
\begin{equation}
z''/z =\mathcal{H}^2\left[\frac{\varepsilon'-\delta'}{\mathcal{H}}+
(1+\varepsilon-\delta)(2-\delta)\right].
\end{equation}

We can consider $\varepsilon',\delta'$ as of second order: For instance, by (5.39)
\[\varepsilon'=\frac{4\pi}{M^2_{Pl}}\frac{2zz'}{a^2}-2\varepsilon\mathcal{H}\]
or
\begin{equation}
\varepsilon'=2\mathcal{H}\varepsilon(\varepsilon-\delta).
\end{equation}

Treating $\varepsilon,\delta$ as constant, eq. (5.41) gives
$\eta=-(1/\mathcal{H})+\varepsilon\eta$, thus
\begin{equation}
\eta=-\frac{1}{\mathcal{H}}\frac{1}{1-\varepsilon}.
\end{equation}
This generalizes (5.19), in which $\varepsilon=1/p$ (see (5.20)). Using this in (5.42) we
obtain to first order
\[ \frac{z''}{z}=\frac{1}{\eta^2}(2+2\varepsilon-3\delta).\]
We write this as (5.16), but with a different $\nu$:
\begin{equation}
\frac{z''}{z}=\left(\nu^2-\frac{1}{4}\right)\frac{1}{\eta^2}~,~~~~
\nu:=\frac{1+\varepsilon-\delta}{1-\varepsilon}+\frac{1}{2}.
\end{equation}

As a result of all this we can immediately write down the power spectrum in the slow-roll
approximation. From the derivation it is clear that the formula (5.33) still holds, and the
same is true for (5.28). Since $\nu$ is close to 3/2 we have $C(\nu)\simeq1$. In sufficient
approximation we thus finally obtain the important result:
\begin{equation}
\fbox{$\displaystyle
P_{\mathcal{R}}(k)=\frac{4}{M^4_{Pl}}\left.\frac{H^4}{(dH/d\varphi)^2}\right|_{k=aH}
=\frac{1}{\pi M^2_{Pl}}\frac{H^2}{\varepsilon} \left(\frac{k}{aH}\right)^{3-2\nu} .$}
\end{equation}

This spectrum is \textit{nearly scale-free}. This is evident if we use the formula (5.31),
from which we get
\begin{equation}
\fbox{$\displaystyle n-1:=\frac{d\ln P_{\mathcal{R}(k)}}{d\ln
k}=3-2\nu=2\delta-4\varepsilon,$}
\end{equation}
so $n$ is \textit{close to unity}.

\paragraph{Exercise.}

Show that (5.47) follows also from (5.46).

\textit{Solution}: In a first step we get
\[n-1=\frac{d}{d\varphi}\ln\left[\left.\frac{H^4}{(dH/d\varphi)^2}\right|_{k=aH}\right]\frac{d\varphi}{d\ln
k}.\] For the last factor we note that $k=aH$ implies
\[d\ln k=\frac{da}{a}+\frac{dH}{H}\Rightarrow \frac{d\ln
k}{d\varphi}=\frac{H}{\dot{\varphi}}+\frac{dH/d\varphi}{H}\] or, with (3.37),
\[\frac{d\ln k}{d\varphi}=\frac{4\pi}{M^2_{Pl}}\frac{H}{dH/d\varphi}\left[\frac{M^2_{Pl}}{4\pi}
\left(\frac{dH/d\varphi}{H}\right)^2-1\right].\] Hence, using (5.34),
\[\frac{d\varphi}{d\ln
k}=\frac{M^2_{Pl}}{4\pi}\frac{dH/d\varphi}{H}\frac{1}{\varepsilon-1}.\] Therefore,
\[n-1=\frac{M^2_{Pl}}{4\pi}\frac{dH/d\varphi}{H}\frac{1}{\varepsilon-1}
\left[4\frac{dH/d\varphi}{H}-2\frac{d^2H/d\varphi^2}{dH/d\varphi}\right]
=\frac{1}{\varepsilon-1}(4\varepsilon-2\delta)\]
by (5.34) and (5.35).

\subsection{Power spectrum for density fluctuations}

Let $P_\Phi(k)$ be the power spectrum for the Bardeen potential $\Phi=D_\chi$. The latter
is related to the density fluctuation $\Delta$ by the Poisson equation (2.3),
\begin{equation}
k^2\Phi=4\pi G\rho a^2\Delta.
\end{equation}
Recall also that for $\Pi=0$ we have $\Phi=-\Psi~(=-A_\chi)$, and according to (4.67) the
following relation for a period with $w=const.$
\begin{equation}
\fbox{$\displaystyle \Phi=\frac{3(w+1)}{3w+5}\mathcal{R},$}
\end{equation}
and thus
\begin{equation}
P^{1/2}_\Phi(k)=\frac{3(w+1)}{3w+5}P^{1/2}_\mathcal{R}(k).
\end{equation}
Inserting (5.46) gives for the \textit{primordial} spectrum on super-horizon scales
\begin{equation}
P_\Phi(k)=\left[\frac{3(w+1)}{3w+5}\right]^2\frac{4}{M^4_{Pl}}\left.\frac{H^4}{(dH/d\varphi)^2}\right|_{k=aH}.
\end{equation}
From (5.48) we obtain
\begin{equation}
\Delta(k)=\frac{2(w+1)}{3w+5} \left(\frac{k}{aH}\right)^2\mathcal{R}(k),
\end{equation}
and thus for the power spectrum of $\Delta$:
\begin{equation}
P_\Delta(k)=\frac{4}{9}\left(\frac{k}{aH}\right)^4P_\Phi(k)=\frac{4}{9}\left[\frac{3(w+1)}{3w+5}\right]^2
\left(\frac{k}{aH}\right)^4P_\mathcal{R}(k).
\end{equation}

During the plasma era until recombination the primordial spectra (5.46) and (5.51) are
modified in a way that will be studied in Part III of these lectures. The modification is
described by the so-called \textit{transfer function}\footnote{For more on this, see Sect.
6.2.4, where the $z$-dependence of $T(k,z)$ is explicitly split off.} $T(k,z)$, normalized
such that $T(k)\simeq 1$ for $(k/aH)\ll 1$. Including this, we have in the (dark) matter
dominated era (in particular at the time of recombination)
\begin{equation}
P_\Delta(k)=\frac{4}{25}\left(\frac{k}{aH}\right)^4P^{prim}_\mathcal{R}(k)T^2(k),
\end{equation}
where $P^{prim}_\mathcal{R}(k)$ denotes the primordial spectrum ((5.46) for our simple
model of inflation).

\paragraph{Remark.} Using the fact that $\mathcal{R}$ is constant on super-horizon scales
allows us to establish the relation between $\Delta_H(k):=\Delta(k,\eta)\mid_{k=aH}$ and
$\Delta(k,\eta)$ on these scales. From (5.52) we see that
\begin{equation}
\Delta(k,\eta)=\left(\frac{k}{aH}\right)^2\Delta_H(k).
\end{equation}
In particular, if $\mid\mathcal{R}(k)\mid\propto k^{n-1}$, thus
$\mid\Delta(k,\eta)\mid^2=Ak^{n+3}$, then
\begin{equation}
\fbox{$\displaystyle \mid\Delta_H(k)\mid^2=Ak^{n-1},$}
\end{equation}
and this is \textit{independent} of $k$ for $n=1$. In this case the density fluctuation for
each mode at horizon crossing has the same magnitude. This explains why the case $n=1$ --
also called the \textit{Harrison-Zel'dovich spectrum} -- is called \textit{scale free}.

\section{Generation of gravitational waves}

In this section we determine the power spectrum of gravitational waves by quantizing tensor
perturbations of the metric.

These are parametrized as follows
\begin{equation}
g_{\mu\nu}=a^2(\eta)[\gamma_{\mu\nu}+2H_{\mu\nu}],
\end{equation}
where $a^2(\eta)\gamma_{\mu\nu}$ is the Friedmann metric
($\gamma_{\mu0}=0,~\gamma_{ij}$:~metric of ($\Sigma,\gamma)$), and $H_{\mu\nu}$ satisfies
the \textit{transverse traceless} (TT) gauge conditions
\begin{equation}
H_{00}=H_{0i}=H^i{}_{i}=H_i{}^j{}_{\mid j}=0.
\end{equation}

The tensor perturbation amplitudes $H_{ij}$ remain invariant under gauge transformations
(1.14). Indeed, as in Sect. 1.14, one readily finds
\begin{eqnarray*}
L_\xi g^{(0)}=2a^2(\eta)\left\{-(\mathcal{H}\xi^0+(\xi^0)')d\eta^2 +(\xi'_i-\xi^0{}_{\mid
i})dx^id\eta \right.\\
\left.+(\mathcal{H}\gamma_{ij}\xi^0+\xi_{i\mid j})dx^idx^j\right\}.
\end{eqnarray*}
Decomposing $\xi^\mu$ into scalar and vector parts gives the scalar and vector
contributions of $L_\xi g^{(0)}$, but there are obviously \textit{no} tensor contributions.

The perturbations of the Einstein tensor belonging to $H_{\mu\nu}$ are derived in the
Appendix to this Chapter. The result is:
\begin{eqnarray}
\delta G^0{}_0 &=& \delta G^0{}_{j}=\delta G^i{}_{0}=0,\nonumber \\
\delta G^i{}_{j} &=&
\frac{1}{a^2}\left[(H^i{}_j)''+2\frac{a'}{a}(H^i{}_j)'+(-\triangle+2K)H^i{}_j\right].
\end{eqnarray}

We claim that the quadratic part of the Einstein-Hilbert action is
\begin{equation}
S^{(2)}=\frac{M^2_{Pl}}{16\pi}\int\left[(H^i{}_k)'(H^k{}_i)'-H^i{}_{k\mid l}H^k{}_i{}^{\mid
l}-2KH^i{}_k H^k{}_i\right]a^2(\eta)d\eta\sqrt{\gamma}d^3x.
\end{equation}
(Remember that the indices are raised and lowered with $\gamma_{ij}$.) Note first that
$\sqrt{-g}d^4x=\sqrt{\gamma}a^4(\eta)d\eta d^3x+$ quadratic terms in $H_{ij}$, because
$H_{ij}$ is traceless. A direct derivation of (5.60) from the Einstein-Hilbert action would
be extremely tedious (see \cite{MFB}). It suffices, however, to show that the variation of
(5.60) is just the linearization of the general variation formula (see Sect. 2.3 of
\cite{NS1})
\begin{equation}
\delta S=-\frac{M^2_{Pl}}{16\pi}\int G^{\mu\nu}\delta g_{\mu\nu}\sqrt{-g}d^4x
\end{equation}
for the Einstein-Hilbert action
\begin{equation}
S=\frac{M^2_{Pl}}{16\pi}\int R\sqrt{-g}d^4x.
\end{equation}
Now, we have after the usual partial integrations,
\[\delta S^{(2)}=-\frac{M^2_{Pl}}{8\pi}\int\left[\frac{(a^2H^i{}_k)')'}{a^2}+(-\triangle+2K)H^i{}_k\right]
\delta H^k{}_i a^2(\eta)d\eta\sqrt{\gamma}d^3x.\] Since $\delta H^k{}_i=\frac{1}{2}\delta
g^k{}_i$ this is, with the expression (5.59), indeed the linearization of (5.61).

We absorb in (5.60) the factor $a^2(\eta)$ by introducing the rescaled perturbation
\begin{equation}
P^i{}_j(x):=\left(\frac{M^2_{Pl}}{8\pi}\right)^{1/2}a(\eta)H^i{}_j(x).
\end{equation}
Then $S^{(2)}$ becomes, after another partial integration,
\begin{equation}
\fbox{$\displaystyle S^{(2)}=\frac{1}{2}\int\left[(P^i{}_k)'(P^k{}_i)'-P^i{}_{k\mid
l}P^k{}_i{}^{\mid l}+\left(\frac{a''}{a}-2K\right)P^i{}_k
P^k{}_i\right]d\eta\sqrt{\gamma}d^3x.$}
\end{equation}

In what follows we take again $K=0$. Then we have the following Fourier decomposition: Let
$\epsilon_{ij}(\mathbf{k},\lambda)$ be the two polarization tensors, satisfying
\begin{eqnarray}
\epsilon_{ij}
=\epsilon_{ji},~~\epsilon^i{}_i&=&0,~~k^i\epsilon_{ij}(\mathbf{k},\lambda)=0,~~
\epsilon_i{}^j(\mathbf{k},\lambda)\epsilon_j{}^i(\mathbf{k},\lambda)^{\ast}=\delta_{\lambda\lambda'},\nonumber \\
\epsilon_{ij}(\mathbf{-k},\lambda)&=&\epsilon^{\ast}_{ij}(\mathbf{k},\lambda),
\end{eqnarray}
then
\begin{equation}
P^i{}_j(\eta,\mathbf{x})=(2\pi)^{-3/2}\int d^3k\sum_\lambda
v_{\mathbf{k},\lambda}(\eta)\epsilon^i{}_j(\mathbf{k},\lambda)e^{i\mathbf{k}\cdot\mathbf{x}}.
\end{equation}
The field is now quantized by interpreting $v_{\mathbf{k},\lambda}(\eta)$ as the operator
\begin{equation}
\hat{v}_{\mathbf{k},\lambda}(\eta)=v_k(\eta)\hat{a}_{\mathbf{k},\lambda}+v^{\ast}_k(\eta)
\hat{a}^{\dagger}_{-\mathbf{k},\lambda},
\end{equation}
where $v_k(\eta)\epsilon_{ij}(\mathbf{k},\lambda)e^{i\mathbf{k}\cdot\mathbf{x}}$ satisfies
the field equation\footnote{We ignore possible tensor contributions to the energy-momentum
tensor} corresponding to the action (5.64), that is (for $K=0$)
\begin{equation}
v''_k+\left(k^2-\frac{a''}{a}\right)v_k=0.
\end{equation}
(Instead of $z''/z$ in (5.6) we now have the ``mass'' $a''/a$.)

In the long-wavelength regime the growing mode now behaves as $v_k\propto a$, hence $v_k/a$
remains constant.

Again we have to impose the normalization (5.7):
\begin{equation}
v^\ast_kv'_k-v_kv'^{\ast}_k=-i,
\end{equation}
and the asymptotic behavior
\begin{equation}
v_k(\eta) \sim\frac{1}{\sqrt{2k}}e^{-ik\eta}~~~~~~(k/aH\gg 1).
\end{equation}

The decomposition (5.66) translates to
\begin{equation}
H^i{}_j(\eta,\mathbf{x})=(2\pi)^{-3/2}\int d^3k\sum_\lambda
\hat{h}_{\mathbf{k},\lambda}(\eta)\epsilon^i{}_j(\mathbf{k},\lambda)e^{i\mathbf{k}\cdot\mathbf{x}},
\end{equation}
where
\begin{equation}
\hat{h}_{\mathbf{k},\lambda}(\eta)=\left(\frac{8\pi}{M^2_{Pl}}\right)^{1/2}\frac{1}{a}
\hat{v}_{\mathbf{k},\lambda}(\eta).
\end{equation}

We define the \textit{power spectrum of gravitational waves} by
\begin{equation}
\frac{2\pi^2}{k^3}P_g(k)\delta^{(3)}(\mathbf{k}-\mathbf{k}')=\sum_\lambda
\langle0|\hat{h}_{\mathbf{k},\lambda}\hat{h}^{\dagger}_{\mathbf{k'},\lambda}|0\rangle
\end{equation}
thus
\begin{equation}
\sum_\lambda
\langle0|\hat{v}_{\mathbf{k},\lambda}\hat{v}^{\dagger}_{\mathbf{k'},\lambda}|0\rangle=\frac{M^2_{Pl}a^2}{8\pi}
\frac{2\pi^2}{k^3}P_g(k)\delta^{(3)}(\mathbf{k}-\mathbf{k}').
\end{equation}
Using (5.67) for the left-hand side we obtain instead of (5.15)\footnote{In the literature
one often finds an expression for $P_g(k)$ which is 4 times larger, because the power
spectrum is defined in terms of $h_{ij}=2H_{ij}$.}
\begin{equation}
\fbox{$\displaystyle P_g(k)=2 \frac{8\pi}{M^2_{Pl}a^2}\frac{k^3}{2\pi^2}|v_k(\eta)|^2.$}
\end{equation}
The factor $2$ on the right is due to the two polarizations.

\subsection{Power spectrum for power-law inflation}

For the modes $v_k(\eta)$ we need $a''/a$. From
\[\frac{a''}{a}=(a\mathcal{H})'/a=\mathcal{H}^2+\mathcal{H}'=2\mathcal{H}^2\left[1-\frac{1}{2}
(1-\mathcal{H}'/\mathcal{H}^2)\right]\] and (5.39) we obtain the generally valid formula
\begin{equation}
\frac{a''}{a}=2\mathcal{H}^2(1-\varepsilon/2).
\end{equation}
For power-law inflation we had $\varepsilon=1/p,~ a(\eta)\propto\eta^{p/(1-p)}$, thus
\[ \mathcal{H}=\frac{p}{p-1}\frac{1}{\eta}\]
and hence
\begin{equation}
\frac{a''}{a}=\left(\mu^2-\frac{1}{4}\right)\frac{1}{\eta^2},~~~~\mu:=\frac{3}{2}+\frac{1}{p-1}.
\end{equation}

This shows that for power-law inflation $v_k(\eta)$ is identical to $u_k(\eta)$. Therefore,
we have by eq. (5.27)
\begin{equation}
|v_k|=C(\mu)\frac{1}{\sqrt{2k}} \left(\frac{k}{aH}\right)^{-\mu+1/2},
\end{equation}
with
\begin{equation}
C(\mu)=2^{\mu-3/2}\frac{\Gamma(\mu)}{\Gamma(3/2)}(1-\varepsilon)^{\mu-1/2}.
\end{equation}
Inserting this in (5.75) gives
\begin{equation}
P_g(k)=\frac{16\pi}{M^2_{Pl}}\frac{k^3}{2\pi^2}\frac{1}{a^2}C^2(\mu)\frac{1}{2k}\left(\frac{k}{aH}\right)^{1-2\mu}.
\end{equation}
or
\begin{equation}
P_g(k)=C^2(\mu)\frac{4}{\pi}\left(\frac{H}{M_{Pl}}\right)^2\left(\frac{k}{aH}\right)^{1-2\mu}.
\end{equation}
Alternatively, we have
\begin{equation}
\fbox{$\displaystyle
P_g(k)=C^2(\mu)\frac{4}{\pi}\left.\frac{H^2}{M^2_{Pl}}\right|_{k=aH}.$}
\end{equation}

\subsection{Slow-roll approximation}

From (5.76) and (5.44) we obtain again the first equation in (5.77), but with a different
$\mu$:
\begin{equation}
\mu=\frac{1}{1-\varepsilon}+\frac{1}{2}.
\end{equation}
Hence $v_k(\eta)$ is equal to $u_k(\eta)$ if $\nu$ is replaced by $\mu$. The formula
(5.82), with $C(\mu)$ given by (5.79), remains therefore valid, but now $\mu$ is given by
(5.83), where $\varepsilon$ is the slow-roll parameter in (5.34) or (5.39). Again
$C(\mu)\simeq1$.

The power index for tensor perturbations,
\begin{equation}
n_T(k):=\frac{d\ln P_g(k)}{d\ln k},
\end{equation}
can be read off from (5.81):
\begin{equation}
\fbox{$\displaystyle n_T\simeq-2\varepsilon,$}
\end{equation}
showing that the \textit{power spectrum is almost flat}\footnote{The result (5.86) can also
be obtained from (5.82). Making use of an intermediate result in the solution of the
Exercise on p.88 and (5.34), we get \[n_T=\frac{d\ln H^2}{d\varphi}\frac{d\varphi}{d\ln
k}=\frac{2\varepsilon}{\varepsilon-1}\simeq-2\varepsilon.\]}.

\subsubsection{Consistency equation}

Let us collect some of the important formulas:
\begin{eqnarray}
A_S(k):&=& \frac{2}{5}P^{1/2}_{\mathcal{R}}(k)=\frac{4}{5}\left.\frac{H^2}{M^2_{Pl}|dH/d\varphi|}\right|_{k=aH},\\
A_T(k):&=& \frac{1}{5}P^{1/2}_g(k)=\frac{2}{5\sqrt{\pi}}\left.\frac{H}{M_{Pl}}\right|_{k=aH},\\
n-1&=& 2\delta-4\varepsilon,\\
n_T&=& -2\varepsilon.
\end{eqnarray}
The relative amplitude of the two spectra (scalar and tensor) is thus given by
\begin{equation}
\fbox{$\displaystyle \frac{A_T^2}{A_S^2}=\varepsilon
~~~\left(\frac{P_g}{P_\mathcal{R}}=4\varepsilon\right).$}
\end{equation}
More importantly, we obtain the \textit{consistency condition}
\begin{equation}
\fbox{$\displaystyle n_T=-2\frac{A_T^2}{A_S^2},$}
\end{equation}
which is characteristic for inflationary models. In principle this can be tested with CMB
measurements, but there is a long way before this can be done in practice.

\subsection{Stochastic gravitational background radiation}

The spectrum of gravitational waves, generated during the inflationary era and stretched to
astronomical scales by the expansion of the Universe, contributes to the background energy
density. Using the results of the previous section we can compute this.

I first recall  a general formula for the effective energy-momentum tensor of gravitational
waves. (For detailed derivations see Sect. 4.4 of \cite{NS1}.)

By `gravitational waves' we mean propagating ripples in curvature on scales much smaller
than the characteristic scales of the background spacetime (the Hubble radius for the
situation under study). For sufficiently high frequency waves it is meaningful to associate
them -- in an \textit{averaged }sense -- an energy-momentum tensor. Decomposing the full
metric $g_{\mu\nu}$ into a background $\bar{g}_{\mu\nu}$ plus fluctuation $h_{\mu\nu}$, the
effective energy-momentum tensor is given by the following expression
\begin{equation}
T^{(GW)}_{\alpha\beta}=\frac{1}{32\pi G}\left\langle
h_{\mu\nu\mid\alpha}h^{\mu\nu}{}_{\mid\beta}\right\rangle,
\end{equation}
if the gauge is chosen such that $h^{\mu\nu}{}_{\mid\nu}=0,~h^\mu{}_\mu=0$. Here, a
vertical stroke indicates covariant derivatives with respect to the background metric, and
$\langle \cdot\cdot\cdot\rangle$ denotes a four-dimensional average over regions of several
wave lengths.

For a Friedmann background we have in the TT gauge for
$h_{\mu\nu}=2H_{\mu\nu}:\\~h_{\mu0}=0,~h_{ij\mid0}=h_{ij,0}$, thus
\begin{equation}
T^{GW)}_{00}=\frac{1}{8\pi G}\left\langle \dot{H}_{ij}\dot{H}^{ij}\right\rangle.
\end{equation}

As in (5.71) we perform (for $K=0$) a Fourier decomposition
\begin{equation}
H_{ij}(\eta,\mathbf{x})=(2\pi)^{-3/2}\int d^3k\sum_\lambda
h_\lambda(\eta,\mathbf{k})\epsilon_{ij}(\mathbf{k},\lambda)e^{i\mathbf{k}\cdot\mathbf{x}}.
\end{equation}
The gravitational background energy density, $\rho_g$, is obtained by taking the space-time
average in (5.93). At this point we regard $h_\lambda(\eta,\mathbf{k})$ as a random field,
indicated by a hat (since it is on macroscopic scales equivalent to the original quantum
field $\hat{h}_\lambda(\eta,\mathbf{k})$), and replace the spatial average by the
\textit{stochastic average} (for which we use the same notation). Clearly, this is only
justified if some \textit{ergodicity} property holds. This issue will appear again in Part
III, and we shall devote Appendix C for some clarifications.

If we adopt this procedure we obtain, anticipating the $\delta$-function in (5.96),
\begin{equation}
\rho_g=\frac{1}{8\pi Ga^2(2\pi)^3}\int d^3k d^3k'
\sum_\lambda\left\langle\left\langle\hat{h}^{'}_\lambda(\eta,\mathbf{k})
\hat{h}^{'\star}_\lambda(\eta,\mathbf{k'})\right\rangle\right\rangle.
\end{equation}
Here, the average on the right includes also an average over several periods. (As always, a
dot denotes the derivative with respect to the cosmic time $t$, thus $\dot{h}=h'/a$.) Using
(5.72) and (5.67) we obtain for the statistical average
\begin{equation}
\sum_\lambda\left\langle\hat{h}^{'}_\lambda(\eta,\mathbf{k})
\hat{h}^{'\ast}_\lambda(\eta,\mathbf{k'})\right\rangle =2
|h'_k(\eta)|^2\delta^{(3)}(\mathbf{k}-\mathbf{k}'),
\end{equation}
where (see (5.72))
\begin{equation}
h_k(\eta)=\left(\frac{8\pi}{M^2_{Pl}}\right)^{1/2}\frac{1}{a}v_k(\eta).
\end{equation}
Thus
\begin{equation}
\rho_g=\frac{2}{8\pi Ga^2(2\pi)^3}\int d^3k \left\langle|h'_k(\eta)|^2\right\rangle,
\end{equation}
where from now on $\langle\cdot\cdot\cdot\rangle$ denotes the \textit{average over several
periods}. For the spectral density this gives
\begin{equation}
k\frac{d\rho_g(k)}{dk}=\frac{k^3}{Ga^2(2\pi)^3}\left\langle|h'_k(\eta)|^2\right\rangle.
\end{equation}
If $\eta_i$ is some early time, we can write
\begin{equation}
|h'_k(\eta)|^2=\left|\frac{h'_k(\eta)}{h_k(\eta_i)}\right|^2 \frac{\pi^2}{ k^3}
P_g(k,\eta_i),
\end{equation}
where $P_g(k,\eta_i)$ is the power spectrum at $\eta_i$ (for which we may take (5.75)).
Hence we obtain
\begin{equation}
\fbox{$\displaystyle k\frac{d\rho_g(k)}{dk}=\frac{M^2_{Pl}}{8\pi
a^2}\left\langle\left|\frac{h'_k(\eta)}{h_k(\eta_i)}\right|^2\right\rangle P_g(k,\eta_i).$}
\end{equation}
When the radiation is well inside the horizon, we can replace $h'_k$ by $kh_k$.

The differential equation (5.68) reads in terms of $h_k(\eta)$
\begin{equation}
h''+2\frac{a'}{a}h'+k^2h=0.
\end{equation}
For the matter dominated era ($a(\eta)\propto\eta^2$) this becomes
\[ h''+\frac{4}{\eta}h'+k^2h=0.\]

Using 9.1.53 of \cite{Abr} one sees that this is satisfied by $j_1(k\eta)/k\eta$.
Furthermore, by 10.1.4 of the same reference, we have $3j_1(x)/x\rightarrow 1$ for
$x\rightarrow0$ and
\[\left(\frac{j_1(x)}{x}\right)'=-\frac{1}{x}j_2(x)~\rightarrow 0~~~~(x\rightarrow 0).\]
So the correct solution is
\begin{equation}
\frac{h_k(\eta)}{h_k(0)}= 3\frac{j_1(k\eta)}{k\eta}
\end{equation}
if the modes cross inside the horizon during the matter dominated era. Note also that
\begin{equation}
j_1(x)=\frac{\sin x}{x^2}-\frac{\cos x}{x}.
\end{equation}
For modes which enter the horizon earlier, we introduce a transfer function  $T_g(k)$ by
\begin{equation}
\frac{h_k(\eta)}{h_k(0)}=:3\frac{j_1(k\eta)}{k\eta}T_g(k),
\end{equation}
that has to be determined numerically from the differential equation (5.102). We can then
write the result (5.101) as
\begin{equation}
k\frac{d\rho_g(k)}{dk}=\frac{M^2_{Pl}}{8\pi}\frac{k^2}{a^2}P_g^{prim}(k)|T_g(k)|^2
\left\langle\left[\frac{3j_1(k\eta)}{k\eta)}\right]^2\right\rangle,
\end{equation}
where $P_g^{prim}(k)$ denotes the primordial power spectrum. This
holds in particular at the present time $\eta_0$ ($a_0=1$). Since
the time average $\langle\cos^2 k\eta\rangle=\frac{1}{2}$, we
finally obtain for $\Omega_g(k):=\rho_g(k)/\rho_{crit}$ (using
$\eta_0=2H_0^{-1}$)
\begin{equation}
\fbox{$\displaystyle \frac{d\Omega_g(k)}{d\ln
k}=\frac{3}{8}P_g^{prim}(k)|T_g(k)|^2\frac{1}{(k\eta_0)^2}.$}
\end{equation}
Here one may insert the inflationary result (5.82), giving
\begin{equation}
\frac{d\Omega_g(k)}{d\ln
k}=\frac{3}{2\pi}\left.\frac{H^2}{M^2_{Pl}}\right|_{k=aH}|T_g(k)|^2\frac{1}{(k\eta_0)^2}.
\end{equation}

\subsubsection{Numerical results}

Since the normalization in (5.82) can not be predicted, it is reasonable to choose it, for
illustration, to be equal to the observed CMB normalization at large scales. (In reality
the tensor contribution is presumably only a small fraction of this; see (5.90).) Then one
obtains the result shown in Fig. 5.1, taken from \cite{TWL}. This shows that the spectrum
of the stochastic gravitational background radiation is predicted to be flat in the
interesting region, with $d\Omega_g/d\ln(k\eta_0)\sim 10^{-14}$. Unfortunately, this is too
small to be detectable by the future LISA interferometer in space.

\begin{figure}
\begin{center}
\includegraphics[height=0.3\textheight]{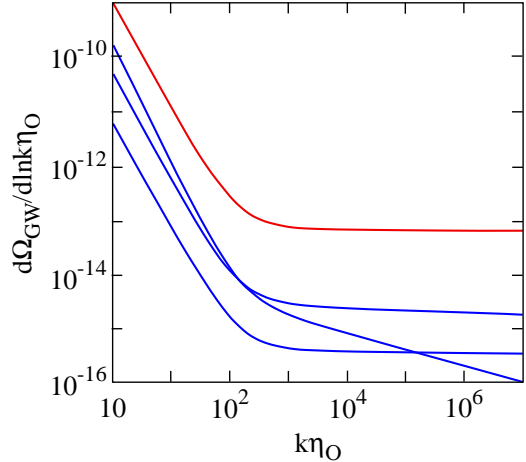}
\caption{Differential energy density (5.108) of the stochastic background of
inflation-produced gravitational waves. The normalization of the dashed curve, representing
the scale-invariant limit, is arbitrary. For the solid curves are normalized to the COBE
quadrupole, and show the result for $n_T=-0.003,~-0.03$, and -0.3. (Adapted from
\cite{TWL}.)} \label{Fig-10}
\end{center}
\end{figure}

------------

\textbf{Exercise}. Consider a massive free scalar field $\phi$ (mass $m$) and discuss the
quantum fluctuations for a de Sitter background (neglecting gravitational back reaction).
Compute the power spectrum as a function of conformal time for $m/H<3/2$.

\textit{Hint:} Work with the field $a\phi$ as a function of conformal time.

\textit{Remark}: This exercise was solved at an astonishingly early time ($\sim$ 1940) by
E. Schr\"{o}dinger.

------------

\section{Appendix to Chapter 5:\\ Einstein tensor for tensor perturbations}

In this Appendix we derive the expressions (5.59) for the tensor perturbations of the
Einstein tensor.

The metric (5.57) is conformal to $\tilde{g}_{\mu\nu}=\gamma_{\mu\nu}+2H_{\mu\nu}$. We
first compute the Ricci tensor $\tilde{R}_{\mu\nu}$ of this metric, and then use the
general transformation law of Ricci tensors for conformally related metrics (see eq.
(2.264) of \cite{NS1}).

Let us first consider the simple case $K=0$, that we considered in Sect. 5.2. Then
$\gamma_{\mu\nu}$ is the Minkowski metric. In the following computation of
$\tilde{R}_{\mu\nu}$ we drop temporarily the tildes.

The Christoffel symbols are immediately found (to first order in $H_{\mu\nu}$)
\begin{eqnarray}
\Gamma^\mu{}_{00} &=& \Gamma^0{}_{0i}=0,~~
\Gamma^0{}_{ij}=H'_{ij},~~\Gamma^i{}_{0j}=(H^i{}_j)',\nonumber \\
\Gamma^i{}_{jk}&=& H^i{}_{j, k}+H^i{}_{k, j}-H_{jk}{}^{, i}.
\end{eqnarray}
So these vanish or are of first order small. Hence, up to higher orders,
\begin{equation}
R_{\mu\nu}=\partial_\lambda\Gamma^\lambda{}_{\nu\mu}-\partial_\nu\Gamma^\lambda{}_{\lambda\mu}.
\end{equation}
Inserting (5.109) and using the TT conditions (5.58) readily gives
\begin{equation}
R_{00}=0,~~~R_{0i}=0,
\end{equation}
\begin{eqnarray*}
R_{ij}&=&\partial_\lambda\Gamma^\lambda{}_{ij}-\partial_j\Gamma^\lambda{}_{\lambda i}=
\partial_0\Gamma^0{}_{ij}+\partial_k\Gamma^k{}_{ij}-\partial_j\Gamma^0{}_{0 i}-
\partial_j\Gamma^k{}_{k i}\nonumber \\
&=&H''_{ij}+(H^k{}_{i,j}+H^k{}_{j,i}-H_{ij}{}^{,k})_{,k}.
\end{eqnarray*}
Thus
\begin{equation}
R_{ij}=H''_{ij}-\triangle H_{ij}.
\end{equation}

Now we use the quoted general relation between the Ricci tensors for two metrics related as
$g_{\mu\nu}=e^f \tilde{g}_{\mu\nu}$. In our case $e^f=a^2(\eta)$, hence
\begin{eqnarray*}
\tilde{\nabla}_\mu f &=& 2\mathcal{H}\delta_{\mu0},~~\tilde{\nabla}_\mu\tilde{\nabla}_\nu
f=\partial_\mu(2\mathcal{H}\delta_{\nu0})-\Gamma^\lambda{}_{\mu\nu}2\mathcal{H}\delta_{\lambda0}\\
& =& 2\mathcal{H}'\delta_{\mu0}\delta_{\nu0}-2\mathcal{H}H'_{\mu\nu},~~\tilde{\triangle}f=
\tilde{g}^{\mu\nu}\tilde{\nabla}_\mu\tilde{\nabla}_\nu f=2\mathcal{H}'.
\end{eqnarray*}
As a result we find
\begin{equation}
R_{\mu\nu}=\tilde{R}_{\mu\nu}+(-2\mathcal{H}'+2\mathcal{H}^2)\delta_{\mu0}\delta_{\nu0}
+(\mathcal{H}'+2\mathcal{H}^2)\tilde{g}_{\mu\nu}+2\mathcal{H}H'_{\mu\nu},
\end{equation}
thus
\begin{eqnarray}
\delta R_{00} &=& \delta R_{0i}=0, \nonumber \\
\delta R_{ij} &=& H''_{ij}-\triangle
H_{ij}+2(\mathcal{H}'+2\mathcal{H}^2)H_{ij}+2\mathcal{H}H'_{ij}.
\end{eqnarray}
From this it follows that
\begin{equation}
\delta R=g^{(0)\mu\nu}\delta R_{\mu\nu}+\delta g^{\mu\nu}R^{(0)}_{\mu\nu}=0.
\end{equation}
The result (5.59) for the Einstein tensor is now easily obtained.

\subsection*{Generalization to $K\neq0$}

The relation (5.113) still holds. For the computation of $\tilde{R}_{\mu\nu}$ we start with
the following general formula for the Christoffel symbols (again dropping tildes).
\begin{equation}
\delta\Gamma^\mu{}_{\alpha\beta}=\gamma^{\mu\nu}(H_{\nu\alpha\mid\beta}+
H_{\nu\beta\mid\alpha}-H_{\alpha\beta\mid\nu})
\end{equation}
(see \cite{NS1}, eq. (2.93)). For the computation of the covariant derivatives
$H_{\alpha\beta\mid\mu}$ with respect to the unperturbed metric $\gamma_{\mu\nu}$, we
recall the unperturbed Christoffel symbols (1.229) with $a\rightarrow 1$,
\begin{equation}
\Gamma^0{}_{00}=\Gamma^0{}_{i0}=\Gamma^i{}_{00}=\Gamma^0{}_{ij}=\Gamma^i{}_{0j}=0,~~\Gamma^i{}_{jk}
=\bar{\Gamma}^i{}_{jk}.
\end{equation}
One readily finds
\begin{equation} H_{\mu0\mid\nu}=0,~~H_{ij\mid0}=H'_{ij},~~H_{ij\mid
k}=H_{ij\parallel k},
\end{equation}
where the double stroke denotes covariant differentiation on $(\Sigma,\gamma)$. Therefore,
\begin{eqnarray}
\delta\Gamma^0{}_{00}&=&\delta\Gamma^0{}_{i0}=\delta\Gamma^i{}_{00}=0,~~\delta\Gamma^0{}_{ij}=H'_{ij},~~
\delta\Gamma^i{}_{0j}=(H^i{}_j)'\nonumber \\
\delta\Gamma^i{}_{jk}&=& H^i{}_{j\parallel k}+H^i{}_{k\parallel j}-H_{jk}{}^{\parallel i}.
\end{eqnarray}

With these expressions we can compute $\delta R_{\mu\nu}$,using the formula (1.249). The
first of the following two equations
\begin{equation}
\delta R_{00}=0,~~\delta R_{0i}=0
\end{equation}
is immediate, while one finds in a first step $\delta R_{0i}=H^k{}_{j\parallel k}$, and
this vanishes because of the TT condition. A bit more involved is the computation of the
remaining components. From (1.249) we have
\begin{eqnarray*}
\delta R_{ij}&=&\partial_\lambda\delta\Gamma^\lambda{}_{ij}-
\partial_j\delta\Gamma^\lambda{}_{\lambda
i} + \delta\Gamma^\sigma{}_{ji}\Gamma^\lambda{}_{\lambda\sigma}+
\Gamma^\sigma{}_{ji}\delta\Gamma^\lambda{}_{\lambda\sigma}-\delta\Gamma^\sigma{}_{\lambda
i}\Gamma^\lambda{}_{j\sigma}-\Gamma^\sigma{}_{\lambda i}\delta\Gamma^\lambda{}_{j\sigma}\\
&=& H''_{ij}+\partial_l\delta\Gamma^l{}_{ij}-\partial_j\delta\Gamma^l{}_{li}
+\delta\Gamma^s{}_{ji}\Gamma^l{}_{ls}+
\Gamma^s{}_{ji}\delta\Gamma^l{}_{ls}-\delta\Gamma^s{}_{l i}\Gamma^l{}_{js}-\Gamma^s{}_{l
i}\delta\Gamma^l{}_{js}.
\end{eqnarray*}
But
\[\delta\Gamma^l{}_{ls}=H^l{}_{l\parallel s}+H^l{}_{s\parallel l}-H_{ls}{}^{\parallel
l}=0,\] so
\[\delta R_{ij}=H''_{ij}+\partial_l\delta\Gamma^l{}_{ij}+\delta\Gamma^s{}_{ji}\Gamma^l{}_{ls}
-\delta\Gamma^s{}_{l i}\Gamma^l{}_{js}-\Gamma^s{}_{l i}\delta\Gamma^l{}_{js}=
H''_{ij}+(\delta\Gamma^l{}_{ij})_{\parallel l}\] or
\begin{equation}
\delta R_{ij}=H''_{ij}+H^l{}_{i\parallel jl}+H^l{}_{j\parallel il}-H_{ij\parallel
l}{}^{\parallel l}.
\end{equation}
In order to impose the TT conditions , we make use of the Ricci identity\footnote{On
$(\Sigma,\gamma)$ we have:\[H^l{}_{i\parallel jl}-H^l{}_{i\parallel
jl}=R^l{}_{slj}H^s{}_i+R_i{}^s{}_{lj}H^l{}_s=3KH_{ij}.\]}
\[H^l{}_{i\parallel jl}=H^l{}_{i\parallel jl}+3KH_{ij},\]
giving
\begin{equation}
\delta R_{ij}=H''_{ij}+6KH_{ij}-\triangle H_{ij}.
\end{equation}

\part{Microwave Background Anisotropies}

\section*{Introduction}

Investigations of the cosmic microwave background have presumably contributed most to the
remarkable progress in cosmology during recent years. Beside its spectrum, which is
Planckian to an incredible degree, we also can study the temperature fluctuations over the
``cosmic photosphere'' at a redshift $z\approx1100$. Through these we get access to crucial
cosmological information (primordial density spectrum, cosmological parameters, etc). A
major reason for why this is possible relies on the fortunate circumstance that the
fluctuations are tiny ($\sim 10^{-5}$ ) at the time of recombination. This allows us to
treat the deviations from homogeneity and isotropy for an extended period of time
perturbatively, i.e., by linearizing the Einstein and matter equations about solutions of
the idealized Friedmann-Lema\^{\i}tre models. Since the physics is effectively {\it
linear}, we can accurately work out the {\it evolution} of the perturbations during the
early phases of the Universe, given a set of cosmological parameters. Confronting this with
observations, tells us a lot about the cosmological parameters as well as the initial
conditions, and thus about the physics of the very early Universe. Through this window to
the earliest phases of cosmic evolution we can, for instance, test general ideas and
specific models of inflation.

Let me add in this introduction some qualitative remarks, before we start with a detailed
treatment. Long before recombination (at temperatures $T>6000 K$, say) photons, electrons
and baryons were so strongly coupled that these components may be treated together as a
single fluid. In addition to this there is also a dark matter component. For all practical
purposes the two interact only gravitationally. The investigation of such a two-component
fluid for small deviations from an idealized Friedmann behavior is a well-studied
application of cosmological perturbation theory, and will be treated in Chapter 6.

At a later stage, when decoupling is approached, this approximate treatment breaks down
because the mean free path of the photons becomes longer (and finally `infinite' after
recombination). While the electrons and baryons can still be treated as a single fluid, the
photons and their coupling to the electrons have to be described by the general
relativistic Boltzmann equation. The latter is, of course, again linearized about the
idealized Friedmann solution. Together with the linearized fluid equations (for baryons and
cold dark matter, say), and the linearized Einstein equations one arrives at a complete
system of equations for the various perturbation amplitudes of the metric and matter
variables. Detailed derivations will be given in Chapter 7. There exist widely used codes
e.g. CMBFAST \cite{S22}, that provide the CMB anisotropies -- for given initial conditions
-- to a precision of about 1\%. A lot of qualitative and semi-quantitative insight into the
relevant physics can, however, be gained by looking at various approximations of the basic
dynamical system.

Let us first discuss the temperature fluctuations. What is observed is the temperature
autocorrelation:
\[
 C(\vartheta ):= \left\langle \frac{\Delta T(\mathbf{n})}{T}\cdot
\frac{\Delta T(\mathbf{n'})}{T}\right\rangle\\
= \sum_{l=2}^\infty \frac{2l+1}{4\pi} C_l P_l(\cos \vartheta) ,
\]
where $\vartheta$ is the angle between the two directions of observation $\mathbf{n},
\mathbf{n'}$, and the average is taken ideally over all sky. The {\it angular power
spectrum} is by definition $\frac{l(l+1)}{2\pi}C_l \; \; versus \; \;l \; \; (\vartheta
\simeq \pi /l ).$

A characteristic scale, which is reflected in the observed CMB anisotropies, is the sound
horizon at last scattering, i.e., the distance over which a pressure wave can propagate
until decoupling. This can be computed within the unperturbed model and subtends about half
a  degree on the sky for typical cosmological parameters. For scales larger than this sound
horizon the fluctuations have been laid down in the very early Universe. These have first
been detected by the COBE satellite. The (gauge invariant brightness) temperature
perturbation $\Theta = \Delta T/T$ is dominated by the combination of the intrinsic
temperature fluctuations and gravitational redshift or blueshift effects. For example,
photons that have to climb out of potential wells for high-density regions are redshifted.
We shall show in Sect. 8.5 that these effects combine for adiabatic initial conditions to
$\frac{1}{3}\Psi$, where $\Psi$ is one of the two gravitational Bardeen potentials. The
latter, in turn, is directly related to the density perturbations. For scale-free initial
perturbations and almost vanishing spatial curvature the corresponding angular power
spectrum of the temperature fluctuations turns out to be nearly flat (Sachs-Wolfe plateau;
see Fig. 8.1 ).

On the other hand, inside the sound horizon before decoupling, acoustic, Doppler,
gravitational redshift, and photon diffusion effects combine to the spectrum of small angle
anisotropies. These result from gravitationally driven synchronized acoustic oscillations
of the photon-baryon fluid, which are damped by photon diffusion (Sect. 8.2).

A particular realization of $\Theta(\mathbf{n})$, such as the one accessible to us (all sky
map from our location), cannot be predicted. Theoretically, $\Theta $ is a random field
$\Theta(\mathbf{x},\eta,\mathbf{n})$, depending on the conformal time $\eta$, the spatial
coordinates, and the observing direction $\mathbf{n}$. Its correlation functions should be
rotationally invariant in $\mathbf{n}$, and respect the symmetries of the background time
slices. If we expand $\Theta$ in terms of spherical harmonics,
\[
\Theta(\mathbf{n}) = \sum_{lm} a_{lm} Y_{lm}(\mathbf{n}),
\]
the random variables $a_{lm}$ have to satisfy\footnote{A formal proof of this can easily be
reduced to an application of Schur's Lemma for the group $SU(2)$ (Exercise).}
\[
\langle a_{lm}\rangle = 0,\;\;\;  \langle a_{lm}^\star a_{l'm'}\rangle =
\delta_{ll'}\delta_{mm'}C_l(\eta),
\]
where the $C_l(\eta)$ depend only on $\eta$. Hence the correlation function at the present
time $\eta_0$ is given by the previous expression with $C_l = C_l(\eta_0)$, and the bracket
now denotes the statistical average. Thus,
\[
C_l = \frac{1}{2l+1}\left\langle\sum_{m=-l}^la_{lm}^\star a_{lm}\right\rangle.
\]
The standard deviations $\sigma(C_l)$ measure a fundamental uncertainty in the knowledge we
can get about the $C_l$'s. These are called {\it cosmic variances}, and are most pronounced
for low $l$. In simple inflationary models the $a_{lm}$ are Gaussian distributed, hence
\[
\frac{ \sigma(C_l)}{C_l} = \sqrt{\frac{2}{2l+1}}.
\]
Therefore, the limitation imposed on us (only one sky in one universe) is small for large
$l$.

\paragraph{Exercise.}
Derive the last equation.

\textit{Solution}: The claim is a special case of the following general fact: Let
$\xi_1,\xi_2,...,\xi_n$ be independent Gaussian random variables with mean 0 and variance
1, and let
\[ \zeta=\frac{1}{n}\sum_{i=1}^n\xi_i^2.\]
Then the variance and standard deviation of $\zeta$ are
\[ var(\zeta)=\frac{2}{n},~~~\sigma(\zeta)=\sqrt{\frac{2}{n}}.\]
To show this, we use the equation of Bienaym\'{e}
\[ var(\zeta)=\frac{1}{n^2}\sum_{i=1}^n var(\xi_i^2),\]
and the following formula for the variance for each $\xi_i^2$:
\[ var(\xi^2)=\langle\xi^4\rangle-\langle\xi^2\rangle^2=1\cdot3-1=2\]
(the even moments of $\xi$ are $m_{2k}=1\cdot3\cdot~\cdot\cdot\cdot(2k-1)$).

Alternatively, we can use the fact that $\sum_{i=1}^n \xi_i^2$ is $\chi_n^2$-distributed,
with distribution function ($p=n/2,~\lambda=1/2$):
\[ f(x)= \frac{\lambda^p}{\Gamma(p)}x^{p-1}e^{-\lambda x}\]
for $x>0$, and 0 otherwise. This gives the same result.

\chapter{Tight Coupling Phase}

Long before recombination, photons, electrons and baryons are so strongly coupled that
these components may be treated as a single fluid, indexed by $r$ in what follows. Beside
this we have to include a CDM component for which we we use the index $d$ (for `dust' or
dark). For practical purposes these two fluids interact only gravitationally.

\section{Basic equations}

We begin by specializing the basic equations, derived in Part I and collected in Sect.1.5.C
to the situation just described. Beside neglecting the spatial curvature ($K=0$), we may
assume $q_\alpha=\Gamma_\alpha=0,~E_\alpha=F_\alpha=0$ (no energy and momentum exchange
between $r$ and $d$). In addition, it is certainly a good approximation to neglect in this
tight coupling era the anisotropic stresses $\Pi_\alpha$. Then $\Psi=-\Phi$ and since
$\Gamma_{int}=0$ the amplitude $\Gamma$ for entropy production is proportional to
\begin{equation}
S:=S_{dr}=\frac{\Delta_{cd}}{1+w_{d}}-\frac{\Delta_{cr}}{1+w_{r}},~~~~
\frac{w}{1+w}\Gamma=\frac{h_d h_r}{h^2}(c^2_d-c^2_r)S.
\end{equation}
We also recall the definition (1.221)
\begin{equation}
c^2_z=\frac{h_r}{h}c^2_d+\frac{h_d}{h}c^2_r.
\end{equation}
The energy and momentum equations are
\begin{eqnarray}
\Delta'-3\frac{a'}{a}w\Delta &=& -k(1+w)V,\\
V'+\frac{a'}{a}V &=& k\Psi+k\frac{c^2_s}{1+w}\Delta+k\frac{w}{1+w}\Gamma.
\end{eqnarray}
By (1.290) the derivative of $S$ is given by
\begin{equation}
S'=-kV_{dr},
\end{equation}
and that of $V_{dr}$ follows from (1.289):
\begin{equation}
V'_{dr}+\frac{a'}{a}(1-3c_z^2)V_{dr}\nonumber \\
=k(c_d^2-c_r)\frac{\Delta}{1+w}+kc_z^2S.
\end{equation}
In the constraint equation (1.261) we use the Friedmann equation for $K=0$,
\begin{equation}
\frac{8\pi G\rho}{3H^2}=1,
\end{equation}
and obtain
\begin{equation}
\Phi=-\Psi=\frac{3}{2}\left(\frac{Ha}{k}\right)^2\Delta.
\end{equation}

It will be convenient to introduce the comoving wave number in units of the Hubble length
$x:=Ha/k$ and the renormalized scale factor $\zeta:=a/a_{eq}$, where $a_{eq}$ is the scale
factor at the `equality time' (see Sect. 0.3.E). Then the last equation becomes
\begin{equation}
\fbox{$\displaystyle \Phi=-\Psi=\frac{3}{2}x^2\Delta .$}
\end{equation}
Using $\zeta'=kx\zeta$ and introducing the operator $D:=\zeta d/d\zeta$ we can write (6.3)
as
\begin{equation}
\fbox{$\displaystyle (D-3w)\Delta=-\frac{1}{x}(1+w)V.$}
\end{equation}
Similarly, (6.4) (together with (6.1)) gives
\begin{equation}
\fbox{$\displaystyle
(D+1)V=\frac{\Psi}{x}+\frac{c^2_s}{x}\frac{\Delta}{1+w}+\frac{1}{x}\frac{h_d
h_r}{h^2}(c^2_d-c^2_r)S.$}
\end{equation}
We also rewrite (6.5) and (6.6)
\begin{equation}
\fbox{$\displaystyle DS=-\frac{1}{x}V_{dr},$}
\end{equation}
\begin{equation}
\fbox{$\displaystyle (D+1-3c_z^2)V_{dr}
=\frac{1}{x}(c_d^2-c_r)\frac{\Delta}{1+w}+\frac{1}{x}c_z^2S.$}
\end{equation}

It will turn out to be useful to work alternatively with the equations of motion for
$V_\alpha$ and
\begin{equation}
X_\alpha:=\frac{\Delta_{c\alpha}}{1+w_\alpha}~~~(\alpha=r,d).
\end{equation}
From (1.288) we obtain
\begin{equation}
V'_\alpha+\frac{a'}{a}V_\alpha=k\Psi +k\frac{c_\alpha^2}{1+w_\alpha}\Delta_{\alpha},
\end{equation}
Here, we replace $\Delta_\alpha$ by $\Delta_{c\alpha}$ with the help of (1.174) and
(1.175), implying (in the harmonic decomposition)
\begin{equation}
\Delta_\alpha=\Delta_{c\alpha}+3(1+w_\alpha)\frac{a'}{a}\frac{1}{k}(V_\alpha-V).
\end{equation}
We then get
\begin{equation}
V'_\alpha+\frac{a'}{a}(1-3c^2_\alpha)V_\alpha=k\Psi+kc^2_\alpha
X_\alpha-3\frac{a'}{a}c^2_\alpha V.
\end{equation}
From (1.287) we find, using (6.1),
\begin{equation}
X'_\alpha=-kV_\alpha+3\frac{a'}{a}c^2_s\frac{\Delta}{1+w}+3\frac{a'}{a}\frac{h_d
h_r}{h^2}(c^2_d-c^2_r)S.
\end{equation}

Rewriting the last two equations as above, we arrive at the system
\begin{eqnarray}
(D+ 1-3c^2_\alpha)V_\alpha &=& \frac{\Psi}{x}+\frac{c^2_\alpha}{x}X_\alpha-3c^2_\alpha V,\\
DX_\alpha &=& -\frac{V_\alpha}{x}+3c^2_s\frac{\Delta}{1+w}+3\frac{h_d
h_r}{h^2}(c^2_d-c^2_r)S.
\end{eqnarray}
This system is closed, since by (6.1), (1.272) and (1.275)
\begin{equation}
S=X_d-X_r,~~ \frac{\Delta}{1+w}=\sum_\alpha\frac{h_\alpha}{h}X_\alpha,~~
V=\sum_\alpha\frac{h_\alpha}{h}V_\alpha.
\end{equation}
Note also that according to (1.220)
\begin{equation}
\frac{\Delta}{1+w}=X_r+\frac{h_d}{h}S=X_d-\frac{h_r}{h}S.
\end{equation}
From these basic equations we now deduce second order equations for the pair ($\Delta,S$),
respectively, for $X_\alpha~(\alpha=r,d)$. For doing this we note that for any function $f,
f'=(a'/a)Df $, in particular (using (1.80) and (1.62))
\begin{equation}
Dx=-\frac{1}{2}(3w+1)x,~~Dw=-3(1+w)(c^2_s-w).
\end{equation}
The result of the somewhat tedious but straightforward calculation is \cite{KS87}:
\begin{eqnarray}
\lefteqn{D^2\Delta+\left[\frac{1-3w}{2}+3c^2_s-6w\right]D\Delta} \nonumber \\ &&
+\left[\frac{c^2_s}{x^2}-3w+9(c^2_s-w)+
\frac{3}{2}(3w^2-1)\right]\Delta=\frac{1}{x^2}\frac{h_r h_d}{\rho h}(c^2_r-c^2_d)S,\nonumber \\
\end{eqnarray}
\begin{equation}
D^2S+\left[\frac{1-3w}{2}-3c^2_z\right]DS+\frac{c^2_z}{x^2}S=\frac{c^2_r-c^2_d}{x^2(1+w)}\Delta
\end{equation}
for the pair $\Delta,S$, and
\begin{eqnarray}
\lefteqn{D^2X_\alpha+\left[\frac{1-3w}{2}-3c^2_\alpha\right]DX_\alpha }\nonumber \\
&& +\left\{\frac{c^2_\alpha}{x^2}-\frac{h_\alpha}{h}\left[\frac{3}{2}(1+w)+
\frac{3}{2}(1-3w)c^2_\alpha+9c^2_\alpha(c^2_s-c^2_\alpha)+3Dc^2_s\right]\right\}X_\alpha
\nonumber \\ && =3\frac{h_\beta}{h}\left[(c^2_\beta-c^2_\alpha)D+
\frac{1+w}{2}+\frac{1-3w}{2}c^2_\beta+3c^2_\beta(c^2_s-c^2_\beta)+Dc^2_\beta\right]X_\beta\nonumber
\\
\end{eqnarray}
for the pair $X_\alpha$.

\subsubsection{Alternative system for tight coupling limit}

Instead of the first order system (6.17), (6.18) one may work with similar equations for
the amplitudes $\Delta_{s\alpha}$ and $V_\alpha$. From (1.291) we obtain instead of (6.17)
for $\Pi_\alpha=F_\alpha=0$
\begin{equation}
V'_\alpha+\frac{a'}{a}(1-3c^2_\alpha)V_\alpha=k\Psi+k\frac{c^2_\alpha}{1+w_\alpha}
\Delta_{s\alpha}.
\end{equation}
Beside this we have eq. (1.286)
\begin{equation}
\left(\frac{\Delta_{s\alpha}}{1+w_\alpha}\right)'=-kV_\alpha-3\Phi'.
\end{equation}
To this we add the following consequence of the constraint equations (1.261), (1.262) and
the relations (1.260), (1.274), (1.275):
\begin{equation}
k^2\Psi=-4\pi
Ga^2\sum_\alpha\left[\rho_\alpha\Delta_{s\alpha}+3\frac{aH}{k}\rho_\alpha(1+w_\alpha)V_\alpha\right].
\end{equation}
Instead one can also use, for instance for generating numerical solutions, the following
first order differential equation that is obtained similarly
\begin{equation}
k^2\Psi +3\frac{a'}{a}(\Psi'+\frac{a'}{a}\Psi)=-4\pi
Ga^2\sum_\alpha\rho_\alpha\Delta_{s\alpha}.
\end{equation}

\subsubsection{Adiabatic and isocurvature perturbations}

These differential equations have to be supplemented with initial conditions. Two linearly
independent types are considered for some very early stage, for instance at the end of the
inflationary era:

\vspace{0.5cm}

 $\bullet$ \textbf{adiabatic} perturbations: all $S_{\alpha\beta}=0$, but $\mathcal{R}\neq
 0$;

\vspace{0.5cm}

$\bullet$ \textbf{isocurvature} perturbations: some $S_{\alpha\beta}\neq 0$, but
$\mathcal{R}=0$.  \vspace{0.5cm}

Recall that $\mathcal{R}$ measures the spatial curvature for the slicing $\mathcal{Q}=0$.
According to the initial definition (1.58) of $\mathcal{R}$ and the eqs. (6.9), (6.10) we
have
\begin{equation}
\mathcal{R}=\Phi-xV=\frac{x^2}{1+w}\left[D+\frac{3}{2}(1-w)\right]\Delta.
\end{equation}

\subsubsection{Explicit forms of the two-component differential equations}

At this point we make use of the equation of state for the two-component model under
consideration. It is convenient to introduce a parameter $c$ by
\begin{equation}
R:=\frac{3\rho_b}{4\rho_\gamma}=\frac{\zeta}{c}~~\Rightarrow
\frac{\Omega_d}{\Omega_b}=\frac{3c}{4}-1.
\end{equation}
We then have for various background quantities
\begin{eqnarray}
\frac{\rho_d}{\rho_{eq}} &=& \frac{1}{2}\left(1-\frac{4}{3c}\right)\frac{1}{\zeta^3},~~
p_d=0,\nonumber \\
\frac{\rho_r}{\rho_{eq}} &=& \frac{2}{3}\frac{\zeta+3c/4}{c}\frac{1}{\zeta^4},~~
\frac{p_r}{\rho_{eq}} = \frac{1}{6}\frac{1}{\zeta^4},\nonumber \\
\frac{\rho}{\rho_{eq}} &=& \frac{1}{2}(\zeta+1)\frac{1}{\zeta^4},~~
\frac{p}{\rho_{eq}}=\frac{1}{6}\frac{1}{\zeta^4},\nonumber \\
\frac{h_r}{h} &=& \frac{4}{3}\frac{\zeta+c}{c(\zeta+4/3)}, ~~
\frac{h_d}{h}=\left(1-\frac{4}{3c}\right)\frac{\zeta}{\zeta+4/3}, \nonumber \\
w &=& \frac{1}{3(\zeta+1)},~~w_r=\frac{c}{4\zeta+3c}, ~~ w_d=0,\nonumber \\
c^2_d &=& 0, ~~ c_r^2=\frac{1}{3}\frac{c}{\zeta+c}, ~~
c^2_s=\frac{4}{9}\frac{1}{\zeta+4/3}, ~~
c^2_z=\frac{1}{3}\frac{(c-4/3)\zeta}{(\zeta+c)(\zeta+4/3)}, \nonumber \\
H^2 &=& H^2_{eq}\frac{\zeta+1}{2}\frac{1}{\zeta^4},
~~x^2=\frac{\zeta+1}{2\zeta^2}\frac{1}{\omega^2}, ~~ \omega:=\frac{1}{x_{eq}}=
\left(\frac{k}{aH}\right)_{eq}.
\end{eqnarray}

Since we now know that the dark matter fraction is much larger than the baryon fraction, we
write the basic equations only in the limit $c\rightarrow\infty$. (For finite $c$ these are
given in \cite{KS87}.) Eq.(6.26) leads to the pair
\begin{eqnarray}
\lefteqn{D^2X_r+\left(\frac{1}{2}\frac{\zeta}{1+\zeta}-1\right)DX_r}\nonumber \\&&
 + \left\{\frac{2}{3}\frac{\omega^2\zeta^2}{1+\zeta}
+\frac{4}{3}\frac{1}{\zeta+4/3}\left[\frac{\zeta}{\zeta+4/3}-2\right]\right\}X_r=
\left[\frac{3}{2}\frac{\zeta}{\zeta+1}-\frac{\zeta}{\zeta+4/3}D\right]X_d,\nonumber \\
\end{eqnarray}
\begin{equation}
\left\{D^2+\frac{1}{2}\frac{\zeta}{1+\zeta}D-\frac{3}{2}\frac{\zeta}{1+\zeta}\right\}X_d
=\frac{4}{3}\frac{1}{\zeta+4/3}\left[D+2-\frac{\zeta}{\zeta+4/3}\right]X_r.
\end{equation}

From (6.24) and (6.25) we obtain on the other hand
\begin{eqnarray}
\lefteqn{
D^2\Delta+\left(-1+\frac{5}{2}\frac{\zeta}{\zeta+1}-\frac{\zeta}{\zeta+4/3}\right)D\Delta}\nonumber
\\ &&
+\left\{-2+\frac{3}{4}\zeta+\frac{1}{2}\left(\frac{\zeta}{\zeta+1}\right)^2-\frac{3\zeta^2}{\zeta+1}
+\frac{9\zeta^2}{4(\zeta+4/3)}\right\}\Delta\nonumber
\\ &&=\frac{8}{9}\omega^2\frac{\zeta^2}{(\zeta+1)^2(\zeta+4/3)} \left[\zeta
S-(\zeta+1)\Delta\right],
\end{eqnarray}
\begin{eqnarray}
\lefteqn{D^2S+\left(\frac{1}{2}\frac{1}{\zeta+1}-\frac{1}{\zeta+4/3}\right)\zeta DS
}\nonumber \\ && +\frac{2}{3}\omega^2\frac{\zeta^3}{(\zeta+1)(\zeta+4/3)} S
=\frac{2}{3}\omega^2 \frac{\zeta^2}{\zeta+4/3}\Delta.
\end{eqnarray}

We also note that (6.31) becomes
\begin{equation}
\mathcal{R}=\frac{1}{2\omega^2}\frac{\zeta+1}{\zeta^2(\zeta+4/3)}
\left[(\zeta+1)D+\frac{3}{2}\zeta+1\right]\Delta.
\end{equation}

We can now define more precisely what we mean by the two types of primordial initial
perturbations by considering solutions of our perturbation equations for $\zeta\ll 1$.

$\bullet$ \textit{adiabatic} (or \textit{curvature}) perturbations: growing mode behaves as
\begin{eqnarray}
\Delta &=&\zeta^2\left[1-\frac{17}{16}\zeta+\cdot\cdot\cdot\right]-
\frac{\omega^2}{15}\zeta^4[1-\cdot\cdot\cdot],\nonumber \\
S &=&
\frac{\omega^2}{32}\zeta^4\left[1-\frac{28}{25}\zeta+\cdot\cdot\cdot\right];~~~\Rightarrow
\mathcal{R}=\frac{9}{8\omega^2}(1+\mathcal{O}(\zeta)).
\end{eqnarray}

$\bullet$ \textit{isocurvature} perturbations: growing mode behaves as
\begin{eqnarray}
\Delta &=& \frac{\omega^2}{6}\zeta^3
\left[1-\frac{17}{10}\zeta+\cdot\cdot\cdot\right],\nonumber \\
S &=& 1-\frac{\omega^2}{18}\zeta^3 \left[1-\cdot\cdot\cdot\right]; ~~~\Rightarrow
\mathcal{R}=\frac{1}{4}\zeta(1+\mathcal{O}(\zeta)).
\end{eqnarray}

From (6.21) and (6.22) we obtain the relation between the two sets of perturbation
amplitudes:
\begin{eqnarray}
X_r &=& \frac{\zeta+1}{\zeta+4/3}\Delta-\frac{\zeta}{\zeta+4/3}S,~~
X_d=\frac{\zeta+1}{\zeta+4/3}\Delta+\frac{4}{3}\frac{1}{\zeta+4/3}S, \\
\Delta &=& \frac{1}{\zeta+1}\left(\frac{4}{3}X_r+\zeta X_d\right),~~S=X_d-X_r.
\end{eqnarray}

\section{Analytical and numerical analysis}

The system of linear differential equations (6.34)-(6.37) has been discussed analytically
in great detail in \cite{KS87}. One learns, however, more about the physics of the
gravitationally coupled fluids in a mixed analytical-numerical approach.

\subsection{Solutions for super-horizon scales}

For super-horizon scales ($x\gg 1$) eq. (6.12) implies that $S$ is
constant. If the mode enters the horizon in the matter dominated
era, then the parameter $\omega$ in (6.33) is small. For $\omega\ll
1$ eq. (6.36) reduces to
\begin{eqnarray}
\lefteqn{
D^2\Delta+\left(-1+\frac{5}{2}\frac{\zeta}{\zeta+1}-\frac{\zeta}{\zeta+4/3}\right)D\Delta}\nonumber
\\ &&
+\left\{-2+\frac{3}{4}\zeta+\frac{1}{2}\left(\frac{\zeta}{\zeta+1}\right)^2-\frac{3\zeta^2}{\zeta+1}
+\frac{9\zeta^2}{4(\zeta+4/3)}\right\}\Delta\nonumber
\\ &&=\frac{8}{9}\omega^2\frac{\zeta^3}{(\zeta+1)^2(\zeta+4/3)}S.
\end{eqnarray}
For \textit{adiabatic} modes we are led to the homogeneous equation already studied in
Sect. 2.1, with the two independent solutions $U_g$ and $U_d$ given in (2.28) and (2.29).
Recall that the Bardeen potentials remain constant both in the radiation and in the matter
dominated eras. According to (2.32) $\Phi$ decreases to 9/10 of the primordial value
$\Phi^{prim}$.

For \textit{isocurvature} modes we can solve (6.41) with the Wronskian method, and obtain
for the growing mode \cite{KS87}
\begin{equation}
\Delta_{iso}=\frac{4}{15}\omega^2S\zeta^3\frac{3\zeta^2+22\zeta+24+4(3\zeta+4)\sqrt{1+\zeta}}
{(\zeta+1)(3\zeta+4)[1+(1+\zeta)^{1/2}]^4}.
\end{equation}
thus
\begin{equation}
\Delta_{iso}\simeq \left\{\begin{array}{r@{\quad:\quad}l} \frac{1}{6}\omega^2S\zeta^3 & \zeta\ll 1\\
\frac{4}{15}\omega^2S\zeta & \zeta\gg 1 .\end{array} \right.
\end{equation}

\subsection{Horizon crossing}

We now study the behavior of adiabatic modes more closely, in particular what happens in
horizon crossing.

\subsubsection{Crossing in radiation dominated era}

When the mode enters the horizon in the radiation dominated phase we can neglect in (6.36)
the term proportional to $S$ for $\zeta<1$. As long as the radiation dominates $\zeta$ is
small, whence (6.36) gives in leading order
\begin{equation}
(D^2-D-2)\Delta=-\frac{2}{3}\omega^2\zeta^2\Delta.
\end{equation}
(This could also be directly obtained from (6.24), setting $c^2_s\simeq1/3,~w\simeq 1/3$.)
Since $D^2-D=\zeta^2d^2/d\zeta^2$ this perturbation equation can be written as
\begin{equation}
\left[\zeta^2\frac{d^2}{d\zeta^2}+\left(\frac{2}{3}\omega^2\zeta^2-2\right)\right]\Delta=0.
\end{equation}

Instead of $\zeta$ we choose as independent variable the comoving sound horizon $r_s$ times
$k$. We have
\[r_s=\int c_sd\eta=\int c_s\frac{d\eta}{d\zeta}d\zeta,\]
with $c_s\simeq1/\sqrt{3},~
d\zeta/d\eta=kx\zeta=aH\zeta=(aH)/(aH)_{eq})(k/\omega)\zeta\simeq (k/\omega\sqrt{2})$, thus
$\zeta\simeq (k/\sqrt{2}\omega)\eta$ and
\begin{equation}
u:=kr_s\simeq\sqrt{\frac{2}{3}}\omega\zeta\simeq k\eta/\sqrt{3}.
\end{equation}
Therefore, (6.45) is equivalent to
\begin{equation}
\fbox{$\displaystyle\left[\frac{d^2}{du^2}+\left(1-\frac{2}{u^2}\right)\right]\Delta=0.$}
\end{equation}
This differential equation is well-known. According to 9.1.49 of \cite{Abr} the functions
$w(x)\propto x^{1/2}\mathcal{C}_\nu(\lambda x),~\mathcal{C}_\nu\propto
H^{(1)}_\nu,H^{(2)}_\nu$, satisfy
\begin{equation}
w''+\left(\lambda-\frac{\nu^2-\frac{1}{4}}{x^2}\right)w=0.
\end{equation}
Since $j_\nu(x)=\sqrt{\pi/2x}J_{\nu+1/2}(x),~n_\nu(x)=\sqrt{\pi/2x}Y_{\nu+1/2}(x)$, we see
that $\Delta$ is a linear combination of $uj_1(u)$ and $un_1(u)$:
\begin{equation}
\Delta(\zeta)=Cuj_1(u)+Dun_1(u);~~~u=\sqrt{\frac{2}{3}}\omega\zeta~~~
(u=kr_s=\frac{k\eta}{\sqrt{3}}).
\end{equation}
Now,
\begin{equation}
xj_1(x)=\frac{1}{x}\sin x-\cos x,~~~xn_1(x)=-\frac{1}{x}\cos x-\sin x.
\end{equation}
On super-horizon scales $u=kr_s\ll 1$, and $uj_1(u)\approx u\propto a$, while
$un_1(u)\approx-1/u\propto 1/a$. Thus the first term in (6.49) corresponds to the growing
mode. If we only keep this, we have
\begin{equation}
\Delta(\zeta)\approx C\left(\frac{1}{u}\sin u-\cos u\right).
\end{equation}
Once the mode is deep within the Hubble horizon only the $\cos$-term survives. This is an
important result, because if this happens long before recombination we can use for
adiabatic modes the \textit{initial condition}
\begin{equation}
\fbox{$\displaystyle \Delta(\eta)\propto cos[kr_s(\eta)].$}
\end{equation}
We conclude that all adiabatic modes are temporally correlated (\textbf{synchroni\-zed}),
while they are spatially uncorrelated (random phases). This is one of the basic reasons for
the appearance of acoustic peaks in the CMB anisotropies. Note also that, as a result of
(6.9) and (6.33), $\Phi\propto \Delta/\zeta^2\propto \Delta/u^2$, i.e.,
\begin{equation}
\Psi=3\Psi^{(prim)}\left[\frac{\sin u-u\cos u}{u^3}\right].
\end{equation}
Thus: If the mode enters the horizon during the radiation dominated era, its
\textit{potential begins to decay}.

As an exercise show that for isocurvature perturbations the $\cos$ in (6.52) has to be
replaced by the $\sin$ (out of phase).

We could have used in the discussion above the system (6.34) and (6.35). In the same limit
it reduces to
\begin{equation}
\left(D^2-D-2+\frac{2}{3}\omega^2\zeta^2\right)X_r\simeq 0,~~~D^2X_d\simeq(D+2)X_r.
\end{equation}
As expected, the equation for $X_r$ is the same as for $\Delta$. One also sees that $X_d$
is driven by $X_r$, and is growing logarithmically for $\omega\gg1$.

The previous analysis can be improved by not assuming radiation domination and also
including baryons (see \cite{KS87}). It turns out that for $\omega\gg 1$ the result (6.54)
is not much modified: The $\cos$-dependence remains, but with the exact sound horizon; only
the amplitude is slowly varying in time $\propto(1+R)^{-1/4}$.

Since the matter perturbation is driven by the radiation, we may use the potential (6.55)
and work out its influence on the matter evolution. It is more convenient to do this for
the amplitude $\Delta_{sd}$ (instead of $\Delta_{cd}$), making use of the equations (6.27)
and (6.28) for $\alpha=d$:
\begin{equation}
\Delta'_{sd} = -kV_d-3\Phi',~~~~ V'_d = -\frac{a'}{a}V_d -k\Phi.
\end{equation}
Let us eliminate $V_d$:
\[\Delta''_{sd}=-V'_d-3\Phi''=\frac{a'}{a}kV_d+k^2\Phi-3\Phi''=\frac{a'}{a}(-\Delta'_{sd}-3\Phi')+k^2\Phi-3\Phi''.\]
The resulting equation
\begin{equation}
\Delta''_{sd}+\frac{a'}{a}\Delta'_{sd}= k^2\Phi-3\Phi''-3\frac{a'}{a}\Phi'
\end{equation}
can be solved with the Wronskian method. Two independent solutions of the homogeneous
equation are $\Delta_{sd}=const.$ and $\Delta_{sd}=\ln(a)$. These determine the Green's
function in the standard manner. One then finds in the radiation dominated regime (for
details, see \cite{Cos4}, p.198)
\begin{equation}
\Delta_{sd}(\eta)=A\Phi^{prim}\ln(Bk\eta),
\end{equation}
with $A\simeq 9.0,~B\simeq0.62$.

\subsubsection{Matter dominated approximation}

As a further illustration we now discuss the matter dominated approximation. For this
($\zeta\gg 1$) the system (6.34),(6.35) becomes

\begin{eqnarray}
\left(D^2-\frac{1}{2}D+\frac{2}{3}\omega^2\zeta\right)X_r &=&
\left(-D+\frac{3}{2}\right)X_d, \\
\left(D^2+\frac{1}{2}D-\frac{3}{2}\right)X_d &=& 0.
\end{eqnarray}
As expected, the equation for $X_d$ is independent of $X_r$, while the radiation
perturbation is driven by the dark matter. The solution for $X_d$ is
\begin{equation}
X_d=A\zeta+B\zeta^{-3/2}.
\end{equation}
Keeping only the growing mode, (6.60) becomes
\begin{equation}
\frac{d}{d\zeta}\left(\zeta\frac{dX_r}{d\zeta}\right)-\frac{1}{2}\frac{dX_r}{d\zeta}
+\frac{2}{3}\omega^2\left(X_r-\frac{3A}{4\omega^2}\right)=0.
\end{equation}
Substituting
\[X_r=:\frac{3A}{4\omega^2}+\zeta^{-3/4}f(\zeta), \]
we get for $f(\zeta)$ the following differential equation
\begin{equation}
f''=-\left(\frac{3}{16}\frac{1}{\zeta^2}+\frac{2}{3}\frac{\omega^2}{\zeta}\right)f.
\end{equation}

For $\omega\gg1$ we can use the WKB approximation
\[f=\frac{\zeta^{1/4}}{\sqrt{\omega}}\exp\left(\pm
i\sqrt{\frac{8}{3}}\omega\zeta^{1/2}\right),
\]
implying the following oscillatory behavior of the radiation
\begin{equation}
X_r=\frac{3A}{4\omega^2}+B\frac{1}{\sqrt{\omega\zeta}}\exp\left(\pm
i\sqrt{\frac{8}{3}}\omega\zeta^{1/2}\right).
\end{equation}

A look at (6.42) shows that this result for $X_d,X_r$ implies the constancy of the Bardeen
potentials in the matter dominated era.

\subsection{Sub-horizon evolution}

For $\omega\gg1$ one may expect on physical grounds that the dark matter perturbation $X_d$
eventually evolves independently of the radiation. Unfortunately, I can not see this from
the basic equations (6.34), (6.35). Therefore, we choose a different approach, starting
from the alternative system (6.27) -- (6.29). This implies
\begin{eqnarray}
\Delta'_{sd} &=& -kV_d-3\Phi', \\
V'_d &=& -\frac{a'}{a}V_d -k\Phi, \\
k^2\Phi &=& 4\pi Ga^2[\rho_d\Delta_{sd} + \cdot\cdot\cdot].
\end{eqnarray}
As an approximation, we drop in the last equation the radiative\footnote{The growth in the
matter perturbations implies that eventually $\rho_d\Delta_{sd}>\rho_r\Delta_{sr}$ even if
$\Delta_{sd}<\Delta_{sr}$.} and velocity contributions that have not been written out. Then
we get a closed system which we again write in terms of the variable $\zeta$:
\begin{eqnarray}
D\Delta_{sd} &=& -\frac{1}{x}V_d-3D\Phi, \\
DV_d &=& -V_d-\frac{1}{x}\Phi, \\
\Phi &\simeq & \frac{3}{4}\frac{1}{\omega^2}\frac{1}{\zeta}\Delta_{sd}.
\end{eqnarray}
In the last equation we used $\rho_d=(\zeta/\zeta+1)\rho$, (6.7) and the expression (6.33)
for $x^2$.

For large $\omega$ we can easily deduce a second order equation for $\Delta_{sd}$: Applying
$D$ to (6.69) and using (6.70) gives
\begin{eqnarray*}
D^2\Delta_{sd}&=&-\frac{1}{x}DV_d+\frac{1}{x^2}(Dx)V_d-3D^2\Phi\\
&=&\frac{1}{x^2}\Phi+\frac{1}{2}(1-3w)\frac{1}{x}V_d -3D^2\Phi\\
&=&\frac{1}{x^2}\Phi-\frac{1}{2}(1-3w)D\Delta_{sd}-\frac{3}{2}(1-3w)D\Phi-3D^2\Phi.
\end{eqnarray*}
Because of (6.71) the last two terms are small, and we end up (using again (6.33)) with
\begin{equation}
\left\{D^2+\frac{1}{2}\frac{\zeta}{1+\zeta}D-\frac{3}{2}\frac{\zeta}{1+\zeta}\right\}\Delta_{sd}
=0,
\end{equation}
known in the literature as the \textit{Meszaros equation}. Note that this agrees, as was to
be expected, with the homogeneous equation belonging to (6.35).

The Meszaros equation can be solved analytically. On the basis of (6.62) one may guess that
one solution is linear in $\zeta$. Indeed, one finds that
\begin{equation}
X_d(\zeta)=D_1(\zeta)=\zeta+2/3
\end{equation}
is a solution. A linearly independent solution can then be found by
quadratures. It is a general fact that $f(\zeta):=
\Delta_{sd}/D_1(\zeta)$ must satisfy a differential equation which
is first order for $f'$. One readily finds that this equation is
\[(1+\frac{3\zeta}{2})f'' +\frac{1}{4\zeta(\zeta+1)}[21\zeta^2+24\zeta+4]f'=0.\]
The solution for $f'$ is
\[ f'\propto(\zeta+2/3)^{-2}\zeta^{-1}(\zeta+1)^{-1/2}.\]
Integrating once more provides the second solution of (6.72)
\begin{equation}
D_2(\zeta)=D_1(\zeta)
\ln\left[\frac{\sqrt{1+\zeta}+1}{\sqrt{1+\zeta}-1}\right]-2\sqrt{1+\zeta}.
\end{equation}
For late times the two solutions approach to those found in (6.62).

The growing and the decaying solutions $D_1,D_2$ have to be superposed such that a match to
(6.59) is obtained.

\subsection{Transfer function, numerical results}

According to (2.31),(2.32) the early evolution of $\Phi$ on super-horizon scales is given
by\footnote{The origin of the factor 9/10 is best seen from the constancy of $\mathcal{R}$
for super-horizon perturbations, and eq. (4.67).}
\begin{equation}
\Phi(\zeta)=\Phi^{(prim)}\frac {9}{10}\frac{\zeta+1}{\zeta^2}U_g \simeq
\frac{9}{10}\Phi^{(prim)}~,~ for~ \zeta\gg1.
\end{equation}

At sufficiently late times in the matter dominated regime all modes evolve identically with
the \textit{growth function } $D_g(\zeta)$ given in (2.37). I recall that this function is
normalized such that it is equal to $a/a_0$ when we can still ignore the dark energy (at
$z>10$, say). The growth function describes the evolution of $\Delta$, thus by the Poisson
equation (2.3) $\Phi$ grows with $D_g(a)/a$. We therefore define the \textit{transfer
function} $T(k)$ by (we choose the normalization $a_0=1$)
\begin{equation}
\Phi(k,a)=\Phi^{(prim)}\frac{9}{10}\frac{D_g(a)}{a}T(k)
\end{equation}
for late times. This definition is chosen such that $T(k)\rightarrow1$ for $k\rightarrow0$,
and does not depend on time.

At these late times $\rho_M=\Omega_M a^{-3}\rho_{crit}$, hence the Poisson equation gives
the following relation between $\Phi$ and $\Delta$
\[\Phi=\left(\frac{a}{k}\right)^24\pi
G\rho_M\Delta=\frac{3}{2}\frac{1}{ak^2}H_0^2\Omega_M\Delta.\] Therefore, (6.76) translates
to
\begin{equation}
\fbox{$\displaystyle
\Delta(a)=\frac{3}{5}\frac{k^2}{\Omega_MH_0^2}\Phi^{(prim)}D_g(a)T(k).$}
\end{equation}

The transfer function can be determined by solving numerically the pair (6.24), (6.25) of
basic perturbation equations. One can derive even a reasonably good analytic approximation
by putting our previous results together (for details see again \cite{Cos4}, Sect. 7.4).
For a CDM model the following accurate fitting formula  to the numerical solution in terms
of the variable $\tilde{q}=k/k_{eq}$, where $k_{eq}$ is defined such that the corresponding
value of the parameter $\omega$ in (6.33) is equal to 1 (i.e.,
$k_{eq}=a_{eq}H_{eq}=\sqrt{2\Omega_M}H_0/\sqrt{a_{eq}}$, using (0.52)) was given in
\cite{BBKS}:
\begin{equation}
T_{BBKS}(\tilde{q})=
\frac{\ln(1+0.171\tilde{q})}{0.171\tilde{q}}[1+0.284\tilde{q}+(1.18\tilde{q})^2
+(0.399\tilde{q})^3+(0.490\tilde{q})^4]^{-1/4}.
\end{equation}
Note that $\tilde{q}$ depends on the cosmological parameters through the
combination\footnote{since $k$ is measured in units of $h_0~Mpc^{-1}$ and
$a_{eq}=4.15\times 10^{-5}/(\Omega_M h_0^2)$.} $\Omega_Mh_0$, usually called the
\textit{shape parameter} $\Gamma$. In terms of the variable $q=k/(\Gamma h_0 Mpc^{-1})$
(6.78) can be written as
\begin{equation}
T_{BBKS}(q)=\frac{\ln(1+2.34q)}{2.34q}[1+3.89q+(16.1q)^2+(5.46q)^3+(6.71q)^4]^{-1/4}.
\end{equation}

This result for the transfer function is based on a simplified analysis. The tight coupling
approximation is no more valid when the decoupling temperature is approached. Moreover,
anisotropic stresses and baryons have been ignored. We shall reconsider the transfer
function after having further developed the basic theory in the next chapter. It will, of
course, be very interesting to compare the theory with available observational data. For
this one has to keep in mind that the linear theory only applies to sufficiently large
scales. For late times and small scales it has to be corrected by numerical simulations for
nonlinear effects.

For a given primordial power spectrum, the transfer function determines the power spectrum
after the `transfer regime' (when all modes evolve with the growth function $D_g$). From
(6.77) we obtain for the power spectrum of $\Delta$
\begin{equation}
P_\Delta(z)=\frac{9}{25}\frac{k^4}{\Omega_M^2H_0^4}P_\Phi^{(prim)}D_g^2(z)T^2(k).
\end{equation}
We choose $P_\Phi^{(prim)}\propto k^{n-1}$ and the amplitude such that
\begin{equation}
\fbox{$\displaystyle P_\Delta(z)=\delta_H^2\left(\frac{k}{H_0}\right)^{3+n}T^2(k)
\left(\frac{D_g(z)}{D_g(0)}\right)^2.$}
\end{equation}
Note that $P_\Delta(0)=\delta_H^2$ for $k=H_0$. The normalization factor $\delta_H$ has to
be determined from observations (e.g. from CMB anisotropies at large scales). Comparison of
(6.80) and (6.81) and use of (5.50) implies
\begin{equation}
P^{(prim)}_{\mathcal{R}}(k)=\frac{9}{4}P^{(prim)}_\Phi(k)=\frac{25}{4}\delta^2_H
\left(\frac{\Omega_M}{D_g(0)}\right)^2\left(\frac{k}{H_0}\right)^{n-1}.
\end{equation}


------------

\textbf{Exercise}. Write the equations (6.27)-(6.30) in explicit
form, using (6.33) in the limit when baryons are neglected
($c\rightarrow \infty$). (For a truncated subsystem this was done in
(6.69) -- (6.71)). Solve the five first order differential equations
(6.27), (6.28) for $\alpha=d,r$ and (6.30) numerically. Determine,
in particular, the transfer function defined in (6.76). (A standard
code gives this in less than a second.)

------------



\chapter{Boltzmann Equation in GR}

For the description of photons and neutrinos before recombination we need the general
relativistic version of the Boltzmann equation.

\section{One-particle phase space, Liouville \\operator for geodesic spray}

For what follows we first have to develop some kinematic and differential geometric tools.
Our goal is to generalize the standard description of Boltzmann in terms of one-particle
distribution functions.

Let $g$ be the metric of the spacetime manifold $M$. On the cotangent bundle $T^\ast
M=\bigcup_{p\in M}T^\ast_pM$ we have the natural symplectic 2-form $\omega$, which is given
in natural bundle coordinates\footnote{If $x^\mu$ are coordinates of $M$ then the $dx^\mu$
form in each point $p\in M$ a basis of the cotangent space $T^\ast_pM$. The \textit{bundle
coordinates} of  $\beta\in T^\ast_p M $ are then $(x^\mu,\beta_\nu)$ if $\beta=\beta_\nu
dx^\nu$ and $x^\mu$ are the coordinates of $p$. With such bundle coordinates one can define
an atlas, by which $T^\ast M$ becomes a differentiable manifold.}$(x^\mu,p_\nu)$ by
\begin{equation}
\omega=dx^\mu\wedge dp_\mu.
\end{equation}
(For an intrinsic description, see Chap. 6 of \cite{Mar}.) So far no metric is needed. The
pair $(T^\ast M,\omega)$ is always a symplectic manifold.

The metric $g$ defines a natural diffeomorphism between the tangent bundle $TM$ and $T^\ast
M$ which can be used to pull $\omega$ back to a symplectic form $\omega_g$ on $TM$. In
natural bundle coordinates the diffeomorphism is given by $(x^\mu,p^\alpha)\mapsto
(x^\mu,p_\alpha=g_{\alpha\beta}p^\beta)$, hence
\begin{equation}
\omega_g=dx^\mu\wedge d(g_{\mu\nu}p^\nu).
\end{equation}

On $TM$ we can consider the ``Hamiltonian function''
\begin{equation}
L=\frac{1}{2}g_{\mu\nu}p^\mu p^\nu
\end{equation} and its associated Hamiltonian vector field $X_g$, determined by the
equation
\begin{equation}
i_{X_g}\omega_g=dL.
\end{equation}
It is not difficult to show that in bundle coordinates
\begin{equation}
X_g=p^\mu\frac{\partial}{\partial x^\mu}-\Gamma^\mu{}_{\alpha\beta}p^\alpha
p^\beta\frac{\partial}{\partial p^\mu}
\end{equation}
(Exercise). The Hamiltonian vector field $X_g$ on the symplectic manifold $(TM,\omega_g)$
is the \textit{geodesic spray}. Its integral curves satisfy the canonical equations:
\begin{eqnarray}
\frac{dx^\mu}{d\lambda} &=& p^\mu, \\
\frac{dp^\mu}{d\lambda} &=& -\Gamma^\mu{}_{\alpha\beta}p^\alpha p^\beta.
\end{eqnarray}
The \textit{geodesic flow} is the flow of the vector field $X_g$.

Let $\Omega_{\omega_g}$ be the volume form belonging to $\omega_g$, i.e., the Liouville
volume
\[\Omega_{\omega_g}= const~\omega_g\wedge\cdot\cdot\cdot\wedge\omega_g,\]
or ($g=\det(g_{\alpha\beta})$)
\begin{eqnarray}
\Omega_{\omega_g} &=&(-g)(dx^0\wedge dx^1\wedge dx^2\wedge dx^3)\wedge (dp^0\wedge
dp^1\wedge dp^2\wedge dp^3) \nonumber\\
 &\equiv & (-g) dx^{0123}\wedge dp^{0123}.
\end{eqnarray}

The \textit{one-particle phase space} for particles of mass $m$ is the following
submanifold of $TM$:
\begin{equation}
\Phi_m=\{v\in TM\mid v~\mbox{future directed},~ g(v,v)=-m^2\}.
\end{equation}
This is invariant under the geodesic flow. The restriction of $X_g$ to $\Phi_m$ will also
be denoted by $X_g$. $\Omega_{\omega_g}$ induces a volume form $\Omega_m$ (see below) on
$\Phi_m$, which is also invariant under $X_g$:
\begin{equation}
L_{X_g}\Omega_m=0.
\end{equation}
$\Omega_m$ is determined as follows (known from Hamiltonian mechanics): Write
$\Omega_{\omega_g}$ in the form
\[ \Omega_{\omega_g}=-dL\wedge\sigma, \]
(this is always possible, but $\sigma$ is not unique), then $\Omega_m$ is the pull-back of
$\Omega_{\omega_g}$ by the injection $i:~\Phi_m\rightarrow TM$,
\begin{equation}
\Omega_m=i^\ast\sigma.
\end{equation}
While $\sigma$ is not unique (one can, for instance, add a multiple of $dL$), the form
$\Omega_m$ is independent of the choice of $\sigma$ (show this). In natural bundle
coordinates a possible choice is
\[ \sigma=(-g)dx^{0123}\wedge\frac{dp^{123}}{(-p_0)},\]
because
\[ -dL\wedge\sigma=[- g_{\mu\nu}p^\mu dp^\nu+ \cdot\cdot\cdot]\wedge
\sigma=(-g)dx^{0123}\wedge g_{\mu0}p^\mu
dp^0\wedge\frac{dp^{123}}{p_0}=\Omega_{\omega_g}.\] Hence,
\begin{equation}
\Omega_m=\eta\wedge\Pi_m,
\end{equation}
where $\eta$ is the volume form of $(M,g)$,
\begin{equation}
\eta=\sqrt{-g}dx^{0123},
\end{equation}
and
\begin{equation}
\Pi_m=\sqrt{-g}\frac{dp^{123}}{|p_0|},
\end{equation}
with $p^0>0$, and $g_{\mu\nu}p^\mu p^\nu=-m^2$.

We shall need some additional tools. Let $\Sigma$ be a hypersurface of $\Phi_m$ transversal
to $X_g$. On $\Sigma$ we can use the volume form
\begin{equation}
vol_\Sigma=i_{X_g}\Omega_m\mid\Sigma.
\end{equation}
Now we note that the 6-form
\begin{equation}
\omega_m:=i_{X_g}\Omega_m
\end{equation}
on $\Phi_m$ is closed,
\begin{equation}
d\omega_m=0,
\end{equation}
because
\[ d\omega_m=di_{X_g}\Omega_m=L_{X_g}\Omega_m=0\]
(we used $d\Omega_m=0$ and (7.10)). From (7.12) we obtain
\begin{equation}
\omega_m=(i_{X_g}\eta)\wedge\Pi_m+\eta\wedge i_{X_g}\Pi_m.
\end{equation}

In the special case when $\Sigma$ is a ``time section'', i.e., in the inverse image of a
spacelike submanifold of $M$ under the natural projection $\Phi_m\rightarrow M$, then the
second term in (7.18) vanishes on $\Sigma$, while the first term is on $\Sigma$ according
to (7.5) equal to $i_p\eta\wedge\Pi_m,~p=p^\mu\partial/\partial x^\mu$. Thus, we have on a
time section\footnote{Note that in Minkowski spacetime we get for a constant time section
$vol_\Sigma=dx^{123}\wedge dp^{123}$.} $\Sigma$
\begin{equation}
\fbox{$\displaystyle vol_\Sigma=\omega_m\mid\Sigma=i_p\eta\wedge\Pi_m.$}
\end{equation}

Let $f$ be a one-particle distribution function on $\Phi_m$, defined such that the number
of particles in a time section $\Sigma$ is
\begin{equation}
N(\Sigma)=\int_\Sigma f\omega_m.
\end{equation}
The particle number current density is
\begin{equation}
n^\mu(x)=\int_{P_m(x)}fp^\mu\Pi_m,
\end{equation}
where $P_m(x)$ is the fiber over $x$ in $\Phi_m$ (all momenta with $\langle
p,p\rangle=-m^2$). Similarly,, one defines the energy-momentum tensor, etc.

Let us show that
\begin{equation}
n^\mu{}_{;\mu}=\int_{P_m}\left(L_{X_g}f\right)\Pi_m.
\end{equation}
We first note that (always in $\Phi_m$)
\begin{equation}
d(f\omega_m)=\left(L_{X_g}f\right)\Omega_m.
\end{equation}
Indeed, because of (7.17) the left-hand side of this equation is
\[ df\wedge\omega_m=df\wedge i_{X_g}\Omega_m=\left(i_{X_g}df\right)\wedge\Omega_m=\left(L_{X_g}f\right)\Omega_m.
\]
Now, let $D$ be a domain in $\Phi_m$ which is the inverse of a domain $\bar{D}\subset M$
under the projection $\Phi_m\rightarrow M$. Then we have on the one hand by (7.18), setting
$i_X\eta\equiv X^\mu\sigma_\mu$,
\[ \int_{\partial D}f\omega_m=\int_{\partial\bar{D}}\sigma_\mu\int_{P_m(x)}p^\mu f\Pi_m=
\int_{\partial\bar{D}}\sigma_\mu n^\mu=\int_{\partial
\bar{D}}i_n\eta=\int_{\bar{D}}(\nabla\cdot n)\eta. \] On the other hand, by (7.23) and
(7.12)
\[ \int_{\partial D}f\omega_m=\int_D d(f\omega_m)=\int_D \left(L_{X_g}f\right)\Omega_m
=\int_{\bar{D}}\eta\int_{P_m(x)}\left(L_{X_g}f\right)\Pi_m.\] Since $\bar{D}$ is arbitrary,
we indeed obtain (7.22).

The proof of the following equation for the energy-momentum tensor
\begin{equation}
T^{\mu\nu}{}_{;\nu}=\int_{P_m}p^\mu\left(L_{X_g}f\right)\Pi_m
\end{equation}
can be reduced to the previous proof by considering instead of $n^\nu$ the vector field
$N^\nu:=v_\mu T^{\mu\nu}$, where $v_\mu$ is geodesic in $x$.

\section{The general relativistic Boltzmann\\ equation}

Let us first consider particles for which collisions can be neglected (e.g. neutrinos at
temperatures much below 1 MeV). Then the conservation of the particle number in a domain
that is comoving with the flow $\phi_s$ of $X_g$ means that the integrals
\[ \int_{\phi_s(\Sigma)}f\omega_m,\]
$\Sigma$ as before a hypersurface of $\Phi_m$ transversal to $X_g$, are independent of $s$.
We now show that this implies the \textit{collisionless Boltzmann equation}
\begin{equation}
\fbox{$\displaystyle L_{X_g}f=0.$}
\end{equation}

The proof of this expected result proceeds as follows. Consider a `cylinder' $\mathcal{G}$,
sweping by $\Sigma$ under the flow $\phi_s$ in the interval $[0,s]$  (see Fig. 7.1), and
the integral
\[ \int_\mathcal{G}L_{X_g}f\Omega_m=\int_{\partial \mathcal{G}}f\omega_m \]
(we used eq. (7.23)). Since $i_{X_g}\omega_m=i_{X_g}(i_{X_g}\Omega_m)=0$, the integral over
the mantle of the cylinder vanishes, while those over $\Sigma$ and $\phi_s(\Sigma)$ cancel
(conservation of particles). Because $\Sigma$ and $s$ are arbitrary, we conclude that
(7.25) must hold.

\begin{figure}
\begin{center}
\includegraphics[height=0.3\textheight]{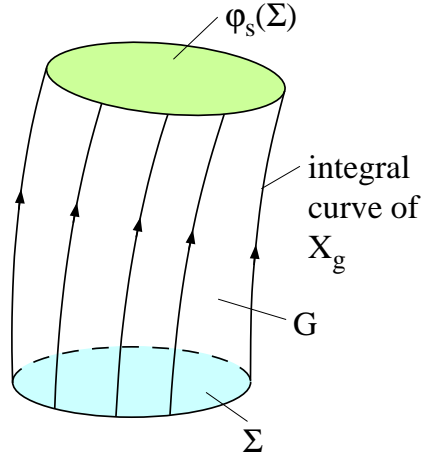}
\caption{Picture for the proof of (7.25).} \label{Fig-12}
\end{center}
\end{figure}

From (7.22) and (7.23) we obtain, as expected, the conservation of the particle number
current density: $n^\mu{}_{;\mu}=0$.

With collisions, the Boltzmann equation has the symbolic form
\begin{equation}
\fbox{$\displaystyle L_{X_g}f=C[f]$},
\end{equation}
where $C[f]$ is the ``collision term''. For the general form of this in terms of the
invariant transition matrix element for a two-body collision, see (B.9). In Appendix B we
also work this out explicitly for photon-electron scattering.

By (7.24) and (7.26) we have
\begin{equation}
T^{\mu\nu}{}_{;\nu}=Q^\mu,
\end{equation}
with
\begin{equation}
Q^\mu=\int_{P_m}p^\mu C[f]\Pi_m.
\end{equation}

\section{Perturbation theory (generalities)}

We consider again small deviations from Friedmann models, and set correspondingly
\begin{equation}
f=f^{(0)}+\delta f.
\end{equation}
How does $\delta f$ change under a gauge transformation? At first sight one may think that
we simply have $\delta f\rightarrow \delta f+L_{T\xi}f^{(0)}$, where $T\xi$ is the lift of
the vector field $\xi$, defining the gauge transformation, to the tangent bundle. (We
recall that $T\xi$ is obtained as follows: Let $\phi_s$ be the flow of $\xi$ and consider
the flow $T\phi_s$ on $TM,~T\phi_s=$ tangent map. Then $T\xi$ is the vector field belonging
to $T\phi_s$.) Unfortunately, things are not quite as simple, because $f$ is only defined
on the one-particle subspace of $TM$, and this is also perturbed when the metric is
changed. One way of getting the right transformation law is given in \cite{DS}. Here, I
present a more pedestrian, but simpler derivation.

First, we introduce convenient independent variables for the distribution function. For
this we choose an adapted orthonormal frame $\{e_{\hat{\mu}},~ \hat{\mu}=0,1,2,3\}$ for the
perturbed metric (1.16), which we recall
\begin{equation}
g=a^2(\eta)\left\{-(1+2A)d\eta^2 - 2B,_idx^id\eta + [(1+2D)\gamma_{ij} + 2E_{\mid
ij}]dx^idx^j\right\}.
\end{equation}
$e_{\hat{0}}$ is chosen to be orthogonal to the time slices $\eta=const$, whence
\begin{equation}
e_{\hat{0}}=\frac{1}{\alpha}\left(\partial_\eta+\beta^i\partial_i\right),~~\alpha=1+A,~\beta_i=B_{,i}.
\end{equation}
This is indeed normalized and perpendicular to $\partial_i$. At the moment we do not need
explicit expressions for the spatial basis $e_{\hat{i}}$ tangential to $\eta=const$.

From
\[p=p^{\hat{\mu}}e_{\hat{\mu}}=p^\mu\partial_\mu \]
we see that $p^{\hat{0}}/\alpha=p^0$. From now on we consider massless particles and
set\footnote{This definition of $q$ is only used in the present subsection. Later, after
eqn. (7.62), $q$ will denote the comoving momentum $aq$.} $q=p^{\hat{0}}$, whence
\begin{equation}
q=a(1+A)p^0.
\end{equation}
Furthermore, we use the unit vector $\gamma^i=p^{\hat{i}}/q$. Then the distribution
function can be regarded as a function of $\eta, x^i, q, \gamma^i$, and this we shall adopt
in what follows. For the case $K=0$, which we now consider for simplicity, the unperturbed
tetrad is $\{\frac{1}{a}\partial_\eta,\frac{1}{a}\partial_i\}$, and for the unperturbed
situation we have $q=ap^0,~p^i=p^0\gamma^i$.

As a further preparation we interpret the Lie derivative as an infinitesimal coordinate
change. Consider the infinitesimal coordinate transformation
\begin{equation}
\bar{x}^\mu=x^\mu-\xi^\mu(x),
\end{equation}
then to first order in $\xi$
\begin{equation}
\left(L_\xi g\right)_{\mu\nu}(x)=\bar{g}_{\mu\nu}(x)-g_{\mu\nu}(x),
\end{equation}
and correspondingly for other tensor fields. One can verify this by a direct comparison of
the two sides. For the simplest case of a function $F$,
\[ \bar{F}(x)-F(x)=F(x+\xi)-F(x)=\xi^\mu\partial_\mu F=L_\xi F.\]

Under the transformation (7.33) and its extension to $TM$ the $p^\mu$ transform as
\[ \bar{p}^\mu=p^\mu-\xi^\mu{}_{,\nu}p^\nu.\]
We need the transformation law for $q$. From
\[ \bar{q}=a(\bar{\eta})[1+\bar{A}(\bar{x})]\bar{p}^0 \]
and the transformation law (1.18) of $A$,
\[ A\rightarrow A+\frac{a'}{a}\xi^0+\xi^{0'}, \]
we get
\[\bar{q}=a(\eta)[1-\mathcal{H}\xi^0][1+A(x)\mathcal{H}\xi^0+\xi^{0'}][p^0-\xi^0{}_{,\nu}p^\nu].\]
The last square bracket is equal to $p^0(1-\xi^{0'}-\xi^0{}_{,i}\gamma^i)$. Using also
(7.32) we find
\begin{equation}
\bar{q}=q-q\xi^0{}_{,i}\gamma^i.
\end{equation}
Since the unperturbed distribution function $f^{(0)}$ depends only on $q$ and $\eta$, we
conclude from this that
\begin{equation}
\delta f\rightarrow\delta f + q\frac{\partial f^{(0)}}{\partial
q}\xi^0{}_{,i}\gamma^i+\xi^0f^{(0)'}.
\end{equation}
Here, we use the equation of motion for $f^{(0)}$. For massless particles this is an
equilibrium distribution that is stationary when considered as a function of the
\textit{comoving momentum} $aq$. This means that
\[ \frac{\partial f^{(0)}}{\partial\eta}+\frac{\partial f^{(0)}}{\partial q}q'=0 \]
for $(aq)'=0$, i.e., $q'=-\mathcal{H}q$. Thus,
\begin{equation}
f^{(0)'}-\mathcal{H}q\frac{\partial f^{(0)}}{\partial q}=0.
\end{equation}
If this is used in (7.36) we get
\begin{equation}
\fbox{$\displaystyle \delta f\rightarrow \delta f+q\frac{\partial f^{(0)}}{\partial
q}[\mathcal{H}\xi^0+\xi^0{}_{,i}\gamma^i]$}.
\end{equation}
Since this transformation law involves only $\xi^0$, we can consider various gauge
invariant distribution functions, such as $(\delta f)_\chi,~(\delta f)_{\mathcal{Q}}$. From
(1.21), $\chi\rightarrow \chi+a\xi^0$, we find
\begin{equation}
\mathcal{F}_s:=(\delta f)_\chi=\delta f-q\frac{\partial f^{(0)}}{\partial
q}[\mathcal{H}(B+E')+\gamma^i(B+E')_{,i}].
\end{equation}
$\mathcal{F}_s$ reduces to $\delta f$ in the longitudinal gauge, and we shall mainly work
with this gauge invariant perturbation. In the literature sometimes $\mathcal{F}_c:=(\delta
f)_{\mathcal{Q}}$ is used. Because of (1.49), $v-B\rightarrow(v-B)-\xi^0$, we obtain
$\mathcal{F}_c$ from (7.39) in replacing $B+E'$ by $-(v-B)$:
\begin{equation}
\mathcal{F}_c:=(\delta f)_\mathcal{Q}=\delta f+q\frac{\partial f^{(0)}}{\partial
q}[\mathcal{H}(v-B)+\gamma^i(v-B)_{,i}].
\end{equation}
Since by (1.56) $(v-B)+(B+E')=V$, we find the relation
\begin{equation}
\mathcal{F}_c=\mathcal{F}_s+q\frac{\partial f^{(0)}}{\partial
q}[\mathcal{H}V+\gamma^iV_{,i}].
\end{equation}

Instead of $v,V$ we could also use the baryon velocities $v_b,V_b$.

\section{Liouville operator in the\\ longitudinal gauge}

We want to determine the action of the Liouville operator $\mathcal{L}:=L_{X_g}$ on
$\mathcal{F}_s$. The simplest way to do this is to work in the longitudinal gauge $B=E=0$.

In this section we do not assume a vanishing $K$. It is convenient to introduce an adapted
orthonormal tetrad
\begin{equation}
e_0=\frac{1}{a(1+A)}\partial_\eta,~~~ e_i=\frac{1}{a(1+D)}\hat{e}_i,
\end{equation}
where $\hat{e}_i$ is an orthonormal basis for the unperturbed space $(\Sigma,\gamma)$. Its
dual basis will be denoted by $\hat{\vartheta}^i$, and that of $e_\mu$ by $\theta^\mu$. We
have
\begin{equation}
\theta^0=(1+A)\bar{\theta}^0,~~~\theta^i=(1+D)\bar{\theta}^i,
\end{equation}
where \begin{equation} \bar{\theta}^0=a(\eta)d\eta,~~~
\bar{\theta}^i=a(\eta)\hat{\vartheta}^i.
\end{equation}

\paragraph{Connection forms.}

The unperturbed connection forms have been obtained in Sect. 0.1.2. In the present notation
they are
\begin{equation}
\bar{\omega}^i{}_0=\bar{\omega}^0{}_i=\frac{a'}{a^2}\bar{\theta}^i,~~~
\bar{\omega}^i{}_j=\hat{\omega}^i{}_j,
\end{equation}
where $\hat{\omega}^i{}_j$ are the connection forms of $(\Sigma,\gamma)$ relative to
$\hat{\vartheta}^i$.

For the determination of the perturbations $\delta\omega^\mu{}_\nu$ of the connection forms
we need $d\theta^\mu$. In the following calculation we make use of the first structure
equations, both for the unperturbed and the actual metric. The former, together with
(7.45), implies that the first term in
\[d\theta^0=(1+A)d\bar{\theta}^0+dA\wedge\bar{\theta}^0\]
vanishes. Using the notation $dA=A'd\eta+A_{\mid
i}\bar{\theta}^i=A_{\mid\mu}\bar{\theta}^\mu$ we obtain
\begin{equation}
d\theta^0=A_{\mid i}\bar{\theta}^i\wedge\bar{\theta}^0.
\end{equation}
Similarly,
\begin{equation}
d\theta^i=(1+D)d\bar{\theta}^i+dD\wedge\bar{\theta}^i=(1+D)[-\bar{\omega}^i{}_j\wedge\bar{\theta}^j
-\bar{\omega}^i{}_0\wedge\bar{\theta}^0]+D_{\mid
j}\bar{\theta}^j\wedge\bar{\theta}^i+D_{\mid 0}\bar{\theta}^0\wedge\bar{\theta}^i.
\end{equation}
On the other hand, inserting
$\omega^\mu{}_\nu=\bar{\omega}^\mu{}_\nu+\delta\omega^\mu{}_\nu$ into
$d\theta^\mu=-\omega^\mu{}_\nu\wedge\theta^\nu$, and comparing first orders, we obtain the
equations
\begin{equation}
-\delta\omega^0{}_i\wedge\bar{\theta}^i-\underbrace{\bar{\omega}^0{}_i\wedge(D\bar{\theta}^i)}_{0}=-
A_{\mid i}\bar{\theta}^0\wedge\bar{\theta}^i,
\end{equation}
\begin{eqnarray}
\lefteqn{-\delta\omega^i{}_0\wedge\bar{\theta}^0-\delta\omega^i{}_j\wedge\bar{\theta}^j
-\bar{\omega}^i{}_0\wedge A\bar{\theta}^0 -\bar{\omega}^i{}_j\wedge D\bar{\theta}^j=} \nonumber \\
& & -D\bar{\omega}^i{}_j\wedge\bar{\theta}^j -D\bar{\omega}^i{}_0\wedge\bar{\theta}^0
+D_{\mid j}\bar{\theta}^j\wedge\bar{\theta}^i+D_{\mid 0}\bar{\theta}^0\wedge\bar{\theta}^i.
\end{eqnarray}
Eq. (7.48) requires
\begin{equation}
\delta\omega^0{}_i=A_{\mid i}\bar{\theta}^0+(\propto\bar{\theta}^i).
\end{equation}
Let us try the guess
\begin{equation}
\delta\omega^i{}_j=-D_{\mid i}\bar{\theta}^j+D_{\mid j}\bar{\theta}^i
\end{equation}
and insert this into (7.49). This gives
\begin{equation}
-\delta\omega^i{}_0\wedge\bar{\theta}^0-A\bar{\omega}^i{}_0\wedge\bar{\theta}^0=
-D\bar{\omega}^i{}_0\wedge\bar{\theta}^0+D_{\mid 0}\bar{\theta}^0\wedge\bar{\theta}^i,
\end{equation}
and this is satisfied if the last term in (7.50) is chosen according to
\begin{equation}
\delta\omega^0{}_i=A_{\mid
i}\bar{\theta}^0-(A-D)\bar{\omega}^0{}_i+\frac{1}{a}D'\bar{\theta}^i.
\end{equation}
Since the first structure equations are now all satisfied (to first order) our guess (7.51)
is correct, and we have determined all $\delta\omega^\mu{}_\nu$.

From (7.45) and (7.53) we get to first order
\begin{equation}
\omega^i{}_0=\left[\frac{a'}{a^2}(1-A)+\frac{1}{a}D'\right]\theta^i+A_{\mid i}\theta^0.
\end{equation}
We shall not need $\omega^i{}_j$ explicitly, except for the property $\omega^i{}_j(e_0)=0$,
which follows from (7.45) and (7.51).

We take the spatial components $p^i$ of the momenta $p$ relative to the orthonormal tetrad
$\{e_\mu\}$ as independent variables of $f$ (beside $x$). Then
\begin{equation}
\fbox{$\displaystyle \mathcal{L}f=p^\mu e_\mu(f)-\omega^i{}_\alpha(p)p^\alpha\frac{\partial
f}{\partial p^i}$}~~~(p=p^\mu e_\mu).
\end{equation}

\paragraph{Derivation.} Eq. (7.55) follows from (7.5) and the result of the following
consideration.

Let $X=\sum_{i=1}^{n+1}\xi^i\partial_i$ be a vector field on a domain of $\mathbf{R}^{n+1}$
and let $\Sigma$ be a hypersurface in $\mathbf{R}^{n+1}$, parametrized by
\[ \varphi:U\subset\mathbf{R}^n\rightarrow \mathbf{R}^{n+1},~~~
(x^1,\cdot\cdot\cdot x^n)\mapsto (x^1,\cdot\cdot\cdot x^n,g(x^1,\cdot\cdot\cdot x^n)),\] to
which $X$ is tangential. Furthermore, let $f$ be a function on $\Sigma$, which we regard as
a function of $x^1,\cdot\cdot\cdot,x^n$. I claim that
\begin{equation}
X(f)=\sum_{i=1}^n \xi^i\frac{\partial(f\circ\varphi)}{\partial x^i}.
\end{equation}
This can be seen as follows: Extend $f$ in some manner to a neighborhood of $\Sigma$ (at
least locally). Then
\begin{equation}
X(f)\mid\Sigma=\left.\sum_{i=1}^n \left(\xi^i\frac{\partial f}{\partial
x^i}+\xi^{n+1}\frac{\partial f}{\partial
x^{n+1}}\right)\right|_{x^{n+1}=g(x^1,\cdot\cdot\cdot x^n)}.
\end{equation}
Now, we have on $\Sigma:~dg-dx^{n+1}=0$ and thus $\langle dg-dx^{n+1},X\rangle=0$ since $X$
is tangential. Using (7.57) this implies
\[ \xi^{n+1}=\sum_{i=1}^n \xi^i\frac{\partial g}{\partial x^i}, \]
whence (7.57) gives by the chain rule
\[ X(f)\mid\Sigma=\sum_{i=1}^n \xi^i\left(\frac{\partial f}{\partial
x^i}+\frac{\partial f}{\partial x^{n+1}}\frac{\partial g}{\partial x^i}\right)=
\sum_{i=1}^n \xi^i\frac{\partial(f\circ\varphi)}{\partial x^i}.\] This fact was used in
(7.55) for the vector field
\begin{equation}
X_g=p^\mu e_\mu-\omega^\mu{}_\alpha(p)p^\alpha\frac{\partial}{\partial p^\mu}.
\end{equation}

\paragraph{$\mathcal{L}f$ to first order.}

For $\mathcal{L}f$ we need
\[p^\mu e_\mu(f)=p^0\frac{1}{a}(1-A)f'+p^i e_i(f)=p^0\frac{1}{a}(1-A)f'+p^i \frac{1}{a}\hat{e}_i(\delta f) \]
and
\begin{eqnarray*}
\omega^i{}_\alpha(p)p^\alpha\frac{\partial}{\partial
p^i}&=&\omega^i{}_0(p)p^0\frac{\partial} {\partial
p^i}+\omega^i{}_j(p)p^j\frac{\partial}{\partial p^i}\nonumber \\
&=& [\omega^i{}_0(e_0)p^0+\omega^i{}_0(\mathbf{p})]p^0\frac{\partial} {\partial p^i}+
[\omega^i{}_j(e_0)p^0+\omega^i{}_j(\mathbf{p})]p^j\frac{\partial} {\partial p^i}.
\end{eqnarray*}
From (7.54) we get $\omega^i{}_0(e_0)=A^{\mid i}$, and
\[\omega^i{}_0(\mathbf{p})=\left[\frac{a'}{a^2}(1-A)+\frac{1}{a}D'\right]p^i.\]
Furthermore, the Gauss equation implies
$\omega^i{}_j(\mathbf{p})=\tilde{\omega}^i{}_j(\mathbf{p})$, where $\tilde{\omega}^i{}_j$
are the connection forms of the spatial metric (see Appendix A of \cite{NS1}).

As an intermediate result we obtain
\begin{eqnarray}
\mathcal{L}f &=& (1-A)\frac{p^0}{a}f'+\frac{p^i}{a}\hat{e}_i(\delta f)\nonumber
\\
&-&\left[\tilde{\omega}^i{}_j(\mathbf{p})p^j+(p^0)^2A^{\mid i}+\frac{p^0}{a}D'p^i
+p^0\frac{a'}{a^2}(1-A)p^i\right]\frac{\partial f} {\partial p^i}.
\end{eqnarray}

From now on we use as independent variables
$\eta,x^i,p,\gamma^i=p^i/p~(p=[\sum_i(p^i)^2]^{1/2})$. We have
\begin{equation}
\frac{\partial f} {\partial p^i}=\frac{p_i}{p}\frac{\partial f} {\partial
p}+\frac{1}{p}\left(\delta^l{}_i-p_ip^l/p^2\right)\frac{\partial f} {\partial \gamma^l}.
\end{equation}
Contracting this with $\tilde{\omega}^i{}_j(\mathbf{p})p^j$, appearing in (7.59), the first
term on the right in (7.60) gives no contribution (antisymmetry of $\tilde{\omega}^i{}_j$),
and since $\partial f/\partial\gamma^l$ is of first order we can replace
$\tilde{\omega}^i{}_j$ by the connection forms of the unperturbed metric $a^2\gamma_{ij}$;
these are the same as the connection forms $\hat{\omega}^i{}_j$ of $\gamma_{ij}$ relative
to $\hat{\vartheta}^i$. What remains is thus
\[\hat{\omega}^i{}_j(\mathbf{p})\frac{p^j}{p}\left(\delta^l{}_i-p_ip^l/p^2\right)
\frac{\partial\delta f} {\partial \gamma^l}=\hat{\omega}^i{}_j(\mathbf{p})\frac{p^j}{p}
\frac{\partial\delta f} {\partial
\gamma^i}=\frac{p}{a}\gamma^j\gamma^k\hat{\Gamma}^i{}_{jk}\frac{\partial\delta f} {\partial
\gamma^i}. \] Inserting this and (7.60) into (7.59) gives in zeroth order for the Liouville
operator
\[ (\mathcal{L}f)^{(0)}=\frac{p^0}{a}\left( f^{(0)'}-\mathcal{H}p\frac{\partial f^{(0)}}{\partial p}\right),\]
and the first order contribution is
\begin{eqnarray*}
-A(\mathcal{L}f)^{(0)}& + &\frac{p^0}{a}(\delta f)'+\frac{p^i}{a}\hat{e}_i(\delta f)  -
\frac{p}{a}\gamma^j\gamma^k\hat{\Gamma}^i{}_{jk}\frac{\partial\delta f} {\partial
\gamma^i}\\
&-& \frac{(p^0)^2}{ap}\hat{e}_i(A)p^i\frac{\partial f^{(0)}}{\partial p}
-\frac{p^0}{a}D'p\frac{\partial f^{(0)}}{\partial
p}-\frac{p^0}{a}\mathcal{H}p\frac{\partial\delta f}{\partial p}.
\end{eqnarray*}
Therefore, we obtain for the Liouville operator, up to first order,
\begin{eqnarray}
\frac{a}{p^0}\mathcal{L}f &=& (1-A)\left( f^{(0)'}-\mathcal{H}p\frac{\partial
f^{(0)}}{\partial p}\right)+(\delta f)'-\mathcal{H}p\frac{\partial\delta f}{\partial
p}\nonumber \\
&+& \frac{p^i}{p^0}\hat{e}_i(\delta
f)-\frac{p}{p^0}\gamma^j\gamma^k\hat{\Gamma}^i{}_{jk}\frac{\partial\delta f} {\partial
\gamma^i}-p\left[D'+\frac{p^0}{p}\gamma^i\hat{e}_i(A)\right]\frac{\partial
f^{(0)}}{\partial p}.\nonumber \\
\end{eqnarray}

As a first application we consider the collisionless Boltzmann equation for $m=0$. In
zeroth order we get the equation (7.37) ($q$ in that equation is our present $p$). The
perturbation equation becomes
\begin{eqnarray}
(\delta f)'-\mathcal{H}p\frac{\partial\delta f}{\partial p}+ \gamma^i\hat{e}_i(\delta
f)-\gamma^j\gamma^k\hat{\Gamma}^i{}_{jk}\frac{\partial\delta f} {\partial
\gamma^i}-\left[D'+\gamma^i\hat{e}_i(A)\right]p\frac{\partial f^{(0)}}{\partial
p}=0.\nonumber \\
\end{eqnarray}
It will be more convenient to write this in terms of the \textit{comoving momentum}, which
we denote by $q,~q=ap$. (This slight change of notation is unfortunate, but should not give
rise to confusions, because the equations at the beginning of Sect. 7.3, with the earlier
meaning $q\equiv p$, will no more be used. But note that (7.38)-(7.41) remain valid with
the present meaning of $q$.) Eq. (7.62) then becomes
\begin{equation}
\fbox{$\displaystyle (\partial_\eta+\gamma^i\hat{e}_i)\delta
f-\hat{\Gamma}^i{}_{jk}\gamma^j\gamma^k\frac{\partial\delta f} {\partial
\gamma^i}-\left[D'+\gamma^i\hat{e}_i(A)\right]q\frac{\partial f^{(0)}}{\partial q}=0.$}
\end{equation}
It is obvious how to write this in gauge invariant form
\begin{equation}
\fbox{$\displaystyle
(\partial_\eta+\gamma^i\hat{e}_i)\mathcal{F}_s-\hat{\Gamma}^i{}_{jk}\gamma^j\gamma^k\frac{\partial
\mathcal{F}_s} {\partial
\gamma^i}=\left[\Phi'+\gamma^i\hat{e}_i(\Psi)\right]q\frac{\partial f^{(0)}}{\partial q}.$}
\end{equation}
(From this the collisionless Boltzmann equation follows in any gauge; write this out.)

In the special case $K=0$ we obtain for the Fourier amplitudes, with $\mu:=
\mbox{\boldmath$\hat{k}\cdot\gamma$}$,
\begin{equation}
\fbox{$\displaystyle \mathcal{F}'_s+i\mu
k\mathcal{F}_s=\left[\Phi'+ik\mu\Psi\right]q\frac{\partial f^{(0)}}{\partial q}.$}
\end{equation}
This equation can be used for neutrinos as long as their masses are negligible (the
generalization to the massive case is easy).

\section{Boltzmann equation for photons}

The collision term for photons due to Thomson scattering on electrons will be derived in
Appendix B. We shall find that in the longitudinal gauge, ignoring polarization effects (to
be discussed later),
\begin{equation}
C[f]=x_en_e\sigma_Tp\left[\langle\delta f\rangle-\delta f-q\frac{\partial f^{(0)}}{\partial
q} \gamma^i\hat{e}_i(v_b)+\frac{3}{4}Q_{ij}\gamma^i\gamma^j\right].
\end{equation}
On the right, $x_e n_e$ is the unperturbed free electron density ($x_e = $ ionization
fraction), $\sigma_T$ the Thomson cross section, and $v_b$ the scalar velocity perturbation
of the baryons. Furthermore, we have introduced the spherical averages
\begin{eqnarray}
\langle\delta f\rangle &= & \frac{1}{4\pi}\int_{S^2}\delta f~ d\Omega_\gamma, \\
Q_{ij} &=& \frac{1}{4\pi} \int_{S^2} [ \gamma_i \gamma_j - \frac{1}{3}\delta_{ij}] \delta
f~ d\Omega_\gamma.
\end{eqnarray}
(Because of the tight coupling of electrons and ions we can take $v_e=v_b$.)

Since the left-hand side of (7.63) is equal to $(a/p_0)\mathcal{L}f$, the linearized
Boltzmann equation becomes
\begin{eqnarray}
(\partial_\eta+\gamma^i\hat{e}_i)\delta f &-&
\hat{\Gamma}^i{}_{jk}\gamma^j\gamma^k\frac{\partial\delta f} {\partial
\gamma^i}-\left[D'+\gamma^i\hat{e}_i(A)\right]q\frac{\partial f^{(0)}}{\partial q}\nonumber
\\
&=& ax_en_e\sigma_T\left[\langle\delta f\rangle-\delta f-q\frac{\partial f^{(0)}}{\partial
q} \gamma^i\hat{e}_i(v_b)+\frac{3}{4}Q_{ij}\gamma^i\gamma^j\right].\nonumber \\
\end{eqnarray}
This can immediately be written in a gauge invariant form, by replacing
\begin{equation}
\delta f\rightarrow \mathcal{F}_s,~~v_b\rightarrow V_b,~~ A\rightarrow \Psi,~~ D\rightarrow
\Phi.
\end{equation}

In our applications to the CMB we work with the gauge invariant {\it brightness
temperature} perturbation
\begin{equation}
\Theta_s(\eta,x^i,\gamma^j) = \int \mathcal{F}_s q^3dq \; \Big / \;\; 4\int f^{(0)}q^3dq.
\end{equation}
(The factor $4$ is chosen because of the Stephan-Boltzmann law, according to which $\delta
\rho / \rho = 4 \delta T / T.$) It is simple to translate the Boltzmann equation for
$\mathcal{F}_s$ to a kinetic equation for $\Theta_s$. Using
\[ \int q\frac{\partial f^{(0)}}{\partial q}q^3dq=-4\int f^{(0)}q^3dq \]
we obtain for the convective part (from the left-hand side of the Boltzmann equation for
$\mathcal{F}_s$)
\[\Theta'_s + \gamma^i \hat{e}_i(\Theta_s ) - \hat{\Gamma}^i{}_{jk}
\gamma^j \gamma^k \frac{\partial\Theta_s}{\partial \gamma^i}+\Phi'+\gamma^i
\hat{e}_i(\Psi). \] The collision term gives
\[ \dot{\tau}(\theta_0 -\Theta_s + \gamma^i \hat{e}_i V_b +
\frac{1}{16}\gamma^i\gamma^j \Pi_{ij}),\] with $\dot{\tau}=x_e n_e \sigma_T a/a_0,\;
\theta_0=\langle\Theta_s\rangle$ (spherical average), and
\begin{equation}
\frac{1}{12} \Pi_{ij}: = \frac{1}{4\pi}\int [\gamma_i\gamma_j - \frac{1}{3}\delta_{ij} ]
\Theta_s \; d\Omega_\gamma.
\end{equation}
The basic equation for $\Theta_s$ is thus
\begin{eqnarray}
\lefteqn{(\Theta_s + \Psi)' + \gamma^i \hat{e}_i(\Theta_s + \Psi ) - \hat{\Gamma}^i{}_{jk}
\gamma^j \gamma^k \frac{\partial}{\partial \gamma^i}(\Theta_s + \Psi ) = } \nonumber \\ & &
(\Psi'-\Phi') + \dot{\tau}(\theta_0 -\Theta_s + \gamma^i \hat{e}_i V_b +
\frac{1}{16}\gamma^i\gamma^j \Pi_{ij}).
\end{eqnarray}

In a mode decomposition we get for $K=0$ (I drop from now on the index $s$ on $\Theta$):
\begin{equation}
\fbox{$\displaystyle\Theta' + ik\mu(\Theta + \Psi) = -\Phi' +\dot{\tau}[ \theta_0 - \Theta
- i\mu V_b -\frac{1}{10}\theta_2 P_2(\mu)]$}
\end{equation}
(recall $V_b\rightarrow-(1/k)V_b$). The last term on the right comes about as follows. We
expand the Fourier modes $\Theta(\eta,k^i,\gamma^j)$ in terms of Legendre polynomials
\begin{equation}
\Theta(\eta,k^i,\gamma^j) = \sum_{l=0}^{\infty }(-i)^l \theta_l(\eta,k) P_l(\mu), \; \; \;
\mu = \mbox{\boldmath$\hat{k}\cdot\gamma$},
\end{equation}
and note that
\begin{equation}
\frac{1}{16}\gamma^i\gamma^j \Pi_{ij}=-\frac{1}{10}\theta_2 P_2(\mu)
\end{equation}
(Exercise). The expansion coefficients $\theta_l(\eta,k)$ in (7.75) are the
\textit{brightness moments}\footnote{In the literature the normalization of the $\theta_l$
is sometimes chosen differently: $\theta_l\rightarrow (2l+1)\theta_l$.}. The lowest three
have simple interpretations. We show that in the notation of Chap. 1:
\begin{equation}
\theta_0=\frac{1}{4}\Delta_{s\gamma},~~\theta_1=V_\gamma,~~\theta_2=\frac{5}{12}\Pi_\gamma.
\end{equation}

\paragraph{Derivation of (7.77).} We start from the general formula (see Sect. 7.1)
\begin{equation}
T^\mu_{(\gamma){}\nu}=\int p^\mu p_\nu f(p)\frac{d^3p}{p^0}=\int p^\mu p_\nu
f(p)pdp~d\Omega_\gamma.
\end{equation}
According to the general parametrization (1.156) we have
\begin{equation}
\delta T^0_{(\gamma){}0}=-\delta\rho_\gamma=-\int p^2 \delta f(p)pdp~d\Omega_\gamma.
\end{equation}
Similarly, in zeroth order
\begin{equation}
T_{(\gamma)}^{(0)0}{}_0=-\rho^{(0)}_\gamma=-\int p^2 f^{(0)}(p)pdp~d\Omega_\gamma.
\end{equation}
Hence,
\begin{equation}
\frac{\delta\rho_\gamma}{\rho^{(0)}_\gamma}=\frac{\int q^3 \delta f~dq~d\Omega_\gamma}{\int
q^3 f^{(0)}dq~d\Omega_\gamma}.
\end{equation}
In the longitudinal gauge we have
$\Delta_{s\gamma}=\delta\rho_\gamma/\rho^{(0)}_\gamma,~\mathcal{F}_s=\delta f$ and thus by
(7.71) and (7.75)
\[ \Delta_{s\gamma}=4\frac{1}{4\pi}\int\Theta~d\Omega_\gamma=4\theta_0.\]

Similarly,
\[ T^i_{(\gamma)0}=-h_\gamma v_\gamma^{\mid i}=\int p^ip_0\delta fp dp~d\Omega_\gamma\]
or
\begin{equation}
v_\gamma^{\mid i}=\frac{3}{4\rho^{(0)}_\gamma}\int\gamma^i\delta f p^3 dp~d\Omega_\gamma.
\end{equation}
With (7.80) and (7.71) we get
\begin{equation}
V_\gamma^{\mid i}=\frac{3}{4\pi}\int\gamma^i\Theta~d\Omega_\gamma.
\end{equation}
For the Fourier amplitudes this gauge invariant equation gives
($V_\gamma\rightarrow-(1/k)V_\gamma$)
\[-iV_\gamma\hat{k}^i=\frac{3}{4\pi}\int\gamma^i\Theta~d\Omega_\gamma \]
or
\[-iV_\gamma =\frac{3}{4\pi}\int\mu\Theta~d\Omega_\gamma.\]
Inserting here the decomposition (7.75) leads to the second relation in (7.77).

For the third relation we start from (1.156) and (7.79)
\[\delta T^i_{(\gamma)j} = \delta p_\gamma\delta^i{}_j+p^{(0)}_\gamma\left(\Pi^{\mid
i}_{\gamma\mid j}-\frac{1}{3}\delta^i{}_j\triangle\Pi_\gamma\right)=\int p^ip_j\delta f p~
dp~d\Omega_\gamma. \] From this and (7.79) we see that $\delta
p_\gamma=\frac{1}{3}\delta\rho_\gamma$, thus $\Gamma_\gamma=0$ (no entropy production with
respect to the photon fluid). Furthermore, since
$p^{(0)}_\gamma=\frac{1}{3}\rho^{(0)}_\gamma$ we obtain with (7.72)
\[ \Pi^{\mid i}_{\gamma\mid j}-\frac{1}{3}\delta^i{}_j\triangle\Pi_\gamma=4\cdot3\frac{1}{4\pi} \int
[\gamma^i\gamma_j - \frac{1}{3}\delta^i{}_j ] \Theta ~ d\Omega_\gamma=\Pi^i{}_j. \] In
momentum space ($\Pi_\gamma\rightarrow(1/k^2)\Pi_\gamma$) this becomes
\[ -(\hat{k}^i \hat{k}_j-\frac{1}{3})\Pi_\gamma =\Pi^i{}_j \]
or, contracting with $\gamma_i\gamma^j$ and using (7.76), the desired result.

\subsection*{Hierarchy for moment equations}

Now we insert the expansion (7.75) into the Boltzmann equation (7.74). Using the recursion
relations for the Legendre polynomials,
\begin{equation}
\mu P_l(\mu)=\frac{l}{2l+1}P_{l-1}(\mu)+\frac{l+1}{2l+1}P_{l+1}(\mu),
\end{equation}
we obtain
\begin{eqnarray*}
\sum_{l=0}^{\infty }(-i)^l \theta'_l P_l &+& ik\sum_{l=0}^{\infty }(-i)^l \theta_l
\left[\frac{l}{2l+1}P_{l-1}+\frac{l+1}{2l+1}P_{l+1}\right] +ik\Psi P_1 \\
&=& -\Phi'P_0-\dot{\tau} \left[ \sum_{l=1}^{\infty }(-i)^l \theta_l
P_l-iV_bP_1-\frac{1}{10}\theta_2P_2\right].
\end{eqnarray*}
Comparing the coefficients of $P_l$ leads to the following hierarchy of ordinary
differential equations for the brightness moments $\theta_l(\eta)$:
\begin{eqnarray}
\theta_0' & = &-\frac{1}{3}k \theta_1 - \Phi',  \\
\theta_1' & = & k\Bigl(\theta_0 + \Psi -\frac{2}{5}\theta_2 \Bigr)
-\dot{\tau}(\theta_1 -V_b),  \\
\theta_2' & = & k\Bigl(\frac{2}{3}\theta_1 -\frac{3}{7}\theta_3\Bigr) -
\dot{\tau} \frac{9}{10}\theta_2, \\
\theta_l' & = & k\Bigl(\frac{l}{2l-1}\theta_{l-1} -\frac{l+1}{2l+3}\theta_{l+1}\Bigr),  \;
\; \; l>2.
\end{eqnarray}

At this point it is interesting to compare the first moment equation (7.86) with the
phenomenological equation (1.212) for $\gamma$:
\begin{equation}
V'_\gamma=k\Psi+\frac{1}{4}\Delta_{s\gamma}-\frac{1}{6}k\Pi_\gamma+\mathcal{H}F_\gamma.
\end{equation}
On the other hand, (7.86) can be written with (7.77) as
\begin{equation}
V'_\gamma=k\Psi+\frac{1}{4}\Delta_{s\gamma}-\frac{1}{6}k\Pi_\gamma
-\dot{\tau}(V_\gamma-V_b).
\end{equation}
The two equations agree if the phenomenological force $F_\gamma$ is given by
\begin{equation}
\fbox{$\displaystyle \mathcal{H}F_\gamma=-\dot{\tau}(V_\gamma-V_b).$}
\end{equation}
From the general relation (1.203) we then obtain
\begin{equation}
F_b=-\frac{h_\gamma}{h_b}F_\gamma=-\frac{4\rho_\gamma}{3\rho_b}F_\gamma.
\end{equation}

\section{Tensor contributions to the Boltzmann equation}

Considering again only the case $K=0$, the metric (5.57) for tensor perturbations becomes
\begin{equation}
g_{\mu\nu}=a^2(\eta)[\eta_{\mu\nu}+2H_{\mu\nu}],
\end{equation}
where the $H_{\mu\nu}$ satisfy the TT gauge conditions (5.58). An adapted orthonormal
tetrad is
\begin{equation}
\theta^0=a(\eta)d\eta,~~~\theta^i=a(\delta^i{}_j+H^i{}_j) dx^j.
\end{equation}
Relative to this the connection forms are (Exercise):
\begin{equation}
\omega^0{}_i=\frac{a'}{a^2}\theta^i +\frac{1}{a}H'_{ij}\theta^j,~~ \omega^i{}_j
=\frac{1}{2a}(H^i{}_{k,j}-H_{jk}{}^{,i})\theta^k.
\end{equation}

For $\mathcal{L}f$ we get from (7.55) to first order
\begin{eqnarray*}
\mathcal{L}f&=& \frac{p^0}{a}f'+p^i \frac{1}{a}\hat{e}_i(f)+
\omega^i{}_0(\mathbf{p})p^0\frac{\partial f} {\partial p^i}+
\omega^i{}_j(\mathbf{p})p^j\frac{\partial f} {\partial p^i}\\
&=& \frac{p^0}{a}\left[f'+\frac{p^i}{p^0}\partial_i f+H'_{ij}p^j\frac{\partial f} {\partial
p^i}\right].
\end{eqnarray*}
Passing again to the variables $\eta,x^i,p,\gamma^i$ we obtain instead of (7.61)
\begin{eqnarray}
\frac{a}{p^0}\mathcal{L}f &=&  f^{(0)'}-\mathcal{H}p\frac{\partial f^{(0)}}{\partial
p}\nonumber \\
&+& (\delta f)'-\mathcal{H}p\frac{\partial\delta f}{\partial p} +\frac{p^i}{p^0}\partial_i
(\delta f)+H'_{ij}\gamma^i\gamma^jp\frac{\partial f^{(0)}} {\partial p}.
\end{eqnarray}
Instead of (7.63) we now obtain the following  collisionless Boltzmann equation
\begin{equation}
\fbox{$\displaystyle (\partial_\eta+\gamma^i\partial_i)\delta f+H'_{ij}\gamma^i\gamma^j
q\frac{\partial f^{(0)}} {\partial q}=0.$}
\end{equation}
For the temperature (brightness) perturbation this gives
\begin{equation}
\fbox{$\displaystyle (\partial_\eta+\gamma^i\partial_i)\Theta= H'_{ij}\gamma^i\gamma^j.$}
\end{equation}

This describes the influence of tensor modes on $\Theta$. The evolution of these tensor
modes is described according to (5.59) by
\begin{equation}
H''_{ij}+2\mathcal{H}H'_{ij}-\triangle H_{ij}=0,
\end{equation}
if we neglect tensor perturbations of the energy-momentum tensor. We shall study the
implications of the last two equations for the CMB fluctuations in Sect. 8.6.

\chapter{The Physics of CMB Anisotropies}

We have by now developed all ingredients for a full understanding of the CMB anisotropies.
In the present chapter we discuss these for the CDM scenario and primordial initial
conditions suggested by inflation (derived in Part II). Other scenarios, involving for
instance topological defects, are now strongly disfavored.

We shall begin by collecting all independent perturbation equations, derived in previous
chapters. There are fast codes that allow us to solve these equations very accurately,
given a set of cosmological parameters. It is, however, instructive to discuss first
various qualitative and semi-quantitative aspects. Finally, we shall compare numerical
results with observations, and discuss what has already come out of this, which is a lot.
In this connection we have to include some theoretical material on polarization effects,
because WMAP has already provided quite accurate data for the so-called E-polarization.

The B-polarization is much more difficult to get, and is left to future missions (Planck
satellite, etc). This is a very important goal, because accurate data will allow us to
determine the power spectrum of the gravity waves.

For further reading I recommend Chap. 8 of \cite{Cos4} and the the two research articles
\cite{HS1}, \cite{HS2}. For a well written review and extensive references, see \cite{Hu}.

\section{The complete system of perturbation\\ equations}

For references in later sections, we collect below the complete system of (independent)
perturbation equations for scalar modes and $K=0$ (see Sects. 1.5.C and 7.5). Let me first
recall and add some notation.

Unperturbed {\it background} quantities: $\rho_\alpha, p_\alpha$ denote the densities and
pressures for the species $\alpha = b$ (baryons and electrons), $\gamma$ (photons), $c$
(cold dark matter); the total density is the sum $\rho = \sum_\alpha \rho_\alpha$, and the
same holds for the total pressure $p$. We also use $w_\alpha = p_\alpha/\rho_\alpha,
w=p/\rho $. The sound speed of the baryon-electron fluid is denoted by $c_b$, and $R$ is
the ratio $3\rho_b/4\rho_\gamma $.

Here is the list of gauge invariant {\it scalar perturbation} amplitudes:
\begin{itemize}
\item $\delta_\alpha:=\Delta_{s\alpha}, \delta:=\Delta_s$ : density perturbations ($\delta
\rho_\alpha /\rho_\alpha, \delta \rho /\rho \;$  in the longitudinal gauge); clearly: $\rho
\; \delta = \sum \rho_\alpha \delta_\alpha$. \item $V_\alpha ,V$ : velocity perturbations;
$\rho(1+w)V=\sum_\alpha \rho_\alpha (1+w_\alpha)V_\alpha.$ \item $\theta_l, N_l$ :
brightness moments for photons and neutrinos. \item $\Pi_\alpha, \Pi$ : anisotropic
pressures;  $\Pi=\Pi_\gamma + \Pi_\nu$. For the lowest moments the following relations
hold:
\begin{equation}
\delta_\gamma = 4\theta_0, \; \; V_\gamma = \theta_1, \; \; \Pi_\gamma = \frac{12}{5}
\theta_2,
\end{equation}
and similarly for the neutrinos. \item $\Psi, \Phi $: Bardeen potentials for the metric
perturbation.
\end{itemize}

As \textit{independent} amplitudes we can choose: $\delta_b, \delta_c, V_b, V_c, \Phi,
\Psi, \theta_l, N_l$. The basic evolution equations consist of three groups.
\begin{itemize}
\item \textit{Fluid equations}:
\begin{eqnarray}
\delta_c' & = & -kV_c - 3\Phi',  \\
V_c' &=& -aHV_c + k\Psi;  \\
\delta_b' &=& -kV_b - 3\Phi',  \\
V_b' &=& -aHV_b + kc^2_b \delta_b + k\Psi + \dot{\tau}(\theta_1 -V_b) / R .
\end{eqnarray}
\item \textit{Boltzmann hierarchies} for photons (eqs. (7.85)-(7.88)) (and the
collisionless neutrinos):
\begin{eqnarray}
\theta_0' & = &-\frac{1}{3}k \theta_1 - \Phi',  \\
\theta_1' & = & k\Bigl(\theta_0 + \Psi -\frac{2}{5}\theta_2 \Bigr)
-\dot{\tau}(\theta_1 -V_b),  \\
\theta_2' & = & k\Bigl(\frac{2}{3}\theta_1 -\frac{3}{7}\theta_3\Bigr) -
\dot{\tau} \frac{9}{10}\theta_2, \\
\theta_l' & = & k\Bigl(\frac{l}{2l-1}\theta_{l-1} -\frac{l+1}{2l+3}\theta_{l+1}\Bigr),  \;
\; \; l>2.
\end{eqnarray}
\item \textit{Einstein equations} : We only need the following algebraic ones for each
mode:
\begin{eqnarray}
k^2\Phi &=& 4\pi G a^2 \rho\Bigl[ \delta +
3\frac{aH}{k}(1+w)V\Bigr ], \\
k^2(\Phi + \Psi) &=& -8\pi G a^2 p \; \Pi.
\end{eqnarray}
\end{itemize}

In arriving at these equations some approximations have been made which are harmless
\footnote{In the notation of Sect. 1.4 we have set $q_\alpha = \Gamma_\alpha =0$, and are
thus ignoring certain intrinsic entropy perturbations within individual components.},
except for one: We have ignored polarization effects in Thomson scattering. For
quantitative calculations these have to be included. Moreover, polarization effects are
highly interesting, as I shall explain later. We shall take up this topic in Sect. 8.7.

\section{Acoustic oscillations}

In this section we study the photon-baryon fluid. Our starting point is the following
approximate system of equations. For the baryons we use (8.4) and (8.5), neglecting the
term proportional to $c_b^2$. We truncate the photon hierarchy, setting $\theta_l=0$ for
$l\geq 3$. So we consider the system of first order equations:
\begin{eqnarray}
\theta_0' & = &-\frac{1}{3}k \theta_1 - \Phi',  \\
\theta_1' & = & k\Bigl(\theta_0 + \Psi -\frac{2}{5}\theta_2 \Bigr)
-\dot{\tau}(\theta_1 -V_b),  \\
\delta_b' &=& -kV_b - 3\Phi',  \\
V_b' &=& -aHV_b + kc^2_b \delta_b + k\Psi + \dot{\tau}(\theta_1 -V_b) / R ,
\end{eqnarray}
and (8.8). This is, of course, not closed ($\Phi$ and $\Psi$ are ``external'' potentials).

As long as the mean free path of photons is much shorter than the wavelength of the
fluctuation, the optical depth through a wavelength $\sim\dot{\tau}/k$ is
large\footnote{Estimate $\dot{\tau}/k$ as a function of redshift $z>z_{rec}$ and
$(aH/k)$.}. Thus the evolution equations may be expanded in the small parameter
$k/\dot{\tau}$.

In lowest order we obtain $\theta_1=V_b,~\theta_l=0$ for $l\geq2$, thus
$\delta'_b=3\theta'_0~(=3\delta'_\gamma/4)$.

Going to first order, we can replace in the following form of (8.15)
\begin{equation}
\theta_1-V_b=\dot{\tau}^{-1}R\left[V'_b+\frac{a'}{a}\theta_1-k\Psi\right]
\end{equation}
on the right $V_b$ by $\theta_1$:
\begin{equation}
\theta_1-V_b=\dot{\tau}^{-1}R\left[\theta'_1+\frac{a'}{a}\theta_1-k\Psi\right].
\end{equation}
We insert this in (8.13), and set in \textit{first order} also $\theta_2=0$:
\begin{equation}
\theta'_1=k(\theta_0+\Psi)-R\left[\theta'_1+\frac{a'}{a}V_b-k\Psi\right].
\end{equation}
Using $a'/a=R'/R$, we obtain from this
\begin{equation}
\theta'_1=\frac{1}{1+R}k\theta_0+k\Psi-\frac{R'}{1+R}\theta_1.
\end{equation}
Combining this with (8.12), we obtain by eliminating $\theta_1$ the \textit{driven
oscillator equation}:
\begin{equation}
\fbox{$\displaystyle
\theta''_0+\frac{R}{1+R}\frac{a'}{a}\theta'_0+c^2_sk^2\theta_0=F(\eta),$}
\end{equation}
with
\begin{equation}
c^2_s=\frac{1}{3(1+R)},~~~F(\eta)=-\frac{k^2}{3}\Psi-\frac{R}{1+R}\frac{a'}{a}\Phi'-\Phi''.
\end{equation}
According to (1.186) and (1.187) $c_s$ is the velocity of sound in the approximation
$c_b\approx 0$. It is suggestive to write (8.20) as ($m_{eff}\equiv1+R$)
\begin{equation}
(m_{eff}\theta'_0)'+\frac{k^2}{3}(\theta_0+m_{eff}\Psi)=-(m_{eff}\Phi')'.
\end{equation}

This equation provides a lot of insight, as we shall see. It may be interpreted as follows:
The change in momentum of the photon-baryon fluid is determined by a competition between
pressure restoring and gravitational driving forces.

Let us, in a first step, ignore the time dependence of $m_{eff}$ (i.e., of the
baryon-photon ratio $R$), then we get the forced harmonic oscillator equation
\begin{equation}
m_{eff}\theta_0'' + \frac{k^2}{3}\theta_0 = -\frac{k^2}{3}m_{eff}\Psi - (m_{eff}\Phi')'.
\end{equation}
The effective mass $m_{eff}=1+R$ accounts for the inertia of baryons. Baryons also
contribute gravitational mass to the system, as is evident from the right hand side of the
last equation. Their contribution to the pressure restoring force is, however, negligible.

We now ignore in (8.23) also the time dependence of the gravitational potentials
$\Phi,\Psi$. With (8.21) this then reduces to
\begin{equation}
\theta_0'' + k^2 c_s^2 \theta_0 = -\frac{1}{3}k^2\Psi.
\end{equation}
This simple harmonic oscillator under constant acceleration provided by gravitational
infall can immediately be solved:
\begin{equation}
\theta_0(\eta) = [ \theta_0(0) + (1+R)\Psi]\cos(kr_s) +
\frac{1}{kc_s}\dot{\theta}_0(0)\sin(kr_s) - (1+R)\Psi,
\end{equation}
where $r_s(\eta)$ is the comoving sound horizon $\int c_s d\eta$.

We know (see (6.54)) that for {\it adiabatic} initial conditions there is only a cosine
term. Since we shall see that the ``effective'' temperature fluctuation is  $\Delta
T=\theta_0+\Psi$, we write the result as
\begin{equation}
\Delta T(\eta,k) = [ \Delta T(0,k) + R\Psi]\cos(kr_s(\eta)) -R\Psi.
\end{equation}

\subsubsection*{Discussion}

In the radiation dominated phase ($R=0$) this reduces to $\Delta T(\eta)
 \propto \cos kr_s(\eta)$, which shows that the oscillation of $\theta_0$ is
displaced by gravity. The zero point corresponds to the state at which gravity and pressure
are balanced. The displacement $-\Psi>0$ yields hotter photons in the potential well since
gravitational infall not only increases the number density of the photons, but also their
energy through gravitational blue shift. However, well after last scattering the photons
also suffer a redshift when climbing out of the potential well, which precisely cancels the
blue shift. Thus the effective temperature perturbation we see in the CMB anisotropies is
indeed $\Delta T = \theta_0 + \Psi$, as we shall explicitely see later.

It is clear from (8.25) that a characteristic wave-number is $k=\pi/r_s(\eta_{dec})\\
\approx \pi/c_s\eta_{dec}$. A spectrum of $k$-modes will produce a sequence of peaks with
wave numbers
\begin{equation}
k_m = m\pi/ r_s(\eta_{dec}), \; \; m = 1,2,...~.
\end{equation}
Odd peaks correspond to the compression phase (temperature crests), whereas even peaks
correspond to the rarefaction phase (temperature troughs) inside the potential wells. Note
also that the characteristic length scale $r_s(\eta_{dec})$, which is reflected in the peak
structure, is determined by the underlying unperturbed Friedmann model. This comoving sound
horizon at decoupling depends on cosmological parameters, but not on $\Omega_\Lambda$. Its
role will further be discussed below.

Inclusion of baryons not only changes the sound speed, but gravitational infall leads to
greater compression of the fluid in a potential well, and thus to a further displacement of
the oscillation zero point (last term in (8.25)). This is not compensated by the redshift
after last scattering, since the latter is not affected by the baryon content. As a result
all peaks from compression are enhanced over those from rarefaction. Hence, the relative
heights of the first and second peak is a \textit{sensitive measure of the baryon content}.
We shall see that the inferred baryon abundance from the present observations is in
complete agreement with the results from big bang nucleosynthesis.

What is the influence of the slow evolution of the effective mass $m_{eff}=1+R$? Well, from
the adiabatic theorem we know that for a slowly varying $m_{eff}$ the ratio
energy/frequency is an adiabatic invariant. If $A$ denotes the amplitude of the
oscillation, the energy is $\frac{1}{2}m_{eff}\omega^2 A^2$. According to (8.21) the
frequency $\omega=kc_s$ is proportional to $m_{eff}^{-1/2}$. Hence $A\propto\omega^{-1/2}
\propto m_{eff}^{1/4}\propto(1+R)^{-1/4}$.
\paragraph{Photon diffusion.}
In \textit{second order} we do no more neglect $\theta_2$ and use in addition (8.8),
\begin{equation}
\theta_2'= k\Bigl(\frac{2}{3}\theta_1 -\frac{3}{7}\theta_3\Bigr) - \dot{\tau}
\frac{9}{10}\theta_2,
\end{equation}
with $\theta_3\simeq 0$. This gives in leading order
\begin{equation}
\theta_2\simeq \frac{20}{27}\dot{\tau}^{-1}k\theta_1.
\end{equation}
If we neglect in the Euler equation for the baryons the term proportional to $a'/a$, then
the first order equation (8.17) reduces to
\begin{equation}
V_b=\theta_1-\dot{\tau}^{-1}R[\theta'_1-k\Psi].
\end{equation}
We use this in (8.16) without the term with $a'/a$, to get
\begin{equation}
\theta_1-V_b=\dot{\tau}^{-1}R[\theta'_1-k\Psi]-\frac{R^2}{\dot{\tau}^2}(\theta''_1-k\Psi').
\end{equation}
This is now used in (8.13) with the approximation (8.29) for $\theta_2$. One finds
\begin{equation}
(1+R)\theta'_1=k[\theta_0+(1+R)\Psi]-\frac{8}{27}\frac{k^2}{\dot{\tau}}+\frac{R^2}{\dot{\tau}}
(\theta''_1-k\Psi').
\end{equation}
In the last term we use the first order approximation of this equation, i.e.,
\[(1+R)(\theta'_1-k\Psi)=k\theta_0, \] and obtain
\begin{equation}
(1+R)\theta'_1=k[\theta_0+(1+R)\Psi]-\frac{8}{27}\frac{k^2}{\dot{\tau}}+\frac{k}{\dot{\tau}}
\frac{R^2}{1+R}\theta'_0.
\end{equation}
Finally, we eliminate in this equation $\theta'_1$ with the help of (8.12). After some
rearrangements we obtain
\begin{equation}
\theta''_0+\frac{k^2}{3\dot{\tau}}\left[\frac{R^2}{(1+R)^2}+\frac{8}{9}\frac{1}{1+R}\right]\theta'_0
+\frac{k^2}{3(1+R)}\theta_0=-\frac{k^2}{3}\Psi-\Phi''-\frac{8}{27}\frac{k^2}{3\dot{\tau}}\frac{1}{1+R}\Phi'.
\end{equation}
The term proportional to $\theta'_0$ in this equation describes the \textit{damping due to
photon diffusion}. Let us determine the characteristic damping scale.

If we neglect in the homogeneous equation the time dependence of all coefficients, we can
make the ansatz $\theta_0\propto\exp(i\int\omega d\eta)$. (We thus ignore variations on the
time scale $a/\dot{a}$ with those corresponding to the oscillator frequency $\omega$.) The
dispersion law is determined by
\[ -\omega^2+i\frac{\omega}{3}\frac{k^2}{\dot{\tau}}\left[\frac{R^2}{(1+R)^2}+\frac{8}{9}\frac{1}{1+R}\right]
+\frac{k^2}{3}\frac{1}{1+R}=0, \] giving
\begin{equation}
\omega=\pm kc_s+i\frac{k^2}{6}\frac{1}{\dot{\tau}}\frac{R^2+\frac{8}{9}(1+R)}{(1+R)^2}.
\end{equation}
So acoustic oscillations are damped as $\exp[-k^2/k^2_D]$, where
\begin{equation}
k^2_D=\frac{1}{6}\int\frac{1}{\dot{\tau}}\frac{R^2+\frac{8}{9}(1+R)}{(1+R)^2}d\eta.
\end{equation}
This is sometimes written in the form
\begin{equation}
k^2_D=\frac{1}{6}\int\frac{1}{\dot{\tau}}\frac{R^2+\frac{4}{5}f_2^{-1}(1+R)}{(1+R)^2}d\eta.
\end{equation}
Our result corresponds to $f_2=9/10$. In some books and papers one finds $f_2=1$. If we
would include polarization effects, we would find $f_2=3/4$. The damping of acoustic
oscillations is now clearly observed.

\paragraph{Sound horizon}

The sound horizon determines according to (8.27) the position of the first peak. We compute
now this important characteristic scale.

The comoving sound horizon at time $\eta$ is
\begin{equation}
r_s(\eta)=\int_0^\eta c_s(\eta')d\eta'.
\end{equation}
Let us write this as a redshift integral, using $1+z=a_0/a(\eta)$, whence by (0.52) for
$K\neq0$
\begin{equation}
d\eta=-\frac{1}{a_0}\frac{dz}{H(z)}=-|\Omega_K|^{1/2}\frac{dz}{E(z)}.
\end{equation}
Thus
\begin{equation}
r_s(z)=|\Omega_K|^{1/2}\int_z^\infty c_s(z')\frac{dz'}{E(z')}.
\end{equation}
This is seen at present under the (small) angle
\begin{equation}
\theta_s(z)=\frac{r_s(z)}{r(z)},
\end{equation}
where $r(z)$ is given by (0.56) and (0.57):
\begin{equation}
r(z) = \mathcal{S}\left(|\Omega_K|^{1/2} \int_0^z \frac{dz'}{E(z')}\right).
\end{equation}

Before decoupling the sound velocity is given by (8.21), with
\begin{equation}
R=\frac{3}{4}\frac{\Omega_b}{\Omega_\gamma}\frac{1}{1+z}.
\end{equation}
We are left with two explicit integrals. For $z_{dec}$ we can neglect in (8.40) the
curvature and $\Lambda$ terms. The integral can then be done analytically, and is in good
approximation proportional to $(\Omega_M)^{-1/2}$ (exercise). Note that (8.42) is closely
related to the angular diameter distance to the last scattering surface (see (0.34) and
(0.60)). A numerical calculation shows that $\theta_s(z_{dec})$ depends mainly on the
curvature parameter $\Omega_K$. For a typical model with $\Omega_\Lambda=2/3,~
\Omega_bh_0^2=0.02,~\Omega_Mh_0^2=0.16,~n=1$ the parameter sensitivity is approximately
\cite{Hu}
\[ \frac{\Delta\theta_s}{\theta_s}\approx 0.24\frac{\Delta(\Omega_Mh_0^2)}{\Omega_Mh_0^2)}
-0.07\frac{\Delta(\Omega_bh_0^2)}{\Omega_bh_0^2}+0.17\frac{\Delta\Omega_\Lambda}{\Omega_\Lambda}
+1.1\frac{\Delta\Omega_{tot}}{\Omega_{tot}}. \]

\section{Formal solution for the moments $\theta_l$}

We derive in this section a useful integral representation for the brightness moments at
the present time. The starting point is the Boltzmann equation (7.74) for the brightness
temperature fluctuations $\Theta(\eta,k,\mu)$,
\begin{equation}
(\Theta+\Psi)' + ik\mu(\Theta + \Psi) =\Psi' -\Phi' +\dot{\tau}[ \theta_0 - \Theta - i\mu
V_b -\frac{1}{10}\theta_2 P_2(\mu)].
\end{equation}
This is of the form of an inhomogeneous linear differential equation
\[ y'+g(x)y=h(x),\]
whose solution can be written as (variation of constants)
\[y(x)=e^{-G(x)}\left\{y_0+\int_{x_0}^x h(x')e^{G(x')}dx'\right\}, \]
with
\[ G(x)=\int_{x_0}^x g(u)du.\]
In our case $g=ik\mu+\dot{\tau},~h=\dot{\tau}[ \theta_0 +\Psi - i\mu V_b
-\frac{1}{10}\theta_2 P_2(\mu)]+\Psi'-\Phi'$. Therefore, the present value of $\Theta+\Psi$
can formally be expressed as
\begin{eqnarray}
\lefteqn{ (\Theta + \Psi)(\eta_0,\mu;k)= }\nonumber \\ & & \int_0^{\eta_0} d\eta\Bigl[
\dot{\tau}(\theta_0 + \Psi -i\mu V_b - \frac{1}{10}\theta_2 P_2)+ \Psi'-\Phi'\Bigr ]
 e^{-\tau(\eta,\eta_0)} e^{ik\mu(\eta-\eta_0)},\nonumber \\
\end{eqnarray}
where
\begin{equation}
\tau(\eta,\eta_0) = \int_\eta^{\eta_0} \dot{\tau} d\eta
\end{equation}
is the {\it optical depth}. The combination $\dot{\tau}e^{-\tau}$ is the (conformal) {\it
time visibility function}. It has a simple interpretation: Let $p(\eta,\eta_0)$ be the
probability that a photon did not scatter between $\eta$ and today ($\eta_0$). Clearly,
$p(\eta-d\eta,\eta_0) = p(\eta,\eta_0)(1-\dot{\tau}d\eta)$. Thus $p(\eta,\eta_0) =
e^{-\tau(\eta,\eta_0)}$, and the visibility function times $d\eta$ is the probability that
a photon last scattered between $\eta$ and $\eta+d\eta$. The visibility function is
therefore {\it strongly peaked} near decoupling. This is very useful, both for analytical
and numerical purposes.

In order to obtain an integral representation for the multipole moments $\theta_l$, we
insert in (8.45) for the $\mu$-dependent factors the following  expansions in terms of
Legendre polynomials:
\begin{eqnarray}
e^{-ik\mu(\eta_0-\eta)}&=&\sum_l(-i)^l(2l+1)j_l(k(\eta_0-\eta))P_l(\mu),\\
-i\mu e^{-ik\mu(\eta_0-\eta)}&=&\sum_l(-i)^l(2l+1)j'_l(k(\eta_0-\eta))P_l(\mu),\\
(-i)^2P_2(\mu)e^{-ik\mu(\eta_0-\eta)}&=&\sum_l(-i)^l(2l+1)\frac{1}{2}[3j''_l+j_l]P_l(\mu).
\end{eqnarray}
Here, the first is well-known. The others can be derived from (8.47) by using the recursion
relations (7.84) for the Legendre polynomials and the following ones for the spherical
Bessel functions
\begin{equation}
lj_{l-1}-(l+1)j_{l+1}=(2l+1)j'_l,
\end{equation}
or by differentiation of (8.47) with respect to $k(\eta_0-\eta)$. Using the definition
(7.75) of the moments $\theta_l$, we obtain for $l\geq 2$ the following useful formula:
\begin{equation} \frac{
\theta_l(\eta_0)}{2l+1} = \int_0^{\eta_0} d\eta e^{-\tau(\eta)}\Bigl[ (\dot{\tau} \theta_0
+ \dot{\tau}\Psi +\Psi'-\Phi' )j_l(k(\eta_0-\eta))
 +\dot{\tau}V_b j_l' + \dot{\tau}\frac{1}{20}\theta_2(3j_l''+ j_l)\Bigr ].
\end{equation}
\paragraph{Sudden decoupling approximation.}
In a reasonably good approximation we can replace the visibility function by the
$\delta$-function, and obtain with $\Delta\eta\equiv \eta_0 -\eta_{dec},~
V_b(\eta_{dec})\simeq \theta_1(\eta_{dec})$ the instructive result
\begin{equation}
\frac{\theta_l(\eta_0,k)}{2l+1}\simeq[ \theta_0 +\Psi](\eta_{dec},k) j_l(k\Delta \eta) +
\theta_1(\eta_{dec},k)j_l'(k\Delta\eta) + ISW + Quad.
\end{equation}
Here, the quadrupole contribution (last term) is not important. ISW denotes the {\it
integrated Sachs-Wolfe effect}:
\begin{equation}
ISW = \int_0^{\eta_0} d\eta(\Psi'-\Phi')j_l(k(\eta_0- \eta)),
\end{equation}
which only depends on the time variations of the Bardeen potentials between recombination
and the present time.

The interpretation of the first two terms in (8.52) is quite obvious: The first describes
the fluctuations of the {\it effective} temperature $\theta_0 +\Psi$ on the cosmic
photosphere, as we would see them for free streaming between there and us, if the
gravitational potentials would not change in time. ($\Psi$ includes blue- and redshift
effects.) The dipole term has to be interpreted, of course, as a Doppler effect due to the
velocity of the baryon-photon fluid. It turns out that the integrated Sachs-Wolfe effect
enhances the anisotropy on scales comparable to the Hubble length at recombination.

In this approximate treatment we have to know -- beside the ISW -- only the effective
temperature $\theta_0 + \Psi$ and the velocity moment $\theta_1$ at decoupling. The main
point is that eq. (8.52) provides a good understanding of the physics of the CMB
anisotropies. Note that the individual terms are all gauge invariant. In gauge dependent
methods interpretations would be ambiguous.

\section{Angular correlations of temperature\\ fluctuations}

The system of evolution equations has to be supplemented by initial conditions. We can not
hope to be able to predict these, but at best their statistical properties (as, for
instance, in inflationary models). Theoretically, we should thus regard the brightness
temperature perturbation $\Theta(\eta,x^i,\gamma^j)$ as a random field. Of special interest
is its angular correlation function at the present time $\eta_0$. Observers measure only
one realization of this, which brings unavoidable {\it cosmic variances} (see the
Introduction to Part III).

For further elaboration we insert (7.75) into the Fourier expansion of $\Theta$, obtaining
\begin{equation}
\Theta(\eta,\mbox{\boldmath$x$},\mbox{\boldmath$\gamma$}) = (2\pi)^{-3/2} \int d^3k\sum_l
\theta_l(\eta,k) G_l(\mbox{\boldmath$x$},\mbox{\boldmath$\gamma$};\mbox{\boldmath$k$}),
\end{equation}
where
\begin{equation}
G_l(\mbox{\boldmath$x$},\mbox{\boldmath$\gamma$};\mbox{\boldmath$k$}) = (-i)^l
P_l(\mbox{\boldmath$\hat{k}\cdot \gamma$}) \exp(i\mbox{\boldmath$k\cdot x$}).
\end{equation}
With the addition theorem for the spherical harmonics the Fourier transform is thus
\begin{equation}
\Theta(\eta,\mbox{\boldmath$k$},\mbox{\boldmath$\gamma$}) = \sum_{lm}
Y_{lm}(\mbox{\boldmath$\gamma$})\frac{4\pi}{2l+1}\theta_l(\eta,k)\; (-i)^l
Y_{lm}^{\ast}(\mbox{\boldmath$\hat{k}$}).
\end{equation}
This has to be regarded as a stochastic field of $\textbf{k}$ (parametrized by
$\mbox{\boldmath$\gamma$}$ ). The randomness is determined by the statistical properties at
an early time, for instance after inflation. If we write $\Theta$ as (dropping $\eta$)
$\mathcal{R}(\mbox{\boldmath$k$})\times(\Theta(\mbox{\boldmath$k$},\mbox{\boldmath$\gamma$})/
\mathcal{R}(\mbox{\boldmath$k$}))$, the second factor evolves deterministically and is
independent of the initial amplitudes, while the stochastic properties are completely
determined by those of $\mathcal{R}(\mbox{\boldmath$k$})$. In terms of the power spectrum
of $\mathcal{R}(\mbox{\boldmath$k$})$,
\begin{equation}
\langle\mathcal{R}(\mbox{\boldmath$k$})\mathcal{R}^\ast(\mbox{\boldmath$k'$})\rangle
=\frac{2\pi^2}{k^3}P_{\mathcal{R}}(k)\delta^3(\mbox{\boldmath$k$}-\mbox{\boldmath$k'$})
\end{equation}
(see (5.14)), we thus have for the correlation function in momentum space
\begin{equation}
\langle\Theta(\mbox{\boldmath$k$},\mbox{\boldmath$\gamma$})
\Theta^\ast(\mbox{\boldmath$k'$},\mbox{\boldmath$\gamma'$})\rangle=
\frac{2\pi^2}{k^3}P_{\mathcal{R}}(k)\delta^3(\mbox{\boldmath$k$}-\mbox{\boldmath$k'$})
\frac{\Theta(k,\mbox{\boldmath$\hat{k}\cdot \gamma$})}{\mathcal{R}(k)}
\frac{\Theta^\ast(k,\mbox{\boldmath$\hat{k}\cdot \gamma'$})}{\mathcal{R}^\ast(k)}.
\end{equation}
Because of the $\delta$-function the correlation function in \textbf{x}-space is
\begin{equation}
\langle\Theta(\mbox{\boldmath$x$},\mbox{\boldmath$\gamma$})
\Theta(\mbox{\boldmath$x$},\mbox{\boldmath$\gamma'$})\rangle=\int\frac{d^3k}{(2\pi)^3} \int
d^3k'\langle\Theta(\mbox{\boldmath$k$},\mbox{\boldmath$\gamma$})
\Theta(\mbox{\boldmath$k'$},\mbox{\boldmath$\gamma'$})\rangle.
\end{equation}
Inserting here (8.56) and (8.58) finally gives
\begin{equation}
\langle\Theta(\mbox{\boldmath$x$},\mbox{\boldmath$\gamma$})
\Theta(\mbox{\boldmath$x$},\mbox{\boldmath$\gamma'$})\rangle=\frac{1}{4\pi}\sum_l(2l+1)C_l
P_l(\mbox{\boldmath$\gamma\cdot\gamma'$}),
\end{equation}
with
\begin{equation}
\fbox{$\displaystyle\frac{(2l+1)^2}{4\pi}C_l = \int_0^\infty
\frac{dk}{k}\left|\frac{\theta_l(k)}{\mathcal{R}(k)}\right|^2 P_{\mathcal{R}}(k).$}
\end{equation}

Instead of $\mathcal{R}(k)$ we could, of course, use another perturbation amplitude. Note
also that we can take $\mathcal{R}(k)$ and $ P_{\mathcal{R}}(k)$ at any time. If we choose
an early time when $ P_{\mathcal{R}}(k)$ is given by its primordial value, $
P^{(prim)}_{\mathcal{R}}(k)$, then the ratios inside the absolute value,
$\theta_l(k)/\mathcal{R}(k)$, are two-dimensional \textit{CMB transfer functions}.

\section{Angular power spectrum for large scales}

The {\it angular power spectrum} is defined as $l(l+1)C_l$ versus $l$. For large scales,
i.e., small $l$, observed first with COBE, the first term in eq. (8.52) dominates. Let us
have a closer look at this so-called Sachs-Wolfe contribution.

For large scales (small $k$) we can neglect in the first equation (8.6) of the Boltzmann
hierarchy the term proportional to $k$: $\theta_0'\approx - \Phi'\approx \Psi'$, neglecting
also $\Pi$ (i.e., $\theta_2$) on large scales. Thus
\begin{equation}
\theta_0(\eta)\approx \theta_0(0) + \Psi(\eta) - \Psi(0).
\end{equation}
To proceed, we need a relation between $\theta_0(0)$ and $\Psi(0)$. This can be obtained by
looking at superhorizon scales in the tight coupling limit, using the results of Sect. 6.1.
(Alternatively, one can investigate the Boltzmann hierarchy in the radiation dominated
era.)

From (7.77) and (1.175) or (1.217) we get (recall $x=Ha/k$)
\[ \theta_0=\frac{1}{4}\Delta_{s\gamma}=\frac{1}{4}\Delta_{c\gamma}-xV.\]
The last term can be expressed in terms of $\Delta$, making use of (6.10) for $w=1/3$,
\[xV=-\frac{3}{4}x^2(D-1)\Delta. \]
Moreover, we have from (6.41)
\[\frac{3}{4}\Delta_{c\gamma}=\frac{\zeta+1}{\zeta+4/3}\Delta-\frac{\zeta}{\zeta+4/3}S. \]
Putting things together, we obtain for $\zeta\ll 1$
\begin{equation}
\theta_0=\frac{3}{4}\left[x^2(D-1)+\frac{1}{4}\right]\Delta-\frac{1}{4}\zeta S,
\end{equation}
thus
\begin{equation}
\theta_0\simeq\frac{3}{4}x^2(D-1)\Delta-\frac{1}{4}\zeta S,
\end{equation}
on superhorizon scales ($x\gg 1$).

For adiabatic perturbations we can use here the expansion (6.39) for $\omega\ll 1$ and get
with (6.9)
\begin{equation}
\theta_0(0)\simeq \frac{3}{4}x^2\Delta=-\frac{1}{2}\Psi(0).
\end{equation}
For isocurvature perturbations, the expansion (6.40) gives
\begin{equation}
\theta_0(0)=\Psi(0)=0.
\end{equation}
Hence, the initial condition for the effective temperature is
\begin{equation}
(\theta_0+\Psi)(0) = \left\{\begin{array}{r@{\quad:\quad}l} \frac{1}{2}\Psi(0) & (\mbox{adiabatic)}\\
0 & (\mbox{isocurvature}) .\end{array} \right.
\end{equation}
If this is used in (8.62) we obtain
\[ \theta_0(\eta)=\Psi(\eta)-\frac{3}{2}\Psi(0)~~~\mbox{for adiabatic perturbations}.\]
On large scales (2.32) gives for $\zeta\gg1$, in particular for $\eta_{rec}$,
\begin{equation}
\Psi(\eta)=\frac{9}{10}\Psi(0).
\end{equation}
Thus we obtain the result (Sachs-Wolfe)
\begin{equation}
\fbox{$\displaystyle(\theta_0 +\Psi)(\eta_{dec}) =
\frac{1}{3}\Psi(\eta_{dec})$}~~~\mbox{for \textit{adiabatic} perturbations}.
\end{equation}

On the other hand, we obtain for isocurvature perturbations with (8.66)
$\theta_0(\eta)=\Psi(\eta)$, thus
\begin{equation}
\fbox{$\displaystyle(\theta_0 +\Psi)(\eta_{dec}) = 2\Psi(\eta_{dec})$}~~~\mbox{for
\textit{isocurvature} perturbations}.
\end{equation}
Note the factor 6 difference between the two cases. The Sachs-Wolfe contribution to the
$\theta_l$ is therefore
\begin{equation}
\frac{\theta^{SW}_l(k)}{2l+1} = \left\{\begin{array}{r@{\quad:\quad}l}
\frac{1}{3}\Psi(\eta_{dec})j_l(k\Delta\eta) & (\mbox{\textit{adiabatic})}\\
2 \Psi(\eta_{dec})j_l(k\Delta\eta)& (\mbox{\textit{isocurvature}}) .\end{array} \right.
\end{equation}

We express at this point $\Psi(\eta_{dec})$ in terms of the primordial values of
$\mathcal{R}$ and $S$. For adiabatic perturbations $\mathcal{R}$ is constant on
superhorizon scales (see (1.138)), and according to (4.67) we have in the matter dominated
era $\Psi=-\frac{3}{5}\mathcal{R}$. On the other hand, for isocurvature perturbations the
entropy perturbation $S$ is constant on superhorizon scales (see Sect. 6.2.1), and for
$\zeta\gg1$ we have according to (6.45) and (6.9) $\Psi=-\frac{1}{5}S$. Hence we find
\begin{equation}
\fbox{$\displaystyle (\theta_0 +\Psi)(\eta_{dec})
=-\frac{1}{5}(\mathcal{R}^{(prim)}+2S^{(prim)}).$}
\end{equation}

The result (8.71) inserted into (8.61) gives the the dominant Sachs-Wolfe contribution to
the coefficients $C_l$ for large scales (small $l$). For adiabatic initial fluctuations we
obtain with (8.72)
\begin{equation}
\fbox{$\displaystyle C^{SW}_l = \frac{4\pi}{25}\int_0^\infty
\frac{dk}{k}\left|j_l(k\Delta\eta)\right|^2 P^{(prim)}_{\mathcal{R}}(k).$}
\end{equation}
Here we insert (6.82) and obtain
\begin{equation}
C_l^{SW}\simeq \pi H_0^{1-n}\delta^2_H \left(\frac{\Omega_M}{D_g(0)}\right)^2 \int_0^\infty
\frac{dk}{k^{2-n}}\left|j_l(k\Delta\eta)\right|^2.
\end{equation}
The integral can be done analytically. Eq. 11.4.34 in \cite{Abr} implies as a special case
\begin{eqnarray}
\int_0^\infty t^{-\lambda}[J_\mu(at)]^2dt &=& \frac{\Gamma(\frac{2\mu-\lambda+1}{2})}
{2^\lambda a^{1-\lambda}\Gamma(\mu+1)\Gamma(\frac{\lambda+1}{2})}\nonumber\\
&\times& \,_{2\!}F_1\left(\frac{2\mu-\lambda+1}{2},\frac{-\lambda+1}{2};\mu+1;1\right).
\end{eqnarray}
Since
\[ j_l(x)=\sqrt{\frac{\pi}{2x}}J_{l+\frac{1}{2}}(x)\]
the integral in (8.74) is of the form (8.75). If we also use Eq. 15.1.20 of the same
reference,
\[\,_{2\!}F_1(\alpha,\beta;\gamma;1)=\frac{\Gamma(\gamma)\Gamma(\gamma-\alpha-\beta)}
{\Gamma(\gamma-\alpha)\Gamma(\gamma-\beta)}, \] we obtain
\begin{equation}
\int_0^\infty
t^{n-2}[j_l(ta)]^2dt=\frac{\pi}{2^{4-n}a^{n-1}}\frac{\Gamma(3-n)}{[\Gamma(\frac{4-n}{2})]^2}
\frac{\Gamma(\frac{2l+n-1}{2})}{\Gamma(\frac{2l+5-n}{2})}
\end{equation}
and thus
\begin{equation}
\fbox{$\displaystyle C_l^{SW}\simeq 2^{n-4}\pi^2 (H_0\eta_0)^{1-n}\delta^2_H
\left(\frac{\Omega_M}{D_g(0)}\right)^2\frac{\Gamma(3-n)}{[\Gamma(\frac{4-n}{2})]^2}
\frac{\Gamma(\frac{2l+n-1}{2})}{\Gamma(\frac{2l+5-n}{2})}.$}
\end{equation}
For a \textit{Harrison-Zel'dovich spectrum} ($n=1$) we get
\begin{equation}
l(l+1)C_l^{SW}=\frac{\pi}{2}\delta^2_H \left(\frac{\Omega_M}{D_g(0)}\right)^2.
\end{equation}
Because the right-hand side is a constant one usually plots the quantity $l(l+1)C_l$ (often
divided by $2\pi$).

\section{Influence of gravity waves on\\ CMB anisotropies}

In this section we study the effect of a stochastic gravitational wave background on the
CMB anisotropies. According to Sect. 5.2  such a background is unavoidably produced in
inflationary models.

\paragraph{A. Basic equations.} We consider only the case $K=0$. Let us first recall some
basic formulae from Sects. 5.2 and 7.6. The metric for tensor modes is of the form
\begin{equation}
g=a^2(\eta)[-d\eta^2+(\delta_{ij}+2H_{ij})dx^idx^j].
\end{equation}
For a mode $H_{ij}\propto\exp(i\mbox{\boldmath$k\cdot x$})$, the tensor amplitudes satisfy
\begin{equation}
H^i{}_i=0,~~~H^i{}_jk^j=0.
\end{equation}
The tensor perturbations of the energy-momentum tensor can be parametrized as follows
\begin{equation}
\delta T^0{}_0=0,~~\delta T^0{}_i=0,~~ \delta T^i{}_j=\Pi^i_{(T)j},
\end{equation}
where $\Pi^i_{(T)j}$ satisfies in $\textbf{k}$-space
\begin{equation}
\Pi^i_{(T)i}=0,~~~\Pi^i_{(T)j}k^j=0.
\end{equation}
According to (5.59) the Einstein equations reduce to
\begin{equation}
H''_{ij}+2\frac{a'}{a}H'_{ij}+k^2H_{ij}=8\pi Ga^2\Pi_{(T)ij}.
\end{equation}
The Boltzmann equation (7.98) becomes in the metric (8.79)
\begin{equation}
\Theta'+ik\mu\Theta=-H'_{ij}\gamma^i\gamma^j.
\end{equation}
The solution of this equation in terms of $H_{ij}$ is
\begin{equation}
\Theta(\eta_0,\mbox{\boldmath$k$},\mbox{\boldmath$\gamma$})
=-\int_0^{\eta_0}H'_{ij}(\eta_0,\mbox{\boldmath$k$})\gamma^i\gamma^je^{-ik\mu(\eta_0-\eta)}d\eta.
\end{equation}
For the photon contribution to $\Pi^i_{(T)j}$ we obtain as in Sect. 7.5
\begin{equation}
\Pi^i_{(T)\gamma j}=12 \int [\gamma^i\gamma_j - \frac{1}{3}\delta^i{}_j ] \Theta ~
\frac{d\Omega_\gamma}{4\pi}.
\end{equation}
To this one should add the neutrino contribution, but in what follows we can safely neglect
the source $\Pi^i_{(T)\gamma j}$ in the Einstein equation (8.83).

\paragraph{B. Harmonic decompositions.} We decompose $H_{ij}$ as in Sect. 5.2:
\begin{equation}
H_{ij}(\eta,\mbox{\boldmath$k$})=\sum_{\lambda=\pm2}h_\lambda(\eta,\mbox{\boldmath$k$})
\epsilon_{ij}(\mbox{\boldmath$k$},\lambda),
\end{equation}
where the polarization tensor satisfies (5.65). If $\textbf{k}=(0,0,k)$ then the x,y
components of $\epsilon_{ij}(\mbox{\boldmath$k$},\lambda)$ are
\begin{equation}
(\epsilon_{ij}(\mbox{\boldmath$k$},\lambda)) = \left(\begin{array}{cc}
1&\mp i\\
\mp i&-1\\
\end{array}\right),~~\lambda=\pm2.
\end{equation}
One easily verifies that for this choice of $\textbf{k}$
\begin{equation}
\epsilon_{ij}(\lambda)(\gamma^i\gamma^j-\frac{1}{3}\delta^{ij})=\frac{4}{\sqrt{2}}\sqrt{\frac{\pi}{15}}
Y_{2\lambda}(\mbox{\boldmath$\gamma$}),~~\lambda=\pm2.
\end{equation}
If we insert this and the expansion
\begin{equation}
e^{-ik\mu(\eta_0-\eta)}=4\pi\sum_{L,M}(-i)^Lj_l(k(\eta_0-\eta))Y_{LM}^{\ast}(\mbox{\boldmath$\hat{k}$})
Y_{LM}(\mbox{\boldmath$\gamma$})
\end{equation}
in (8.85) we obtain for each polarization $\lambda$ the expansion (dropping the variable
$\eta$)
\begin{equation}
\Theta_\lambda(\mbox{\boldmath$k$},\mbox{\boldmath$\gamma$})=\sum_{l,m}a^{(\lambda)}_{lm}(k)Y_{lm}
(\mbox{\boldmath$\gamma$}),
\end{equation}
with
\begin{eqnarray}
a^{(\lambda)}_{lm}(k)&=&\int Y^\ast_{lm}
(\mbox{\boldmath$\gamma$})\Theta_\lambda(\mbox{\boldmath$k$},\mbox{\boldmath$\gamma$})d\Omega_\gamma\nonumber
\\ &=& -\int_0^{\eta_0}d\eta h'_\lambda(\eta,k)4\pi\sum_{L,M}(-i)^Lj_l(k(\eta_0-\eta))
Y_{LM}^{\ast}(\mbox{\boldmath$\hat{k}$})\nonumber
\\ &\times& \frac{4}{\sqrt{2}}\sqrt{\frac{\pi}{15}}
\int Y^\ast_{lm} (\mbox{\boldmath$\gamma$})Y_{2\lambda}(\mbox{\boldmath$\gamma$})Y_{LM}
(\mbox{\boldmath$\gamma$})d\Omega_\gamma.
\end{eqnarray}
Since $\textbf{k}$ points in the 3-direction we have
$Y_{LM}^{\ast}(\mbox{\boldmath$\hat{k}$})=\delta_{M0}\sqrt{\frac{2L+1}{4\pi}}$. If we also
use the spherical integral
\[\int Y^\ast_{lm} Y_{2\lambda}Y_{L0}d\Omega=\left[\frac{(2l+1)5(2L+1)}{4\pi}\right]^{1/2}
\left(\begin{array}{ccc}
l&2&L\\
0&0&0\\
\end{array}\right)(-1)^m
\left(\begin{array}{ccc}
l&2&L\\
-m&\lambda&0\\
\end{array}\right)
\]
we obtain
\[
a^{(\lambda)}_{lm}=-\sqrt{\frac{8\pi}{3}}\int_0^{\eta_0}d\eta
h'_\lambda(\eta,k)(2l+1)^{1/2}\sum_{L=l,l\pm2}j_L(k(\eta_0-\eta))(-i)^lX_{L,\lambda}\delta_{m\lambda},
\]
where
\[ (-i)^{l}X_{L,\lambda} :=(-i)^{L}(2L+1)\left(\begin{array}{ccc}
l&2&L\\
0&0&0\\
\end{array}\right)
\left(\begin{array}{ccc}
l&2&L\\
-m&\lambda&0\\
\end{array}\right).\]
Note that this is invariant under $\lambda\rightarrow-\lambda$. With a table of
Clebsch-Gordan coefficients one readily finds
\begin{eqnarray*}
X_{l,\lambda}&=&-\sqrt{\frac{3}{2}}\left[(l+2)(l+1)l(l-1)\right]^{1/2}\frac{1}{(2l+3)(2l-1)},\\
X_{l+2,\lambda}&=& -\sqrt{\frac{3}{8}}[\cdot\cdot\cdot]^{1/2}\frac{1}{(2l+3)(2l+1)},\\
X_{l-2,\lambda}&=& -\sqrt{\frac{3}{8}}[\cdot\cdot\cdot]^{1/2}\frac{1}{(2l+1)(2l-1)},
\end{eqnarray*}
and thus
\begin{eqnarray*}
\sum_{L=l,l\pm2}j_L X_{L,\lambda}&=&
-\sqrt{\frac{3}{8}}\left[\frac{(l+2)!}{(l-2)!}\right]^{1/2}\\
&\times&
\left[\frac{j_{l+2}}{(2l+3)(2l+1)}+2\frac{j_l}{(2l+3)(2l-1)}+\frac{j_{l-2}}{(2l+1)(2l-1)}\right].
\end{eqnarray*}
Using twice the recursion relation
\[\frac{j_l(x)}{x}=\frac{1}{2l+1}(j_{l-1}+j_{l+1}), \]
shows that the last square bracket is equal to $j_l(k(\eta_0-\eta))/[k(\eta_0-\eta)]^2$.
Thus we find
\begin{eqnarray}
a^{(\lambda)}_{lm}(k)&=&\sqrt{\pi}(-i)^l
\left[\frac{(l+2)!}{(l-2)!}\right]^{1/2}\nonumber\\
&\times& \int_0^{\eta_0}d\eta h'_\lambda(\eta,k)(2l+1)^{1/2}\frac{
j_l(k(\eta_0-\eta))}{[k(\eta_0-\eta)]^2}~\delta_{m\lambda}.\nonumber\\
\end{eqnarray}
Recall that so far the wave vector is assumed to point in the 3-direction. For an arbitrary
direction $a^{(\lambda)}_{lm}(\mbox{\boldmath$k$})$ is determined by (see (8.91) and use
the fact that $a^{(\lambda)}_{lm}(k)$ is proportional to $\delta_{m\lambda}$)
\[ \sum_{m}a^{(\lambda)}_{lm}(\mbox{\boldmath$k$})Y_{lm}
(\mbox{\boldmath$\gamma$})=a^{(\lambda)}_{l\lambda}(k)Y_{l\lambda}
(R^{-1}(\mbox{\boldmath$\hat{k}$})\mbox{\boldmath$\gamma$}), \] where
$R(\mbox{\boldmath$\hat{k}$})$ rotates (0,0,1) to $\mbox{\boldmath$\hat{k}$}$. Let
$D^l_{m\lambda}(\mbox{\boldmath$\hat{k}$})$ be the corresponding representation
matrices\footnote{The Euler angles are $(\varphi,\vartheta,0)$,where $(\vartheta,\varphi)$
are the polar angles of $\mbox{\boldmath$\hat{k}$}$.}. Since
\[Y_{l\lambda}
(R^{-1}(\mbox{\boldmath$\hat{k}$})\mbox{\boldmath$\gamma$})=\sum_m
D^l_{m\lambda}(\mbox{\boldmath$\hat{k}$})Y_{lm} (\mbox{\boldmath$\gamma$}), \] we obtain
\begin{equation}
a^{(\lambda)}_{lm}(\mbox{\boldmath$k$})=a^{(\lambda)}_{l\lambda}(k)D^l_{m\lambda}(\mbox{\boldmath$\hat{k}$}),
\end{equation}
where $a^{(\lambda)}_{l\lambda}(k)$ is given by (8.93) for $m=\lambda$.

\paragraph{C. The coefficients $C_l$ for tensor modes.}

For the computation of the $C_l$'s due to gravitational waves we proceed as in Sect. 8.4
for scalar modes. On the basis (8.91) and (8.94) we can write
\begin{equation}
\Theta_\lambda(\eta,\mbox{\boldmath$k$},\mbox{\boldmath$\gamma$})=h_\lambda(\eta_i,\mbox{\boldmath$k$})
\sum_{l,m}\frac{a^{(\lambda)}_{lm}(k)}{h_\lambda(\eta_i,k)}D^l_{m\lambda}(\mbox{\boldmath$\hat{k}$})
Y_{lm}(\mbox{\boldmath$\gamma$}),
\end{equation}
where $\eta_i$ is some very early time, e.g., at the end of inflation. A look at (8.93)
shows that the factor $a^{(\lambda)}_{lm}(k)/h_\lambda(\eta_i,k)$ involves only
$h'_\lambda(\eta,k)/h_\lambda(\eta_i,k)$, and is thus independent of the initial amplitude
of $h_\lambda$ and also independent of $\lambda$ (see paragraph D below). The stochastic
properties are entirely located in the first factor of (8.95). Its correlation function is
given in terms of the primordial power spectrum $P^{(prim)}_g(k)$ of the gravitational
waves:
\begin{equation}
\sum_\lambda\langle
h_\lambda(\eta_i,\mbox{\boldmath$k$})h^{\ast}_\lambda(\eta_i,\mbox{\boldmath$k'$})\rangle
=\frac{2\pi^2}{k^3}P^{(prim)}_g(k)\delta^3(\mbox{\boldmath$k$}-\mbox{\boldmath$k'$})
\end{equation}
(see also (5.73)). With this and the orthogonality properties of the representation
matrices
\begin{equation}
\int d\Omega_{\textbf{k}}D^l_{m\lambda}(\mbox{\boldmath$\hat{k}$})
D^{l'\ast}_{m'\lambda'}(\mbox{\boldmath$\hat{k}$})=\frac{2l+1}{4\pi}\delta_{ll'}\delta_{mm'}
\delta_{\lambda\lambda'},
\end{equation}
we obtain at the present time
\begin{equation}
\langle\Theta(\mbox{\boldmath$x$},\mbox{\boldmath$\gamma$})
\Theta(\mbox{\boldmath$x$},\mbox{\boldmath$\gamma'$})\rangle=\frac{1}{4\pi}\sum_l(2l+1)C^{GW}_l
P_l(\mbox{\boldmath$\gamma\cdot\gamma'$}),
\end{equation}
with
\[ C^{GW}_l=\frac{4\pi}{2l+1}\int_0^\infty \frac{dk k^2}{(2\pi)^3}\frac{2\pi^2}{k^3}P^{(prim)}_g(k)
\left|\frac{a^{(\lambda)}_{lm}(k)}{h_\lambda(\eta_i,k)}\right|^2. \] Finally, inserting
here (8.93) gives our main result
\begin{equation}
\fbox{$\displaystyle C^{GW}_l=\pi\frac{(l+2)!}{(l-2)!}\int_0^\infty
\frac{dk}{k}P^{(prim)}_g(k)\left|\int_{\eta_i\approx0}^{\eta_0} d\eta
\frac{h'(\eta,k)}{h(\eta_i,k)} \frac{j_l(k(\eta_0-\eta))}{[k(\eta_0-\eta)]^2}\right|^2.$}
\end{equation}

Note that the tensor modes (8.95) are in $\mbox{\boldmath$\hat{k}$}$-space orthogonal to
the scalar modes, which are proportional to $D^l_{m0}(\mbox{\boldmath$\hat{k}$})$.

\paragraph{D. The modes $h_\lambda(\eta,k)$.}

In the Einstein equations (8.83) we can safely neglect the anisotropic stresses
$\Pi_{(T)ij}$. Then $h_\lambda(\eta,k)$ satisfies the homogeneous linear differential
equation
\begin{equation}
h''+2\frac{a'}{a}h'+k^2h=0.
\end{equation}

At very early times, when the modes are still far outside the Hubble horizon, we can
neglect the last term in (8.100), whence $h$ is \textit{frozen}. For this reason we solve
(8.100) with the initial condition $h'(\eta_i,k)=0$. Moreover, we are only interested in
growing modes.

This problem was already discussed in Sect. 5.2.3. For modes which enter the horizon during
the matter dominated era we have the analytic solution (5.103),
\begin{equation}
\frac{h_k(\eta)}{h_k(0)}= 3\frac{j_1(k\eta)}{k\eta}.
\end{equation}
For modes which enter the horizon earlier, we use again a transfer function  $T_g(k)$:
\begin{equation}
\frac{h_k(\eta)}{h_k(0)}=:3\frac{j_1(k\eta)}{k\eta}T_g(k),
\end{equation}
that has to be determined by solving the differential equation numerically.

On large scales (small $l$), larger than the Hubble horizon at decoupling, we can use
(8.101). Since
\begin{equation}
\left(\frac{j_1(x)}{x}\right)'=-\frac{1}{x}j_2(x),
\end{equation}
we then have
\begin{equation}
\frac{h'(\eta,k)}{3h(0,k)}=-k\frac{j_2(x)}{x},~~ x:=k\eta.
\end{equation}
Using this in (8.99) gives
\begin{equation}
C^{GW}_l=9\pi\frac{(l+2)!}{(l-2)!}\int_0^\infty \frac{dk}{k}P^{(prim)}_g(k)I^2_l(k),
\end{equation}
with
\begin{equation} I_l(k)=\int_0^{x_0} dx\frac{j_l(x_0-x)j_2(x)}{(x_0-x)^2x},~~
x_0:=k\eta_0.
\end{equation}

\textbf{Remark.} Since the power spectrum is often defined in terms of $2H_{ij}$, the
pre-factor in (8.105) is the 4 times smaller.

For inflationary models we obtained for the power spectrum eq. (5.82),
\begin{equation}
P_g(k)\simeq \frac{4}{\pi}\left.\frac{H^2}{M^2_{Pl}}\right|_{k=aH},
\end{equation}
and the power index
\begin{equation}
n_T\simeq-2\varepsilon.
\end{equation}

For a flat power spectrum the integrations in (8.105) and (8.106) can perhaps be done
analytically, but I was not able to do achieve this.

\paragraph{E. Numerical results}

A typical theoretical CMB spectrum is shown in Fig. 8.1. Beside the scalar contribution in
the sense of cosmological perturbation theory, considered so far, the tensor contribution
due to gravity waves is also plotted.

\begin{figure}
\begin{center}
\includegraphics[height=0.45\textheight]
{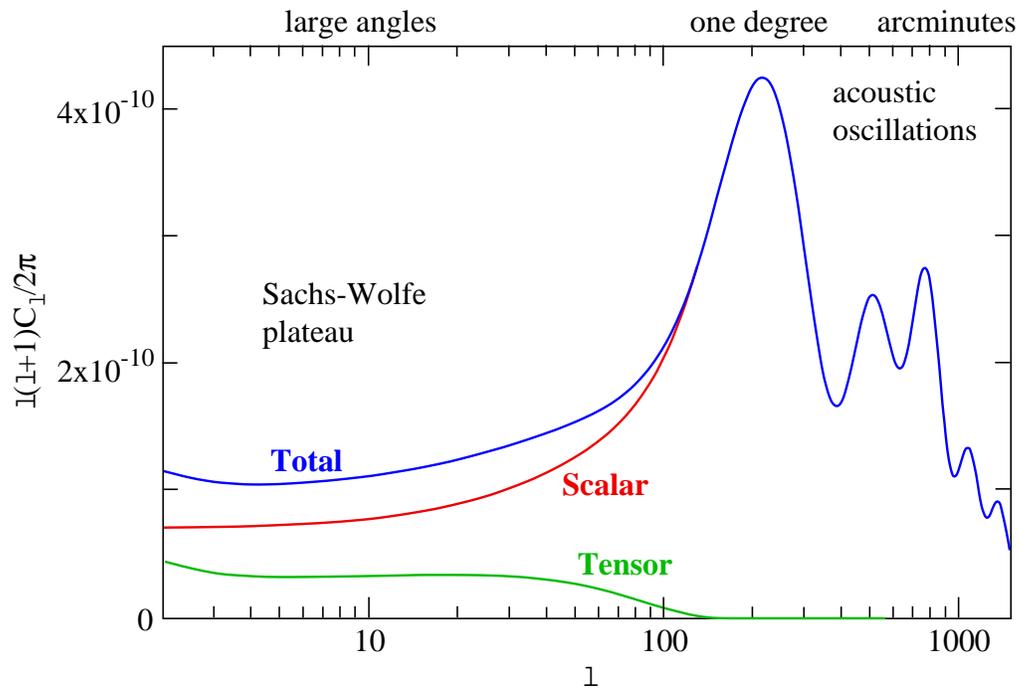}
\caption{Theoretical angular power spectrum for adiabatic initial
perturbations and typical cosmological parameters. The scalar and tensor contributions to
the anisotropies are also shown.} \label{Fig-13}
\end{center}
\end{figure}

Parameter dependences are discussed in detail in \cite{Teg} (see especially Fig. 1 of this
reference).

\section{Polarization}

A polarization map of the CMB radiation provides important additional information to that
obtainable from the temperature anisotropies. For example, we can get constraints about the
epoch of reionization. Most importantly, future polarization observations may reveal a
stochastic background of gravity waves, generated in the very early Universe. In this
section we give a brief introduction to the study of CMB polarization.

The mechanism which partially polarizes the CMB radiation is similar
to that for the scattered light from the sky. Consider first
scattering at a single electron of unpolarized radiation coming in
from all directions. Due to the familiar polarization dependence of
the differential Thomson cross section, the scattered radiation is,
in general, polarized. It is easy to compute the corresponding
Stokes parameters. Not surprisingly, they are not all equal to zero
if and only if the intensity distribution of the incoming radiation
has a non-vanishing quadrupole moment. The Stokes parameters $Q$ and
$U$ are proportional  to the overlap integral with the combinations
$Y_{2,2} \pm Y_{2,-2}$ of the spherical harmonics, while $V$
vanishes.) This is basically the reason why a CMB  polarization map
traces (in the tight coupling limit) the quadrupole temperature
distribution on the last scattering surface.

The polarization tensor of an all sky map of the CMB radiation can be parametrized in
temperature fluctuation units, relative to the orthonormal basis $\{d\vartheta,
\sin\vartheta\; d\varphi\}$ of the two sphere, in terms of the Pauli matrices as
$\Theta\cdot 1 + Q\sigma_3 + U\sigma_1 + V\sigma_2$. The Stokes parameter $V$ vanishes (no
circular polarization). Therefore, the polarization properties can be described by the
following symmetric trace-free tensor on $S^2$:
\begin{equation}
(\mathcal{P}_{ab}) = \left(\begin{array}{cc}
Q&U\\
U&-Q\\
\end{array}\right).
\end{equation}

As for gravity waves, the components $Q$ and $U$ transform under a rotation of the 2-bein
by an angle $\alpha$ as
\begin{equation}
Q \pm iU \rightarrow e^{\pm 2i\alpha}(Q\pm iU),
\end{equation}
and are thus of spin-weight 2. $\mathcal{P}_{ab}$ can be decomposed uniquely into
\emph{`electric'} and \emph{`magnetic'} parts:
\begin{equation}
\mathcal{P}_{ab} = E_{;ab} - \frac{1}{2}g_{ab} \Delta E + \frac{1}{2}(\varepsilon_a{}^c
B_{;bc} +\varepsilon_b{}^c B_{;ac}).
\end{equation}
Expanding here the scalar functions $E$ and $B$ in terms of spherical harmonics, we obtain
an expansion of the form
\begin{equation}
\mathcal{P}_{ab} = \sum_{l=2}^{\infty} \sum_{m} \left[a^E_{(lm)}Y^E_{(lm)ab} +
a^B_{(lm)}Y^B_{(lm)ab} \right]
\end{equation}
in terms of the tensor harmonics:
\begin{equation}
Y^E_{(lm)ab}: = N_l(Y_{(lm);ab} -\frac{1}{2}g_{ab}Y_{(lm);c}{}^c),\;\; Y^B_{(lm)ab}: =
\frac{1}{2}N_l(Y_{(lm);ac} \varepsilon^c{}_b + a\leftrightarrow b),
\end{equation}
where $l\geq 2$ and
\[ N_l \equiv \left(\frac{2(l-2)!}{(l+2)!}\right)^{1/2}.\]
Equivalently, one can write this as
\begin{equation}
Q+iU = \sqrt{2}\sum_{l=2}^\infty\sum_{m}\left[a^E_{(lm)}+ ia^B_{(lm)}\right] \,_{2\!}Y_l^m,
\end{equation}
where $\,_{s\!}Y_l^m$ are the spin-s harmonics:
\[\,_{s\!}Y_l^m=\sqrt{\frac{2l+1}{4\pi}}D^l_{-s,m}(\vartheta,\varphi,0).\]

The multipole moments $a^E_{(lm)}$ and $a^B_{(lm)}$ are random variables, and we have
equations analogous to those of the temperature fluctuations, with
\begin{equation}
C_l^{TE} = \frac{1}{2l+1}\sum_{m} \;\langle a_{lm}^{\Theta \star} a_{lm}^E \rangle ,\;\;
etc.
\end{equation}
(We have now put the superscript $\Theta$ on the $a_{lm}$ of the temperature fluctuations.)
The $C_l$'s determine the various angular correlation functions. For example, one easily
finds
\begin{equation}
\langle\Theta(\mbox{\boldmath$n$})Q(\mbox{\boldmath$n'$})\rangle = \sum_{l}\;
C_l^{TE}\frac{2l+1}{4\pi}N_l P_l^2(\cos \vartheta)
\end{equation}
(the last factor is the associated Legendre function $P^m_l$ for $m=2$).

For the space-time dependent Stokes parameters $Q$ and $U$ of the radiation field we can
perform a normal mode decomposition analogous to
\begin{equation}
\Theta(\eta,\mbox{\boldmath$x$},\mbox{\boldmath$\gamma$}) = (2\pi)^{-3/2} \int d^3k\sum_l
\theta_l(\eta,k) G_l(\mbox{\boldmath$x$},\mbox{\boldmath$\gamma$};\mbox{\boldmath$k$}),
\end{equation}
where
\begin{equation}
G_l(\mbox{\boldmath$x$},\mbox{\boldmath$\gamma$};\mbox{\boldmath$k$}) = (-i)^l
P_l(\mbox{\boldmath$\hat{k}\cdot \gamma$}) \exp(i\mbox{\boldmath$k\cdot x$}).
\end{equation}
If, for simplicity, we again consider only scalar perturbations this reads
\begin{equation}
Q\pm iU = (2\pi)^{-3/2} \int d^3k\sum_l(E_l \pm iB_l)\,_{\pm 2\!}G^0_l,
\end{equation}
where
\begin{equation}
~~~\,_{s\!}G_l^m(\mbox{\boldmath$x$},\mbox{\boldmath$\gamma$};\mbox{\boldmath$k$}) =
(-i)^l\left(\frac{2l+1}{4\pi}\right)^{1/2} \,_{s\!}Y_l^m(\mbox{\boldmath$ \gamma$})
\exp(i\mbox{\boldmath$k\cdot x$}),
\end{equation}
if the mode vector $\mathbf{k}$ is chosen as the polar axis. (Note that $G_l$ in (8.118) is
equal to $\,_{0\!}G_l^0$.)

The Boltzmann equation implies a coupled hierarchy for the moments $\theta_l, E_l$, and
$B_l$ \cite{23}, \cite{24}. It turns out that the $B_l$ vanish for scalar perturbations.
Non-vanishing magnetic multipoles would be a unique signature for a spectrum of gravity
waves. We give here, without derivation, the equations for the $E_l$:
\begin{equation}
E'_l=k\left\{\frac{(l^2-4)^{1/2}}{2l-1}E_{l-1}-\frac{[(l+1)^2-4]^{1/2}}{2l+1}E_{l+1}\right\}
-\dot{\tau}(E_l+\sqrt{6}P\delta_{l,2}),
\end{equation}
where
\begin{equation}
P=\frac{1}{10}[\theta_2-\sqrt{6}E_2].
\end{equation}
The analog of the integral representation (8.51)is
\begin{equation}
\frac{E_l(\eta_0)}{2l+1} = -\frac{3}{2}\sqrt{\frac{(l+2)!}{(l-2)!}}\int_0^{\eta_0} d\eta
e^{-\tau(\eta)}\dot{\tau} P(\eta)\frac{j_l(k(\eta_0-\eta))}{(k(\eta_0-\eta))^2} .
\end{equation}

For large scales the first term in (8.122) dominates, and the $E_l$ are thus determined by
$\theta_2$.

For large $l$ we may use the tight coupling approximation in which
$E_2=-\sqrt{6}\Rightarrow P=\theta_2/4$. In the sudden decoupling approximation, the
present electric multipole moments can thus be expressed in terms of the brightness
quadrupole moment on the last scattering surface and spherical Bessel functions as
\begin{equation}
\frac{E_l(\eta_0,k)}{2l+1}\simeq \frac{3}{8}\theta_2(\eta_{dec},k)
\frac{l^2j_l(k\eta_0)}{(k\eta_0)^2}.
\end{equation}
Here one sees how the observable $E_l$'s trace the quadrupole temperature anisotropy on the
last scattering surface. In the tight coupling approximation the latter is proportional to
the dipole moment $\theta_1$.

\section{Observational results and cosmological\\ parameters}

In recent years several experiments gave clear evidence for multiple
peaks in the angular temperature power spectrum at positions
expected on the basis of the simplest inflationary models and big
bang nucleosynthesis \cite{25}. These results have been confirmed
and substantially improved by the first year WMAP data \cite{26},
\cite{27}. Fortunately, the improved data after three years of
integration are now available \cite{Sperg2}. Below we give a brief
summary of some of the most important results.

\begin{figure}
\begin{center}
\includegraphics[height=0.5\textheight]{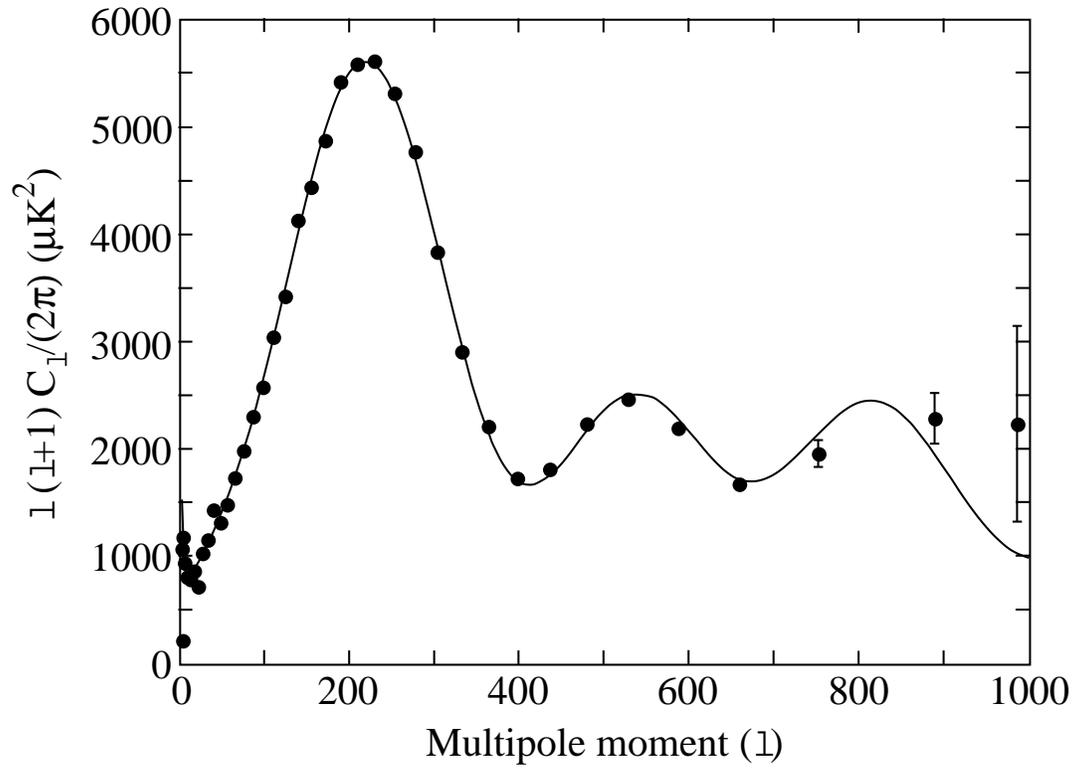}
\caption{Three-year WMAP data for the temperature-temperature (TT)
power spectrum. The black line is the best fit $\Lambda$CDM model
for the three-year WMAP data. (Adapted from Figure 2 of Ref.
\cite{Sperg2}.)} \label{cosmp:Fig-1}
\end{center}
\end{figure}

\begin{figure}
\begin{center}
\includegraphics[height=0.4\textheight]{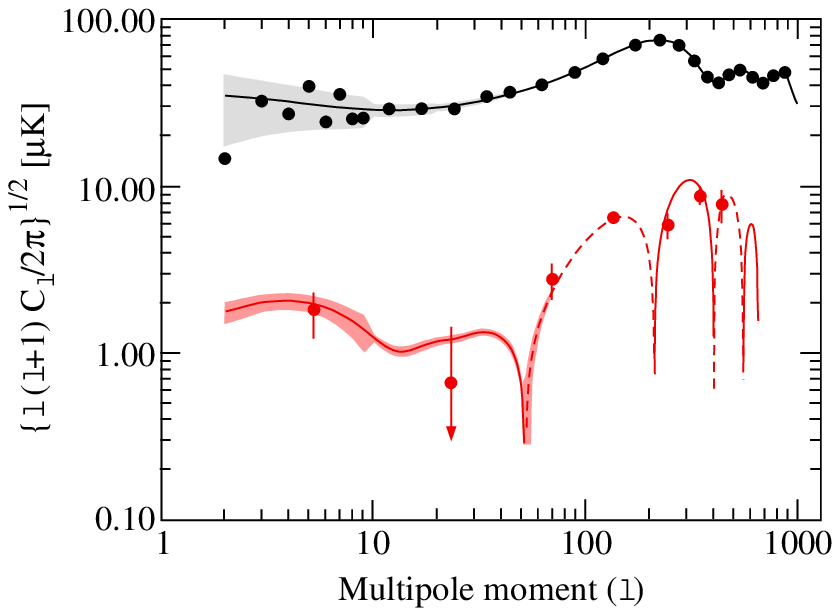}
\caption{WMAP data for the temperature-polarization TE power
spectrum. The best fit $\Lambda$CDM model is also shown. (Adapted
from Figure 25 of Ref. \cite{Pag}.)} \label{cosmp:Fig-2}
\end{center}
\end{figure}

Figure \ref{cosmp:Fig-1} shows the 3 year data of WMAP for the TT
angular power spectrum, and the best fit (power law) $\Lambda$CDM
model. The latter is a spatially flat model and involves the
following six parameters: $\Omega_bh_0^2,~\Omega_Mh_0^2,~H_0$,
amplitude of fluctuations, $\sigma_8$, optical depth $\tau$, and the
spectral index, $n_s$, of the primordial scalar power spectrum (see
Sect. 5.2). Figure \ref{cosmp:Fig-2} shows in addition the TE
polarization data \cite{Pag}. There are now also EE data that lead
to a further reduction of the allowed parameter space. The first
column in Table 1 shows the best fit values of the six parameters,
using only the WMAP data.

\begin{figure}
\begin{center}
\includegraphics[height=0.3\textheight]{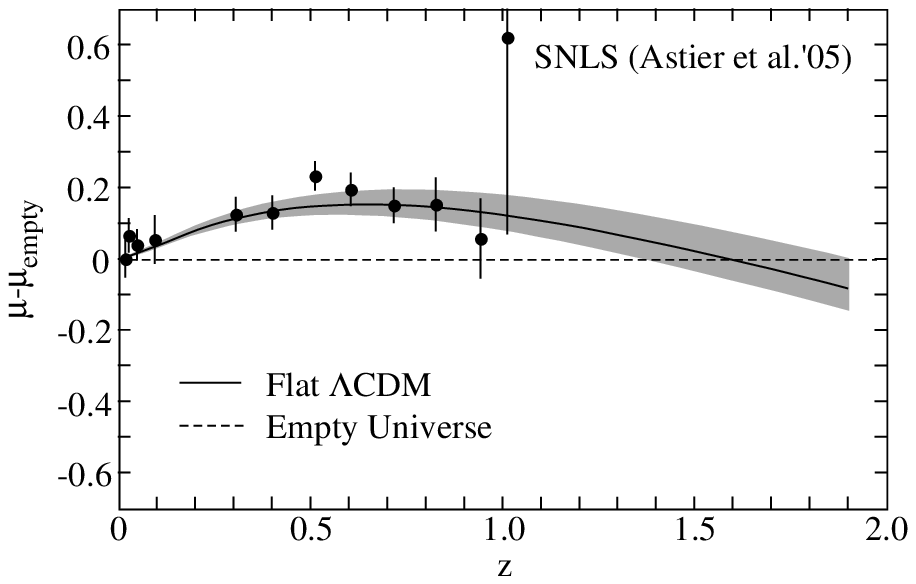}
\caption{Prediction for the luminosity-redshift relation from the
$\Lambda$CDM model model fit to the WMAP data only. The ordinate is
the deviation of the distance modulous from the empty universe
model. The prediction is compared to the SNLS data \cite{Leg}. (From
Figure 8 of Ref. \cite{Sperg2}.)} \label{cosmp:Fig-3}
\end{center}
\end{figure}

Figure \ref{cosmp:Fig-3} shows the prediction of the model for the
luminosity-redshift relation, together with the SLNS data \cite{Leg}
mentioned in Sect. 5.3. For other predictions and corresponding data
sets, see \cite{Sperg2}.

\begin{figure}
\begin{center}
\includegraphics[height=0.5\textheight]{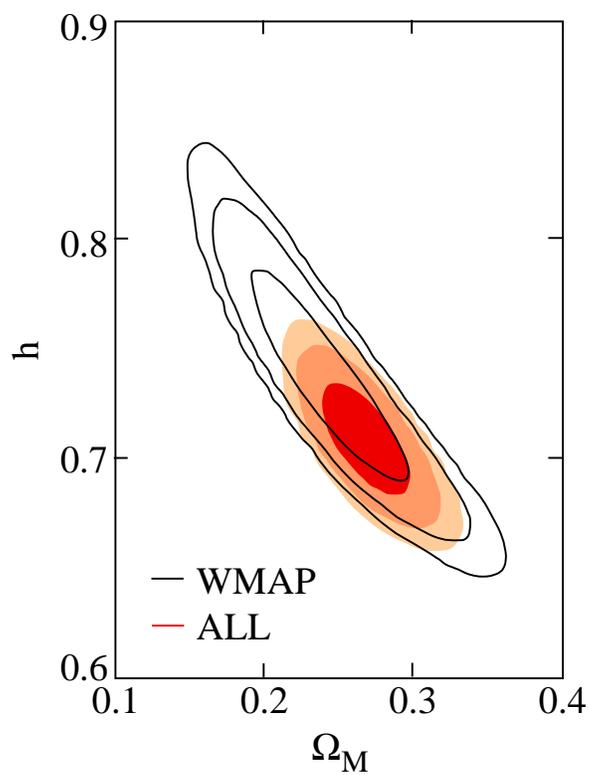}
\caption{Joint marginalized contours (68\% and 95\% confidence
levels) in the ($\Omega_M,h_0$)-plane for WMAP only (solid lines)
and additional data (filled red) for the power-law $\Lambda$CDM
model. (From Figure 10 in \cite{Sperg2}.)} \label{cosmp:Fig-4}
\end{center}
\end{figure}

Combining the WMAP results with other astronomical data reduces the
uncertainties for some of the six parameters. This is illustrated in
the second column which shows the 68\% confidence ranges of a joint
likelihood analysis when the power spectrum from the completed
2dFGRS \cite{Col} is added. In \cite{Sperg2} other joint constraints
are listed (see their Tables 5,6). In Figure \ref{cosmp:Fig-4} we
reproduce one of many plots in \cite{Sperg2} that shows the joint
marginalized contours in the ($\Omega_M,h_0$)-plane.

The parameter space of the cosmological model can be extended in
various ways. Because of intrinsic degeneracies, the CMB data alone
no more determine unambiguously the cosmological model parameters.
We illustrate this for non-flat models. For these the WMAP data (in
particular the position of the first acoustic peak) restricts the
curvature parameter $\Omega_K$ to a narrow region around the
degeneracy line $\Omega_K=-0.3040+0.4067~\Omega_\Lambda$. This does
not exclude models with $\Omega_\Lambda=0$. However, when for
instance the Hubble constant is restricted to an acceptable range,
the universe must be nearly flat. For example, the restriction
$h_0=0.72\pm 0.08$ implies that $\Omega_K=-0.003^{+0.013}_{-0.017}$
and $\Omega_\Lambda= 0.758^{+0.035}_{-0.058}$. Other strong limits
are given in Table 11 of \cite{Sperg2}, assuming that $w=-1$. But
even when this is relaxed, the combined data constrain $\Omega_K$
and $w$ significantly (see Figure 17 of \cite{Sperg2}). The
marginalized best fit values are
$w=-1.062^{+0.128}_{-0.079},~\Omega_K=-0.024^{+0.016}_{-0.013}$ at
the 68\% confidence level.

\begin{figure}
\begin{center}
\includegraphics[height=0.4\textheight]{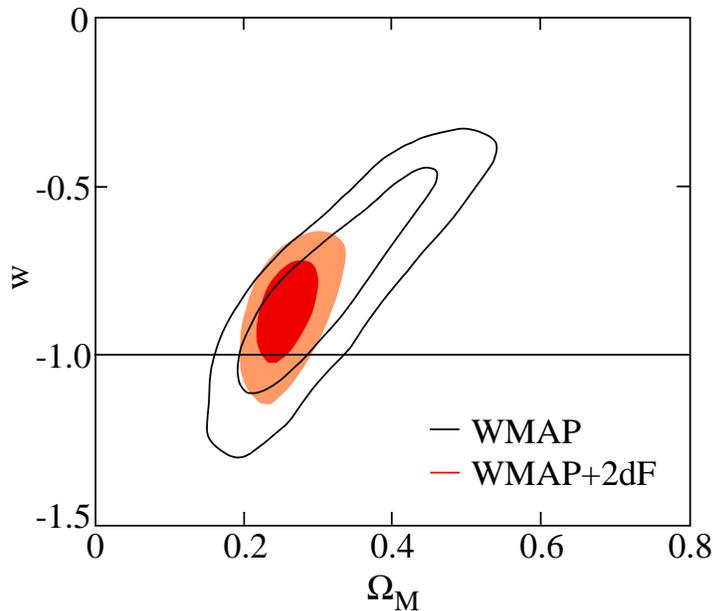}
\caption{Constraints on the equation of state parameter $w$ in a
flat universe model when WMAP data are combined with the 2dFGRS
data. (From Figure 15 in \cite{Sperg2}.)} \label{cosmp:Fig-5}
\end{center}
\end{figure}

The restrictions on $w$ -- assumed to have no $z$-dependence -- for
a flat model are illustrated in Figure \ref{cosmp:Fig-5} .

\vspace{1cm}

\begin{tabular}{|l||c|r|}
\multicolumn{3}{c}{\textbf{Table 1.}} \\  \hline Parameter & WMAP
alone & WMAP + 2dFGRS\\ \hline\hline 
$100\Omega_b h_0^2$ & $2.233^{+0.072}_{-0.091}$ &
$2.223^{+0.066}_{-0.083}$\\
$\Omega_M h_0^2$ & $0.1268^{+0.0072}_{-0.0095}$   & $0.1262^{+0.0045}_{-0.0062}$   \\
$h_0 $ &   $0.734^{+0.028}_{-0.038}$   &  $0.732^{+0.018}_{-0.025}$  \\
$\Omega_M$  &  $0.238^{+0.030}_{-0.041}$   &  $0.236^{+0.016}_{-0.029}$  \\
$\sigma_8$  &  $0.744^{+0.050}_{-0.060}$   &  $0.737^{+0.033}_{-0.045}$ \\
$\tau$      &  $0.088^{+0.028}_{-0.034}$   &  $0.083^{+0.027}_{-0.031}$ \\
$n_s$       &  $0.951^{+0.015}_{-0.019}$   &  $0.948^{+0.014}_{-0.018}$ \\

\hline
\end{tabular}

\vspace{1cm}

Another interesting result is that reionization of the Universe has
set in at a redshift of $z_r = 10.9^{+2.7}_{-2.3}$. Later we shall
add some remarks on what has been learnt about the primordial power
spectrum.

Before the new results possible admixtures of isocurvature modes
were not strongly constraint. But now the measured
temperature-polarization correlations imply that the primordial
fluctuations were primarily \textit{adiabatic}. Admixtures of
isocurvature modes do not improve the fit.

It is most remarkable that a six parameter cosmological model is
able to fit such a rich body of astronomical observations. There
seems to be little room for significant modifications of the
successful $\Lambda$CDM model.

WMAP has determined the amplitude of the primordial power spectrum:
\begin{equation}
P^{(prim)}_{\mathcal{R}}(k)\simeq 2.95\times 10^{-9}~A,~~ A=0.6 - 1
\end{equation}
(depending on the model). Using (5.46) this implies
\begin{equation}
\frac{1}{\pi M^2_{Pl}}\frac{H^2}{\varepsilon}\simeq(2 -
3)\times10^{-9},
\end{equation}
hence the Hubble parameter during inflation is
\begin{equation}
H\simeq(0.9 - 1.2)\times10^{15} \varepsilon^{1/2}~GeV.
\end{equation}
With (5.36) this gives
\begin{equation}
U^{1/4}\simeq (6.3 - 7.1)\times10^{16}\varepsilon^{1/4}~GeV.
\end{equation}

For a comparison with observations the power index, $n_s$, for
scalar perturbations is of particular interest. In terms of the
slow-roll parameters and it is given by (see (5.88))
\begin{equation}
n_s-1= 2\delta-4\varepsilon \simeq-6\varepsilon_U+2\eta_U.
\label{eq:Inf60}
\end{equation}

The WMAP data constrain the ratio $P_g/P_{\mathcal{R}}$, and hence
by (5.90) also $\varepsilon:~\varepsilon < 0.08$. Therefore, we can
conclude that the energy scale of inflation has to satisfy the bound
\begin{equation}
U^{1/4}< 3.8 \times10^{16}~GeV.
\end{equation}
A positive detection of the $B$ - mode in the CMB polarization would
provide a lower bound for $U^{1/4}$.

It is most remarkable that the WMAP data match the basic
inflationary predictions, and are even well fit by the simplest
model $U\propto \varphi^2$.

 \section{Concluding remarks}

In these lectures we have discussed some of the wide range of
astronomical data that support the following `concordance model':
The Universe is spatially flat and dominated by a Dark Energy
component and weakly interacting cold dark matter. Furthermore, the
primordial fluctuations are adiabatic, nearly scale invariant and
Gaussian, as predicted in simple inflationary models. It is very
likely that the present concordance model will survive
phenomenologically.

A dominant Dark Energy component with density parameter $\simeq 0.7$
is so surprising that many authors have examined whether this
conclusion is really unavoidable. On the basis of the available data
one can now say with considerable confidence that if general
relativity is assumed to be also valid on cosmological scales, the
existence of such a dark energy component that dominates the recent
universe is almost inevitable. The alternative possibility that
general relativity has to be modified on distances comparable to the
Hubble scale is currently discussed a lot. It  turns out that
observational data are restricting theoretical speculations more and
more. Moreover, some of the recent proposals have serious defects on
a fundamental level (ghosts, acausalities, superluminal
fluctuations).

\begin{center}
* \quad * \quad *
\end{center}

The dark energy problems will presumably stay with us for a long
time. Understanding the nature of DE is widely considered as one of
the main goals of cosmological research for the next decade and
beyond.

\begin{appendix}
\chapter{Random fields, power spectra, filtering}

Let $\xi(\mathbf{x})$  a random field on $\mathbb{R}^3$, and $\hat{\xi}(\mathbf{k})$ its
Fourier transform, normalized according to
\begin{equation}
\xi(\mathbf{x})=(2\pi)^{-3/2}\int\hat{\xi}(\mathbf{k})e^{i\mathbf{k}\cdot\mathbf{x}}d^3k.
\end{equation}
In our applications $\xi(\mathbf{x})$ will be, for instance, the field of density
fluctuations $\delta(\mathbf{x})$ at a fixed time.

In practice $\hat{\xi}(\mathbf{k})$ will be distributional (generalized random field).

\subsection*{Correlation function and power spectrum}

In our cosmological applications we shall often assume that the different \textbf{k}- modes
are uncorrelated:
\begin{equation}
\langle\hat{\xi}(\mathbf{k})\hat{\xi}(\mathbf{k})^{\ast}\rangle=\delta^{(3)}(\mathbf{k}-\mathbf{k}')
\mathcal{P}(\mathbf{k}).
\end{equation}
Note that $\left|\hat{\xi}(\mathbf{k})\right|^2$ is not defined. (One might, therefore,
prefer to work in a finite volume with periodic boundary conditions.)

The function $\mathcal{P}(\mathbf{k})$ is the \textit{power spectrum} belonging to
$\xi(\mathbf{x})$. This is also the Fourier transform of the correlation function:
\begin{equation}
C_\xi(\mathbf{x}-\mathbf{x}')=\langle\xi(\mathbf{x})\xi(\mathbf{x'})\rangle
=\frac{1}{(2\pi)^3}\int\mathcal{P}(\mathbf{k})e^{i\mathbf{k}\cdot(\mathbf{x}-\mathbf{x}')}d^3k.
\end{equation}

\subsection*{Filtering}

Let $W$ be a window function (filter) and define the filtered $\xi$ by
\begin{equation}
\xi_W=\xi\star W.
\end{equation}
With our convention we have for the Fourier transforms
\begin{equation}
\hat{\xi}_W=(2\pi)^{3/2}\hat{\xi}~\hat{W}.
\end{equation}
Therefore,
\begin{equation}
\mathcal{P}_{\xi_W}(\mathbf{k})=(2\pi)^3\left|\hat{W}(\mathbf{k})\right|^2
\mathcal{P}_\xi(\mathbf{k}).
\end{equation}
With (A.3) this gives, in particular,
\begin{equation}
\langle\xi^2_W(\mathbf{x})\rangle=\int\left|\hat{W}(\mathbf{k})\right|^2
\mathcal{P}_\xi(\mathbf{k})d^3k.
\end{equation}

\subsubsection*{Example}

For $W$ we choose a top-hat:
\begin{equation}
W(\mathbf{x})=\frac{1}{V}\theta(R-|\mathbf{x}|),~~ V=\frac{4\pi}{3}R^3,
\end{equation}
where $\theta$ is the Heaviside function. The Fourier transform is readily found to be
\begin{equation}
\hat{W}(\mathbf{k})=(2\pi)^{-3/2}\tilde{W}(kR), ~~~ \tilde{W}(kR):=\frac{3(\sin kR-kR\cos
kR)}{(kR)^3}.
\end{equation}
Thus,
\begin{equation}
\mathcal{P}_{\xi_W}(\mathbf{k})=\left|\tilde{W}(kR)\right|^2 \mathcal{P}_\xi(\mathbf{k}).
\end{equation}
For a spherically symmetric situation we get from (A.7)
\begin{equation}
\langle\xi^2_W(\mathbf{x})\rangle=\frac{1}{2\pi^2}\int\left|\tilde{W}(kR)\right|^2
\mathcal{P}_\xi(k)k^2dk
\end{equation}
(independent of \textbf{x}).

For this reason one often works with the following definition of the power spectrum
\begin{equation}
P_\xi(k):=\frac{k^3}{2\pi^2}\mathcal{P}_\xi(k).
\end{equation}
Then the last equation becomes
\begin{equation}
\langle\xi^2_W(\mathbf{x})\rangle=\int\left|\tilde{W}(kR)\right|^2
P_\xi(k)\frac{dk}{k}.
\end{equation}

If $\xi$ is the density fluctuation field $\delta(\mathbf{x})$, the filtered fluctuation
$\sigma_R^2$ on the scale $R$ is
\begin{equation}
\sigma^2_R=\int\left|\tilde{W}(kR)\right|^2 P_\delta(k)\frac{dk}{k}.
\end{equation}

\chapter{Collision integral for Thomson scattering}

The main goal of this Appendix is the derivation of equation (7.66) for the collision
integral in the Thomson limit.

When we work relative to an orthonormal tetrad the collision integral has the same form as
in special relativity. So let first consider this case.

\subsection*{Collision integral for two-body scattering}

In SR the Boltzmann equation (7.26) reduces to
\begin{equation}
p^{\mu}\partial_\mu f=C[f]
\end{equation}
or
\begin{equation}
\partial_t f+v^i\partial_i f=\frac{1}{p^0}C[f].
\end{equation}
In order to find the explicit expression for $C[f]$ things become easier if the following
non-relativistic normalization of the one-particle states $|p,\lambda\rangle$ is adopted:
\begin{equation}
\langle p',\lambda'|p,\lambda\rangle =
(2\pi)^3\delta_{\lambda,\lambda'}\delta^{(3)}(\mathbf{p}'-\mathbf{p}).
\end{equation}
(Some readers may even prefer to discretize  the momenta by using a finite volume with
periodic boundary conditions.) Correspondingly, the one-particle distribution functions $f$
are normalized according to
\begin{equation}
\int f(p)\frac{gd^3p}{(2\pi)^3}=n,
\end{equation}
where $g$ is the statistical weight (= 2 for electrons and photons), and $n$ is the
particle number density.

The $S$-matrix element for a 2-body collision $p,q\rightarrow p',q'$ has the form
(suppressing polarization indices)
\begin{equation}
\langle p',q'|S-1|p,q\rangle=-i(2\pi)^4\delta^{(4)}(p'+q'-p-q)\langle p',q'|T|p,q\rangle.
\end{equation}
Because of our non-invariant normalization we introduce the Lorentz invariant matrix
element $M$ by
\begin{equation}
\langle p',q'|T|p,q\rangle= \frac{M}{\left(2p^02q^02p'^02q'^0\right)^{1/2}}.
\end{equation}
The transition probability per unit time and unit volume is then (see, e.g., Sect. 64 of
\cite{LL})
\begin{equation}
dW=(2\pi)^4\frac{1}{2p^02q^0}|M|^2\delta^{(4)}(p'+q'-p-q)\frac{d^3p'}{(2\pi)^32p'^0}\frac{d^3q'}{(2\pi)^32q'^0}.
\end{equation}
Since we ignore in the following polarization effects, we average $|M|^2$ over all
polarizations (helicities) of the initial and final particles. This average is denoted by
$\overline{|M|^2}$. Per polarization we still have the formula (B.7), but with $|M|^2$
replaced by $\overline{|M|^2}$. From time reversal invariance we conclude that
$\overline{|M|^2}$ remains invariant under $p,q\leftrightarrow p',q'$.

With the standard arguments we can now write down the collision integral. For definiteness
we consider Compton scattering $\gamma(p)+e^{-}(q)\rightarrow\gamma(p')+e^{-}(q')$ and
denote the distribution functions of the photons and electrons by $f(p)$ and $f_{(e)}(q)$,
respectively. In the following expression we neglect the Pauli suppression factors
$1-f_{(e)}$, since in our applications the electrons are highly non-degenerate. Explicitly,
we have
\begin{eqnarray}
\frac{1}{p^0}C[f]&=&\frac{1}{2p^0}\int\frac{2d^3q}{(2\pi)^32q'^0}\frac{2d^3q'}{(2\pi)^32q'^0}
\frac{2d^3p'}{(2\pi)^32p'^0}(2\pi)^4\overline{|M|^2}\delta^{(4)}(p'+q'-p-q)\nonumber \\
&\times&\left\{\left(1+f(p)\right)f(p')f_{(e)}(q')-\left(1+f(p')\right)f(p)f_{(e)}(q)\right\}.
\end{eqnarray}

At this point we return to the normalization of the one-particle distributions adopted in
Sect. 7.1. This amounts to the substitution $f\rightarrow 4\pi^3f$. Performing this in
(B.1) and (B.8) we get for the collision integral
\begin{eqnarray}
C[f]&=&\frac{1}{16\pi^2}\int\frac{d^3q}{q^0}\frac{d^3q'}{q'^0}
\frac{d^3p'}{p'^0}\overline{|M|^2}\delta^{(4)}(p'+q'-p-q)\nonumber \\
&\times&\left\{\left(1+4\pi^3f(p)\right)f(p')f_{(e)}(q')-\left(1+4\pi^3f(p')\right)f(p)f_{(e)}(q)\right\}.
\nonumber\\
\end{eqnarray}
The invariant function $\overline{|M|^2}$ is explicitly known, and can for instance be
expressed in terms of the Mandelstam variables $s,t,u$ (see Sect. 86 of \cite{LL}).

The integral with respect to $d^3q'$ can trivially be done
\begin{equation}
C[f]=\frac{1}{16\pi^2}\int\frac{d^3q}{q^0}\frac{1}{q'^0}
\frac{d^3p'}{p'^0}\delta(p'^0+q'^0-p^0-q^0)\overline{|M|^2}\times\{\cdot\cdot\cdot\}.
\end{equation}
The integral with respect to $\mathbf{p}'$ can most easily be evaluated by going to the
rest frame of $q^\mu$. Then
\[\int d^3p'\frac{1}{p'^0q'^0}\delta(p'^0+q'^0-p^0-q^0)\cdot\cdot\cdot=\int d\Omega_{\hat{\mathbf{p}}'}
\int d|\mathbf{p}'|\frac{|\mathbf{p}'|}{q'^0}\delta(m+q'^0-p^0-q^0)\cdot\cdot\cdot .\] We
introduce the following notation: With respect to the rest system of $q^\mu$ let
$\omega:=p^0=|\mathbf{p}|,~ \omega':=p'^0=|\mathbf{p}'|,~ E'=\sqrt{\mathbf{q}'^2+m^2}$.
Then the last integral is equal to
\[\frac{\omega'}{E'}\frac{1}{|1+\partial E'/\partial\omega'|}=\frac{\omega'^2}{m\omega}.\]
In getting the last expression we have used energy and momentum conservation.

So far we are left with
\begin{equation}
C[f]=\frac{1}{16\pi^2m}\int\frac{d^3q}{q^0}\int
d\Omega_{\hat{\mathbf{p}}'}\frac{\omega'^2}{\omega}\overline{|M|^2}\times\{\cdot\cdot\cdot\}.
\end{equation}
In the rest system of $q^\mu$ the following expression for $\overline{|M|^2}$ can be found
in many books (for a derivation, see \cite{NS5})
\begin{equation}
\overline{|M|^2}=3\pi
m^2\sigma_T\left[\frac{\omega'}{\omega}+\frac{\omega}{\omega'}-\sin\vartheta\right],
\end{equation}
where $\vartheta$ is the scattering angle in that frame. For an arbitrary frame, the
combination $d\Omega_{\hat{\mathbf{p}}'}\frac{\omega'^2}{\omega}\overline{|M|^2}$ has to be
treated as a Lorentz invariant object.

At this point we take the non-relativistic limit $\omega/m\rightarrow0$, in which
$\omega'\simeq\omega$ and $C[f]$ reduces to the simple expression
\begin{equation}
C[f]=\frac{3}{16\pi}\sigma_T\omega n_e\int d\Omega_{\hat{\mathbf{p}}'}(1+\cos^2\vartheta)
[f(p')-f(p)].
\end{equation}

\subsection*{Derivation of (7.66)}

In Sect. 7.4 the components $p^\mu$ of the four-momentum $p$ refer to the tetrad $e_\mu$
defined in (7.42). Relative to this\footnote{Without specifying the gauge one can easily
generalize the following relative to the tetrad defined by (7.31).} we introduced the
notation $p^\mu=(p,p\gamma^i)$. The electron four-velocity is according to (1.156) given to
first order by
\begin{equation}
u_{(e)}=\frac{1}{a}(1-A)\partial_\eta+\frac{1}{a}\gamma^{ij}v_{(e)\mid j}\partial_j
=e_0+v_{(e)}^i e_i; ~~v_{(e)}^i=v_{(e)i}=\hat{e}_i(v_{(e)}).
\end{equation}
Now $\omega$ in (B.13) is the energy of the four-momentum $p$ in the rest frame of the
electrons, thus
\begin{equation}
\omega=-\langle p,u_{(e)}\rangle=p[1-\hat{e}_i(v_{(e)})\gamma^i].
\end{equation}
Similarly,
\begin{equation}
\omega'=-\langle p',u_{(e)}\rangle=p'[1-\hat{e}_i(v_{(e)})\gamma'^i].
\end{equation}
Since in the non-relativistic limit $\omega'=\omega$, we obtain the relation
\begin{equation}
p'[1-\hat{e}_i(v_{(e)})\gamma'^i]=p[1-\hat{e}_i(v_{(e)})\gamma^i].
\end{equation}
Therefore, to first order
\begin{eqnarray}
f(p',\gamma'^i) &=& f^{(0)}(p')+ \delta f(p',\gamma'^i)\nonumber\\
&=& f^{(0)}(p)+\frac{\partial f^{(0)}}{\partial p}(p'-p)+\delta f(p,\gamma'^i)\nonumber\\
&=& f^{(0)}(p)+p\frac{\partial f^{(0)}}{\partial
p}\hat{e}_i(v_{(e)})(\gamma'^i-\gamma^i)+\delta f(p,\gamma'^i).
\end{eqnarray}
Remember that the surface element $d\Omega_{\hat{\mathbf{p}}'}$ in (B.13) also refers to
the rest system. This is related to the surface element $d\Omega_{\gamma'}$
by\footnote{Under a Lorentz transformation, the surface element for photons transforms as
\[d\Omega=(\omega'/\omega)^2d\Omega'\]
(exercise).}
\begin{equation}
d\Omega_{\hat{\mathbf{p}}'}=\left(\frac{p'}{\omega'}\right)^2d\Omega_{\gamma'}=
[1+2\hat{e}_i(v_{(e)})\gamma'^i]d\Omega_{\gamma'}.
\end{equation}
Inserting (B.18) and (B.19) into (B.13) gives to first order, with the notation of Sect.
7.5,
\begin{equation}
C[f]=n_e\sigma_Tp\left[\langle\delta f\rangle-\delta f-p\frac{\partial f^{(0)}}{\partial
p}\hat{e}_i(v_{(e)})\gamma^i+\frac{3}{4}Q_{ij}\gamma^i\gamma^j\right],
\end{equation}
that is the announced equation (7.66).

This approximation suffices completely for our applications. The first order corrections to
the Thomson limit have also been worked out \cite{DoJ}.

\chapter{Ergodicity for (generalized) random fields}

In Sect. 5.2.3 we have replaced a spatial average by a stochastic average. Since this is
often done in cosmology, we add some remarks about what is behind this procedure.

\subsection*{Mathematical remarks on generalized random fields}

Let $\phi$ be a generalized random field. Each `smeared' $\phi(f)$ is a random variable on
some probability space $(\Omega,\mathcal{F},\mu)$. Often one can choose
$\Omega=\mathcal{S}'(\mathbb{R}^D),~ \mathcal{F}:~\sigma$-algebra generated by cylindrical
sets, and $\phi(f)$ the `coordinate function'
\[\phi(f)(\omega)=\langle\omega,f\rangle,~~
\omega\in\mathcal{S}'(\mathbb{R}^D),~f\in\mathcal{S}(\mathbb{R}^D).\]

\textbf{Notation}: We use the letter $\phi$ for elements of $\Omega$ and interpret
$\phi(f)$ as the coordinate function: $\phi\mapsto\langle\phi,f\rangle$.

Let $\tau_a$ denote the translation of $\mathbb{R}^D$ by $a$. This induces translations of
$\Omega$, as well as random variables such as $A=\phi(f_1)\cdot\cdot\cdot\phi(f_n)$, which
we all denote by the same symbol $\tau_a$. Assume that $\mu$ is an invariant measure on
$(\Omega,\mathcal{F})$ which is also \textit{ergodic}: For any measurable subset
$M\in\Omega$ which is invariant under translations $\mu(M)$ equals 0 or 1. Then the
following \textbf{Birkhoff ergodic theorem} holds:  \textit{``spatial average (of
individual realization)= stochastic average''}, i.e., $\mu$-almost always
\begin{equation}
\lim_{\Lambda\uparrow\mathbf{R}^D}\frac{1}{|\Lambda|}\int_\Lambda\tau_a A~ da=\langle
A\rangle_\mu,
\end{equation}
where $\Lambda$ is a finite hypercube, and the right-hand side denotes the stochastic
average of the random variable $A$.

\paragraph{Generalized random fields on a torus.} Often it is convenient to work on
a ``big'' torus $T^D$ with volume $V=L^D$. Then $\Omega=\mathcal{D}'(T^D)$ (periodic
distributions), etc. The Fourier transform and cotransform are topological isomorphisms
between $\mathcal{D}(T^D)$ and $\mathcal{S}(\Delta^D),~\Delta^D:=(2\pi/L)^D\mathbb{Z}^D$,
the rapidly decreasing (tempered) sequences\footnote{For proofs of this and some other
statements below, see W. Schempp and B. Dressler,  \textit{Einf\"{u}rung in die harmonische
Analyse} (Teubner, 1980), Sect. I.8.}. These provide, in turn, isomorphisms between
$\mathcal{D}'(T^D)$ and $\mathcal{S}'(\Delta^D)$. Each (periodic) distribution $S\in
\mathcal{D}'(T^D)$ can be expanded in a convergent Fourier series
\begin{equation}
S=\frac{1}{\sqrt{V}}\sum_{k\in\Delta^D}c_k(S)\chi_k,~~~\chi_k(x):=
\frac{1}{\sqrt{V}}e^{ik\cdot x}
\end{equation}
($\chi_k$ regarded as a distribution), where
\begin{equation}
c_k(S)=\langle S,e^{-ik\cdot x}\rangle.
\end{equation}
Written symbolically,
\begin{equation}
S(x)=\frac{1}{V}\sum_{k\in\Delta^D}S_k e^{ik\cdot x},~~~S_k=\int S(x)e^{-ik\cdot x}~dx.
\end{equation}

Let us consider the correlation functions $\langle\phi(f)\phi(g)\rangle_\mu$. In terms of
the Fourier expansion for $\phi(x)$, we have
\[\langle\phi(x)\phi(y)\rangle_\mu=\frac{1}{V^2}\sum_{k,k'}\langle\phi_k\phi_k'^{\ast}\rangle
e^{i(k\cdot x-k'\cdot y)}. \] This is only translationally invariant if the tempered
sequences $\phi_k$ are uncorrelated,
\[ \langle\phi_k\phi_k'^{\ast}\rangle=\delta_{kk'}\langle|\phi_k|^2\rangle. \]
Then
\begin{equation}
\langle\phi(x)\phi(y)\rangle_\mu=\frac{1}{V}\sum_{k}\frac{\langle|\phi_k|^2\rangle}{V}
e^{ik\cdot(x-y)}.
\end{equation}
By definition, the power spectrum $P_\phi(k)$ of the generalized random field $\phi$ is the
Fourier transform of the correlation function (distribution)
\begin{equation}
\langle\phi(x)\phi(y)\rangle_\mu=\frac{1}{V}\sum_{k\in\Delta^D}P_\phi(k)e^{ik\cdot(x-y)}.
\end{equation}
Therefore,
\begin{equation}
P_\phi(k)=\frac{1}{V}\langle|\phi_k|^2\rangle_\mu.
\end{equation}

If the measure is ergodic with respect to translations $\tau_a$, we obtain $\mu$-almost
always the same result if we take for a particular realization of $\phi(x)$ its
\textit{spatial average}. This follows from Birkhoff's ergodic theorem, stated above,
together with the following well-known theorem of H. Weyl:

\paragraph{Theorem (H. Weyl).} Let $f$ be a continuous function on the torus $T^D$, then
\begin{equation}
\lim_{\Lambda\uparrow\mathbf{R}^D}\frac{1}{|\Lambda|}\int_\Lambda f\circ\tau_a ~ da
=\int_{T^D}f~d\lambda,
\end{equation}
where $\lambda$ is the invariant normalized measure on $T^D$.

For a proof I refer to Arnold's ``Mathematical methods of classical mechanics'', Sect. 51.

\subsection*{A discrete example for ergodic random fields}

Proving ergodicity is usually very difficult. Below we give an example of a discrete random
Gaussian field, for which this can be established without much effort.

Let $\Omega=\mathbb{R}^{\mathbb{Z}^D}$, and consider the discrete random field
$\phi_x(\omega)=\omega_x$, where $\omega:\mathbb{Z}^D\rightarrow\mathbb{R}$, and $\omega_x$
denotes the value of $\omega$ at site $x\in\mathbb{Z}^D$. We assume that the random field
$\phi_x$ is Gaussian, and that the underlying probability measure $\mu$ is invariant under
translations. Then the correlation function $C(x-y)=\langle\phi_x\phi_y\rangle$ depends
only on the difference $x-y$. Being of positive type, we have by the Bochner-Herglotz
theorem a representation of the form
\begin{equation}
C(x)=\int_{T^D} e^{ik\cdot x}~d\sigma(k),
\end{equation}
where $\sigma$ is a positive measure.

Now we can formulate an interesting fact:

\paragraph{Theorem (Fomin, Maruyama).} (1) The random field $\phi_x$ is ergodic (i.e., the
probability measure $\mu$ is ergodic relative to discrete translations $\tau_a$), if and
only if the measure $\sigma$ is nonatomic. (2) The translations are mixing if $\sigma$ is
absolutely continuous with respect to $\lambda$.

For a proof, see Cornfeld, Fomin, and Sinai, \textit{Ergodic Theory }, Springer
(Grundlehren, 245), Sect.14.2. (I was able to simplify this proof somewhat.)

\end{appendix}

\end{document}